\documentclass[12pt, a4paper]{article}
\usepackage[T2A]{fontenc}
\usepackage[utf8]{inputenc}
\usepackage{cite}
\usepackage{mathtext}
\usepackage{amssymb,amsthm,amsmath}
\usepackage[mathscr]{eucal}
\usepackage{mathrsfs}
\usepackage{colortbl}
\usepackage{bbm}
\usepackage[mathscr]{eucal}
\usepackage{array}
\usepackage[english]{babel}
\usepackage[title,titletoc]{appendix}
\usepackage{graphicx}
\usepackage{float}
\usepackage{upgreek}

\begin{document}
\author{\bf Yu.A.\,Markov$\!\,$\thanks{e-mail:markov@icc.ru},\,
M.A.\,Markova$\!\,$\thanks{e-mail:markova@icc.ru},\,
N.Yu.\,Markov$\!\:$\thanks{e-mail:NYumarkov@gmail.com}}

\title{Hamiltonian formalism for Bose excitations\\ in a plasma with a non-Abelian interaction I:\\ plasmon -- hard particle scattering}
%
%
\date{\it\normalsize
\begin{itemize}
\item[]Matrosov Institute for System Dynamics and Control Theory, Siberian Branch, Russian Academy of Sciences, Irkutsk, 664033 Russia
\vspace{-0.3cm}
%
\end{itemize}}
%
\thispagestyle{empty}
  \maketitle{}
\def\theequation{\arabic{section}.\arabic{equation}}
\vspace{-0.2cm}
{
\[
\mbox{\bf Abstract}
\]
Hamiltonian theory for collective longitudinally polarized gluon excitations (plasmons) interacting with classical high-energy test color-charged particle propagating through a high-temperature gluon plasma is developed. A generalization of the Lie-Poisson bracket to the case of a continuous medium involving bosonic normal field variable $a^{\phantom{\ast}\!\!a}_{\hspace{0.03cm}{\bf k}}$ and a non-Abelian color charge $Q^{\hspace{0.03cm}a}$ is performed and the corresponding Hamilton equations are presented. The canonical transformations including simultaneously both bosonic degrees of freedom of the soft collective excitations and degree of freedom of hard test particle connecting with its color charge in the hot gluon plasma are written out. A complete system of the canonicity conditions for these transformations is derived. The notion of the plasmon number density ${\mathcal N}^{\;a\hspace{0.03cm}a^{\prime}_{\phantom{1}}\!}_{\hspace{0.02cm}{\bf k}}$, which is a nontrivial matrix in the color space, is introduced. An explicit form of the effective fourth-order Hamiltonian describing the elastic scattering of a plasmon off a hard color particle is found and the self-consistent system of Boltzmann-type kinetic equations taking into account the time evolution of the mean value of the color charge of the hard particle is obtained. On the basis of these equations, a model problem of the interaction of two infinitely narrow wave packets is considered. A system of nonlinear first-order ordinary differential equations defining the dynamics of the interaction of the colorless $N^{\hspace{0.03cm}l}_{\bf k}$ and color $W^{\hspace{0.03cm}l}_{\bf k}$ components of the plasmon number density is derived. The problem of determining the third- and fourth-order coefficient functions entering into the canonical transformations of the original bosonic variable $a^{\phantom{\ast}\!\!a}_{\hspace{0.03cm}{\bf k}}$ and color charge $Q^{\hspace{0.03cm}a}$ is discussed. With the help of the coefficient functions obtained, a complete effective amplitude of the elastic scattering of plasmon off hard test particle is written out. 

}


\newpage

\section{Introduction}
\setcounter{equation}{0}

\indent In two previous papers \cite{markov_2020, markov_2023}, a step forward for the construction of the classical Hamiltonian formalism for the description of the nonlinear interaction processes for soft Bose and Fermi collective excitations in a hot weakly coupled quark-gluon plasma (QGP) was taken. In these papers the simplest cases of interactions, namely, the interaction of collective longitudinal-polarized colorless gluon excitations (plasmons) and collective quark-antiquark excitations with an abnormal relation between chirality and helicity (plasminos) in the QGP were considered. Within the framework of a general Hamiltonian approach to the derivation of the wave theory in nonlinear media with dispersion \cite{zakharov_1971, zakharov_1974, zakharov_1985, zakharov_book_1992, zakharov_1997, krasitskii_1990} we have defined in an explicit form the special canonical transformations up to third-order terms in the fermionic and bosonic normal variables $\bigl(b_{\bf q}^{\,i},\,b_{\bf q}^{\,\ast\hspace{0.03cm}i}\bigr)$ and $\bigl(a^{\phantom{\ast}\!\!a}_{{\bf k}},\,a^{\ast\ \!\!a}_{{\bf k}}\bigr)$. The former take its values in a Grassmann algebra. The systems of the canonicity conditions connecting among themselves the highest and lowest coefficient functions in the integrands of the expansion terms of the canonical transformations were defined. By virtue of the three-wave non-decay nature of the dispersion relations for the colorless plasminos and plasmons, the canonical transformations enabled us to eliminate the third-order Hamiltonians $H_{3}$ in powers of the normal variables 
$\bigl(b_{\bf q}^{\,i},\,b_{\bf q}^{\,\ast\hspace{0.03cm}i}\bigr)$ and $\bigl(a^{\phantom{\ast}\!\!a}_{{\bf k}},\,a^{\ast\ \!\!a}_{{\bf k}}\bigr)$.
Excluding ``nonessential'' (in the terminology of V.E. Zakharov \cite{zakharov_1974}) interaction Hamiltonians $H_{3}$ gave us the possibility of obtaining new effective fourth-order Hamilto\-nians $\widetilde{H}_{4}$, the integrand of which contains the gauge-covariant scattering amplitudes defining the simplest tree-level elastic scattering processes: the scattering of two plasmons off each other, the scattering of plasmino off plasmon and the scattering of plasmino off plasmino. The developed Hamiltonian approach was used for the construction of Boltzmann-type kinetic equations, which describe a change in the number density of colorless plasmons $N^{\,l}_{\bf k}$ and the number density of colorless plasminos $n^{-}_{\bf q}$ in a weakly inhomogeneous and weakly non-stationary quark-gluon plasma by virtue of the elastic scattering processes mentioned above and the so-called nonlinear Landau damping.\\
\indent In the present paper, we enlarge the Hamiltonian analysis of
dynamics of fermion and boson excitations in the hot QCD-medium at the soft momentum scale, carried out in \cite{markov_2020, markov_2023} to the hard sector of the QGP plasma excitations. Here, we focus our research on the study of the scattering processes of soft plasma waves off hard particles within a real time formalism based on kinetic equations for soft modes. The nonlinear Landau damping process studied previously \cite{markov_2000, markov_2001} is a simple example of this type of the scattering processes. For a sufficiently high energy level of the soft plasma excitations the scattering processes of plasmino and plasmon off a hard particle will give contributions of the same order to the right-hand side of the corresponding Boltzmann equations as contributions from the plasmon-plasmon, plasmino-plasmon and plasmino-plasmino scattering processes.\\
\indent The scattering of hard charged particles off plasma excitations has been extensively studied in the context of ordinary electron-ion plasma (see, e.g., \cite{kovrizhnykh_1965, tsytovich_1970, tsytovich_1972}). The generalization to a plasma with a non-Abelian type of interaction provides fundamentally new features to this scattering process. In particular, this is manifested in the structure of the corresponding scattering amplitu\-des and in the form of kinetic equations (see section \ref{section_10}). The main reason for this is that at the leading order in the strong coupling constant, the basic mechanism of one of the possible
interactions of a classical hard color-charged particle and soft gluon excitations is caused not by the spatial oscillations of the charged particle (the normal Thomson scattering), as occurs in an electromagnetic plasma, but it is induced by a precession of the so-called color vector $Q = (Q^{\hspace{0.03cm}a})$ of a color-charged particle in a field of soft gluon wave. In other words, the QED-like part of the induced radiation is suppressed and only the dominant non-Abelian contribution survives. In the classical pattern of the QGP description \cite{heinz_1983, heinz_1985, heinz_1986, elze_1989, mrowczynski_1989, kelly_1994, markov_1995, litim_1999, jalilian_2000, jeon_2004, mueller_2019}, the particle states are characterized in addition to position $x_{\mu}$ and momentum $p_{\mu}$, by the color vector (non-Abelian charge) $Q = (Q^{\hspace{0.03cm}a}),\, a = 1,\,\ldots\,,
N^{\hspace{0.03cm}2}_{c} - 1$ also (see the discussion just below).\\ 
\indent The other interaction mechanism is the scattering of a longitudinal wave off dressing (scree\-ning) ``cloud'' generated by a charged test particle placed in the medium. The interaction of the two fields, the field of the screening cloud and the wave field, is a nonlinear effect. Here, in contrast to the Abelian plasma, the nonlinear scattering is produced not by the oscillation of the screening polarization cloud around the color charge as a result of interaction with the incident scattering wave, but as a consequence of the fact that it is induced by the precession of color vectors of particles forming this cloud in the incident wave field with frequency $\omega^{\hspace{0.03cm}l}_{\hspace{0.03cm}{\bf k}}$. We shall emphasize once again that this process of transition scattering is a purely collective effect. For calculation of its probability it is necessary to solve the kinetic equation describing the motion of color charges in a screening cloud in the field that is equal to the sum of fields of the incident wave and a ``central'' charge producing the screening cloud.\\ 
\indent Our ultimate objective is to construct the (pseudo)classical Hamiltonian formalism for a complete and self-consistent description of the nonlinear scattering processes of soft collective excitations of both bosonic and fermionic types off hard particles with integer and half-integer spins in a hot plasma with non-Abelian interaction. In the present work, we restrict ourselves mainly to research of the processes connected with the interaction of soft boson pure collective excitations (plasmons) with hard thermal particles carrying integer spin (hard test thermal or external gluon). To achieve the stated aim, as the guiding principle, we again use a general Hamiltonian approach to the description of wave processes in nonlinear media with nondecay dispersion laws developed by Zakharov. We generalize the classical Hamiltonian formalism to the case of systems with distributed parameters for the description of the interaction processes of collective excitations obeying Bose statistics with a hard color-charged particle and construct specific canonical transformations that simultaneously account for both hard and soft boson degrees of freedom of the system under consideration. It enables us to eliminate the third-order terms in the normal boson variables $\bigl(a^{\phantom{\ast}\!\!a}_{{\bf k}},\,a^{\ast\ \!\!a}_{{\bf k}}\bigr)$ and the variables $Q^{\hspace{0.03cm}a}$ describing the color degree of freedom of a hard particle, from the initial interaction Hamiltonian and thus to define a new effective fourth-order Hamiltonian. We will obtain an explicit form of the vertex function in the effective amplitude of scattering and construct a self-consistent system of kinetic equations of the Boltzmann type describing the scattering processes of plasmon off the hard color particle.\\
\indent To define the effective amplitude, we must take into account the influence of the soft random gauge field on the state of the color test particle. The particle motion in a field of incident soft-gluon plasma wave is described by
the following system of equations 
\begin{equation}
\begin{split}
	&m\frac{d^{\hspace{0.03cm}2}x^{\hspace{0.02cm}\mu}}
	{d\hspace{0.03cm}\tau^{\hspace{0.02cm}2}} =
	g\hspace{0.03cm}Q^{\hspace{0.03cm}a}
	F^{\hspace{0.03cm}a\hspace{0.03cm}\mu\nu}(x)
	\frac{d\hspace{0.03cm}x_{\hspace{0.02cm}\nu}}{d\hspace{0.02cm}\tau},
	\\[1ex]
&\frac{d\hspace{0.03cm}Q^{\hspace{0.03cm}a}}{d\hspace{0.02cm}\tau} =
-\hspace{0.03cm}gf^{\hspace{0.03cm}a\hspace{0.03cm}b\hspace{0.03cm}c}
\hspace{0.04cm}\frac{d\hspace{0.03cm}x^{\mu}}{d\hspace{0.02cm}\tau}\,
A_{\mu}^b(x)\hspace{0.03cm}Q^{\hspace{0.03cm}c}.
\end{split}
\label{eq:1q}
\end{equation}
Above, $\tau$ is the proper time and the vector $Q = (Q^{\hspace{0.03cm}a})$ transforms under the adjoint representation of the internal-symmetry group $SU(N_{c})$. Equations (\ref{eq:1q}) are the famous Wong equations \cite{wong_1970}. As a color current of the test Yang-Mills charge moving in a 4-dimensional space-time $(t,{\bf x})$ along a space-time path $x^{\mu}(\tau)$, we have to use the following expression:
\begin{equation}
	j^{\hspace{0.02cm}a\hspace{0.02cm}\mu}(x) = g\!\int\!\frac{d\hspace{0.02cm}x^{\hspace{0.03cm}\mu}}{d\hspace{0.03cm}\tau}\,
	Q^{\hspace{0.03cm}a}(\tau)\,\delta^{(4)}(x-x(\tau))\,
	d\hspace{0.02cm}\tau.
	\label{eq:1w}
\end{equation}
\indent The system of equations (\ref{eq:1q}) is, in a general case, very complicated since it describes both the Abelian contribution to radiation connected with the change of trajectory and momentum of the test particle due to interaction with the soft fluctuating gauge field, and the non-Abelian part of the radiation induced by precession of the color spin in the field of incident wave. Our interest is only with the leading hard thermal loop (HTL) contribution with respect to the coupling constant in the theory under consideration. This makes it possible to simplify the treatment and consider the hard test massless particle as moving along a straight line with a constant velocity
\[
{\bf x} = {\bf x}_{0} + {\bf v}(t - t_{0}),
\] 
where ${\bf x}_{0}$ and ${\bf v}$ are the initial position and velocity of the hard particle, respectively. In HTL-approximation, the QED-like part of the Compton scattering is suppressed, and only the dominant specific non-Abelian contribution survives. The processes of plasmon emission and absorption are defined by a ``rotation'' of the color vector $Q^{\hspace{0.03cm}a} = Q^{\hspace{0.03cm}a}(\tau)$ in effective color space. Thus, in the limit of the accepted accuracy of calculation we need only a minimal extension of the color current of the hard test particle, instead of (\ref{eq:1w})
\begin{equation}
	j_Q^{\hspace{0.02cm}a\hspace{0.02cm}\mu} = g\hspace{0.03cm}v^{\hspace{0.02cm}\mu}\hspace{0.02cm}
	Q^{\hspace{0.03cm}a}(t)	\hspace{0.03cm}
	{\delta}^{(3)}\bigl({\bf x} - {\bf x}_{0} - {\bf v}(t - t_{0})\bigr),\quad
	v^{\hspace{0.02cm}\mu} = (1,{\bf v}),
\label{eq:1e}
\end{equation}
where the color charge $Q^{\hspace{0.03cm}a}(t)$ satisfies the following equation:
\begin{equation}
	\frac{d\hspace{0.03cm}Q^{\hspace{0.03cm}a}(t)}{d\hspace{0.03cm}t} = gf^{\hspace{0.03cm}a\hspace{0.03cm}b\hspace{0.03cm}c}\hspace{0.02cm}
	(v\cdot A^{b}(x))\hspace{0.03cm}Q^{\hspace{0.03cm}c}(t),
	\qquad
	Q^{\hspace{0.03cm}a}(t)\hspace{0.02cm}
	|_{\hspace{0.02cm}t\hspace{0.02cm}=\hspace{0.02cm}t_{0}}
	=
	Q^{\hspace{0.03cm}a}_{\hspace{0.03cm}0}.
\label{eq:1r}
\end{equation}
Here, we turn to a coordinate time $t$ by the simple rule: $\tau = \sqrt{1 - {\bf v}^{\hspace{0.03cm}2}}\,t$.\\
\indent In the development of the Hamiltonian formalism we are interested in, an important part is the definition of an appropriate Poisson bracket. For a system of soft bosonic excitations in \cite{markov_2020}, such a bracket has been written out for the normal variables $\bigl(a^{\phantom{\ast}\!\!a}_{{\bf k}},\,a^{\ast\ \!\!a}_{{\bf k}}\bigr)$. Now we need to include in our consideration a new degree of freedom of the system associated with the color charge $Q^{\hspace{0.03cm}a}$ of a hard particle. We can consider the components $Q^{\hspace{0.03cm}a}$ of the color charge as coordinates of point $Q$ in the space $\mathfrak{su}^{*}(N_{c})$ the dual of the Lie algebra $\mathfrak{su}(N_{c})$ of the Lie group $SU(N_{c})$, i.e., the space of linear functionals on $\mathfrak{su}(N_{c})$. The space $\mathcal{F}(\mathfrak{su}^{*}(N_{c}))$ of
smooth functions on $\mathfrak{su}^{*}(N_{c})$ has a natural Poisson structure. Such a generalization for ordinary classical finite-dimensional
mechanics has been extensively studied before (see, e.g., \cite{kirillov_1976, guillemin_1980, gibbons_1982, weinstein_1983, perelomov_1990, dufour_2005, laurent_2013, esposito_2015}).\\
\indent The natural setting for Hamiltonian systems is symplectic manifolds. These may be described as manifolds carrying a Poisson structure that is locally isomorphic to the standard one on an ${\bf R}^{2n}$; the coordinates $(q_{1},q_{2},\dots,q_{n},p_{1},\ldots,p_{n})$ are then called canonical variables. Let the system under consideration also have (internal) degrees of freedom described by some additional coordinates
$x^{\hspace{0.03cm}a}$, which are local coordinates on ${\bf R}^{r}$. Then the generalization of the classical Poisson bracket operation defined on functions on ${\bf R}^{2n}$ and ${\bf R}^{r}$, is
\begin{equation}
\{F,G\} = \sum_{i\hspace{0.02cm}=\hspace{0.02cm}1}^{n} \,\biggl(\frac{\partial F}{\partial\hspace{0.03cm} q_{i}}\,\frac{\partial\hspace{0.03cm} G}{\partial\hspace{0.03cm} p_{i}}
\,-\,
\frac{\partial G}{\partial\hspace{0.03cm}q_{i}}\,\frac{\partial F}{\partial\hspace{0.03cm} p_{i}}
\biggr)
+
\sum_{a,\hspace{0.03cm}b\hspace{0.03cm}=\hspace{0.03cm}1}^{d_{A}}\! w^{\hspace{0.03cm}a\hspace{0.02cm}b}(x)\,\frac{\partial F}{\partial\hspace{0.03cm}x^{\hspace{0.02cm}a}}\frac{\partial\hspace{0.03cm} G}{\partial\hspace{0.03cm}x^{\hspace{0.02cm}b}}\hspace{0.03cm},
\label{eq:1t}
\end{equation}
where $w^{\hspace{0.03cm}a\hspace{0.02cm}b}(x)$ are the components of a skew-symmetric covariant two-tensor called the Poisson tensor. The
rank of $w^{\hspace{0.03cm}a\hspace{0.02cm}b}$ is no longer constant. Otherwise, the last contribution on the right-hand side could be reduced (locally) to the standard form that the first contribution has. The functions $w^{\hspace{0.03cm}a\hspace{0.02cm}b}(x)$ satisfy the identity
\begin{equation}
w^{\hspace{0.03cm}c\hspace{0.03cm}b}\,\frac{\partial\hspace{0.03cm} w^{\hspace{0.03cm}a\hspace{0.02cm}d}}{\partial\hspace{0.03cm} x^{\hspace{0.02cm}c}}
+
w^{\hspace{0.03cm}c\hspace{0.03cm}a}\,\frac{\partial\hspace{0.03cm} w^{\hspace{0.03cm}d\hspace{0.03cm}b}}{\partial\hspace{0.03cm} x^{\hspace{0.02cm}c}}
+
w^{\hspace{0.03cm}c\hspace{0.03cm}d}\,\frac{\partial\hspace{0.03cm} w^{\hspace{0.03cm}b\hspace{0.03cm}a}}{\partial\hspace{0.03cm} x^{\hspace{0.02cm}c}} = 0,
\label{eq:1y}
\end{equation}
which is a consequence of the requirement for satisfying the Jacobi identity.\\
\indent Of special interest is the case where the metric elements $w^{\hspace{0.03cm}a\hspace{0.02cm}b}$
are linearly dependent on the coordinates as follows:
\begin{equation}
w^{\hspace{0.03cm}a\hspace{0.02cm}b}(x) = f^{\hspace{0.03cm}a\hspace{0.03cm}b\hspace{0.03cm}c}x^{\hspace{0.02cm}c}.
\label{eq:1u}
\end{equation}
From the condition (\ref{eq:1y}), it now follows that the $f^{\hspace{0.03cm}a\hspace{0.03cm}b\hspace{0.03cm}c}$ are subject to the relations
\[
f^{\hspace{0.02cm}a\hspace{0.03cm}b\hspace{0.03cm}c}
f^{\hspace{0.03cm}c\hspace{0.03cm}d\hspace{0.03cm}e} 
+ 
f^{\hspace{0.02cm}a\hspace{0.02cm}d\hspace{0.03cm}c}
f^{\hspace{0.03cm}b\hspace{0.03cm}c\hspace{0.03cm}e} 
+ 
f^{\hspace{0.02cm}b\hspace{0.03cm}d\hspace{0.03cm}c}
f^{\hspace{0.03cm}c\hspace{0.03cm}a\hspace{0.03cm}e} = 0.
\]
The numbers $f^{\hspace{0.03cm}a\hspace{0.02cm}b\hspace{0.02cm}c}$ are the structure constants of some Lie algebra and $d_{A}$ in (\ref{eq:1t}) is its dimension. Thus, every Lie algebra has a canonical Poisson structure, which we call, following Weinstein \cite{weinstein_1983}, the {\it Lie-Poisson structure} (the {\it local Lie algebra} in the terminology of A.A. Kirillov \cite{kirillov_1976}). The case (\ref{eq:1u}) was first considered by Berezin \cite{berezin_1967} (see also \cite{vergne_1972}).\\
\indent It is clear that in our case, we need to take the color charge $Q^{\hspace{0.02cm}a}$ as the coordinate $x^{\hspace{0.02cm}a}$. Then, in the representation (\ref{eq:1u}) we have, instead of (\ref{eq:1t}), the Lie-Poisson bracket\footnote{\hspace{0.03cm}Strictly speaking, only the second sum on the right-hand side of (\ref{eq:1i}) is usually called the Lie-Poisson bracket, but for convenience we will understand the whole expression (\ref{eq:1i}) by this term.}
\begin{equation}
	\{F,G\} = \sum_{i = 1}^{n} \,\biggl(\frac{\partial F}{\partial\hspace{0.03cm} q_{\hspace{0.02cm}i}}\,\frac{\partial\hspace{0.03cm} G}{\partial\hspace{0.03cm} p_{\hspace{0.02cm}i}}
	\,-\,
	\frac{\partial G}{\partial\hspace{0.03cm}q_{\hspace{0.02cm}i}}\,\frac{\partial F}{\partial\hspace{0.03cm}p_{\hspace{0.02cm}i}}
	\biggr)
	+
	\sum_{a,\hspace{0.03cm}b\hspace{0.03cm}=\hspace{0.03cm}1}^{d_{A}} f^{\hspace{0.03cm}a\hspace{0.03cm}b\hspace{0.03cm}c}\hspace{0.03cm}
	Q^{\hspace{0.03cm}c}\, 
	\frac{\partial F}{\partial\hspace{0.03cm}Q^{\hspace{0.03cm}a}}
	\hspace{0.03cm}
	\frac{\partial\hspace{0.03cm}G}{\partial\hspace{0.03cm}
	Q^{\hspace{0.03cm}b}}\hspace{0.03cm}.
\label{eq:1i}
\end{equation}
The variables $q_{\hspace{0.02cm}i},\,p_{\hspace{0.02cm}i}$ and $Q^{\hspace{0.03cm}a}$ subject to the canonical equations
\begin{equation}
\{q_{\hspace{0.02cm}i},p_{\hspace{0.02cm}i}\} = \delta_{\hspace{0.02cm}ij}, \qquad 
\{Q^{\hspace{0.03cm}a},Q^{\hspace{0.03cm}b}\} = 
f^{\hspace{0.03cm}a\hspace{0.03cm}b\hspace{0.03cm}c}\hspace{0.03cm}
Q^{\hspace{0.03cm}c}.
\label{eq:1o}
\end{equation}
The expression (\ref{eq:1i}) was first written out in \cite{linden_1995}, where it is derived from somewhat different considerations (see also \cite{balachandran_1977, barducci_1977, montgomery_1984}). The basis here is putting into conside\-ra\-tion, instead of the Yang-Mills charges $Q^{\,a}$, the Grassmann-valued color charges $\theta^{\hspace{0.03cm}\ast\hspace{0.03cm}i}$ and $\theta^{\,i}$, $i = 1,\ldots,N_{c}$, belonging to the defining representation of the $SU(N_c)$ group that are in involution with respect to the operation $\ast$ relative to each other or the Grassmann-valued real color charges $\vartheta^{\hspace{0.03cm}a}$ belonging to the adjoint representation of the same group. These charges are related to the initial ones $Q^{\hspace{0.03cm}a}$ by the relations 
\begin{equation}
Q^{\hspace{0.03cm}a} = \theta^{\dagger\,i}(t^{\hspace{0.03cm}a})^{i\hspace{0.02cm}j}
\hspace{0.01cm}\theta^{\,j}
\quad
\mbox{or}
\quad
Q^{\hspace{0.03cm}a} = \frac{1}{2}\,\vartheta^{\hspace{0.03cm}b}(T^{\,a})^{b\hspace{0.03cm}c}
\hspace{0.03cm}\vartheta^{\hspace{0.02cm}c}, 
\label{eq:1p} 
\end{equation}  
respectively. In this case, the proof of the last relation in (\ref{eq:1o}) is reduced to using the fundamental Poisson brackets between the Grassmann-valued color charges. In the second part of our work  \cite{markov_II_2023}, the representations (\ref{eq:1p}) will be actively used. The approach presented in \cite{markov_II_2023}, is somewhat more
rigorous than the approach proposed in the present work, which is more heuristic in nature. However, the disadvantage of the work \cite{markov_II_2023} is that it is somewhat
cumbersome and complex. This circumstance has motivated us to propose as a first step a physically simpler and clearer construction that, in principle, makes it possible to advance further in the analysis of the scattering processes under study.\\
\indent Finally, let us mention the approach of Bak {\it et  al.} \cite{bak_1994} in the straightforward construction of a symplectic structure for the non-Abelian charges $Q^{\hspace{0.03cm}a}$.The basis of this construction is the use for the $Q^{\hspace{0.03cm}a}$ charges the representation, proposed by Balachandran {\it et al.} \cite{balachandran_1978, balachandran_1983} (see also \cite{kerner_1968, hogreve_1983, abbott_1982, barnich_2002}), namely
\begin{equation}
Q^{\hspace{0.03cm}a} 
= 2\,{\rm tr}(t^{\hspace{0.03cm}a}g\hspace{0.02cm}K\hspace{0.02cm}g^{-1}),
\label{eq:1a} 
\end{equation}
where $g$ is a time-dependent $SU(N_{c})$ group element and 
$K = K^{a}\hspace{0.02cm}t^{\hspace{0.03cm}a}$ is a fixed element of the Lie algebra $\mathfrak{su}(N_{c})$. That is, elements of the group appear as dynamical variables. To determine the classical phase space, the group element $g$ is parametrized by an independent set of variables (local coordinates) $\xi(t) = (\xi^{\hspace{0.03cm}a}(t))$. The fundamental Poisson brackets between $F$ and $G$ as functions of variable $(p,q,\xi)$, can be defined, instead of (\ref{eq:1i}), also as
\[
\{F,G\} = \sum_{i = 1}^{n} \,\biggl(\frac{\partial F}{\partial\hspace{0.03cm} q_{i}}\,\frac{\partial\hspace{0.03cm} G}{\partial\hspace{0.03cm} p_{i}}
\,-\,
\frac{\partial G}{\partial\hspace{0.03cm}q_{i}}\,\frac{\partial F}{\partial\hspace{0.03cm} p_{i}}
\biggr)
+
\sum_{a,\hspace{0.03cm}b\hspace{0.03cm}=\hspace{0.03cm}1}^{d_{A}}
\frac{\partial F}{\partial\hspace{0.03cm} \xi^{\hspace{0.02cm}a}}\,\omega^{\hspace{0.03cm}a\hspace{0.02cm}b}
\hspace{0.03cm}
\frac{\partial \hspace{0.03cm}G}{\partial\hspace{0.03cm} \xi^{\hspace{0.02cm}b}},
\]
where $\omega^{\hspace{0.03cm}a\hspace{0.02cm}b}$ is the symplectic two-form, the explicit form of which was calculated in \cite{bak_1994}. It
was shown there that the last relation in (\ref{eq:1o}) follows from this bracket for the representation (\ref{eq:1a}). The representation (\ref{eq:1a}) gives the possibility of an elegant Lagrangian (and Hamiltonian) formulation of dynamics for a color-charged particle moving in an external Yang-Mills field without using Grassmann-valued color charges. The Euler-Lagrangian equations (\ref{eq:1q}) are determined by varying over the dynamic variables $x_{\mu}$ and $g$. However, in terms of practical
applications and calculations, this representation is not very convenient, and therefore, we will not use it within the present work.\\
\indent In the next section we will give a generalization of (\ref{eq:1i}) to the case of an infinite number of degrees of freedom, when instead of the canonical variables $(q_{\hspace{0.02cm}i},\,p_{\hspace{0.02cm}i})$ for a finite-dimensional mechanical system, we introduce bosonic normal field variables $\bigl(a^{\phantom{\ast}\!\!a}_{{\bf k}},\,a^{\ast\ \!\!a}_{{\bf k}}\bigr)$ describing the (random) wave field of a hot gluon plasma.\\
\indent A few words should be said about the nature of the color charge $Q^{\hspace{0.03cm}a}$ as a function of time. In our consideration, we represent the fluctuating gluon plasma field as an ensemble of interacting Bose-excitations that at a low nonlinearity level have random phases. This allows us to describe the system under consideration statistically by introdu\-cing the corresponding bosonic correlation functions of the normal variables $\bigl(a^{\phantom{\ast}\!\!a}_{{\bf k}},\,a^{\ast\ \!\!a}_{{\bf k}}\bigr)$. Because the gluon field gauge potential $A^{a}_{\mu}(x)$ enters Wong's equation (\ref{eq:1r}), the values of the color charge $Q^{\hspace{0.03cm}a}(t)$ of the hard test particle will also have a random character. In this case, however, we will assume that the
initial value of the charge $Q^{\hspace{0.03cm}a}_{\hspace{0.03cm}0}$ in (\ref{eq:1r}), which the high-energy particle possessed at the initial
time moment, is deterministic\hspace{0.02cm}\footnote{\hspace{0.03cm}Otherwise,  $Q^{\hspace{0.03cm}a}_{\hspace{0.03cm}0}$ would be a random variable given by its distribution  $w(Q^{\hspace{0.03cm}a}_{\hspace{0.03cm}0})\hspace{0.02cm}d\hspace{0.02cm}
Q^{\hspace{0.03cm}a}_{\hspace{0.03cm}0}$, and the further calculated conditional moments (i.e., moments at a fixed value of $Q^{\hspace{0.03cm}a}_{\hspace{0.03cm}0}$) would have to be averaged over this distribution as well.} (nonrandom). In the statistical description of the combined system under consideration -- a hot gluon plasma plus a hard color-charged particle -- it is inevitable to introduce into consideration correlation functions including both the normal field variables $\bigl(a^{\phantom{\ast}\!\!a}_{{\bf k}},\,a^{\ast\ \!\!a}_{{\bf k}}\bigr)$, and the color charge $Q^{\hspace{0.03cm}a}$ of the test particle. Examples of constructing solutions of stochastic differential equations of the type (\ref{eq:1r}) in one- and two-dimensional cases are considered, for example, in the textbook by $\O$ksendal \cite{oksendal_2010}. These solutions depend strongly on whether we consider these stochastic equations in the It\^{o} interpretation or the Stratonovich interpretation. The consequence of this is qualitatively different behavior of the averaged value of the solutions
of these stochastic equations. Thus, in the case of the stochastic Wong equation (\ref{eq:1r}), if we assume the gauge field in the system to be completely random, i.e.,
\[
\bigl\langle\hspace{0.01cm}A^{a}_{\mu}(x)\bigr\rangle = 0,
\]
then we would expect, according to the It\^{o} interpretation, that for the mean value of the solution of the Wong equation (\ref{eq:1r}) to be 
\[
\langle\hspace{0.01cm}Q^{\hspace{0.03cm}a}(t)\rangle
=
Q^{\hspace{0.03cm}a}_{\hspace{0.03cm}0}.
\]
However, as shown in section \ref{section_9}, this is not the case: the average value of the color charge $\langle\hspace{0.01cm}Q^{\hspace{0.03cm}a}(t)\rangle$ increases exponentially, as is the case in the Stratonovich interpretation.\\ 
\indent The notion of random classical colored charges has already been used in the literature for various physical systems. For example, in \cite{akamatsu_2015,akamatsu_2022} the model Hamiltonian of heavy quarkonium bound states in finite temperature QCD matter includes a potential with a function $\theta^{\hspace{0.03cm}a}$, which is a white noise with color. The random color rotation by the stochastic potential is a unique feature in the quark-gluon plasma. Further, random classical colored charges arise in the currently actively developed worldline quantum field theory formalism for the classical gravitational scattering of massive charged point particles coupled to biadjoint scalar field theory, Yang-Mills theory, and dilaton-gravity theory \cite{goldberger_2017_1, goldberger_2017_2, shi_2022}. In particular, in \cite{shi_2022}, the classical color charge of a point particle is defined (in the authors' notations) as $C^{\hspace{0.03cm}a}(\tau) = \Psi^{\dagger\hspace{0.03cm}\alpha}(T^{\,a})_{\alpha}^{\;\beta}\hspace{0.03cm}\Psi_{\beta},\,\alpha,\beta = 1,\ldots,N^{2}_{c} - 1$, where the ``color wave function'' $\Psi_{\alpha}$ is an auxiliary field carrying the color degrees of the particle. The field $\Psi_{\alpha}$ decomposes into the background part $\psi = \Psi(-\infty) = {\rm const}$ and the fluctuation one. The color charge $C^{\hspace{0.03cm}a}(\tau)$ correspondingly decomposes in the background color charge $c^{\hspace{0.03cm}a} = \psi^{\dagger}\hspace{0.03cm}T^{\hspace{0.03cm}a}\hspace{0.03cm}\psi$ and in the fluctuation color part. The latter circumstance is most closely related to the task posed in our paper.\\ 
\indent As one of the possible applications of the developed Hamiltonian theory of soft and hard excitations of high-temperature plasma with non-Abelian interaction, we propose to consider the problem of calculating the energy losses of high-energy external color particles passing through a hot QCD medium. As is well known, the energy losses are one of the most important methods for diagnostics of the quark-gluon plasma in ultrarelativistic heavy-ion collisions. This problem will be considered in detail in our second part of this paper.\\
%
%
\indent The paper is organized as follows. In section \ref{section_2}, the general form of the decomposition into plane waves of the gauge field potentail is written out and the expectation value of the product of two bosonic amplitudes are given. In the same section a generalization of the Lie-Poisson bracket including the Yang-Mills charge $Q^{\hspace{0.03cm}a}$ to the case of a continuous media is presented, the corresponding Hamilton equations are defined and the most general structure of the third- and fourth-order interaction Hamiltonians with respect to the color charge $Q^{\hspace{0.03cm}a}$ of a hard particle and the normal variable $a^{\phantom{\ast}\!\!a}_{\hspace{0.03cm}{\bf k}}$ of Bose field of a hot gluon plasma is written out. In section \ref{section_3}, the canonical transformations including both bosonic and color charge degrees of freedom of the soft collective excitations and hard test particle in the gluon plasma are discussed. Two systems of the canonicity conditions for these transformations based on the Lie-Poisson bracket are derived. The most general structure of the canonical transformations in the form of integro-power series in the new normal field variable $c^{\phantom{\ast}\!\!a}_{\hspace{0.03cm}{\bf k}}$ and new color charge ${\mathcal Q}^{\hspace{0.03cm}a}$ up to the terms of sixth order is presented. Algebraic relations for the second-order coefficient functions of the canonical transformations are written out.\\
\indent In section~\ref{section_4}, making use of the above-mentioned canonical transformations the problem of excluding the ``non-essential'' third-order Hamiltonian $H^{(3)}$ is considered. The explicit expressions for the coefficient functions in the linear and quadratic in $c^{\!\phantom{\ast}a}_{\hspace{0.03cm}{\bf k}}$ and ${\mathcal Q}^{\hspace{0.03cm}a}$ terms of the canonical transformations are obtained. An explicit form of the effective amplitude $\widetilde{T}^{\,(2)\,a_{1}\hspace{0.03cm}a_{2}\, a}_{\,{\bf k}_{1},\, {\bf k}_{2}}$ describing the elastic scattering process of plasmon off a hard color particle in leading tree-level order is given and the corresponding effective fourth-order Hamiltonian ${\mathcal H}^{(4)}_{g\hspace{0.02cm}G\hspace{0.02cm}\rightarrow\hspace{0.02cm} g\hspace{0.02cm}G}$ is written out. A simple diagrammatic interpretation of the individual terms in the effective amplitude is given. Section \ref{section_5} is concerned with the calculation of fourth- and sixth-order correlation functions in the new normal field variable $c^{\phantom{\ast}\!\!a}_{\hspace{0.03cm}{\bf k}}$ and non-Abelian charge ${\mathcal Q}^{\hspace{0.03cm}a}$. The notion of the plasmon number density ${\mathcal N}^{\;a\hspace{0.03cm}a^{\prime}_{\phantom{1}}\!}_{\hspace{0.02cm}
{\bf k}}$\!, which is a nontrivial matrix in color space, is intro\-duced. On the basis of Hamilton's equation of motion with the Lie-Poisson bracket a differential equation to which the fourth-order correlation function obeys is obtained.\\ 
\indent In section \ref{section_6} an approximate solution of the equation for the fourth-order correlator, accounting for the deviation of the four-point correlation function from the Gaussian approximation for a low nonlinearity level of interacting Bose-excitations is found. On the basis of this solution, a matrix kinetic equation for the number density of color plasmons describing an elastic scattering process of collective gluon excitations off hard test color-charged particle with allowance for the Landau linear damping effect is constructed. In section \ref{section_7} the color decomposition of the matrix function ${\mathcal N}^{\;a\hspace{0.03cm}
a^{\prime}_{\phantom{1}}\!}_{\hspace{0.02cm}{\bf k}}$ is written out and the first moment about color of the matrix kinetic equation, defining scalar kinetic equation for colorless part $N^{\hspace{0.03cm}l}_{\bf k}$ of this decomposition, is calculated. Section \ref{section_8} is devoted to determining the second moment about color of the matrix kinetic equation, defining scalar kinetic equation for color part $W^{\hspace{0.03cm}l}_{\bf k}$ of the decomposition of the matrix number density ${\mathcal N}^{\;a\hspace{0.03cm}a^{\prime}_{\phantom{1}}\!}_{
\hspace{0.02cm}{\bf k}}$. Two special cases $SU(2_{c})$ and $SU(3_{c})$ of the color group are discussed.\\
\indent In section \ref{section_9} the derivation of the equation of motion for the expected value of color charge ${\mathcal Q}^{\hspace{0.03cm}a}$ is considered and the time dependence of colorless combinations of the second and fourth orders with respect to the mean value $\langle{\mathcal Q}^{\hspace{0.03cm}a}\rangle$ was determined. In section \ref{section_10} a complete self-consistent system of kinetic equations for soft gluon excitations taking into account the time evolution of the mean value of the color charge of a hard test particle is written out. Section \ref{section_11} is concerned with the model problem of interaction of two infinitely narrow wave packets. The system of two nonlinear first-order ordinary differential equations defining the dynamics of the interaction of colorless $N^{\hspace{0.03cm}l}_{\bf k}$ and color $W^{\hspace{0.03cm}l}_{\bf k}$ components of the plasmon number density is defined. At a certain relation between the constants entering into the nonlinear equations, the exact solution of this system in parametric form is obtained.\\
\indent In section \ref{section_12} the problem of the construction of the third-order coefficient functions $\widetilde{V}^{\,(1)\,a_{1}\hspace{0.03cm}a_{2}\,a}_{\ {\bf k}_{1},\, 
{\bf k}_{2}}$ and $\widetilde{V}^{\,(2)\,a_{1}\hspace{0.03cm}a_{2}\,a}_{\ {\bf k}_{1},\,{\bf k}_{2}}$ entering into the canonical transformation of the original bosonic variable $a^{\,a}_{\hspace{0.03cm}{\bf k}}$ is considered. Based on the requirement of vanishing the so-called non-resonant terms in new effective interaction Hamiltonian ${\mathcal H}^{(4)}_{g\hspace{0.02cm}G\hspace{0.02cm}\rightarrow\hspace{0.02cm} g\hspace{0.02cm}G}$, the explicit form of the coefficient $\widetilde{V}^{\,(1)\,a_{1}\hspace{0.02cm}a_{2}\,a}_{\ {\bf k}_{1},\, {\bf k}_{2}}$ is defined. To determine an explicit form of the coefficient function $\widetilde{V}^{\,(2)\,a_{1}\hspace{0.02cm}a_{2}\,a}_{\ {\bf k}_{1},\, {\bf k}_{2}}$ a functional equation to which this function obeys, is solved. With the help of the function $\widetilde{V}^{\,(2)\,a_{1}\hspace{0.02cm}a_{2}\,a}_{\ {\bf k}_{1},\, {\bf k}_{2}}$, a {\it complete} effective amplitude of elastic scattering of plasmon off hard test particle satisfying a certain symmetry condition is written out. In the concluding section \ref{section_13} the key points of our work are specified. This section briefly discusses several interesting issues which are very close to the subject of the present research but have not been touched in the paper. In particular, they concern both the consideration of higher-order scattering processes and so-called soft ``one-loop''  corrections to the processes of tree-level elastic scattering of plasmon off a hard particle.\\
\indent In Appendix \ref{appendix_A} we give all of the basic expressions for the effective three- and four-gluon vertex functions and the effective gluon propagator within the framework of the hard thermal loop approximation. In Appendix \ref{appendix_B} a complete system of independent relations of the canonicity conditions connecting the lowest and highest coefficient functions in the canonical transformations among themselves is given. In Appendix \ref{appendix_C} an explicit form of the coefficient functions $M^{\hspace{0.02cm}(1)\hspace{0.03cm}a\,a_{1}
\hspace{0.03cm}a_{2}\,a_{3}}_{\ {\bf k}_{1},\,{\bf k}_{2}}$ and $\hspace{0.02cm}M^{(2)\hspace{0.03cm}a\,a_{1}\hspace{0.03cm}a_{2}\, a_{3}}_{\ {\bf k}_{1},\,{\bf k}_{2}}$, which enter into the canonical transformation of the original color charge $Q^{\hspace{0.03cm}a}$ is written out. In Appendix \ref{appendix_D} the necessary traces for generators in adjoint representation of the color group $SU(N_{c})$ up to the fifth order and some useful relations between these generators are given. In Appendix \ref{appendix_E} the construction of an exact solution of the nonlinear Abel differential equation of the second kind is presented and the corresponding exact solution in parametric form of the original evolution system for two interacting infinitely narrow wave packets is recovered.


\section{Interaction Hamiltonian of plasmons and hard particle}
\setcounter{equation}{0}
\label{section_2}

\begin{figure}[hbtp]
\begin{center}
\includegraphics[width=0.45\textwidth]{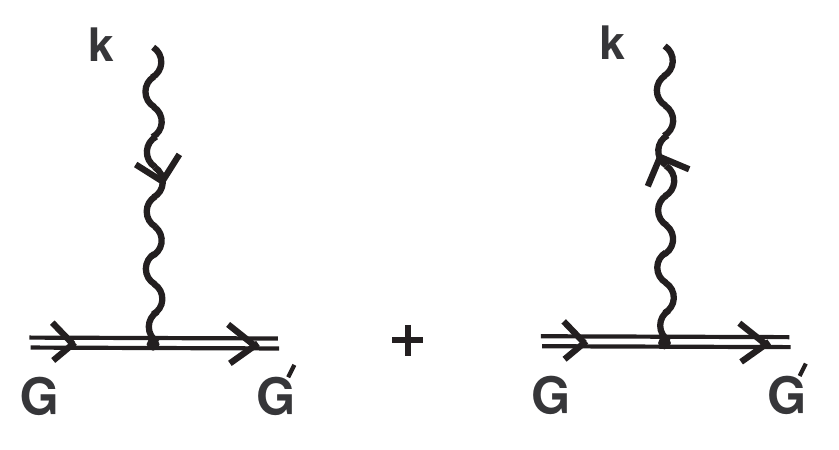}
\end{center}
\vspace{-0.5cm}
\caption{\small Elementary interaction vertices of soft boson excitations with a hard test color-charged particle $\mathrm{G}$. The double line denotes a hard particle carrying a color charge $Q^{\hspace{0.03cm}a}$. The interaction vertices for the incoming and outgoing wave lines (plasmons) are defined by the functions ${\upphi}^{\phantom{\ast}}_{\hspace{0.02cm}{\bf k}}$ and ${\upphi}^{\hspace{0.01cm}\ast}_{\hspace{0.02cm}{\bf k}}$ in the Hamiltonian $H^{(3)}$}
\label{fig1}
\end{figure}

As is well known \cite{kalashnikov_1980}, in an equilibrium hot pure gluon plasma, there exist two types of physical soft bosonic fields -- transverse- and longitudinal-polarized fields. Let us consider the gauge field potential $A_{\mu}^{a}(x)$ in the form of the decomposition into plane waves\footnote{\,The color indices $a,\,b,\,c,\,\ldots$ run through values $1,2,\,\ldots\,,N^{\hspace{0.02cm}2}_{c}-1$, while the vector indices $\mu,\,\nu,\,\lambda,\,\ldots$ run through values $0,1,2,3$. In this article, we imply summation over repeated indices and use the system of units with $\hbar = c = 1$.} \cite{blaizot_1994(1), hakim_book_2011}
\begin{equation}
\begin{split}
A^{a}_{\mu}(x) = &\int\!d\hspace{0.02cm}{\bf k}\left(\frac{Z_{l}({\bf k})}
{2\hspace{0.03cm}\omega^{\hspace{0.03cm}l}_{\hspace{0.02cm}{\bf k}}}\right)^{\!\!1/2}\!\!
\left\{\epsilon^{\hspace{0.03cm}l}_{\mu}({\bf k})\hspace{0.03cm} a^{\phantom{\ast}\!\!a}_{\hspace{0.02cm}{\bf k}}\ \!e^{-i\hspace{0.03cm}\omega^{\hspace{0.03cm}l}_{\hspace{0.02cm}{\bf k}}\hspace{0.02cm}t\hspace{0.03cm} +\hspace{0.03cm} i\hspace{0.03cm}{\bf k}\hspace{0.02cm}\cdot\hspace{0.02cm} {\bf x}}
+
\epsilon^{\ast\, l}_{\mu}({\bf k})\, a^{\ast\ \!\!a}_{\hspace{0.02cm}{\bf k}}\ \!e^{\hspace{0.02cm}i\hspace{0.03cm}\omega^{\hspace{0.03cm}l}_{\hspace{0.02cm}{\bf k}}\hspace{0.02cm}t\hspace{0.03cm} -\hspace{0.03cm} i\hspace{0.03cm}{\bf k}\hspace{0.02cm}\cdot\hspace{0.02cm} {\bf x}}
\right\}
\\[1.3ex]
+ 
\sum_{\zeta\hspace{0.02cm} =\hspace{0.02cm} 1,\hspace{0.02cm} 2}\hspace{0.03cm} 
&\int\!d\hspace{0.02cm}{\bf k}\left(\frac{Z_{t}({\bf k})}
{2\hspace{0.03cm}\omega^{t}_{\hspace{0.02cm}{\bf k}}}\right)^{\!\!1/2}\!\!
\left\{\epsilon^{\hspace{0.03cm}t}_{\mu}({\bf k}, \zeta)\hspace{0.03cm} a^{\phantom{\ast}\!\!a}_{\hspace{0.02cm}{\bf k}}(\zeta)\ \!e^{-i\hspace{0.03cm}\omega^{t}_{\hspace{0.02cm}{\bf k}}\hspace{0.02cm}t\hspace{0.03cm} +\hspace{0.03cm} i\hspace{0.03cm}{\bf k}\hspace{0.02cm}\cdot\hspace{0.02cm} {\bf x}}
+
\epsilon^{\ast\, t}_{\mu}({\bf k},\zeta)\, a^{\ast\ \!\!a}_{\hspace{0.02cm}{\bf k}}(\zeta)\ \!e^{\hspace{0.02cm}i\hspace{0.03cm}\omega^{t}_{\hspace{0.02cm}{\bf k}}\hspace{0.02cm}t\hspace{0.03cm} -\hspace{0.03cm} i\hspace{0.03cm}{\bf k}\hspace{0.02cm}\cdot\hspace{0.02cm} {\bf x}}\right\},
\end{split}
\label{eq:2q}
\end{equation}
where $\epsilon^{\hspace{0.03cm}l}_{\mu} ({\bf k})$ is the polarization vector of a longitudinal mode (${\bf k}$ is the wave vector) and $\epsilon^{\ \! t}_{\mu} ({\bf k},\zeta)$ is the polarization vector of a transverse mode. The symbol $\zeta = 1, 2$ stands for two possible transverse polarization states, and the asterisk $\ast$ denotes the complex conjugation. The factors $Z_{l}({\bf k})$ and $Z_{t}({\bf k})$ are the residues of the effective gluon propagator at the longitudinal and transverse mode poles, respectively. Finally, $\omega^{\ \! l}_{\hspace{0.02cm}{\bf k}}$ and $\omega^{\ \! t}_{\hspace{0.02cm}{\bf k}}$ are the dispersion relations of the corresponding modes. We consider the amplitudes for longitudinal $a^{\phantom{\ast}\!\!a}_{\hspace{0.02cm}{\bf k}}$ and transverse $a^{\phantom{\ast}\!\!a}_{\hspace{0.02cm}{\bf k}}(\zeta)$ excitations to be ordinary (complex) random functions, and the expectation values of the products of two bosonic amplitudes are
\[
\bigl\langle\hspace{0.01cm}a^{\ast\hspace{0.03cm}a}_{\hspace{0.02cm}{\bf k}}\hspace{0.03cm} a^{\phantom{\ast}\!\!b}_{\hspace{0.02cm}
{\bf k}^{\prime}}\bigr\rangle
=
\delta^{\hspace{0.03cm}a\hspace{0.01cm}b}\hspace{0.03cm}\delta({\bf k} - {\bf k}^{\prime})\hspace{0.05cm}{\mathcal N}^{\hspace{0.04cm}l}_{\bf k},
\qquad
\bigl\langle\hspace{0.03cm}a^{\ast\hspace{0.03cm}a}_{\hspace{0.02cm}
{\bf k}}(\zeta)\hspace{0.03cm}
a^{\phantom{\ast}\!\!b}_{\hspace{0.02cm}
{\bf k}^{\prime}}(\zeta^{\hspace{0.02cm}\prime})\bigr\rangle
=
\delta^{\hspace{0.03cm}a\hspace{0.01cm}b}\delta_{\zeta\hspace{0.01cm} \zeta^{\prime}}\hspace{0.03cm}\delta({\bf k} - {\bf k}^{\prime})\hspace{0.05cm}{\mathcal N}^{\hspace{0.04cm}t}_{\bf k},
\]
where ${\mathcal N}^{\hspace{0.04cm}l}_{\bf k}$ and ${\mathcal N}^{\hspace{0.04cm}t}_{\bf k}$ are the number densities of the longitudinal and transverse plasma waves. Note that these correlation functions are written for a hot gluon plasma without external color fields or high-energy color-charged particles penetrating into the plasma from outside.\\
\indent For simplicity, we confine our analysis only to processes involving longitudinally polarized plasma excitations, which are known as {\it plasmons}. The dispersion relation $\omega^{\ \! l}_{\hspace{0.02cm}{\bf k}}$ for plasmons satisfies the following dispersion equation \cite{kalashnikov_1980}:
\begin{equation}
{\rm Re}\ \!\varepsilon^{\hspace{0.02cm}l}(\omega,{\bf k})=0\ \!,
\label{eq:2e}
\end{equation}
where
\[
\varepsilon^{\hspace{0.02cm}l}(\omega,{\bf k})=1+\frac{3\hspace{0.02cm}\omega^{\hspace{0.02cm}2}_{pl}}{{\bf k}^{\hspace{0.01cm}2}}
\biggl[1-F\biggl(\frac{\omega}{|{\bf k}|^{2}}\biggr)\biggr],
\quad
F(x) = \frac{x}{2}\left[\hspace{0.03cm}\ln\left|\frac{1+x}{1-x}\right|-i\pi\theta(1-|x|)\right]
\]
is the longitudinal permittivity, $\omega^{2}_{pl}=g^{2}N_{c}T^{2}/9$ is the plasma frequency squared of the gluon plasma excitations, $T$ is the temperature of the system, and $g$ is the strong coupling constant.\\
\indent The amplitudes $a^{\phantom{\ast}\!\!a}_{\hspace{0.02cm}{\bf k}}$ and $a^{\ast\ \!\!a}_{\hspace{0.02cm}{\bf k}}$ in the expansion for the longitudinal part of the gauge field potential (\ref{eq:2q}) satisfy the following Lie-Poisson bracket $({\rm LPB})$ relations
\begin{equation}
\bigl\{a^{\phantom{\ast}\!\!a}_{\hspace{0.02cm}{\bf k}},\,a^{\phantom{\ast}\!\!b}_{\hspace{0.02cm}{\bf k}^{\prime}}\bigr\}_{\rm LPB} = 0,
\quad\!
\bigl\{a^{\ast\ \!\!a}_{\hspace{0.02cm}{\bf k}},\,a^{\ast\ \!\!b}_{\hspace{0.02cm}{\bf k}^{\prime}}\bigr\}_{\rm LPB} = 0, 
\quad\!
\bigl\{a^{\phantom{\ast}\!\!a}_{\hspace{0.02cm}{\bf k}},\,a^{\ast\ \!\!b}_{\hspace{0.02cm}{\bf k}^{\prime}}\bigr\}_{\rm LPB}
=
\delta^{\hspace{0.02cm} ab}\hspace{0.02cm}\delta({\bf k} - {\bf k}^{\prime}).
\label{eq:2r}
\end{equation}
On the other hand, when we consider the color charge $Q^{\hspace{0.03cm}a}$ of a hard test particle, the same Lie-Poisson bracket, on the strength of (\ref{eq:1o}), must have the following form:
\begin{equation}
\hspace{0.04cm}
\bigl\{Q^{\,a},\hspace{0.03cm}Q^{\,b}\hspace{0.03cm} \bigr\}_{\rm LPB} = \,f^{\hspace{0.03cm}a\hspace{0.03cm}b\hspace{0.03cm}c}\hspace{0.03cm}Q^{\hspace{0.03cm}c}.
\label{eq:2t}
\end{equation}
For the case of continuous media, we take the following expression as the definition of the Lie-Poisson bracket:
\begin{equation}
\bigl\{F,\,G\bigr\}_{\rm LPB} 
=
\int\! d\hspace{0.02cm}{\bf k\hspace{0.01cm}}'\!\hspace{0.02cm}
\left\{\frac{\delta\hspace{0.01cm} F}{\delta\hspace{0.01cm} a^{\phantom{\ast}\!\!c}_{{\bf k}'}}
\hspace{0.03cm}\frac{\delta\hspace{0.01cm}  G}{\delta\hspace{0.01cm} a^{\ast\ \!\!c}_{{\bf k}'}}
\,-\,
\frac{\delta\hspace{0.01cm}F}{\delta\hspace{0.01cm} a^{\ast\ \!\!c}_{{\bf k}'}}\hspace{0.03cm}
\frac{\delta\hspace{0.01cm}G}{\delta\hspace{0.01cm} a^{\phantom{\ast}\!\!c}_{{\bf k}'}}\right\}
\,+\,
\frac{\partial F}{\,\partial\hspace{0.03cm} Q^{\,a}}\hspace{0.03cm}
\frac{\partial\hspace{0.03cm}G}{\,\partial\hspace{0.03cm} Q^{\hspace{0.03cm}b}}
\,f^{\hspace{0.03cm}a\hspace{0.03cm}b\hspace{0.03cm}c}\hspace{0.03cm}
Q^{\hspace{0.03cm}c}.
\label{eq:2y}
\end{equation}
The first term is the standard canonical bracket. Next, for the sake of simplicity of notation, the abbreviation ${\rm LPB}$ will be omitted, thereby suggesting that by the braces $\{\,,\}$ we always mean the Lie-Poisson bracket (\ref{eq:2y}).\\
\indent Let us write the Hamilton equations for the functions  $a^{\phantom{\ast}\!\!a}_{\hspace{0.02cm}{\bf k}},\,a^{\ast\ \!\!a}_{\hspace{0.02cm}{\bf k}}$ and $Q^{\,a}$  
\begin{equation}
\frac{\partial\hspace{0.02cm}a^{\phantom{\ast}\!\!a}_{\hspace{0.02cm}{\bf k}}}{\partial\hspace{0.02cm} t}
=
-\hspace{0.03cm}i\hspace{0.03cm}\bigl\{a^{\phantom{\ast}\!\!a}_{\hspace{0.02cm}{\bf k}}, H\bigr\} \equiv  -i\,\frac{\delta H}{\delta\hspace{0.01cm} a^{\ast\ \!\!a}_{\hspace{0.02cm}{\bf k}}},
\qquad
\frac{\partial\hspace{0.02cm}a^{\ast\ \!\!a}_{\hspace{0.02cm}{\bf k}}}{\partial\hspace{0.02cm} t}
=
-\hspace{0.03cm}i\hspace{0.03cm}\bigl\{a^{\ast\ \!\!a}_{\hspace{0.02cm}{\bf k}}, H\bigr\} \equiv  i\,\frac{\delta H}{\delta\hspace{0.01cm} a^{\phantom{\ast}\!\!a}_{\hspace{0.02cm}{\bf k}}},
\label{eq:2u}
\end{equation}
\begin{equation}
\frac{d\hspace{0.02cm}Q^{\hspace{0.03cm}a}}{d\hspace{0.02cm} t}
=
\left\{Q^{\hspace{0.03cm}a}, H\right\} =  
\frac{\!\partial H}{\partial\hspace{0.03cm} Q^{\hspace{0.03cm}b}}\,f^{\hspace{0.03cm}a\hspace{0.03cm}b\hspace{0.03cm}c}
\hspace{0.03cm}Q^{\hspace{0.03cm}c},
\quad 
Q^{\hspace{0.03cm}a}|_{t\hspace{0.02cm}=\hspace{0.02cm}t_{0}} = Q^{\hspace{0.03cm}a}_{0}.
\label{eq:2i}
\end{equation}
Here, the function $H$ represents a Hamiltonian for the system of plasmons and a hard test particle, which is equal to a sum $H =  H^{(0)} + H_{int}$, where
\begin{equation}
H^{(0)} =  \!\int\!d\hspace{0.02cm}{\bf k}\, \omega^{\hspace{0.03cm}l}_{\hspace{0.02cm}{\bf k}}\ \!
a^{\ast\hspace{0.03cm}a}_{\hspace{0.02cm}{\bf k}}\hspace{0.03cm}a^{\!\!\phantom{\ast}a}_{\hspace{0.02cm}{\bf k}}
\label{eq:2o}
\end{equation}
is the Hamiltonian of noninteracting plasmons and ${H}_{int}$ is the interaction Hamiltonian of plasmons and the hard color-charged particle. We assume that a given (test) particle moves in a gluon plasma with a constant velocity ${\bf v}$. According to \cite{tsytovich_1970}, the simplest way to account for the presence of a moving particle in this medium is to go over to a frame of reference in which this particle is at rest after performing the replacement $a^{\!\!\phantom{\ast}a}_{\hspace{0.02cm}{\bf k}}$ by $a^{\!\!\phantom{\ast}a}_{\hspace{0.02cm}{\bf k}}\hspace{0.03cm}{\rm e}^{-i\hspace{0.02cm}{\bf k}\cdot{\bf v}\hspace{0.02cm}(t - t_{0})}$ in the gauge field potential (\ref{eq:2q}). In this case, instead of the Hamiltonian (\ref{eq:2o}) we now have
\begin{equation}
H^{(0)} =  
\!\int\!d\hspace{0.02cm}{\bf k}\hspace{0.04cm}(\omega^{\hspace{0.03cm}l}_{\hspace{0.02cm}{\bf k}} - {\mathbf v}\cdot {\mathbf k})\ \!
a^{\ast\hspace{0.03cm}a}_{\hspace{0.02cm}{\bf k}}\hspace{0.03cm}a^{\!\!\phantom{\ast}a}_{\hspace{0.02cm}{\bf k}}.
\label{eq:2p}
\end{equation}
Thus, $\tilde{\omega}^{l}_{\hspace{0.02cm}{\bf k}} \equiv \omega^{\hspace{0.03cm}l}_{\hspace{0.02cm}{\bf k}} - {\mathbf v}\cdot {\mathbf k}$ is the frequency of the soft gluon field $A^{\hspace{0.03cm}a}_{\mu}$ in a coordinate system in which the charge is at rest, i.e., the frequency seen by the moving charge (by the Doppler effect). It is necessary to specifically note an important fact for future consideration. Owing to the specific character of the dispersion equation for soft (longitudinal) bosonic excitations (\ref{eq:2e}) in a hot gluon plasma, the factor $(\omega^{\hspace{0.03cm}l}_{\hspace{0.02cm}{\bf k}} - {\mathbf v}\cdot {\mathbf k})$ in the integrand (\ref{eq:2p}) never turns to zero, i.e.,
\begin{equation}
\omega^{\hspace{0.03cm}l}_{\hspace{0.02cm}{\bf k}} - {\mathbf v}\cdot {\mathbf k} \neq 0	
\label{eq:2a}
\end{equation}
for arbitrary values of the wave vector ${\bf k}$. In other words, linear Landau damping (Cherenkov emission) is kinematically forbidden in hot gluon plasma.\\ 
\indent The more rigorous approach to the solving the problem posed in the present work will be realized by us in the second part of our work \cite{markov_II_2023}, when it will not be necessary to introduce somewhat artificial operation -- the frequency shift in the Hamiltonian (\ref{eq:2p}). Nevertheless, the methodology proposed in this paper is fully self-consistent and self-sufficient, allowing us to obtain the required results more easily, quickly, and efficiently and to advance somewhat further at the cost of less effort in analyzing the complex interaction dynamics of a hard color particle with soft boson excitations of a hot gluon plasma.\\ 
\indent In our first work \cite{markov_2020} on the construction of the Hamiltonian formalism, we have considered the process of elastic scattering of two colorless plasmons off each other. This scattering process dominates when the gauge field amplitude $A_{\mu}(x)$ has order \cite{markov_2002}
\[
\vert A_{\mu}(x)\vert\sim\sqrt{g}\hspace{0.04cm}T
\]
and, accordingly, the plasmon number density is of the order
\[
\;N^{\,l}_{\bf k}\,\sim\,\displaystyle\frac{1}{g}\ \!,
\]
which in fact corresponds to the level of thermal fluctuations in a hot gluon plasma. For this value of the gauge field amplitude at $g\ll 1$, plasmon number density $N_{\bf k}^l$ is high, and the application of the purely classical description is justified. In the case of weakly nonlinear waves the Hamiltonian can be expanded as an integer-degree series in $a^{\phantom{\ast}\!\!a}_{\hspace{0.03cm}{\bf k}}$ and $a^{\ast\,a}_{\hspace{0.03cm}{\bf k}}$, which can be limited to a finite number of terms up to the quartic terms, inclusive. Within the present work we suggest that the gauge field amplitude has the same order and the process of plasmon elastic scattering off a hard particle is dominant. For this reason we also restrict the expansion of interaction Hamiltonian to the fourth order terms\footnote{\,It should be noted that the situation changes qualitatively when the system is strongly excited. For high intensity of excitations in a gluon plasma, it is necessary to consider next terms in the expansion of $H_{int}$. In the limiting case of strong excitations when 
\[
\vert A_{\mu}(x)\vert\sim T\,\;\mbox{and, accordingly,}
\;N_{\bf k}^l\,\sim\,\displaystyle\frac{1}{g^2\,},
\]
the expansion of $H_{int}$ must contain an infinite number of terms of any order in $a^{\phantom{\ast}\!\!a}_{\hspace{0.03cm}{\bf k}}$ and $a^{\ast\,a}_{\hspace{0.03cm}{\bf k}}$. In turn, this necessitates the inclusion of all higher-order plasmon elastic scattering processes off the hard particle. Here, we arrive at the truly nonlinear theory of interaction of soft gluon excitations in a plasma with a non-Abelian type of interaction.
For strongly excited states, when we are dealing with an infinite number of terms, a more adequate qualitatively new apparatus is required. If, however, the amplitude $a^{\phantom{\ast}\!\!a}_{\hspace{0.03cm}{\bf k}}$ is small, the wave dynamics may be described in general terms by expanding the Hamiltonian in terms of canonical variables.}.\\
\indent In the approximation of small amplitudes the interaction Hamiltonian can be presented in the form of a formal integro-power series in the bosonic functions ${a}^{a}_{\hspace{0.02cm}{\bf k}}$ and ${a}^{\ast\hspace{0.03cm} a}_{\hspace{0.02cm}{\bf k}}$, and in the color charge $Q^{\hspace{0.03cm}a}$:
\[
H_{int} = H^{(3)} + H^{(4)} + \, \ldots\,\,,
\]
where the third-order interaction Hamiltonian has the following structure:
\begin{align}
H^{(3)} 
=  &\int\!d\hspace{0.02cm}{\bf k}\hspace{0.03cm}
\Bigl[\hspace{0.02cm}{\upphi}^{\phantom{\ast}}_{\hspace{0.03cm} {\bf k}}\, {a}^{\,a}_{\hspace{0.02cm}{\bf k}}\,Q^{\hspace{0.03cm}a}\hspace{0.03cm} 
+
{\upphi}^{\hspace{0.01cm}\ast}_{\hspace{0.02cm}{\bf k}}\, {a}^{\hspace{0.03cm}\ast\hspace{0.03cm} a}_{\hspace{0.02cm}{\bf k}}\, Q^{\hspace{0.03cm} a}\hspace{0.03cm}\Bigr]
\label{eq:2s}\\[0.7ex]
+ &\int\!d\hspace{0.02cm}{\bf k}\, d\hspace{0.02cm}{\bf k}_{1}\hspace{0.03cm} d\hspace{0.02cm}{\bf k}_{2}\hspace{0.03cm}
\Bigl\{\hspace{0.02cm}{\mathcal V}^{\; a\, a_{1}\hspace{0.03cm} a_{2}}_{{\bf k},\, {\bf k}_{1},\, {\bf k}_{2}}\, a^{\ast\hspace{0.03cm}  a}_{\hspace{0.02cm}{\bf k}}\,
a^{\,a_{1}}_{\hspace{0.03cm}{\bf k}_{1}}\, a^{\,a_{2}}_{\hspace{0.03cm}{\bf k}_{2}}
\,+\,
{\mathcal V}^{\,*\,a\, a_{1}\hspace{0.03cm} a_{2}}_{\, {\bf k},\, {\bf k}_{1},\, {\bf k}_{2}}\, 
a^{\!\!\phantom{\ast}a}_{\hspace{0.02cm}{\bf k}}\,a^{\ast\,a_{1}}_{\hspace{0.03cm}{\bf k}_{1}}\hspace{0.03cm}a^{\ast\,a_{2}}_{\hspace{0.03cm}{\bf k}_{2}}
\Bigr\}\hspace{0.04cm}
\delta({\bf k} - {\bf k}_{1} - {\bf k}_{2}) 
\notag\\[0.7ex]
+\, \frac{1}{3}&\int\!d\hspace{0.02cm}{\bf k}\, d\hspace{0.02cm}{\bf k}_{1}\hspace{0.03cm} d\hspace{0.02cm}{\bf k}_{2}\hspace{0.03cm}
\Bigl\{\hspace{0.02cm}{\mathcal U}^{\; a\, a_{1}\hspace{0.03cm} a_{2}}_{\,{\bf k},\, 
{\bf k}_{1},\, {\bf k}_{2}}\, a^{a}_{\hspace{0.02cm}{\bf k}}\, a^{a_{1}}_{\hspace{0.03cm}{\bf k}_{1}}\hspace{0.03cm}
a^{a_{2}}_{\hspace{0.03cm}{\bf k}_{2}}
\,+\,
{\mathcal U}^{\,*\,a\, a_{1}\hspace{0.03cm} a_{2}}_{\; {\bf k},\,
{\bf k}_{1},\, {\bf k}_{2}}\, a^{\ast\ \!a}_{\hspace{0.02cm}{\bf k}}\hspace{0.03cm}
a^{\ast\,a_{1}}_{\hspace{0.03cm}{\bf k}_{1}}\hspace{0.03cm}a^{\ast\,a_{2}}_{\hspace{0.03cm}{\bf k}_{2}}
\Bigr\}\hspace{0.04cm}
\delta({\bf k} + {\bf k}_{1} + {\bf k}_{2}). 
\notag
\end{align}
Correspondingly, the fourth-order interaction Hamiltonian is
\begin{align}
H^{(4)} 
\!=  \frac{1}{2} 
&\int\!\!d\hspace{0.02cm}{\bf k}\,d{\bf k}_{1}\hspace{0.03cm}
\Bigl\{T^{\,\ast\,(1)\,a\,a_{1}\hspace{0.03cm}a_{2}}_{\,{\bf k},\, 
{\bf k}_{1}}\hspace{0.03cm}
a^{\,a}_{\hspace{0.02cm}{\bf k}}\hspace{0.03cm}
a^{\,a_{1}}_{\hspace{0.03cm}{\bf k}_{1}}\, Q^{\,a_{2}}\hspace{0.03cm} 
\!+
T^{\,(1)\,a\,a_{1}\hspace{0.03cm}a_{2}}_{\,{\bf k},\, {\bf k}_{1}}\, {a}^{\hspace{0.03cm}\ast\, a}_{\hspace{0.02cm}{\bf k}}\hspace{0.03cm}
a^{\hspace{0.03cm}\ast\,a_{1}}_{\hspace{0.03cm}{\bf k}_{1}} Q^{\,a_{2}}\hspace{0.03cm} 
\!\Bigr\}
\label{eq:2d}\\[1.3ex]
+\,i\! 
&
\int\!d\hspace{0.02cm}{\bf k}\,d{\bf k}_{1}\,
T^{\,(2)\, a\, a_{1}\hspace{0.03cm} a_{2}}_{\,{\bf k},\,{\bf k}_{1}}\hspace{0.03cm} {a}^{\,\ast\,a}_{\hspace{0.02cm}{\bf k}}\hspace{0.03cm} a^{\,a_{1}}_{\hspace{0.03cm}{\bf k}_{1}}\hspace{0.03cm} Q^{\,a_{2}}\hspace{0.03cm} 
+
\Xi\, Q^{\hspace{0.03cm}a}\hspace{0.03cm}Q^{\hspace{0.03cm}a}.
\notag
\end{align}
It should be especially explained that in spite of the fact that, for example, the Hamiltonian $H^{(3)}$ contains mixed contributions that are formally quadratic in the variables $a^{\,a}_{\,{\bf k}}$ and $Q^{\,a}$, and contributions that are cubic in $a^{\,a}_{\,{\bf k}}$, nevertheless, the whole expression (\ref{eq:2s}) is a third-order interaction Hamiltonian. Thus, we assign the degree of nonlinearity {\it two} to the color charge  $Q^{\hspace{0.03cm}a}$, considering it as a constituent object, as presented, for example, by the expression \cite{barducci_1977, balachandran_1977}
\[
Q^{\,a} = \theta^{\dagger\,i}(t^{a})^{ij}\theta^{\,j}, 
\] 
where $\theta^{\dagger\hspace{0.03cm}i}$ and $\theta^{\,i}$, $i = 1,\ldots,N_{c}$ are Grassmann-valued color charges belonging to the defining representation of the $SU(N_c)$ group. We refer to the two terms in square brackets in the first line in (\ref{eq:2s}) as elementary vertices of the interaction between soft gluon excitations and a hard color-charged particle, as shown in fig.\ref{fig1}. The vertex functions ${\mathcal V}^{\; a\, a_{1}\hspace{0.03cm} a_{2}}_{{\bf k},\, {\bf k}_{1},\, {\bf k}_{2}}$ and ${\mathcal U}^{\; a\, a_{1}\hspace{0.03cm} a_{2}}_{{\bf k},\, {\bf k}_{1},\, {\bf k}_{2}}$ in the remaining contributions in (\ref{eq:2s}) determine the processes of three-plasmon interaction.\\  
\indent The same applies to the Hamiltonian $H^{(4)}$, which contains quadratic contributions in the variable $Q^{\hspace{0.03cm}a}$ and cubic mixed contributions in the variables $a^{\,a}_{\,{\bf k}}$ and $Q^{\hspace{0.03cm}a}$. However, all these contributions form exactly the fourth-order interaction Hamiltonian if one assumes that the degree of nonlinearity of the color charge is two.\\ 
\indent The vertex functions ${\mathcal V}^{\; a\, a_{1}\hspace{0.03cm} a_{2}}_{{\bf k},\, {\bf k}_{1},\, {\bf k}_{2}},\, {\mathcal U}^{\; a\, a_{1}\hspace{0.03cm} a_{2}}_{{\bf k},\, {\bf k}_{1},\, {\bf k}_{2}}$ and $T^{\,(1)\,a\, a_{1}\hspace{0.03cm}a_{2}}_{\,{\bf k},\, {\bf k}_{1}}$ satisfy the ``conditions of natural symmetry'', which specify that the integrals in Eqs.\,(\ref{eq:2s}) and (\ref{eq:2d}) are unaffected by relabeling of the dummy color indices and integration variables. These conditions have the following form:
\begin{equation}
{\mathcal V}^{\;a\,a_{1}\hspace{0.03cm}a_{2}}_{{\bf k},\, {\bf k}_{1},\, {\bf k}_{2}} = {\mathcal V}^{\; a\, a_{2}\, a_{1}}_{{\bf k},\, {\bf k}_{2},\, {\bf k}_{1}},
\quad\;
{\mathcal U}^{\,a\,a_{1}\hspace{0.03cm}a_{2}}_{\hspace{0.03cm}{\bf k},\, {\bf k}_{1},\,{\bf k}_{2}} 
= 
{\mathcal U}^{\,a\,a_{2}\,a_{1}}_{\hspace{0.03cm}{\bf k},\, {\bf k}_{2},\, {\bf k}_{1}}
= 
{\mathcal U}^{\,a_{1}\hspace{0.03cm}a\,a_{2}}_{\hspace{0.03cm}{\bf k}_{1},\, {\bf k},\, {\bf k}_{2}},
\quad\;
T^{\,(1)\,a\, a_{1}\hspace{0.03cm}a_{2}}_{\,{\bf k},\, {\bf k}_{1}} = 
T^{\,(1)\,a_{1}\hspace{0.03cm}a\,a_{2}}_{\,{\bf k}_{1},\, {\bf k}}.
\label{eq:2f}
\end{equation}
The real nature of the Hamiltonian (\ref{eq:2s}) is obvious. A  reality of the Hamiltonian (\ref{eq:2d}) entails a validity of
additional relation for the vertex function $T^{\,(2)\,a\,a_{1}\hspace{0.03cm}a_{2}}_{\,{\bf k},\, {\bf k}_{1}}$:
\[
T^{\,(2)\,a\,a_{1}\hspace{0.03cm}a_{2}}_{\,{\bf k},\, {\bf k}_{1}}
=
-\hspace{0.03cm}
T^{\,\ast\,(2)\,a_{1}\hspace{0.03cm}a\,a_{2}}_{\,{\bf k}_{1},\, {\bf k}}.
\]
%
\indent The vertex functions in the Hamiltonians $H^{(3)}$ and $H^{(4)}$ are defined by specific properties of the system under study, in our case, by a high-temperature gluon plasma. An explicit form of the three-point amplitudes ${\mathcal V}^{\; a\, a_{1}\hspace{0.03cm} a_{2}}_{{\bf k},\, {\bf k}_{1},\, {\bf k}_{2}}$ and ${\mathcal U}^{\; a\, a_{1}\hspace{0.03cm} a_{2}}_{{\bf k},\, {\bf k}_{1},\, {\bf k}_{2}}$ within the hard thermal loop approximation was obtained in \cite{markov_2020}. They have the following color and momentum structures: 
\begin{equation}
{\mathcal V}^{\ \!a\,a_{1}\hspace{0.03cm}a_{2}}_{{\bf k},\, {\bf k}_{1},\,{\bf k}_{2}}
=
f^{\hspace{0.03cm}a\,a_{1}\hspace{0.03cm}a_{2}\,}\hspace{0.02cm}{\mathcal V}_{\, {\bf k},\, {\bf k}_{1},\, {\bf k}_{2}},
\qquad
{\mathcal U}^{\ \! a\, a_{1}\hspace{0.03cm} a_{2}}_{{\bf k},\, {\bf k}_{1},\, {\bf k}_{2}}
=
f^{\hspace{0.03cm}a\,a_{1}\hspace{0.03cm}a_{2}\,}\hspace{0.02cm}{\mathcal U}_{\, {\bf k},\, {\bf k}_{1},\, {\bf k}_{2}},
\label{eq:2h}
\end{equation}
where
\begin{equation}
{\mathcal V}_{\, {\bf k},\, {\bf k}_{1},\, {\bf k}_{2}} = 
\frac{1}{2^{3/4}}\,g\hspace{0.03cm}
\Biggl(\frac{\epsilon^{\hspace{0.03cm}l}_{\mu}({\bf k})}{\sqrt{2\hspace{0.03cm}\omega^ {l}_{\hspace{0.03cm}
			{\bf k}_{\phantom{1}}}}}\Biggr)\!
\Biggl(\frac{\epsilon^{\hspace{0.03cm}l}_{\mu_{1}}({\bf k}_{1})}{\sqrt{2\hspace{0.03cm}\omega^{\hspace{0.03cm}l}_{\hspace{0.03cm}
			{\bf k}_{1}}}}\Biggr)\!
\Biggl(\frac{\epsilon^{\hspace{0.03cm}l}_{\mu_{2}}({\bf k}_{2})}{\sqrt{2\hspace{0.03cm}\omega^{\hspace{0.03cm}l}_{\hspace{0.03cm}
			{\bf k}_{2}}}}\Biggr)\!
\,^{\ast}\Gamma^{\mu\mu_1\mu_2}(k,- k_{1},- k_{2})\Bigr|_{\rm \,on-shell}
\hspace{0.4cm} 
\label{eq:2j}
\end{equation}
and
\begin{equation}
{\mathcal U}_{\, {\bf k},\, {\bf k}_{1},\, {\bf k}_{2}} =
\frac{1}{2^{3/4}}\,g\hspace{0.03cm}
\Biggl(\frac{ \epsilon^{\hspace{0.03cm}l}_{\mu}({\bf k})}{\sqrt{2\hspace{0.03cm}\omega^{\hspace{0.03cm}l}_{\hspace{0.03cm}
			{\bf k}_{\phantom{1}}}}}\Biggr)\!
\Biggl(\frac{\epsilon^{\hspace{0.03cm}l}_{\mu_{1}}({\bf k}_{1})}{\sqrt{2\hspace{0.03cm}\omega^{\hspace{0.03cm}l}_{\hspace{0.03cm}
			{\bf k}_{1}}}}\Biggr)\!
\Biggl(\frac{\epsilon^{\hspace{0.03cm}l}_{\mu_{2}}({\bf k}_{2})}{\sqrt{2\hspace{0.03cm}\omega^{\hspace{0.03cm}l}_{\hspace{0.03cm}
			{\bf k}_{2}}}}\Biggr)\!
\,^{\ast}\Gamma^{\mu\mu_1\mu_2}(- k,- k_{1},- k_{2})\Bigr|_{\rm \,on-shell}.
\!
\label{eq:2k}
\end{equation}
The explicit form of the effective three-gluon vertex $\,^{\ast}\Gamma^{\mu\mu_1\mu_2}(k, k_{1}, k_{2})$ on the right-hand side of these expressions is defined by formulae (\ref{ap:A1})\,--\,(\ref{ap:A3}) in Appendix \ref{appendix_A}. By virtue of the color decomposition (\ref{eq:2h}) and of the symmetry properties (\ref{eq:2f}) for the effective vertices (\ref{eq:2j}) and (\ref{eq:2k}), the following permutation relations 
\begin{equation}
{\mathcal V}_{\, {\bf k},\, {\bf k}_{1},\, {\bf k}_{2}}
=
-\hspace{0.02cm}{\mathcal V}_{\, {\bf k},\, {\bf k}_{2},\, {\bf k}_{1}},
\qquad
{\mathcal U}_{\, {\bf k},\, {\bf k}_{1},\, {\bf k}_{2}}
=
-\,{\mathcal U}_{\, {\bf k},\, {\bf k}_{2},\, {\bf k}_{1}}
=
-\,{\mathcal U}_{\, {\bf k}_{1},\, {\bf k},\, {\bf k}_{2}}
\label{eq:2l}
\end{equation}
are satisfied.\\
\indent In closing this section we define an explicit form of the vertex function ${\upphi}^{\phantom{\ast}}_{\,{\bf k}}$, which enters into the third-order interaction Hamiltonian (\ref{eq:2s}). It is simpler to define this function on the basis of the interaction Hamiltonian of a classical color-charged particle with an external gauge field $A^{a}_{\mu}(x)$:
\[
H_{int} = \!\int\!\frac{d\hspace{0.02cm}{\bf x}}{(2\pi)^{3}}\,j^{a\hspace{0.03cm}\mu}_{Q}(x)A^{a}_{\mu}(x),
\]
where the color current of hard particle $j^{a\hspace{0.03cm}\mu}_{Q}(x)$ is defined by expression  (\ref{eq:1e}). We substitute this current in $H_{int}$ and integrate over ${\bf x}$. As the gauge potential $A^{a}_{\mu}(x)$ we use the first term with the longitudinal polarization on the right-hand side of the expansion (\ref{eq:2q}). As a result, we get for $t_{0} = 0$ and ${\bf x}_{0} = 0$
\[
H_{int} = g\hspace{0.03cm}v^{\hspace{0.02cm}\mu}\hspace{0.03cm}Q^{\hspace{0.03cm}a}(t)\hspace{0.03cm}
A^{a}_{\mu}\bigl(t,\,{\bf v}\hspace{0.02cm}t\bigr) 
\]  
\begin{align}
	= \int\!d\hspace{0.02cm}{\bf k}\,
	\biggl[\hspace{0.03cm}&g
	\left(\frac{Z_{l}({\bf k})}
	{2\hspace{0.03cm}\omega^{\hspace{0.03cm}l}_{\hspace{0.03cm}{\bf k}}}\right)^{\!\!1/2}\!\!\!
	(v\cdot\epsilon^{\hspace{0.03cm}l}({\bf k}))\, a^{\phantom{\ast}\!\!a}_{\hspace{0.03cm}{\bf k}}\hspace{0.03cm}Q^{\hspace{0.03cm}a}
	\ \!e^{-i\hspace{0.03cm}(\omega^{\hspace{0.03cm}l}_{\hspace{0.03cm}{\bf k}} - {\bf k}\cdot{\bf v})\hspace{0.02cm}t}
	\notag \\[1ex]
	+\,&g
	\left(\frac{Z_{l}({\bf k})}
	{2\hspace{0.03cm}\omega^{\hspace{0.03cm}l}_{\hspace{0.03cm}{\bf k}}}\right)^{\!\!1/2}\!\!\!
	(v\cdot\epsilon^{\hspace{0.03cm}\ast\,l}({\bf k}))\, a^{\hspace{0.03cm}\ast\hspace{0.03cm}a}_{\hspace{0.03cm}{\bf k}}\hspace{0.03cm}Q^{\hspace{0.03cm}a}
	\ \!e^{\hspace{0.03cm}i\hspace{0.03cm}(\omega^{\hspace{0.03cm}l}_{\hspace{0.03cm}{\bf k}} - {\bf k}\cdot{\bf v})\hspace{0.02cm}t}\biggr].
	\notag
\end{align}
Comparing this expression with Hamiltonian $H^{(3)}$, Eq.\,(\ref{eq:2s}), we find the following identification:
\begin{equation}
	{\upphi}^{\phantom{\ast}}_{\hspace{0.03cm}{\bf k}}
	=
	g
	\left(\frac{Z_{l}({\bf k})}
	{2\hspace{0.03cm}\omega^{\hspace{0.03cm}l}_{\hspace{0.03cm}
			{\bf k}}}\right)^{\!\!1/2}\!\!\!
	(v\cdot\epsilon^{\hspace{0.03cm}l}({\bf k})).
	\label{eq:4s}
\end{equation}


\section{Canonical transformation}
\setcounter{equation}{0}
\label{section_3}

It is intuitively clear that in the case of a nondecay dispersion law, the terms with the vertex functions ${\mathcal V}^{\; a\, a_{1}\hspace{0.03cm} a_{2}}_{{\bf k},\, {\bf k}_{1},\, {\bf k}_{2}}$ and ${\mathcal U}^{\; a\, a_{1}\hspace{0.03cm} a_{2}}_{{\bf k},\, {\bf k}_{1},\, {\bf k}_{2}}$ in the Hamiltonian $H^{(3)}$ describing three-wave processes may turn out to be irrelevant in some respect. Indeed, due to the specific character of the dispersion equation for soft bosonic longitudinal excitation (\ref{eq:2e}) in a hot gluon plasma, the resonance conditions for three-wave processes involving plasmons
\begin{equation}
\left\{
\begin{array}{ll}
	{\bf k} = {\bf k}_{1} + {\bf k}_{2}, \\[1.5ex]
	\omega^{\hspace{0.03cm}l}_{\hspace{0.03cm}{\bf k}} = \omega^{\hspace{0.03cm}l}_{\hspace{0.03cm}{\bf k}_{1}} + \omega^{\hspace{0.03cm}l}_{\hspace{0.03cm}{\bf k}_{2}},
\end{array}
\right.
\quad
\left\{
\begin{array}{ll}
	{\bf k} + {\bf k}_{1} + {\bf k}_{2} = 0, \\[1.5ex]
	\omega^{\hspace{0.03cm}l}_{\hspace{0.03cm}{\bf k}} + \omega^{\hspace{0.03cm}l}_{\hspace{0.03cm}{\bf k}_{1}} + \omega^{\hspace{0.03cm}l}_{\hspace{0.03cm}{\bf k}_{2}} = 0,
\end{array}\
\right.
\label{eq:3qq}
\end{equation}
have no solutions. Furthermore, the terms in $H^{(3)}$ with vertex function ${\upphi}^{\phantom{\ast}}_{\hspace{0.03cm}{\bf k}}$ describe the processes of Cherenkov emission (absorption) by a particle moving in a medium. These processes are kinematically forbidden due to the condition (\ref{eq:2a}). We shall show that in this case one can go over to new canonical variables ($c^{\,a}_{\hspace{0.02cm}\bf k}$, $c^{\hspace{0.03cm}\ast\ \!\!a}_{{\hspace{0.02cm}\bf k}}$) and ${\mathcal Q}^{\hspace{0.03cm}a}$ such that in new variables the third-order interaction Hamiltonian $H^{(3)} = 0$. In this procedure of excluding the Hamiltonian $H^{(3)}$ a canonical transformation is derived by sequentially excluding the terms for which the resonance conditions (\ref{eq:3qq}) and (\ref{eq:2a}) are not satisfied.\\
\indent Let us consider the transformation from the normal boson variable $a^{a}_{\bf k}$ and classical color charge ${Q}^{\hspace{0.03cm}a}$ to the new field variable $c^{\,a}_{\hspace{0.02cm}\bf k}$ and color charge ${\mathcal Q}^{\hspace{0.03cm}a}$:
\begin{align}
&a^{a}_{\bf k} = a^{\,a}_{\hspace{0.02cm}{\bf k}}\hspace{0.01cm}(c^{\,a}_{\hspace{0.02cm}\bf k},\, 
c^{\hspace{0.03cm}\ast\ \!\!a}_{{\hspace{0.02cm}\bf k}}\!,\, 
{\mathcal Q}^{\hspace{0.03cm}a}),
\label{eq:3q}\\[0.8ex]
&Q^{\,a} = Q^{\,a}\hspace{0.01cm}(\hspace{0.02cm}c^{\,a}_{\hspace{0.02cm}\bf k},\, 
c^{\hspace{0.03cm}\ast\ \!\!a}_{{\hspace{0.02cm}\bf k}}\!,\, {\mathcal Q}^{\hspace{0.03cm}a}\hspace{0.02cm}).
\label{eq:3w}
\end{align}
We will demand that the Hamilton equations in terms of the new variables have the form (\ref{eq:2u}) and (\ref{eq:2i}) with the same Hamiltonian $H$. Because we consider those transformations that are not explicitly time  
dependent, the old and the new Hamiltonians are numerically equal, but generally differ in functional form because they are written in different variables. Straightforward but rather cumbersome calculations result in two systems of integral relations. The first of them has the following form:
\begin{subequations} 
\label{eq:3e}
\begin{align}
&\int\! d\hspace{0.02cm}{\bf k\hspace{0.01cm}}'\!\hspace{0.01cm}
\left\{\frac{\delta\hspace{0.01cm}  a^{\phantom{\ast}\!\!a}_{\hspace{0.02cm}{\bf k}}}{\delta c^{\phantom{\ast}\!\!c}_{\hspace{0.02cm}{\bf k}'}}
\hspace{0.03cm}\frac{\delta\hspace{0.01cm}  a^{\ast\ \!\!b}_{\hspace{0.02cm}{\bf k}''}}{\delta c^{\hspace{0.03cm}\ast\ \!\!c}_{\hspace{0.02cm}{\bf k}'}}
\,-\,
\frac{\delta\hspace{0.01cm}  a^{\phantom{\ast}\!\!a}_{\hspace{0.02cm}{\bf k}}}
{\delta c^{\hspace{0.03cm}\ast\ \!\!c}_{\hspace{0.02cm}{\bf k}'}}\hspace{0.03cm}
\frac{\delta\hspace{0.01cm}  a^{\ast\ \!\!b}_{\hspace{0.02cm}{\bf k}''}}
{\delta c^{\phantom{\ast}\!\!c}_{\hspace{0.02cm}{\bf k}'}}\right\}
+
i\,\frac{\partial a^{\phantom{\ast}\!\!a}_{\hspace{0.02cm}{\bf k}}}{\,\partial\hspace{0.03cm} {\mathcal Q}^{\hspace{0.03cm}c}}\hspace{0.03cm}
\frac{\partial\hspace{0.03cm}a^{\ast\ \!\!b}_{\hspace{0.02cm}{\bf k}''}}{\,\partial\hspace{0.03cm} {\mathcal Q}^{\hspace{0.03cm}c^{\hspace{0.02cm}\prime}}}
\,f^{\hspace{0.03cm}c\hspace{0.03cm}c^{\hspace{0.02cm}\prime}\hspace{0.02cm}d}\hspace{0.03cm}{\mathcal Q}^{\hspace{0.03cm}d}
\!=
\delta^{\hspace{0.03cm}a\hspace{0.02cm}b}\delta ({\bf k}-{\bf k}\!\ ''),
\label{eq:3ea}
\\[0.8ex]
&\int\! d\hspace{0.02cm}{\bf k\hspace{0.01cm}}'\!\hspace{0.01cm}\left\{\frac{\delta\hspace{0.02cm}  a^{\phantom{\ast}\!\!a}_{\hspace{0.02cm}{\bf k}}}
{\delta c^{\phantom{\ast}\!\!c}_{\hspace{0.02cm}{\bf k}^{\prime}}}
\hspace{0.03cm}\frac{\delta\hspace{0.02cm}  a^{\phantom{\ast}\!\!b}_{\hspace{0.02cm}{\bf k}''}}
{\delta c^{\hspace{0.03cm}\ast\ \!\!c}_{\hspace{0.02cm}{\bf k}^{\prime}}}
\,-\,
\frac{\delta\hspace{0.02cm}  a^{\phantom{\ast}\!\!a}_{\hspace{0.02cm}{\bf k}}}
{\delta c^{\hspace{0.03cm}\ast\ \!\!c}_{\hspace{0.02cm}{\bf k}^{\prime}}}\hspace{0.03cm}
\frac{\delta\hspace{0.02cm}  a^{\phantom{\ast}\!\!b}_{\hspace{0.02cm}{\bf k}''}}
{\delta c^{\phantom{\ast}\!\!c}_{\hspace{0.02cm}{\bf k}^{\prime}}}\right\}
+
i\,\frac{\partial a^{\phantom{\ast}\!\!a}_{\hspace{0.02cm}{\bf k}}}{\,\partial\hspace{0.03cm}{\mathcal Q}^{\hspace{0.03cm}c}}\hspace{0.03cm}
\frac{\partial\hspace{0.03cm}a^{\phantom{\ast}\!\!b}_{\hspace{0.02cm}{\bf k}''}}{\,\partial\hspace{0.03cm} {\mathcal Q}^{\hspace{0.03cm}c^{\hspace{0.02cm}\prime}}}
\,f^{\hspace{0.03cm}c\hspace{0.03cm}c^{\hspace{0.02cm}\prime}\hspace{0.02cm}d}
\hspace{0.03cm}{\mathcal Q}^{\hspace{0.03cm}d} = 0,
\label{eq:3eb}
\\[0.8ex]
&\int\! d\hspace{0.02cm}{\bf k\hspace{0.01cm}}'\!\hspace{0.01cm}\left\{\frac{\delta\hspace{0.02cm}  
a^{\phantom{\ast}\!\!a}_{\hspace{0.02cm}{\bf k}}}
{\delta c^{\phantom{\ast}\!\!c}_{\hspace{0.02cm}{\bf k}^{\prime}}}
\hspace{0.03cm}\frac{\delta\hspace{0.02cm}  Q^{\,b}}
{\delta c^{\hspace{0.03cm}\ast\ \!\!c}_{\hspace{0.02cm}{\bf k}^{\prime}}}
\,-\,
\frac{\delta\hspace{0.02cm}  a^{\phantom{\ast}\!\!a}_{\hspace{0.02cm}{\bf k}}}
{\delta c^{\hspace{0.03cm}\ast\ \!\!c}_{\hspace{0.02cm}{\bf k}^{\prime}}}\hspace{0.03cm}
\frac{\delta\hspace{0.02cm} Q^{\,b}}
{\delta c^{\phantom{\ast}\!\!c}_{\hspace{0.02cm}{\bf k}^{\prime}}}\right\}
+
i\,\frac{\partial a^{\phantom{\ast}\!\!a}_{\hspace{0.02cm}{\bf k}}}{\,\partial\hspace{0.02cm}{\mathcal Q}^{\hspace{0.03cm}c}}\hspace{0.03cm}
\frac{\partial\hspace{0.04cm} Q^{\hspace{0.03cm}b}}{\,\partial\hspace{0.02cm} 
{\mathcal Q}^{\hspace{0.03cm}c^{\hspace{0.02cm}\prime}}}
\,f^{\hspace{0.03cm}c\hspace{0.03cm}c^{\hspace{0.02cm}\prime}\hspace{0.03cm}d}
\hspace{0.03cm}{\mathcal Q}^{\hspace{0.03cm}d} = 0.
\label{eq:3ec}
\end{align}
\end{subequations}
Correspondingly, the second system is
\begin{subequations} 
\label{eq:3r}
\begin{align}
&\int\! d\hspace{0.02cm}{\bf k\hspace{0.01cm}}'\!\hspace{0.01cm}
\left\{\frac{\delta\hspace{0.01cm}  
Q^{\phantom{\ast}\!\!a}}{\delta c^{\phantom{\ast}\!\!c}_{\hspace{0.02cm}{\bf k}'}}
\hspace{0.03cm}\frac{\delta\hspace{0.01cm}  a^{\ast\ \!\!b}_{\hspace{0.02cm}{\bf k}''}}{\delta c^{\hspace{0.03cm}\ast\ \!\!c}_{\hspace{0.02cm}{\bf k}'}}
\,-\,
\frac{\delta\hspace{0.01cm}  Q^{\phantom{\ast}\!\!a}}
{\delta c^{\hspace{0.03cm}\ast\ \!\!c}_{\hspace{0.02cm}{\bf k}'}}\hspace{0.03cm}
\frac{\delta\hspace{0.01cm}  a^{\ast\ \!\!b}_{\hspace{0.02cm}{\bf k}''}}
{\delta c^{\phantom{\ast}\!\!c}_{\hspace{0.02cm}{\bf k}'}}\right\}
+
i\,\frac{\partial\hspace{0.03cm}   Q^{\phantom{\ast}\!\!a}}{\,\partial\hspace{0.04cm} {\mathcal Q}^{\,c}}\hspace{0.03cm}
\frac{\partial\hspace{0.03cm}  a^{\ast\ \!\!b}_{\hspace{0.02cm}{\bf k}''}}{\,\partial\hspace{0.03cm} {\mathcal Q}^{\hspace{0.03cm}c^{\hspace{0.02cm}\prime}}}
\,f^{\hspace{0.03cm}c\hspace{0.03cm}c^{\hspace{0.02cm}\prime}\hspace{0.02cm}d}\hspace{0.03cm}{\mathcal Q}^{\hspace{0.03cm}d}
\!= 0,
\label{eq:3ra}
\\[0.8ex]
&\int\! d\hspace{0.02cm}{\bf k\hspace{0.01cm}}'\!\hspace{0.01cm}\left\{\frac{\delta\hspace{0.02cm}  
Q^{\phantom{\ast}\!\!a}}{\delta c^{\phantom{\ast}\!\!c}_{\hspace{0.02cm}{\bf k}^{\prime}}}
\hspace{0.03cm}\frac{\delta\hspace{0.02cm}a^{\phantom{\ast}\!\!b}_{\hspace{0.02cm}{\bf k}''}}
{\delta c^{\hspace{0.03cm}\ast\ \!\!c}_{\hspace{0.02cm}{\bf k}^{\prime}}}
\,-\,
\frac{\delta\hspace{0.02cm} Q^{\phantom{\ast}\!\!a}}
{\delta c^{\hspace{0.03cm}\ast\ \!\!c}_{\hspace{0.02cm}{\bf k}^{\prime}}}\,
\frac{\delta\hspace{0.02cm}a^{\phantom{\ast}\!\!b}_{\hspace{0.02cm}{\bf k}''}}
{\delta c^{\phantom{\ast}\!\!c}_{\hspace{0.02cm}{\bf k}^{\prime}}}\right\}
+
i\,\frac{\partial\hspace{0.03cm}  Q^{\phantom{\ast}\!\!a}}{\,\partial\hspace{0.04cm} {\mathcal Q}^{\,c}}\hspace{0.03cm}
\frac{\partial\hspace{0.03cm}  a^{\phantom{\ast}\!\!b}_{\hspace{0.02cm}{\bf k}''}}{\,\partial\hspace{0.03cm} 
{\mathcal Q}^{\hspace{0.03cm}c^{\hspace{0.02cm}\prime}}}
\,f^{\hspace{0.03cm}c\hspace{0.03cm}c^{\hspace{0.02cm}\prime}\hspace{0.02cm}d}
\hspace{0.03cm}{\mathcal Q}^{\hspace{0.03cm}d} = 0,
\label{eq:3rb}
\\[0.8ex]
&\int\! d\hspace{0.02cm}{\bf k\hspace{0.01cm}}'\!\hspace{0.01cm}\left\{\frac{\delta\hspace{0.02cm}  
Q^{\phantom{\ast}\!\!a}}
{\delta c^{\phantom{\ast}\!\!c}_{\hspace{0.02cm}{\bf k}^{\prime}}}
\hspace{0.03cm}\frac{\delta\hspace{0.02cm}  Q^{\,b}}
{\delta c^{\hspace{0.03cm}\ast\ \!\!c}_{\hspace{0.02cm}{\bf k}^{\prime}}}
\,-\,
\frac{\delta\hspace{0.02cm} Q^{\phantom{\ast}\!\!a}}
{\delta c^{\hspace{0.03cm}\ast\ \!\!c}_{\hspace{0.02cm}{\bf k}^{\prime}}}\hspace{0.03cm}
\frac{\delta\hspace{0.02cm} Q^{\,b}}
{\delta c^{\phantom{\ast}\!\!c}_{\hspace{0.02cm}{\bf k}^{\prime}}}\right\}
+
i\,\frac{\partial\hspace{0.03cm}  Q^{\phantom{\ast}\!\!a}}{\,\partial\hspace{0.03cm}
{\mathcal Q}^{\hspace{0.03cm}c}}\hspace{0.03cm}
\frac{\partial\hspace{0.03cm} Q^{\hspace{0.03cm}b}}{\,\partial\hspace{0.03cm} 
{\mathcal Q}^{\hspace{0.03cm}c^{\hspace{0.02cm}\prime}}}
\,f^{\hspace{0.03cm}c\hspace{0.03cm}c^{\hspace{0.02cm}\prime}\hspace{0.02cm}d}
\hspace{0.03cm} {\mathcal Q}^{\hspace{0.03cm}d} = i\hspace{0.02cm} 
f^{\hspace{0.03cm}a\hspace{0.03cm}b\hspace{0.03cm}d}\hspace{0.03cm} {Q}^{\hspace{0.03cm}d}.
\label{eq:3rc}
\end{align}
\end{subequations}
These canonicity conditions can be written in a very compact form if we make use of the definition of the Lie-Poisson bracket (\ref{eq:2y}) and replace the variation variables by the new ones: $a^{\,a}_{\bf k}\rightarrow c^{\,a}_{\hspace{0.02cm}\bf k}$ and $Q^{\,b} \rightarrow {\mathcal Q}^{\,b}$. In this case the Lie-Poisson bracket for the original variables $a^{\,a}_{\bf k}$ and $Q^{\,b}$, Eqs.\,(\ref{eq:2r}) and (\ref{eq:2t}), turns to the canonicity conditions (\ref{eq:3e}) and (\ref{eq:3r}), which impose certain restrictions on the functional dependencies (\ref{eq:3q}) and (\ref{eq:3w}). Let us present the canonical transformations (\ref{eq:3q}) and (\ref{eq:3w}) in the form of integro-power series in normal variable $c^{\,a}_{\hspace{0.02cm}\bf k}$ and color charge ${\mathcal Q}^{\,a}$. In this case the transformation (\ref{eq:3q}) up to the terms of the sixth order\footnote{\hspace{0.03cm}Recall again that we consider a degree of nonlinearity of the color charge to be two.} in $c^{\,a}_{\hspace{0.02cm}\bf k}$ and ${\mathcal Q}^{\hspace{0.03cm}a}$ has the following form:
\begin{equation}
a^{a}_{\hspace{0.02cm}{\bf k}} = c^{a}_{\hspace{0.02cm}{\bf k}}\,+ {F}_{\hspace{0.03cm} \bf k}\hspace{0.02cm}
{\mathcal Q}^{\hspace{0.03cm}a}
\label{eq:3t}
\end{equation}
\[
+ \int\!d\hspace{0.02cm}{\bf k}_{1}\hspace{0.02cm} d\hspace{0.02cm}{\bf k}_{2}\! 
\left[V^{\hspace{0.02cm}(1)\,a\,a_{1}\hspace{0.03cm}a_{2}}_{\ {\bf k},\, {\bf k}_{1},\, {\bf k}_{2}}\hspace{0.03cm}c^{a_{1}}_{\hspace{0.02cm}{\bf k}_{1}}\hspace{0.03cm}c^{a_{2}}_{\hspace{0.02cm}{\bf k}_{2}}
\,+\,
V^{\hspace{0.02cm}(2)\,a\,a_{1}\hspace{0.03cm}a_{2}}_{\ {\bf k},\, {\bf k}_{1},\, {\bf k}_{2}}\hspace{0.03cm}c^{\hspace{0.03cm}\ast\, a_{1}}_{\hspace{0.02cm}{\bf k}_{1}}\hspace{0.03cm}c^{\phantom{\ast}\!\!a_{2}}_{\hspace{0.02cm}{\bf k}_{2}}
\,+\,
V^{\hspace{0.02cm}(3)\,a\,a_{1}\hspace{0.03cm}a_{2}}_{\ {\bf k},\, {\bf k}_{1},\, {\bf k}_{2}}\, 
c^{\hspace{0.03cm}\ast\, a_{1}}_{\hspace{0.02cm}{\bf k}_{1}} c^{\hspace{0.03cm}\ast\, a_{2}}_{\hspace{0.02cm}{\bf k}_{2}}\right] 
 \]
\[
+ \int\!d\hspace{0.02cm}{\bf k}_{1}\! 
\left[\hspace{0.03cm}\widetilde{V}^{\hspace{0.02cm}(1)\,a\,a_{1}\hspace{0.03cm}a_{2}}_{\ {\bf k},\, {\bf k}_{1}}\hspace{0.03cm}c^{\hspace{0.03cm}\ast\, a_{1}}_{\hspace{0.02cm}{\bf k}_{1}}\hspace{0.03cm}{\mathcal Q}^{\hspace{0.03cm}a_{2}}
\,+\,
\widetilde{V}^{\hspace{0.02cm}(2)\,a\,a_{1}\hspace{0.03cm}a_{2}}_{\ {\bf k},\,
{\bf k}_{1}}\hspace{0.03cm}c^{a_{1}}_{\hspace{0.02cm}{\bf k}_{1}}\hspace{0.03cm} 
{\mathcal Q}^{\hspace{0.03cm}a_{2}}\hspace{0.03cm}
\right] 
\]
\[
+\! \int\!d\hspace{0.02cm}{\bf k}_{1}\hspace{0.02cm} d\hspace{0.02cm}{\bf k}_{2}
\Bigl[\hspace{0.04cm}W^{\hspace{0.02cm}(1)\,a\,a_{1}\hspace{0.03cm}a_{2}\, a_{3}}_{\ {\bf k},\, {\bf k}_{1},\, {\bf k}_{2}}\, c^{a_{1}}_{\hspace{0.02cm}{\bf k}_{1}}\hspace{0.03cm} c^{a_{2}}_{\hspace{0.02cm}{\bf k}_{2}}\hspace{0.02cm}{\mathcal Q}^{\hspace{0.03cm}a_{3}}
\,+\,
W^{\hspace{0.02cm}(2)\,a\,a_{1}\hspace{0.03cm}a_{2}\,a_{3}}_{\ {\bf k},\, {\bf k}_{1},\, {\bf k}_{2}}\hspace{0.03cm}c^{\hspace{0.03cm}\ast\, a_{1}}_{\hspace{0.02cm}{\bf k}_{1}}\hspace{0.03cm}c^{a_{2}}_{\hspace{0.02cm}{\bf k}_{2}}\hspace{0.02cm}{\mathcal Q}^{\hspace{0.03cm}a_{3}}
+
W^{\hspace{0.02cm}(3)\,a\,a_{1}\hspace{0.03cm}a_{2}\,a_{3}}_{\ {\bf k},\, {\bf k}_{1},\, {\bf k}_{2}}\hspace{0.03cm}c^{\hspace{0.03cm}\ast\, a_{1}}_{\hspace{0.02cm}{\bf k}_{1}}\hspace{0.03cm}c^{\hspace{0.03cm}\ast\, a_{2}}_{\hspace{0.02cm}{\bf k}_{2}}\hspace{0.02cm}{\mathcal Q}^{\hspace{0.03cm}a_{3}}\hspace{0.03cm} 
\Bigr]
\]
\[
+\! \int\!d\hspace{0.02cm}{\bf k}_{1}
\Bigl[\hspace{0.04cm}\widetilde{W}^{\hspace{0.02cm}(1)\,a\,a_{1}\hspace{0.03cm}a_{2}\, a_{3}}_{\ {\bf k},\, {\bf k}_{1}}\hspace{0.03cm}c^{\hspace{0.03cm}\ast\, a_{1}}_{\hspace{0.02cm}{\bf k}_{1}}\hspace{0.01cm}{\mathcal Q}^{\hspace{0.03cm}a_{2}}\hspace{0.01cm}{\mathcal Q}^{\hspace{0.03cm}a_{3}}
\,+\,
\widetilde{W}^{\hspace{0.02cm}(2)\,a\,a_{1}\hspace{0.03cm}a_{2}\,a_{3}}_{\ {\bf k},\, {\bf k}_{1}}\hspace{0.03cm}c^{a_{1}}_{\hspace{0.02cm}{\bf k}_{1}}\hspace{0.02cm}{\mathcal Q}^{\hspace{0.03cm}a_{2}}\hspace{0.02cm}{\mathcal Q}^{\hspace{0.03cm}a_{3}}
\hspace{0.03cm}
\Bigr] +\,\ldots
\vspace{0.15cm}
\]
\[
+\, 
\bigl(G^{\, a\, a_{1}\hspace{0.02cm} a_{2}}_{\bf k}\hspace{0.02cm}{\mathcal Q}^{\hspace{0.03cm}a_{1}}\hspace{0.01cm}{\mathcal Q}^{\hspace{0.03cm}a_{2}}
\,+\, 
G^{\, a\, a_{1}\hspace{0.02cm} a_{2}\hspace{0.02cm} a_{3}}_{\bf k}\hspace{0.02cm}{\mathcal Q}^{\hspace{0.03cm}a_{1}}\hspace{0.01cm}{\mathcal Q}^{\hspace{0.03cm}a_{2}}\hspace{0.01cm}{\mathcal Q}^{\hspace{0.03cm}a_{3}}
\,+\,\ldots\,\bigr)\, _{.}
\]
Similarly, the most common power-series expansion for the transformation (\ref{eq:3w}) up to the terms of the sixth order  is
\begin{equation}
Q^{\hspace{0.03cm}a} = {\mathcal Q}^{\hspace{0.03cm}a}\,
+ 
\int\!d\hspace{0.02cm}{\bf k}_{1}
\left[\hspace{0.03cm}M^{\,a\,a_{1}\hspace{0.03cm}a_{2}}_{\; {\bf k}_{1}}\, c^{a_{1}}_{\hspace{0.02cm}{\bf k}_{1}}\hspace{0.02cm}{\mathcal Q}^{\hspace{0.03cm}a_{2}}
\,+\,
M^{\hspace{0.02cm}\ast\,a\,a_{1}\hspace{0.03cm}a_{2}}_{\; {\bf k}_{1}}\, c^{\hspace{0.03cm}\ast\, a_{1}}_{\hspace{0.02cm}{\bf k}_{1}}\hspace{0.02cm}{\mathcal Q}^{\hspace{0.03cm}a_{2}}\hspace{0.03cm} 
\right] 
\label{eq:3y}
\end{equation}
\[
+\!\int\!d\hspace{0.02cm}{\bf k}_{1}\hspace{0.02cm} d\hspace{0.02cm}{\bf k}_{2}
\Bigl[\hspace{0.04cm}M^{\hspace{0.03cm}(1)\,a\,a_{1}\hspace{0.03cm}a_{2}\,
a_{3}}_{\ {\bf k}_{1},\, {\bf k}_{2}}\hspace{0.03cm}c^{a_{1}}_{\hspace{0.02cm}{\bf k}_{1}}\hspace{0.03cm} c^{a_{2}}_{\hspace{0.02cm}{\bf k}_{2}}\hspace{0.02cm}{\mathcal Q}^{\hspace{0.03cm}a_{3}}
+
M^{\hspace{0.03cm}(2)\,a\,a_{1}\hspace{0.03cm}a_{2}\,a_{3}}_{\ {\bf k}_{1},\, {\bf k}_{2}}\hspace{0.03cm}c^{\hspace{0.03cm}\ast\, a_{1}}_{\hspace{0.02cm}{\bf k}_{1}}\hspace{0.03cm} c^{a_{2}}_{\hspace{0.02cm}{\bf k}_{2}}\hspace{0.02cm}{\mathcal Q}^{\hspace{0.03cm}a_{3}}
+
M^{\hspace{0.03cm}\ast\,(1)\,a\,a_{1}\hspace{0.03cm}a_{2}\,a_{3}}_{\ {\bf k}_{1},\, {\bf k}_{2}}\, c^{\hspace{0.03cm}\ast\, a_{1}}_{\hspace{0.02cm}{\bf k}_{1}}\hspace{0.03cm}c^{\hspace{0.03cm}\ast\, a_{2}}_{\hspace{0.02cm}{\bf k}_{2}}\hspace{0.02cm}{\mathcal Q}^{\hspace{0.03cm}a_{3}}\hspace{0.03cm} 
\Bigr]
\]
\[
+\!\int\!d\hspace{0.02cm}{\bf k}_{1}
\Bigl[\hspace{0.04cm}\widetilde{M}^{\,a\,a_{1}\hspace{0.03cm}a_{2}\, a_{3}}_{\ {\bf k}_{1}}\hspace{0.03cm} c^{a_{1}}_{\hspace{0.02cm}{\bf k}_{1}}
\hspace{0.02cm}{\mathcal Q}^{\hspace{0.03cm}a_{2}}
{\mathcal Q}^{\hspace{0.03cm}a_{3}}
\,+\,
\widetilde{M}^{\hspace{0.02cm}\ast\,a\,a_{1}\hspace{0.03cm}a_{2}\,a_{3}}_{\ {\bf k}_{1}}\hspace{0.03cm} c^{\hspace{0.03cm}\ast\, a_{1}}_{\hspace{0.02cm}{\bf k}_{1}}\hspace{0.01cm}
{\mathcal Q}^{\hspace{0.03cm}a_{2}}
{\mathcal Q}^{\hspace{0.03cm}a_{3}}
\hspace{0.03cm}
\Bigr] +\,\ldots
\vspace{0.15cm}
\]
\[
+\; 
F^{\, a\, a_{1}\hspace{0.02cm} a_{2}}\hspace{0.02cm}{\mathcal Q}^{\hspace{0.03cm}a_{1}}{\mathcal Q}^{\hspace{0.03cm}a_{2}}
\,+\,
F^{\, a\, a_{1}\hspace{0.02cm} a_{2}\hspace{0.02cm} a_{3}}\hspace{0.02cm}{\mathcal Q}^{\hspace{0.03cm}a_{1}}{\mathcal Q}^{\hspace{0.03cm}a_{2}}{\mathcal Q}^{\hspace{0.03cm}a_{3}}
+\,\ldots\ _{.}
\]
In the transformation (\ref{eq:3t}) we have not written out cubic terms such as $c^{\,a_{1}}_{\hspace{0.02cm}{\bf k}_{1}}\hspace{0.03cm} c^{\,a_{2}}_{\hspace{0.02cm}{\bf k}_{2}}\hspace{0.03cm} c^{\,a_{3}}_{\hspace{0.02cm}{\bf k}_{3}}$, since they do not give any contributions in the approximation we are considering. In the transformation (\ref{eq:3y}), the contributions containing only the normal variables $c^{\,a}_{\hspace{0.02cm}{\bf k}}$ and $c^{\hspace{0.03cm}\ast\, a}_{\hspace{0.02cm}{\bf k}}$ (i.e., without the color charges ${\mathcal Q}^{\hspace{0.03cm}a}$) are omitted. This can be interpreted so that the color charges are given physical entities that cannot be induced by the soft gauge field of the system under consideration. However, the fact that these contributions in any case turn to zero can be verified by direct calculations. In addition, in the transformation (\ref{eq:3y}) the requirement of a reality of the color charge is taken into account in particular, it also leads to the conditions
\[
M^{\hspace{0.03cm}\ast\hspace{0.03cm}(2)\,a\,a_{1}\hspace{0.03cm}a_{2}\, a_{3}}_{\ {\bf k}_{1},\, {\bf k}_{2}}
=
M^{\hspace{0.03cm}(2)\,a\,a_{2}\,a_{1}\hspace{0.03cm}a_{3}}_{\ {\bf k}_{2},\, {\bf k}_{1}},
\quad
F^{\,\ast\, a\, a_{1}\hspace{0.02cm} a_{2}} = F^{\, a\, a_{1}\hspace{0.02cm} a_{2}},
\quad
F^{\,\ast\, a\, a_{1}\hspace{0.02cm} a_{2}\hspace{0.02cm} a_{3}} = F^{\, a\, a_{1}\hspace{0.02cm} a_{2}\hspace{0.02cm} a_{3}}.
\]
In addition we note that the coefficient functions $V^{\hspace{0.02cm}(1)\, a\, a_{1}\, a_{2}}_{\ {\bf k},\, {\bf k}_{1},\, {\bf k}_{2}}\!$, $V^{\hspace{0.02cm}(3)\, a\, a_{1}\, a_{2}}_{\ {\bf k},\, {\bf k}_{1},\, {\bf k}_{2}}\!$, $W^{\hspace{0.02cm}(1,3)\, a\, a_{1}\, a_{2}\, a_{3}}_{\ {\bf k},\, {\bf k}_{1},\, {\bf k}_{2}}\!$, $\widetilde{W}^{\hspace{0.02cm}(1,2)\, a\, a_{1}\, a_{2}\, a_{3}}_{\ {\bf k},\,{\bf k}_{1}}\!$, $M^{\hspace{0.02cm}(1)\, a\, a_{1}\, a_{2}\, a_{3}}_{\ {\bf k}_{1},\, {\bf k}_{2}}$ and $\widetilde{M}^{\, a\, a_{1}\, a_{2}\, a_{3}}_{\ {\bf k}_{1}}$ must satisfy the natural symmetry conditions:
\begin{equation}
\begin{array}{llll}
&V^{\hspace{0.02cm}(1)\,a\,a_{1}\hspace{0.03cm}a_{2}}_{\ {\bf k},\, {\bf k}_{1},\, {\bf k}_{2}} = V^{\hspace{0.02cm}(1)\,a\,a_{2}\,a_{1}}_{\ {\bf k},\, {\bf k}_{2},\, {\bf k}_{1}},
\qquad
&V^{\hspace{0.02cm}(3)\,a\,a_{1}\hspace{0.03cm}a_{2}}_{\ {\bf k},\, {\bf k}_{1},\, {\bf k}_{2}} = V^{\hspace{0.02cm}(3)\,a\,a_{2}\,a_{1}}_{\ {\bf k},\, {\bf k}_{2},\, {\bf k}_{1}}, 
\\[2.5ex]
&W^{\hspace{0.02cm}(1)\,a\,a_{1}\hspace{0.03cm}a_{2}\, a_{3}}_{\ {\bf k},\, {\bf k}_{1},\, {\bf k}_{2}} = W^{\hspace{0.02cm}(1)\,a\,a_{2}\,a_{1}\hspace{0.03cm}a_{3}}_{\ {\bf k},\, {\bf k}_{2},\, {\bf k}_{1}},
\qquad
&W^{\hspace{0.02cm}(3)\,a\,a_{1}\hspace{0.03cm}a_{2}\,a_{3}}_{\ {\bf k},\, {\bf k}_{1},\, {\bf k}_{2}} = W^{\hspace{0.02cm}(3)\, a\, a_{2}\, a_{1}\hspace{0.03cm}a_{3}}_{\ {\bf k},\, {\bf k}_{2},\, {\bf k}_{1}},\\[3ex]
&\widetilde{W}^{\hspace{0.02cm}(1)\,a\,a_{1}\hspace{0.03cm}a_{2}\,a_{3}}_{\ {\bf k},\, 
{\bf k}_{1}} 
= 
\widetilde{W}^{\hspace{0.02cm}(1)\,a\,a_{1}\hspace{0.03cm}a_{3}\, a_{2}}_{\ {\bf k},\, 
{\bf k}_{1}},
\quad
&\widetilde{W}^{\hspace{0.02cm}(2)\,a\,a_{1}\hspace{0.03cm}a_{2}\,a_{3}}_{\ {\bf k},\, 
{\bf k}_{1}} 
= 
\widetilde{W}^{\hspace{0.02cm}(2)\,a\,a_{1}\hspace{0.03cm}a_{3}\,a_{2}}_{\ {\bf k},\, 
{\bf k}_{1}}, 
\\[2.5ex]
&M^{\hspace{0.02cm}(1)\,a\,a_{1}\hspace{0.03cm}a_{2}\,a_{3}}_{\ {\bf k}_{1},\, {\bf k}_{2}} 
= 
M^{\hspace{0.02cm}(1)\,a\,a_{2}\hspace{0.03cm}a_{1}\,a_{3}}_{\ {\bf k}_{2},\, {\bf k}_{1}},
\quad
&\widetilde{M}^{\,a\,a_{1}\hspace{0.03cm}a_{2}\,a_{3}}_{\ {\bf k}_{1}} 
= 
\widetilde{M}^{\,a\,a_{1}\hspace{0.03cm}a_{3}\,a_{2}}_{\ {\bf k}_{1}}.
\end{array}
\label{eq:3u}  
\end{equation}
\indent Furthermore, substituting the expansions (\ref{eq:3t}) and  (\ref{eq:3y}) into a system of the canonicity conditions (\ref{eq:3e}) and (\ref{eq:3r}), we obtain rather nontrivial integral relations connecting various coefficient functions among themselves. A complete list of the integral relations connecting the coefficient functions of the third and fourth orders is given in Appendices \ref{appendix_B} and \ref{appendix_C}. Here, we have provided only algebraic relations for the lowest second-order coefficient functions:
\begin{equation}
V^{\hspace{0.02cm}(2)\,a\,a_{1}\hspace{0.03cm}a_{2}}_{\ {\bf k},\, {\bf k}_{1},\, {\bf k}_{2}} = -\hspace{0.01cm}2\hspace{0.03cm}V^{\,\ast\hspace{0.03cm}(1)\, a_{2}\,a_{1}\hspace{0.03cm}a}_{\ {\bf k}_{2},\, {\bf k}_{1},\, {\bf k}},
\quad
V^{\hspace{0.02cm}(3)\,a\,a_{1}\hspace{0.03cm}a_{2}}_{\ {\bf k},\, {\bf k}_{1},\, {\bf k}_{2}} = V^{\hspace{0.02cm}(3)\,a_{1}\hspace{0.03cm}a\,a_{2}}_{\ {\bf k}_{1},\, {\bf k},\, 
{\bf k}_{2}},
\label{eq:3i}
\end{equation}
\begin{equation}
M^{\,a\,a_{1}\hspace{0.03cm}a_{2}}_{\; {\bf k}} + i\hspace{0.02cm} F^{\hspace{0.03cm}\ast}_{\, {\bf k}}\hspace{0.03cm} f^{\,a\, a_{1}\hspace{0.03cm}a_{2}} = 0,
\label{eq:3o}
\end{equation}
\begin{align}
&\widetilde{V}^{\hspace{0.03cm}(1)\,a\,a_{1}\hspace{0.03cm}a_{2}}_{\ {\bf k},\, {\bf k}_{1}} 
-
\widetilde{V}^{\hspace{0.03cm}(1)\,a_{1}\hspace{0.03cm}a\,a_{2}}_{\ {\bf k}_{1},\, {\bf k}} 
-
i\hspace{0.02cm}F^{\hspace{0.03cm}\phantom{\ast}}_{\, {\bf k}}\hspace{0.01cm} F^{\hspace{0.03cm} \phantom{\ast}}_{\, {\bf k}_{1}} f^{\,a\,a_{1}\hspace{0.03cm}a_{2}} = 0,
\label{eq:3p}\\[1ex]
&\widetilde{V}^{\hspace{0.03cm}(2)\,a\, a_{1}\hspace{0.03cm}a_{2}}_{\ {\bf k},\, {\bf k}_{1}} 
+
\widetilde{V}^{\,\ast\hspace{0.03cm}(2)\,a_{1}\hspace{0.03cm}a\,a_{2}}_{\ {\bf k}_{1},\, {\bf k}} 
+
i\hspace{0.02cm}F^{\hspace{0.03cm}\phantom{\ast}}_{\, {\bf k}}\hspace{0.01cm} F^{\hspace{0.03cm}\ast}_{\, {\bf k}_{1}} f^{\,a\, a_{1}\hspace{0.03cm}a_{2}}  = 0.
\label{eq:3a}
\end{align}


\section{Eliminating ``nonessential'' Hamiltonian $H^{(3)}$. Effective fourth-order Hamiltonian}
\setcounter{equation}{0}
\label{section_4}

\begin{figure}[hbtp]
\begin{center}
\includegraphics*[scale=0.9]{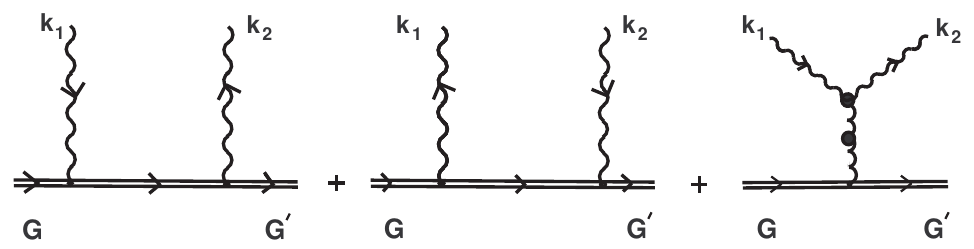}
\end{center}
\vspace{-0.5cm}
\caption{\small The effective amplitude $\widetilde{T}^{\,(2)\,a_{1}\hspace{0.03cm}a_{2}\, a}_{\,{\bf k}_{1},\, 
{\bf k}_{2}}$ for the elastic scattering process of plasmon off a hard color particle. The blob stands for HTL resummation and the double line denotes the hard  particle}
\label{fig2}
\end{figure}

\begin{figure}[hbtp]
\begin{center}
\includegraphics*[scale=0.65]{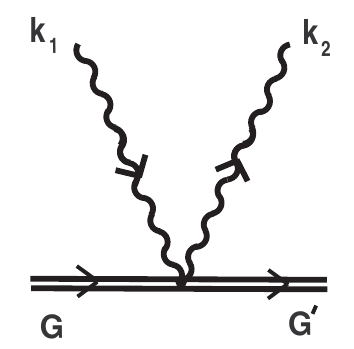}
\end{center}
\vspace{-0.5cm}
\caption{\small Direct interaction of two plasmons with a hard particle}
\label{fig3}
\end{figure}

The next step in constructing the effective theory is the procedure of eliminating the third-order interaction Hamiltonian $H^{(3)}$, Eq.\,(\ref{eq:2s}), upon switching from the original bosonic function $a^{\hspace{0.03cm}a}_{\bf k}$ and the color charge $Q^{\hspace{0.03cm}a}$ to the new function $c^{\hspace{0.03cm}a}_{\hspace{0.02cm}{\bf k}}$ and color charge ${\mathcal Q}^{\hspace{0.03cm}a}$ as a result of the canonical transformations (\ref{eq:3t}) and (\ref{eq:3y}). We have already performed such an elimination in \cite{markov_2020}, considering the interaction of purely soft bosonic excitations among themselves. In this paper, a new element is the consideration of the interaction of soft bosonic excitations with a hard test particle. Therefore, here we will focus in more detail on this new aspect.\\
\indent To eliminate the third-order interaction Hamiltonian $H^{(3)}$, we substitute the expansions (\ref{eq:3t}) and (\ref{eq:3y}) into the free-field Hamiltonian $H^{(0)}$ given by expression (\ref{eq:2p}) and keep only the terms that have a degree of nonlinearity of two or three in the new variables $c^{\!\phantom{\ast}a}_{\hspace{0.02cm}{\bf k}}$ and ${\mathcal Q}^{\hspace{0.03cm}a}$. Then in the third-order Hamiltonian $H^{(3)}$, Eq.\,(\ref{eq:2s}), we perform the simple replacement of variables: $a^{a}_{\bf k}\rightarrow c^{\,a}_{\hspace{0.02cm}{\bf k}}$ and $Q^{\hspace{0.03cm}a} \rightarrow {\mathcal Q}^{\hspace{0.03cm}a}$. Adding
the expression thus obtained to the expression that follows from the free-field Hamiltonian $H^{(0)}$ and collecting similar terms, finally we obtain 
\begin{equation}
H^{(0)} + H^{(3)}
=  
\!\int\!d\hspace{0.02cm}{\bf k}\, 
(\omega^{\hspace{0.03cm}l}_{\hspace{0.03cm}{\bf k}} - {\mathbf v}\cdot {\mathbf k})\ \!
c^{\ast\hspace{0.03cm}a}_{\hspace{0.02cm}{\bf k}}\hspace{0.03cm} c^{\!\!\phantom{\ast} a}_{\hspace{0.02cm}{\bf k}}
\label{eq:4q}
\end{equation}
\[
+
\int\!d\hspace{0.02cm}{\bf k}\,
\Bigl\{\bigl[\hspace{0.03cm}(\omega^{\hspace{0.03cm}l}_{\hspace{0.03cm}{\bf k}} - {\mathbf v}\cdot {\mathbf k}){F}^{\,\ast}_{\hspace{0.03cm}\bf k} + 
{\upphi}^{\phantom{\ast}}_{\hspace{0.03cm}{\bf k}}\bigr]
\hspace{0.03cm}
c^{\,a}_{\hspace{0.02cm}{\bf k}}\hspace{0.03cm}
{\mathcal Q}^{\hspace{0.03cm}a}
\hspace{0.03cm}+\hspace{0.03cm}
\bigl[\hspace{0.03cm}(\omega^{\hspace{0.03cm}l}_{\hspace{0.03cm}{\bf k}} - {\mathbf v}\cdot {\mathbf k}){F}^{\,\phantom{\ast}}_{\hspace{0.03cm}\bf k} + 
{\upphi}^{\hspace{0.01cm}\ast}_{\hspace{0.03cm}{\bf k}}\bigr]
\hspace{0.03cm}
{c}^{\,\ast\, a}_{\hspace{0.02cm}{\bf k}}\hspace{0.03cm}
{\mathcal Q}^{\hspace{0.03cm}a}\hspace{0.03cm} 
\Bigr\}.
\]
Requiring that the expression in curly brackets on the right-hand side of (\ref{eq:4q}) be turned to zero, we obtain an explicit form of the coefficient function ${F}^{\,\phantom{\ast}}_{\hspace{0.03cm}{\bf k}}$ in the canonical transformation (\ref{eq:3t}) in terms of the vertex function ${\upphi}^{\phantom{\ast}}_{\hspace{0.03cm}{\bf k}}$:
\begin{equation}
F^{\,\phantom{\ast}}_{\hspace{0.03cm}\bf k} 
=
-\hspace{0.03cm}\frac{{\upphi}^{\hspace{0.01cm}\ast}_{\hspace{0.03cm}{\bf k}}}
{\omega^{\hspace{0.03cm}l}_{\hspace{0.03cm}{\bf k}} - {\mathbf v}\cdot {\mathbf k}}\,.
\label{eq:4w}
\end{equation}
The relation (\ref{eq:4w}) has a meaning due to the condition (\ref{eq:2a}). Making use of (\ref{eq:4w}), from (\ref{eq:3o}) we immediately find the explicit form of the coefficient function $M^{\,a\,a_{1}\hspace{0.03cm}a_{2}}_{\;{\bf k}}$ entering into the canonical transformation of color charge $Q^{\hspace{0.03cm}a}$, Eq.\,(\ref{eq:3y}):
\begin{equation}
M^{\,a\,a_{1}\hspace{0.03cm}a_{2}}_{\; {\bf k}} = i\hspace{0.02cm}f^{\,a\, a_{1}\hspace{0.03cm}a_{2}}\, \frac{{\upphi}_{\hspace{0.03cm}{\bf k}}}
{\omega^{\hspace{0.03cm}l}_{\hspace{0.03cm}{\bf k}} - {\mathbf v}\cdot {\mathbf k}}. 
\label{eq:4e}
\end{equation}
\indent Furthermore, the requirement to exclude third-order terms in the Hamiltonian $H^{(3)}$, Eq.\,(\ref{eq:2s}), containing the vertex functions ${\mathcal V}^{\,a\,a_{1}\hspace{0.03cm}a_{2}}_{\ {\bf k},\,{\bf k}_{1},\, {\bf k}_{2}}$ and ${\mathcal U}^{\,a\,a_{1}\hspace{0.03cm}a_{2}}_{\ {\bf k},\,{\bf k}_{1},\,{\bf k}_{2}}$ leads to the already known expressions \cite{markov_2020} for the coefficient functions $V^{\,(1,3)\,a\,a_{1}\hspace{0.03cm}a_{2}}_{\ {\bf k},\,{\bf k}_{1},\,
{\bf k}_{2}}$ in the canonical transformation (\ref{eq:3t}):
\begin{equation}
\begin{array}{ll}
&V^{\,(1)\,a\,a_{1}\hspace{0.03cm}a_{2}}_{\ {\bf k},\,{\bf k}_{1},\, 
{\bf k}_{2}} 
		=
		-\hspace{0.02cm}\displaystyle\frac{{\mathcal V}^{\,a\,a_{1}\hspace{0.03cm}a_{2}}_{\ {\bf k},\,{\bf k}_{1},\, 
		{\bf k}_{2}}}
		{\omega^{\hspace{0.03cm}l}_{\hspace{0.03cm}{\bf k}} - \omega^{\hspace{0.03cm}l}_{\hspace{0.03cm}{\bf k}_{1}} - \omega^{\hspace{0.03cm}l}_{\hspace{0.03cm}{\bf k}_{2}}}\,
		\delta({\bf k} - {\bf k}_{1} - {\bf k}_{2}), 
		\\[4ex]
&V^{\,(3)\,a\,a_{1}\hspace{0.03cm}a_{2}}_{\ {\bf k},\,{\bf k}_{1},\, 
{\bf k}_{2}}
= 
-\hspace{0.02cm}\displaystyle\frac{{\mathcal U}^{\hspace{0.03cm}*\,a\, a_{1}\hspace{0.03cm}a_{2}}_{\ {\bf k},\,{\bf k}_{1},\, 
{\bf k}_{2}}}{\omega^{\hspace{0.03cm}l}_{\hspace{0.03cm}{\bf k}} + \omega^{\hspace{0.03cm}l}_{\hspace{0.03cm}{\bf k}_{1}} + \omega^{\hspace{0.03cm}l}_{\hspace{0.03cm}{\bf k}_{2}}}\,
\delta({\bf k} + {\bf k}_{1} + {\bf k}_{2}).
\end{array}
\label{eq:4r}
\end{equation}
The coefficient $V^{\,(2)\, a\, a_{1}\, a_{2}}_{\ {\bf k},\, {\bf k}_{1},\, {\bf k}_{2}} $ is found from Eq.\,(\ref{eq:3i}). These expressions\footnote{\hspace{0.02cm}Strictly speaking, due to the eigenfrequency shift in the free Hamiltonian $H^{(0)}$, Eq.\,(\ref{eq:2p}), for example, instead of the first expression in (\ref{eq:4r}), we must write
\[
V^{\,(1)\,a\,a_{1}\hspace{0.03cm}a_{2}}_{\ {\bf k},\, {\bf k}_{1},\, {\bf k}_{2}} 
=
-\hspace{0.02cm}\displaystyle\frac{{\mathcal V}^{\,a\,a_{1}\hspace{0.03cm}a_{2}}_{\ {\bf k},\, {\bf k}_{1},\, 
{\bf k}_{2}}}
{\omega^{\hspace{0.03cm}l}_{\hspace{0.03cm}{\bf k}} - \omega^{\hspace{0.03cm}l}_{\hspace{0.03cm}{\bf k}_{1}} - \omega^{\hspace{0.03cm}l}_{\hspace{0.03cm}{\bf k}_{2}}
-
{\bf v}\cdot({\bf k} - {\bf k}_{1} - {\bf k}_{2})}\,
\delta({\bf k}-{\bf k}_{1}-{\bf k}_{2}),
\]
but it is clear that by virtue of the $\delta$-function we still return to the old expression (\ref{eq:4r}).} have a meaning due to the fact that the resonance conditions (\ref{eq:3qq}) for three-plasmon processes have no solutions.\\ 
%
%
\indent Thus, instead of the sum of the initial Hamiltonians $H^{(0)} + H^{(3)}$, Eq.\,(\ref{eq:4q}), we now obtain a new free-field Hamiltonian ${\mathcal H}^{(0)}$ for noninteracting plasmons in terms of the new normal variables $c^{\ast\hspace{0.03cm}a}_{\hspace{0.02cm}{\bf k}}$ and $c^{\!\!\phantom{\ast} a}_{\hspace{0.03cm}{\bf k}}$:
\begin{equation}
{\mathcal H}^{(0)} =  
\!\int\!d\hspace{0.02cm}{\bf k}\, (\omega^{\hspace{0.03cm}l}_{\hspace{0.03cm}{\bf k}} - {\mathbf v}\cdot {\mathbf k})\ \!
c^{\ast\hspace{0.03cm}a}_{\hspace{0.02cm}{\bf k}}\hspace{0.03cm} c^{\!\!\phantom{\ast} a}_{\hspace{0.03cm}{\bf k}}.
\label{eq:4t}
\end{equation}
Hereinafter,\,the\,Hamiltonians\,in\,the\,new\,variables\,will\,be designated by the calligraphic letter~${\mathcal H}$.\\ 
\indent Furthermore, we can move to the construction of an explicit form of effective fourth-order Hamilto\-nian ${\mathcal H}^{(4)}$, which describes the elastic scattering of plasmon off hard particle. For this purpose, we need to collect all contributions proportional to the product $c^{\ast\ \!\!a_{1}}_{\hspace{0.03cm}{\bf k}_{1}} c^{\hspace{0.03cm}a_{2}}_{\hspace{0.03cm}{\bf k}_{2}}\hspace{0.03cm} \mathcal{Q}^{\,a}$ from the free-field Hamiltonian $H^{(0)}$, Eq.\,(\ref{eq:2p}), and from the interaction Hamiltonians $H^{(3)}$ and $H^{(4)}$, Eqs.\,(\ref{eq:2s}) and  (\ref{eq:2d}), to be generated by the canonical transformations (\ref{eq:3t}) and (\ref{eq:3y}). Thereby we obtain the effective fourth-order Hamiltonian describing the elastic scattering process of plasmon off a hard color-charged particle:  
\begin{equation}
{\mathcal H}^{(4)}_{g\hspace{0.02cm}G\hspace{0.02cm}\rightarrow
\hspace{0.02cm} g\hspace{0.02cm}G} 
=
i\!\int\!d\hspace{0.02cm}{\bf k}_{1}\hspace{0.03cm} d\hspace{0.02cm}{\bf k}_{2}\hspace{0.04cm}
\mathscr{T}^{\hspace{0.03cm}(2)\hspace{0.03cm}a\,a_{1}\hspace{0.03cm}a_{2}}_{\; {\bf k}_{1},\, {\bf k}_{2}}\,
c^{\ast\ \!\!a_{1}}_{\hspace{0.02cm}{\bf k}_{1}} c^{\hspace{0.03cm}a_{2}}_{\hspace{0.02cm}{\bf k}_{2}}
\mathcal{Q}^{\,a},
\label{eq:4y}
\end{equation}
where the {\it complete effective amplitude} $\mathscr{T}^{\hspace{0.03cm}(2)\hspace{0.03cm}a\; a_{1}\hspace{0.03cm}a_{2}}_{\;{\bf k}_{1},\,{\bf k}_{2}}$ has the following structure:
\begin{equation}
i\hspace{0.02cm}
\mathscr{T}^{\hspace{0.03cm}(2)\hspace{0.03cm}a\,a_{1}\hspace{0.03cm}a_{2}}_{\; {\bf k}_{1},\, {\bf k}_{2}}
=
i\hspace{0.04cm}
T^{\hspace{0.03cm}(2)\hspace{0.03cm}a\,a_{1}\hspace{0.03cm}a_{2}}_{\; {\bf k}_{1},\, {\bf k}_{2}}
+\,
\Bigl[\hspace{0.03cm}\bigl(\omega^{\hspace{0.03cm}l}_{\hspace{0.03cm}{\bf k}_{2}} - {\mathbf v}\cdot {\mathbf k}_{2}\bigr)\,
\widetilde{V}^{\,\ast\hspace{0.03cm}(2)\,a_{2}\hspace{0.04cm}a_{1}\,a}_{\ {\bf k}_{2},\, {\bf k}_{1}}
+
\bigl(\omega^{\hspace{0.03cm}l}_{\hspace{0.03cm}{\bf k}_{1}} - {\mathbf v}\cdot {\mathbf k}_{1}\bigr)\,
\widetilde{V}^{\,(2)\,a_{1}\hspace{0.03cm}a_{2}\,a}_{\ {\bf k}_{1},\, {\bf k}_{2}}
\Bigr]
\label{eq:4u}
\end{equation}
\[
+\!
\int\!d\hspace{0.02cm}{\bf k}\,(\omega^{\hspace{0.03cm}l}_{\hspace{0.03cm}{\bf k}} - {\mathbf v}\cdot {\mathbf k})\,
\Bigl\{F^{\,\ast}_{\hspace{0.03cm}\bf k}\hspace{0.03cm} 
V^{\hspace{0.03cm}(2)\,a\,a_{1}\hspace{0.03cm}a_{2}}_{\ {\bf k},\, {\bf k}_{1},\, {\bf k}_{2}}
+
F^{\phantom{\ast}}_{\hspace{0.03cm}\bf k}\hspace{0.03cm} 
V^{\,\ast\hspace{0.03cm}(2)\, a\, a_{2}\, a_{1}}_{\ {\bf k},\, {\bf k}_{2},\, {\bf k}_{1}}\Bigr\}
\]
\[
+\!
\int\!d\hspace{0.02cm}{\bf k}\, 
\Bigl\{\upphi^{\phantom{\ast}}_{\hspace{0.03cm}\bf k}\hspace{0.03cm} 
V^{\hspace{0.03cm}(2)\,a\,a_{1}\hspace{0.03cm}a_{2}}_{\ {\bf k},\, {\bf k}_{1},\, {\bf k}_{2}}
+\hspace{0.03cm}
\upphi^{\hspace{0.03cm}\ast}_{\hspace{0.03cm}\bf k}\,
V^{\,\ast\hspace{0.03cm}(2)\, a\, a_{2}\, a_{1}}_{\ {\bf k},\, {\bf k}_{2},\, {\bf k}_{1}}\Bigr\}
\]
\[
+\,
\Bigl[\upphi^{\phantom{\ast}}_{\hspace{0.03cm}{\bf k}_{2}}\hspace{0.02cm}
M^{\,\ast\hspace{0.03cm}a_{2}\,a_{1}\,a}_{\; {\bf k}_{1}}
\,+\,
\upphi^{\hspace{0.03cm}\ast}_{\hspace{0.03cm}{\bf k}_{1}}\hspace{0.01cm}
M^{\hspace{0.03cm}a_{1}\hspace{0.03cm}a_{2}\,a}_{\; {\bf k}_{2}}
\Bigr]
\]
\[
+\,
2\!
\int\!d\hspace{0.02cm}{\bf k}\,
\Bigl\{\hspace{0.02cm}{\mathcal V}^{\;a_{1}\hspace{0.03cm}a_{2}\,a}_{\hspace{0.03cm}{\bf k}_{1},\, {\bf k}_{2},\, {\bf k}}\, F^{\phantom{\ast}}_{\hspace{0.03cm}\bf k}
\hspace{0.04cm}
\delta({\bf k}_{1} - {\bf k}_{2} - {\bf k})
\,+\,
{\mathcal V}^{\,*\,a_{2}\, a_{1}\, a}_{\hspace{0.03cm}{\bf k}_{2},\, {\bf k}_{1},\, {\bf k}}\, 
F^{\,\ast}_{\hspace{0.03cm}\bf k}\hspace{0.04cm}
\delta({\bf k}_{2} - {\bf k}_{1} - {\bf k})
\Bigr\}.
\]
Note that the complete effective amplitude $\mathscr{T}^{\hspace{0.03cm}(2)\hspace{0.03cm}a\,a_{1}\hspace{0.03cm}a_{2}}_{\; {\bf k}_{1},\, {\bf k}_{2}}$ must satisfy the following symmetry condition 
\begin{equation}
\mathscr{T}^{\hspace{0.03cm} (2)\hspace{0.03cm}a\,a_{1}\hspace{0.03cm}a_{2}}_{\; {\bf k}_{1},\, {\bf k}_{2}}
=
-\hspace{0.03cm}
\mathscr{T}^{\,\ast\hspace{0.03cm}(2)\,a\,a_{2}\,a_{1}}_{\; {\bf k}_{2},\, {\bf k}_{1}}.
\label{eq:4i}
\end{equation}
This relation is a consequence of the requirement of reality for the effective Hamiltonian (\ref{eq:4y}). We already faced this condition for the vertex function $T^{\,(2)\,a\,a_{1}\hspace{0.03cm}a_{2}}_{\,{\bf k},\, {\bf k}_{1}}$ in section \ref{section_2}.\\ 
\indent Let us analyze the structure $\mathscr{T}^{\hspace{0.03cm}(2)\hspace{0.03cm}a\,a_{1}\hspace{0.03cm}a_{2}}_{\; {\bf k}_{1},\, {\bf k}_{2}}$ in more detail. The first step is to use the relation (\ref{eq:3a}) to get rid of the coefficient function $\widetilde{V}^{\hspace{0.03cm}\ast\hspace{0.03cm}(2)\, a_{2}\, a_{1}\, a}_{\ {\bf k}_{2},\, {\bf k}_{1}}$ in the first line of (\ref{eq:4u}). Further, note that the expressions in the second and third lines are mutually reduced with the use of (\ref{eq:4w}). For the coefficient function $M^{\,a\,a_{1}\hspace{0.03cm}a_{2}}_{\; {\bf k}_{1}}$ in the next fourth line we make use of its connection with the coefficient function $F^{\hspace{0.03cm} \ast}_{\, {\bf k}_{1}}$, Eq.\,(\ref{eq:3o}). Finally, in the last contribution in (\ref{eq:4u}), with the vertex function ${\mathcal V}^{\;a_{1}\hspace{0.03cm}a_{2}\,a}_{\hspace{0.03cm}{\bf k}_{1},\, {\bf k}_{2},\, {\bf k}}$, we perform an integration with respect to ${\bf k}$, take into account color decomposition (\ref{eq:2h}) and pass from the coefficient function  $F^{\phantom{\ast}}_{\hspace{0.03cm}\bf k}$ to the ``physical'' vertex function $\upphi^{\phantom{\ast}}_{\hspace{0.03cm}{\bf k}}$ by the rule  (\ref{eq:4w}). As a result, instead of (\ref{eq:4u}), we finally obtain 
\begin{equation}
\mathscr{T}^{\hspace{0.03cm}(2)\hspace{0.03cm}a\,a_{1}\hspace{0.03cm} a_{2}}_{\; {\bf k}_{1},\,{\bf k}_{2}}
=
-\hspace{0.03cm}i\hspace{0.04cm}
\bigl[\hspace{0.03cm}\omega^{\hspace{0.03cm}l}_{\hspace{0.03cm}{\bf k}_{1}} - \omega^{\hspace{0.03cm}l}_{\hspace{0.03cm}{\bf k}_{2}}
-
{\mathbf v}\cdot ({\mathbf k}_{1} - {\mathbf k}_{2})\bigr]\,
\widetilde{V}^{\,(2)\,a_{1}\hspace{0.03cm}a_{2}\,a}_{\ {\bf k}_{1},\, {\bf k}_{2}}
+\,
\widetilde{T}^{\,(2)\,a_{1}\hspace{0.03cm}a_{2}\, a}_{\,{\bf k}_{1},\, 
{\bf k}_{2}}, 
\label{eq:4o}
\end{equation}
where the effective amplitude $\widetilde{T}^{\,(2)\,a_{1}\hspace{0.03cm}a_{2}\, a}_{\,{\bf k}_{1},\, 
{\bf k}_{2}}$ has the following structure:
\begin{equation}
\widetilde{T}^{\,(2)\,a_{1}\hspace{0.03cm}a_{2}\,a}_{\,{\bf k}_{1},\, 
{\bf k}_{2}}
=
T^{\hspace{0.03cm}(2)\hspace{0.03cm}a\,a_{1}\hspace{0.03cm}a_{2}}_{\; {\bf k}_{1},\, {\bf k}_{2}}
\label{eq:4p}
\end{equation}
\[
+\,
f^{\hspace{0.03cm}a_{1}\hspace{0.03cm}a_{2}\,a}\hspace{0.03cm}
\biggl\{\frac{{\upphi}^{\hspace{0.02cm}\ast}_{\,
{\bf k}_{1}}\hspace{0.03cm}{\upphi}^{\phantom{\ast}}_{\,{\bf k}_{2}}}
{\omega^{\hspace{0.02cm} l}_{\hspace{0.03cm}{\bf k}_{2}} - {\bf v}\cdot {\bf k}_{2}}
\,+\,
2\hspace{0.03cm}i\hspace{0.02cm}\biggl(\,
\frac{{\mathcal V}^{\phantom{\ast}}_{\,{\bf k}_{1},\, {\bf k}_{2},\, {\bf k}_{1} - {\bf k}_{2}} 
{\upphi}^{\hspace{0.02cm}{\ast}}_{\,{\bf k}_{1} - {\bf k}_{2}}}
{\omega^{\hspace{0.02cm} l}_{\hspace{0.03cm}{\bf k}_{1} - {\bf k}_{2}}\! - {\bf v}\cdot ({\bf k}_{1} - {\bf k}_{2})}
\,-\,
\frac{{\mathcal V}^{\,{\ast}}_{\,{\bf k}_{2},\, {\bf k}_{1},\, {\bf k}_{2} - {\bf k}_{1}} 
{\upphi}^{\phantom{\ast}}_{\,{\bf k}_{2} - {\bf k}_{1}}}
{\omega^{\hspace{0.02cm} l}_{\hspace{0.03cm}{\bf k}_{2} - {\bf k}_{1}}\! - {\bf v}\cdot ({\bf k}_{2} - {\bf k}_{1})}\biggr)\!\biggr\}.
\]
We can simplify this amplitude somewhat more by taking into account the conjugation property for the three-plasmon vertex:
\[
{\mathcal V}^{\hspace{0.03cm}\ast}_{\, {\bf k}_{2},\, {\bf k}_{1},\, {\bf k}_{2} - {\bf k}_{1}}
=
{\mathcal V}_{\, {\bf k}_{1},\, {\bf k}_{2},\, {\bf k}_{1}-{\bf k}_{2}},
\]  
but we won't do it. The first term on the right-hand side of (\ref{eq:4o}) has the resonance factor $\bigl[\hspace{0.03cm}\omega^{\hspace{0.03cm}l}_{\hspace{0.03cm}{\bf k}_{1}} - \omega^{\hspace{0.03cm}l}_{\hspace{0.03cm}{\bf k}_{2}}
- {\mathbf v}\cdot ({\mathbf k}_{1} - {\mathbf k}_{2})\bigr]$, which in fact represents a consequence of the momentum and energy conservation laws in the scattering process under investigation. If this resonance condition is approximately satisfied, then the contribution of this term to the effective Hamiltonian can be completely neglected. In section \ref{section_12} we will discuss in more detail the case when the ``resonance frequency difference'' 
\begin{equation}
\Delta\hspace{0.02cm}\omega_{\hspace{0.03cm}{\mathbf k}_{1},\hspace{0.03cm}{\mathbf k}_{2}} 
\equiv
\omega^{\hspace{0.02cm}l}_{\hspace{0.03cm}{\bf k}_{1}} - \omega^{\hspace{0.02cm}l}_{\hspace{0.03cm}{\mathbf k}_{2}}
- {\mathbf v}\cdot (\hspace{0.03cm}{\mathbf k}_{1} - {\mathbf k}_{2})
\label{eq:4a}
\end{equation}
can be arbitrary and not necessarily small.\\ 
\indent Figure\,\ref{fig2} gives the diagrammatic interpretation of different terms in curly brackets in the effective amplitude (\ref{eq:4p}).
The first two graphs represent the Compton scattering of soft boson excitations off a hard test particle induced by the first term in curly brackets of the expression (\ref{eq:4p}). The incoming and outgoing wave lines in fig.\,\ref{fig2} correspond
to the normal variables $c^{\hspace{0.03cm}a_{1}}_{\hspace{0.03cm}{\bf k}_{1}}$ and $c^{\ast\ \!\!a_{2}}_{\hspace{0.03cm}{\bf k}_{2}}$, respectively, and the horizontal double line between two interaction vertices corresponds to the ``propagator'' of the hard particle
\begin{equation}
\frac{\!\!1}{\omega^{\hspace{0.02cm}l}_{\hspace{0.03cm}{\bf k}_{1}} - {\bf v}\cdot {\bf k}_{1}}.
\label{eq:4aa}
\end{equation}
The interaction vertices correspond to the functions ${\upphi}^{\hspace{0.02cm}\ast}_{\,{\bf k}_{1}}$ or ${\upphi}^{\phantom{\ast}}_{\,{\bf k}_{2}}$. The remaining graph is connected with the interaction of hard particle with plasmons through the three-plasmon vertex function ${\mathcal V}^{\;a\,a_{1}\hspace{0.03cm}a_{2}}_{\;{\bf k},\, {\bf k}_{1},\, {\bf k}_{2}}$ with intermediate ``virtual'' oscillation to which the factor 
\[
\frac{1}{\omega^{\hspace{0.02cm} l}_{\hspace{0.03cm}{\bf k}_{1} - {\bf k}_{2}}\! - {\bf v}\cdot ({\bf k}_{1} - {\bf k}_{2})}
\]
in (\ref{eq:4p}) corresponds.
Note that this factor can also be written in a slightly different form 
\[
\frac{1}{\omega^{\hspace{0.02cm} l}_{\hspace{0.03cm}{\bf k}_{1} - {\bf k}_{2}}\! - \omega^{\hspace{0.03cm}l}_{\hspace{0.03cm}{\bf k}_{1}} + \omega^{\hspace{0.03cm}l}_{\hspace{0.03cm}{\bf k}_{2}}},
\]
if the resonance frequency difference (\ref{eq:4a}) is exactly zero. The last expression represents (up to a multiplier) an approximation of the effective gluon propagator $^{\ast}\widetilde{\cal D}_{\mu\hspace{0.02cm} \nu }(k)$, Eqs.\,(\ref{ap:A7})--(\ref{ap:A10}), at the plasmon
pole $\omega\sim\omega^{\hspace{0.03cm}l}_{\hspace{0.03cm}{\bf k}}$ in a spirit of the works by Weldon \cite{weldon_1998} and Blaizot and Iancu \cite{blaizot_1994(1)}. We considered such an approximation in \cite{markov_2020, markov_2023}.\\
\indent Finally, the first term $T^{\hspace{0.03cm}(2)\hspace{0.03cm}a\,a_{1}\hspace{0.03cm}a_{2}}_{\; {\bf k}_{1},\, {\bf k}_{2}}$ on the right-hand side of (\ref{eq:4p}) should be associated with the process of direct interaction of two plasmons with a hard particle, as depicted in fig.\,\ref{fig3}. 
In the particular physical system under consideration there is no a double contact vertex function that would describe this scattering process\footnote{\hspace{0.03cm}However, one may recall here the second order formalism for fermions \cite{brown_1958, morgan_1995, martinez_2012, acosta_2015}, within which  the interaction vertex of similar type arises. It is not clear whether such a vertex can be defined within our purely classical formalism, without fermions. Besides it is possible to mention a double photon vertex function in the theory of a complex (pseudo)scalar field $\phi$ interacting with the electromagnetic field $A_{\mu}$ \cite{peskin_1995, volkov_2003} (scalar QED). With a minimal generalization to color degrees of freedom (scalar QCD) perhaps this will give us a nontrivial representation for the $T^{\hspace{0.03cm}(2)\hspace{0.03cm}a\,a_{1}\hspace{0.03cm}a_{2}}_{\; {\bf k}_{1},\, {\bf k}_{2}}$-\,vertex.} and therefore we should just assume that
\[
T^{\hspace{0.03cm}(2)\hspace{0.03cm}a\,a_{1}\hspace{0.03cm}a_{2}}_{\; {\bf k}_{1},\, {\bf k}_{2}} \equiv 0.
\]
Taking into account the  last circumstance and
substituting the vertex function ${\upphi}^{\phantom{\ast}}_{\,{\bf k}}$, Eq.\,(\ref{eq:4s}), into (\ref{eq:4p}), we find the final form for the effective amplitude $\hspace{0.03cm} \widetilde{T}^{\,(2)\,a_{1}\hspace{0.03cm}a_{2}\,a}_{\,{\bf k}_{1},\,{\bf k}_{2}}$. The expression thus obtained coincides with the effective amplitude derived in the framework of high-temperature chromodynamics in \cite{markov_2004}.


\section{\bf Fourth-order correlation function}
\label{section_5}
\setcounter{equation}{0}

Now we turn to the construction of a kinetic equation describing the elastic scattering process of a plasmon off a hard color particle. As the free Hamiltonian here, we take the new free Hamiltonian ${\mathcal H}^{(0)} $, Eq.\,(\ref{eq:4t}), and as the interaction Hamiltonian we consider the effective Hamiltonian ${\mathcal H}^{(4)}_{g\hspace{0.02cm}G\hspace{0.02cm}\rightarrow\hspace{0.02cm} g\hspace{0.02cm}G}$, Eq.\,(\ref{eq:4y}). The equations of motion for the bosonic normal variables $c^{\phantom{\hspace{0.03cm}\ast} \!\!a^{\prime}}_{\hspace{0.02cm}{\bf k}^{\prime}}$ and $c^{\hspace{0.03cm}\ast\,a}_{\hspace{0.02cm}{\bf k}}$ and the color charge $\mathcal{Q}^{\,a}$ are defined by the corresponding Hamilton equations. For soft Bose-excitations we find
\begin{align}
&\frac{\partial \hspace{0.02cm}c^{\phantom{\hspace{0.03cm}\ast} \!\!a^{\prime}}_{\hspace{0.02cm}{\bf k}^{\hspace{0.02cm}\prime}}}{\partial\hspace{0.03cm} t}
	=
	-\hspace{0.03cm}i\hspace{0.05cm}\Bigl\{c^{\phantom{\hspace{0.03cm}\ast} \!\!a^{\prime}}_{\hspace{0.02cm}{\bf k}^{\hspace{0.02cm}\prime}}\hspace{0.03cm},\hspace{0.03cm} {\mathcal H}^{(0)\!} + {\mathcal H}^{(4)}_{g\hspace{0.02cm}G\hspace{0.02cm}\rightarrow\hspace{0.02cm} g\hspace{0.02cm}G}\Bigr\}
	=
	-\hspace{0.03cm}i\hspace{0.03cm}
	\bigl(\omega^{\hspace{0.02cm}l}_{\hspace{0.02cm}{\bf k}^{\hspace{0.02cm}\prime}}
	- {\bf v}\cdot{\bf k}^{\hspace{0.02cm}\prime}\bigr)
	\hspace{0.03cm}c^{\phantom{\hspace{0.03cm}\ast} \!\!a^{\prime}}_{\hspace{0.02cm}{\bf k}^{\hspace{0.02cm}\prime}}
	+\!
	\int\!d\hspace{0.03cm} {\bf k}_{1}\hspace{0.03cm}  
	\mathscr{T}^{\,a^{\prime}\hspace{0.03cm}a_{1}\hspace{0.03cm} d}_{\; {\bf k}^{\hspace{0.02cm}\prime},\, {\bf k}_{1}}\, 
	c^{\hspace{0.03cm}a_{1}}_{\hspace{0.02cm}{\bf k}_{1}}
	\hspace{0.03cm}\mathcal{Q}^{\,d}\hspace{0.03cm},
\label{eq:5q}\\[1ex]	
&\frac{\partial \hspace{0.02cm}c^{\hspace{0.03cm}\ast\,a}_{\hspace{0.02cm}{\bf k}}}{\partial\hspace{0.03cm} t}
	=
	-\hspace{0.03cm}i\hspace{0.05cm}\Bigl\{c^{\hspace{0.03cm}\ast\,a}_{\hspace{0.02cm}{\bf k}}\hspace{0.03cm},\hspace{0.03cm} {\mathcal H}^{(0)\!} + {\mathcal H}^{(4)}_{g\hspace{0.02cm}G\hspace{0.02cm}\rightarrow\hspace{0.02cm} g\hspace{0.02cm}G}\Bigr\}
	=
	i\hspace{0.03cm}
	\bigl(\omega^{\hspace{0.02cm}l}_{\hspace{0.02cm}{\bf k}}
	- {\bf v}\cdot{\bf k}\bigr)\hspace{0.03cm} 
	c^{\hspace{0.03cm}\ast\,a}_{\hspace{0.02cm}{\bf k}}
	-\!
	\int\!d\hspace{0.03cm} {\bf k}_{1}\hspace{0.03cm}  
	\mathscr{T}^{\,\ast\hspace{0.04cm}a\,a_{1}\hspace{0.03cm}d}_{\; {\bf k},\, {\bf k}_{1}}\, 
	c^{\hspace{0.03cm}\ast\,a_{1}}_{\hspace{0.02cm}{\bf k}_{1}}\hspace{0.03cm}\mathcal{Q}^{\,d}\hspace{0.03cm}, 
\label{eq:5w}	
\end{align}
and, respectively, for the classical color charge we get  
\begin{equation}
\frac{d \hspace{0.01cm}\mathcal{Q}^{\,d}}{d\hspace{0.03cm} t}
=
\Bigl\{\mathcal{Q}^{\,d}\hspace{0.03cm},\hspace{0.03cm} {\mathcal H}^{(0)\!} + {\mathcal H}^{(4)}_{g\hspace{0.02cm}G\hspace{0.02cm}\rightarrow\hspace{0.02cm} g\hspace{0.02cm}G}\Bigr\}
=
\frac{\partial\hspace{0.03cm} \bigl({\mathcal H}^{(0)\!} + 
{\mathcal H}^{(4)}_{g\hspace{0.02cm}G\hspace{0.02cm}\rightarrow\hspace{0.02cm} g\hspace{0.02cm}G}\bigr)}{\,\partial\hspace{0.02cm} {\mathcal Q}^{\hspace{0.03cm}d^{\hspace{0.02cm}\prime}}}
\,f^{\hspace{0.03cm}d\hspace{0.03cm}d^{\hspace{0.02cm}\prime}
\hspace{0.01cm}e}\hspace{0.03cm} {\mathcal Q}^{\hspace{0.03cm}e}
\label{eq:5e}
\end{equation}
\[
=
i\hspace{0.02cm}
f^{\hspace{0.03cm}d\hspace{0.03cm}d^{\hspace{0.02cm}\prime}\hspace{0.01cm}e}
\!\!\int\!d\hspace{0.03cm} {\bf k}_{1}\hspace{0.02cm}d\hspace{0.03cm} 
{\bf k}_{2}\, 
\mathscr{T}^{\,a_{1}\hspace{0.03cm}a_{2}\,d^{\hspace{0.02cm}\prime}}_{\; {\bf k}_{1},\, {\bf k}_{2}}\hspace{0.03cm} 
c^{\hspace{0.03cm}\ast\,a_{1}}_{\hspace{0.02cm}{\bf k}_{1}}c^{\hspace{0.03cm}a_{2}}_{\hspace{0.02cm}{\bf k}_{2}}\hspace{0.03cm}\mathcal{Q}^{\,e}\hspace{0.03cm}. 
\]
Here and hereafter, for the sake of simplicity, we have omitted the tag ``(2)'' in the designation of the complete effective amplitude $\mathscr{T}^{\hspace{0.03cm}(2)\hspace{0.03cm}a\;a_{1}\hspace{0.03cm}a_{2}}_{\; {\bf k}_{1},\, {\bf k}_{2}}$. If the ensemble of interacting Bose-excitations at a low nonlinearity level has random phases, then it can be statistically described by introducing the bosonic correlation function of the following form:
\begin{equation}
\bigl\langle\hspace{0.03cm}c^{\ast\ \!\!a}_{\hspace{0.02cm}{\bf k}}\hspace{0.03cm} c^{\phantom{\ast}\!\!a^{\prime}}_{\hspace{0.02cm}{\bf k}^{\hspace{0.02cm}\prime}}\hspace{0.03cm}\bigr\rangle
=
{\mathcal N}^{\;a\hspace{0.03cm}a^{\prime}_{\phantom{1}}\!}_{\hspace{0.02cm}{\bf k}}
\delta({\bf k} - {\bf k}^{\hspace{0.02cm}\prime}),
\label{eq:5r}
\end{equation}
where ${\mathcal N}^{\;a\hspace{0.03cm}a^{\prime}_{\phantom{1}}\!}_{\hspace{0.02cm}{\bf k}}$ is the plasmon number density.\\ 
\indent Let us define the kinetic equation for the  plasmon number density  employing the Hamilton equations (\ref{eq:5q}) and (\ref{eq:5w}), and the definition (\ref{eq:5r}). Using precisely the same reasoning as in \cite{markov_2020, markov_2023}, we get
\begin{equation}
\delta({\bf k} - {\bf k}\!\ ')\,
\frac{\partial\hspace{0.02cm}{\mathcal N}^{\;a\hspace{0.03cm}a^{\prime}_{\phantom{1}}\!}_{\hspace{0.02cm}{\bf k}}}{\partial\hspace{0.03cm}t}
=
\!\int\!d\hspace{0.02cm}{\bf k}_{1}\hspace{0.03cm} 
\Bigl\{\mathscr{T}^{\,\ast\,a\,a_{1}\hspace{0.03cm}d}_{\;{\bf k},\, {\bf k}_{1}}
\hspace{0.03cm}
\bigl\langle\hspace{0.03cm}c^{\ast\ \!\!a_{1}}_{\hspace{0.02cm}{\bf k}_{1}}\hspace{0.02cm} 
c^{\phantom{\ast}\!\!a^{\prime}}_{\hspace{0.02cm}{\bf k}^{\hspace{0.02cm}\prime}}
\hspace{0.03cm}\mathcal{Q}^{\,d}\hspace{0.03cm}\bigr\rangle
+
\mathscr{T}^{\,a^{\prime}\hspace{0.03cm}a_{1}\hspace{0.03cm}d}_{\, {\bf k}^{\hspace{0.02cm}\prime},\, {\bf k}_{1}}\hspace{0.03cm} 
\bigl\langle\hspace{0.03cm}c^{\ast\ \!\!a_{\phantom{1}}\!\!\!}_{\hspace{0.02cm}{\bf k}}\hspace{0.03cm} c^{\phantom{\ast}\!\!a_{1}}_{\hspace{0.02cm}{\bf k}_{1}}\hspace{0.03cm}\mathcal{Q}^{\,d}\hspace{0.03cm}\bigr\rangle
\Bigr\}.
\label{eq:5t}
\end{equation}
Further, by differentiating the fourth-order correlation function $\bigl\langle\hspace{0.03cm}c^{\ast\ \!\!a_{\phantom{1}}\!\!\!}_{\hspace{0.02cm}{\bf k}}\hspace{0.03cm} c^{\phantom{\ast}\!\!a_{1}}_{\hspace{0.02cm}{\bf k}_{1}}\hspace{0.03cm}\mathcal{Q}^{\,d}\hspace{0.03cm}\bigr\rangle$ with respect to $t$ with allowance made for (\ref{eq:5q})\,--\,(\ref{eq:5e}), we derive the equation the right-hand side of which will contain the fourth- and sixth-order correlation functions in the variables  $c^{\hspace{0.03cm}\ast\,a}_{\hspace{0.02cm}{\bf k}},\,c^{\phantom{\hspace{0.03cm}\ast} \!\!a}_{\hspace{0.02cm}{\bf k}}$ and $\mathcal{Q}^{\,d}$
\begin{equation}
\frac{\partial\hspace{0.03cm} \bigl\langle\hspace{0.03cm}c^{\ast\ \!\!a_{\phantom{1}}\!\!\!}_{\hspace{0.02cm}{\bf k}}\hspace{0.02cm} c^{\phantom{\ast}\!\!a_{1}}_{\hspace{0.02cm}{\bf k}_{1}}\hspace{0.02cm}\mathcal{Q}^{\,d}\hspace{0.03cm}\bigr\rangle}{\partial\hspace{0.03cm} t}
=
i\hspace{0.03cm}\bigl[\,\omega^{\hspace{0.02cm}l}_{\hspace{0.02cm}{\bf k}} - 
\omega^{\hspace{0.02cm}l}_{\hspace{0.02cm}{\bf k}_{1}} - {\bf v}\cdot({\bf k} - {\bf k}_{1})\hspace{0.03cm}\bigr]
\, \bigl\langle\hspace{0.03cm}c^{\ast\ \!\!a_{\phantom{1}}\!\!\!}_{\hspace{0.02cm}{\bf k}}\hspace{0.02cm} c^{\phantom{\ast}\!\!a_{1}}_{\hspace{0.02cm}{\bf k}_{1}}\hspace{0.02cm}\mathcal{Q}^{\,d}\hspace{0.03cm}\bigr\rangle
\label{eq:5y}
\end{equation}
\[
+
\int\!d\hspace{0.03cm}{\bf k}^{\hspace{0.02cm}\prime}_{1}\hspace{0.03cm}
\Bigl\{\mathscr{T}^{\,\ast\, a\, a^{\prime}_{1}\hspace{0.03cm} e}_{\, {\bf k},\, {\bf k}^{\hspace{0.02cm}\prime}_{1}}\, 
\bigl\langle\hspace{0.03cm}c^{\ast\ \!\!a^{\prime}_{1}}_{\hspace{0.02cm}{\bf k}^{\hspace{0.02cm}\prime}_{1}}\hspace{0.03cm} c^{\phantom{\ast}\!\!a_{1}}_{\hspace{0.02cm}
{\bf k}^{\phantom{\prime}}_{1}}\hspace{0.03cm}\mathcal{Q}^{\,d}
\hspace{0.03cm}\mathcal{Q}^{\,e}
\hspace{0.03cm}\bigr\rangle
\,+\,
\mathscr{T}^{\,a_{1}\hspace{0.03cm}a^{\prime}_{1}\hspace{0.03cm}e}_{\, {\bf k}_{1},\, {\bf k}^{\hspace{0.02cm}\prime}_{1}}\, 
\bigl\langle\hspace{0.03cm}c^{\ast\ \!\!a^{\phantom{\prime}}_{\phantom{1}}}_{\hspace{0.02cm}{\bf k}^{\phantom{\prime}}_{\phantom{1}}}\!\! c^{\phantom{\ast}\!\!a^{\prime}_{1}}_{\hspace{0.02cm}{\bf k}^{\hspace{0.02cm}\prime}_{1}}\hspace{0.03cm}\mathcal{Q}^{\,d}\hspace{0.03cm}
\mathcal{Q}^{\,e}\hspace{0.03cm}\bigr\rangle
\Bigr\}
\]
\[
+\; i\hspace{0.03cm}
f^{\hspace{0.03cm}d\hspace{0.03cm}d^{\hspace{0.02cm}\prime}\hspace{0.01cm}e}\! \!\int\!d\hspace{0.02cm}{\bf k}^{\hspace{0.02cm}\prime}_{1}\hspace{0.03cm} 
d\hspace{0.02cm}{\bf k}^{\hspace{0.02cm}\prime}_{2}\,
\mathscr{T}^{\, a^{\prime}_{1}\, a^{\prime}_{2}\, d^{\hspace{0.02cm}\prime}}_{\, {\bf k}^{\hspace{0.02cm}\prime}_{1},\, {\bf k}^{\hspace{0.02cm}\prime}_{2}}
\hspace{0.03cm} 
\bigl\langle
c^{\ast\ \!\!a^{\phantom{\prime}}_{\phantom{1}}}_{\hspace{0.02cm}{\bf k}^{\phantom{\prime}}_{\phantom{1}}}\!\!
c^{\ \!\!a^{\phantom{\prime}}_{1}}_{\hspace{0.02cm}{\bf k}^{\phantom{\prime}}_{1}}
c^{\ast\ \!\!a^{\prime}_{1}}_{\hspace{0.02cm}{\bf k}^{\hspace{0.02cm}\prime}_{1}}\, 
c^{\hspace{0.03cm}a^{\prime}_{2}}_{\hspace{0.02cm}{\bf k}^{\hspace{0.02cm}\prime}_{2}}
\,\mathcal{Q}^{\,e}\bigr\rangle.
\]
We close the chain of equations by expressing the sixth-order correlation functions in terms of the pair correlation functions for the normal variables $c^{\hspace{0.03cm}\ast\,a}_{\hspace{0.02cm}{\bf k}}$ and $c^{\phantom{\hspace{0.03cm}\ast} \!\!a}_{\hspace{0.02cm}{\bf k}}$, and the mean value of the color charge $\mathcal{Q}^{\,d}$. We keep only the terms that will give the proper contributions to the required kinetic equation:
\begin{equation}
\begin{split}
&\bigl\langle\hspace{0.03cm}
c^{\ast\ \!\!a^{\prime}_{1}}_{{\bf k}^{\hspace{0.02cm}\prime}_{1}}\hspace{0.02cm} c^{\phantom{\ast}\!\!a^{\phantom{\prime}}_{1}}_{{\bf k}^{\phantom{\prime}}_{1}}\hspace{0.02cm}
\mathcal{Q}^{\,d}\hspace{0.02cm}\mathcal{Q}^{\,e}
\hspace{0.03cm}\bigr\rangle
\,\simeq\,
\delta({\bf k}^{\hspace{0.02cm}\prime}_{1} - {\bf k}^{\phantom{\prime}}_{1})\,
{\mathcal N}^{\;a^{\prime}_{1}\hspace{0.03cm}a^{\phantom{\prime}}_{1}}_{\hspace{0.02cm}{\bf k}_{1}}
\hspace{0.01cm}\bigl\langle\hspace{0.03cm}\mathcal{Q}^{\hspace{0.03cm}d}
\hspace{0.03cm}\bigr\rangle
\hspace{0.03cm}\bigl\langle\hspace{0.03cm}\mathcal{Q}^{\hspace{0.03cm}e}
\hspace{0.03cm}\bigr\rangle,
\\[1.5ex]
&\bigl\langle\hspace{0.03cm}
c^{\ast\ \!\!a^{\phantom{\prime}}_{\phantom{1}}}_{\hspace{0.02cm}{\bf k}^{\phantom{\prime}}_{\phantom{1}}}\!\! c^{\phantom{\ast}\!\!a^{\prime}_{1}}_{{\bf k}^{\hspace{0.02cm}\prime}_{1}}\hspace{0.02cm}
\mathcal{Q}^{\,d}\hspace{0.02cm}\mathcal{Q}^{\,e}
\hspace{0.03cm}\bigr\rangle
\simeq\,
\delta({\bf k}^{\hspace{0.02cm}\prime}_{1} - {\bf k}^{\phantom{\prime}})\,
{\mathcal N}^{\;a^{\phantom{\prime}}\! a^{\prime}_{1}}_{\hspace{0.02cm}{\bf k}}
\hspace{0.01cm}\bigl\langle\hspace{0.03cm}\mathcal{Q}^{\hspace{0.03cm}d}
\hspace{0.03cm}\bigr\rangle
\hspace{0.03cm}\bigl\langle\hspace{0.03cm}\mathcal{Q}^{\hspace{0.03cm}e}
\hspace{0.03cm}\bigr\rangle,
\\[1.5ex]
&\bigl\langle
c^{\ast\ \!\!a^{\phantom{\prime}}_{\phantom{1}}}_{\hspace{0.02cm}{\bf k}^{\phantom{\prime}}_{\phantom{1}}}\!\!
c^{\phantom{\ast}\!\!a^{\phantom{\prime}}_{1}}_{{\bf k}^{\phantom{\prime}}_{1}}\hspace{0.02cm}
c^{\ast\ \!\!a^{\prime}_{1}}_{{\bf k}^{\hspace{0.02cm}\prime}_{1}}\hspace{0.02cm} 
c^{\hspace{0.03cm}a^{\prime}_{2}}_{{\bf k}^{\hspace{0.02cm}\prime}_{2}}
\hspace{0.02cm}\mathcal{Q}^{\,e}
\bigr\rangle
\simeq
\delta({\bf k}^{\hspace{0.02cm}\prime}_{2} - {\bf k})
\hspace{0.03cm}
\delta({\bf k}^{\hspace{0.02cm}\prime}_{1} - {\bf k}^{\phantom{\prime}}_{1})\,
{\mathcal N}^{\;a^{\phantom{\prime}}\!a^{\prime}_{2}}_{{\bf k}^{\phantom{\prime}}}
{\mathcal N}^{\;a^{\prime}_{1}\hspace{0.02cm}a^{\phantom{\prime}}_{1}}_{\hspace{0.02cm}{\bf k}_{1}}
\hspace{0.01cm}\bigl\langle\hspace{0.03cm}\mathcal{Q}^{\hspace{0.03cm}e}
\hspace{0.03cm}\bigr\rangle.
\end{split}
\label{eq:5u}
\end{equation}
We set for the complete effective amplitude $\mathscr{T}^{\hspace{0.03cm}a\,a_{1}\hspace{0.03cm}a_{2}}_{\; {\bf k}_{1},\, {\bf k}_{2}}$ the following color and momentum decomposition
\begin{equation}
\mathscr{T}^{\hspace{0.03cm}a\,a_{1}\hspace{0.03cm}a_{2}}_{\; {\bf k}_{1},\, {\bf k}_{2}} 
=
f^{\hspace{0.03cm}a\,a_{1}\hspace{0.03cm}a_{2}\hspace{0.03cm}} \mathscr{T}^{\hspace{0.03cm} \phantom{a}}_{\; {\bf k}_{1},\, {\bf k}_{2}}.
\label{eq:5i}
\end{equation}
It should be mentioned that the function $\mathscr{T}^{\hspace{0.03cm} \phantom{a}}_{\; {\bf k}_{1},\, {\bf k}_{2}}$, due to the requirement (\ref{eq:4i}), must meet the condition
\begin{equation}
\mathscr{T}^{\hspace{0.03cm} \phantom{a}}_{\; {\bf k}_{1},\, {\bf k}_{2}} 
=
\mathscr{T}^{\hspace{0.03cm}\ast \phantom{a}}_{\; {\bf k}_{2},\, {\bf k}_{1}}.
\label{eq:5o}
\end{equation} 
\indent Substituting the expressions (\ref{eq:5u}) and (\ref{eq:5i}) into the right-hand side of (\ref{eq:5y}) and considering the symmetry condition (\ref{eq:5o}) for the scattering amplitude, instead of (\ref{eq:5y}), we derive the equation for the fourth-order correlation function 
\begin{equation}
	\frac{\partial\hspace{0.03cm} \bigl\langle\hspace{0.03cm}c^{\ast\ \!\!a_{\phantom{1}}\!\!\!}_{\hspace{0.02cm}{\bf k}}\hspace{0.02cm} c^{\phantom{\ast}\!\!a_{1}}_{\hspace{0.02cm}{\bf k}_{1}}\hspace{0.02cm}\mathcal{Q}^{\,d}\hspace{0.03cm}\bigr\rangle}{\partial\hspace{0.03cm} t}
	\,=\,
	i\hspace{0.03cm}\bigl[\,\omega^{\hspace{0.02cm}l}_{\hspace{0.02cm}{\bf k}} - 
	\omega^{\hspace{0.02cm}l}_{\hspace{0.02cm}{\bf k}_{1}} - {\bf v}\cdot({\bf k} - {\bf k}_{1})\hspace{0.03cm}\bigr]
	\, \bigl\langle\hspace{0.03cm}c^{\ast\ \!\!a_{\phantom{1}}\!\!\!}_{\hspace{0.02cm}{\bf k}}\hspace{0.02cm} c^{\phantom{\ast}\!\!a_{1}}_{\hspace{0.02cm}{\bf k}_{1}}\hspace{0.02cm}\mathcal{Q}^{\,d}\hspace{0.03cm}\bigr\rangle	\label{eq:5p}
\end{equation}
\[
+\;i\hspace{0.03cm}
\mathscr{T}^{\hspace{0.03cm}\ast\,\phantom{a}}_{\; {\bf k},\, {\bf k}_{1}}
\bigl[
\bigl(\hspace{0.03cm}T^{\,e}{\mathcal N}_{\hspace{0.02cm}{\bf k}_{1}}\bigr)^{a\hspace{0.03cm}a_{1}}
-
\bigl({\mathcal N}_{\hspace{0.02cm}{\bf k}}\,T^{\,e}\bigr)^{a\hspace{0.03cm} a_{1}}
\bigr]\hspace{0.03cm} 
\hspace{0.01cm}\bigl\langle\hspace{0.03cm}\mathcal{Q}^{\hspace{0.03cm}d}
\hspace{0.03cm}\bigr\rangle
\hspace{0.03cm}\bigl\langle\hspace{0.03cm}\mathcal{Q}^{\hspace{0.03cm}e}
\hspace{0.03cm}\bigr\rangle
\vspace{0.2cm}
\,+\,
\mathscr{T}^{\hspace{0.03cm}\ast\,\phantom{a}}_{\; {\bf k},\, {\bf k}_{1}}\hspace{0.03cm}
\bigl(\hspace{0.03cm}{\mathcal N}_{\hspace{0.02cm}{\bf k}}\hspace{0.03cm}T^{\,d^{\hspace{0.03cm}\prime}\!}
{\mathcal N}_{\hspace{0.02cm}{\bf k}_{1}}\bigr)^{a\hspace{0.03cm}a_{1}}
\hspace{0.03cm}\bigl\langle\hspace{0.03cm}\mathcal{Q}^{\hspace{0.03cm}e}
\hspace{0.03cm}\bigr\rangle
f^{\hspace{0.03cm}d\hspace{0.03cm}d^{\hspace{0.02cm}\prime}\hspace{0.01cm}e}.
\]
Here, $(\hspace{0.01cm}T^{\,a})^{\hspace{0.01cm}b\hspace{0.03cm}c} \equiv -i\hspace{0.03cm}f^{\hspace{0.03cm}a\hspace{0.02cm}b\hspace{0.03cm}c}$ are generators in the adjoint representation and we have introduced a matrix notation for the plasmon number density ${\mathcal N}^{\phantom{a_{1}}\!\!\! }_{\hspace{0.02cm}{\bf k}} = \bigl({\mathcal N}^{\,a\hspace{0.03cm}a_{1}}_{{\bf k}^{\phantom{\prime}}}\bigr)$.


\section{\bf Kinetic equation for soft gluon excitations}
\label{section_6}
\setcounter{equation}{0}

The self-consistent system of equations (\ref{eq:5t}) and (\ref{eq:5p}), when it is supplemented by an equation for the averaged color charge (see section \ref{section_9}) determines, in principle, the time evolution of the plasmon number density ${\mathcal N}^{\;a\hspace{0.03cm}a^{\prime}_{\phantom{1}}\!}_{\hspace{0.02cm}{\bf k}}$. However, one more simplification is introduced: in 
Eq.\,(\ref{eq:5p}) we disregard the term with a time derivative compared to the term containing the resonance factor $\Delta\hspace{0.02cm}\omega_{\,{\bf k},\hspace{0.03cm}{\bf k}_{1}}$, Eq.\,(\ref{eq:4a}). Instead of equation (\ref{eq:5p}), we now have
\begin{equation}
\bigl\langle\hspace{0.03cm}c^{\ast\ \!\!a_{\phantom{1}}\!\!\!}_{{\bf k}}\hspace{0.03cm} c^{\phantom{\ast}\!\!a_{1}}_{\hspace{0.02cm}{\bf k}_{1}}\hspace{0.02cm}\mathcal{Q}^{\,d}\hspace{0.03cm}\bigr\rangle
\simeq
\delta({\bf k} - {\bf k}_{1})\,
{\mathcal N}^{\;a\hspace{0.03cm}a_{1}\!}_{\hspace{0.02cm}{\bf k}}
\bigl\langle\hspace{0.03cm}\mathcal{Q}^{\hspace{0.03cm}d}
\hspace{0.03cm}\bigr\rangle
\vspace{0.15cm}
\label{eq:6q}
\end{equation}
\[
-\; \frac{1}{\Delta\hspace{0.02cm}\omega_{\,{\bf k},\hspace{0.03cm}
{\bf k}_{1}}\! + i\hspace{0.03cm}0}\,
\mathscr{T}^{\hspace{0.03cm}\ast\,\phantom{a}}_{\; {\bf k},\, {\bf k}_{1}}
\bigl[
\bigl(T^{\,e}{\mathcal N}_{{\bf k}_{1}}\bigr)^{a\hspace{0.03cm}a_{1}}
-
\bigl({\mathcal N}_{{\bf k}}\,T^{\,e}\bigr)^{a\hspace{0.03cm}a_{1}}
\bigr]\hspace{0.03cm} 
\hspace{0.01cm}\bigl\langle\hspace{0.03cm}\mathcal{Q}^{\hspace{0.03cm}d}
\hspace{0.03cm}\bigr\rangle
\hspace{0.03cm}\bigl\langle\hspace{0.03cm}\mathcal{Q}^{\hspace{0.03cm}e}
\hspace{0.03cm}\bigr\rangle
\]
\[
-\,
\frac{i}{\Delta\hspace{0.02cm}\omega_{\,{\bf k},\hspace{0.03cm}
{\bf k}_{1}}\! + i\hspace{0.03cm}0}\,
\mathscr{T}^{\hspace{0.03cm}\ast\,\phantom{a}}_{\; {\bf k},\, {\bf k}_{1}}
\hspace{0.03cm}
\bigl({\mathcal N}_{{\bf k}}\,T^{\,d^{\hspace{0.03cm}\prime}\!}{\mathcal N}_{{\bf k}_{1}}\bigr)^{a\hspace{0.03cm}a_{1}}\hspace{0.03cm} 
\hspace{0.03cm}
\bigl\langle\hspace{0.03cm}\mathcal{Q}^{\hspace{0.03cm}e}
\hspace{0.03cm}\bigr\rangle
f^{\hspace{0.03cm}d\hspace{0.03cm}d^{\hspace{0.02cm}\prime}\hspace{0.01cm}e}.
\]
Here, the first term on the right-hand side, which corresponds to the absence of interaction between soft gluon excitations and a hard color particle is the solution to the homogeneous equation for the fourth-order correlation function
$\bigl\langle\hspace{0.03cm}c^{\ast\ \!\!a_{\phantom{1}}\!\!\!}_{{\bf k}}\hspace{0.03cm} c^{\phantom{\ast}\!\!a_{1}}_{{\bf k}_{1}}\hspace{0.03cm}\mathcal{Q}^{\,d}\hspace{0.03cm}\bigr\rangle$. The other terms determine the deviation of the three-point correlator from the free interaction case.\\ 
\indent We substitute the first term from (\ref{eq:6q}) into the right-hand side of Eq.\,(\ref{eq:5t}) for ${\mathcal N}^{\;a\hspace{0.03cm}a^{\prime}_{\phantom{1}}\!}_{\hspace{0.02cm}{\bf k}}$. As a result we obtain
\begin{equation}
-\hspace{0.03cm}i\,\Bigl\{
\mathscr{T}^{\,\phantom{a}}_{\; {\bf k}^{\prime},\,{\bf k}}\hspace{0.03cm}
\bigl(\hspace{0.03cm}{\mathcal N}_{\hspace{0.02cm}{\bf k}}\,T^{\,d}\hspace{0.03cm}\bigr)^{a\hspace{0.03cm}a^{\prime}}
-\,
\mathscr{T}^{\hspace{0.03cm}\ast\,\phantom{a}}_{\; {\bf k},\,{\bf k}^{\prime}}
\hspace{0.03cm}
\bigl(\hspace{0.03cm}T^{\,d}{\mathcal N}_{\hspace{0.02cm}{\bf k}^{\prime}}\bigr)^{a\hspace{0.03cm}a^{\prime}}
\Bigr\}
\hspace{0.03cm}\bigl\langle\hspace{0.03cm}\mathcal{Q}^{\hspace{0.03cm}d}
\hspace{0.03cm}\bigr\rangle.
\label{eq:6e}
\end{equation}
Further, we substitute the second and third terms from (\ref{eq:6q}) into the right-hand side of Eq.\,(\ref{eq:5t}). Simple algebraic transformations, in view of the symmetry condition (\ref{eq:5o}), lead us to
\begin{equation}
-\hspace{0.01cm} \!\int\!d\hspace{0.02cm}
{\bf k}_{1}\biggl(\mathscr{T}^{\ast}_{\,{\bf k},\,{\bf k}_{1}}\,
\mathscr{T}^{\phantom{\ast}}_{\,{\bf k}^{\hspace{0.02cm}\prime}\!,\,{\bf k}_{1}}\times
\label{eq:6r}
\end{equation}
\begin{align}
\times
\biggl\{&\frac{1}{\Delta\hspace{0.02cm}\omega_{\hspace{0.03cm}{\bf k},\hspace{0.03cm}{\bf k}_{1}}\! +\hspace{0.03cm} i\hspace{0.03cm}0}
\Bigl(
\bigl({\mathcal N}_{{\bf k}}\hspace{0.02cm} T^{\,d^{\hspace{0.03cm}\prime}}\!{\mathcal N}_{{\bf k}_{1}} T^{\,d}\hspace{0.03cm}\bigr)^{a\hspace{0.03cm}a^{\prime}}
\!f^{\hspace{0.03cm}d\,d^{\prime}\hspace{0.02cm}e\hspace{0.03cm}}
\bigl\langle\hspace{0.03cm}\mathcal{Q}^{\hspace{0.03cm}e}
\hspace{0.03cm}\bigr\rangle
+ i\hspace{0.04cm}
\Bigl[
\bigl({\mathcal N}_{{\bf k}}\hspace{0.02cm} T^{\,e}\hspace{0.03cm}T^{\,d}\hspace{0.03cm}\bigr)^{a\hspace{0.03cm}a^{\prime}}
\!-\,
\bigl(T^{\,e} {\mathcal N}_{{\bf k}_{1}} T^{\,d}\hspace{0.03cm}\bigr)^{a\hspace{0.03cm}a^{\prime}}
\Bigr]
\bigl\langle\hspace{0.03cm}\mathcal{Q}^{\hspace{0.03cm}d}
\hspace{0.03cm}\bigr\rangle
\bigl\langle\hspace{0.03cm}\mathcal{Q}^{\hspace{0.03cm}e}
\hspace{0.03cm}\bigr\rangle
\Bigr)
\notag\vspace{0.12cm}\\[1ex]
-\,
&\frac{1}{\Delta\hspace{0.02cm}\omega_{\hspace{0.03cm}{\bf k}^{\hspace{0.02cm}\prime},\hspace{0.03cm}{\bf k}_{1}} \!-\hspace{0.03cm} i\hspace{0.03cm}0}
\Bigl(
\bigl(T^{\,d^{\hspace{0.03cm}\prime}}\!{\mathcal N}_{{\bf k}_{1}} T^{\,d}{\mathcal N}_{{\bf k}^{\hspace{0.02cm}\prime}}\hspace{0.03cm}\bigr)^{a\hspace{0.03cm}a^{\prime}}
\!f^{\hspace{0.03cm}d\, d^{\hspace{0.02cm}\prime}\hspace{0.03cm}e\hspace{0.03cm}}
\bigl\langle\hspace{0.03cm}\mathcal{Q}^{\hspace{0.03cm}e}
\hspace{0.03cm}\bigr\rangle
+ i\hspace{0.04cm}
\Bigl[
\bigl(T^{\,d}\hspace{0.03cm}T^{\,e}{\mathcal N}_{{\bf k}^{\hspace{0.02cm}\prime}} \hspace{0.03cm}\bigr)^{a\hspace{0.03cm}a^{\prime}}
\!-
\bigl(T^{\,d} {\mathcal N}_{{\bf k}_{1}} T^{\,e}\hspace{0.03cm}\bigr)^{a\hspace{0.03cm}a^{\prime}}
\Bigr]
\bigl\langle\hspace{0.03cm}\mathcal{Q}^{\hspace{0.03cm}d}
\hspace{0.03cm}\bigr\rangle
\bigl\langle\hspace{0.03cm}\mathcal{Q}^{\hspace{0.03cm}e}
\hspace{0.03cm}\bigr\rangle
\Bigr)\!\biggr\}.
\notag
\end{align}
In the second term in braces, by virtue of the definition of (\ref{eq:4a}), we took into account that
\[
\frac{1}{\Delta\hspace{0.02cm}\omega_{\hspace{0.03cm}{\bf k}_{1},\hspace{0.03cm}{\bf k}^{\hspace{0.02cm}\prime}} +\hspace{0.03cm} 
i\hspace{0.03cm}0}
=
-\frac{1}{\Delta\hspace{0.02cm}\omega_{\hspace{0.03cm}{\bf k}^{\hspace{0.02cm}\prime} ,\hspace{0.03cm}{\bf k}_{1}} \!-\hspace{0.03cm} i\hspace{0.03cm}0}.
\]
Considering the expressions obtained (\ref{eq:5t}), (\ref{eq:6e}), and (\ref{eq:6r}) and setting ${\bf k} = {\bf k}^{\hspace{0.02cm}\prime}$, we arrive at the following kinetic equation for the plasmon number density ${\mathcal N}^{\;a\hspace{0.03cm}a^{\prime}_{\phantom{1}}\!}_{\hspace{0.02cm}
{\bf k}}$, instead of (\ref{eq:5t}):
\begin{equation}
\frac{\partial\hspace{0.03cm}{\mathcal N}^{\;a\hspace{0.03cm}a^{\prime}_{\phantom{1}}\!}_{\hspace{0.02cm}{\bf k}}}{\!\partial\hspace{0.03cm} t}
=
-\hspace{0.03cm}i\,\Bigl\{
\mathscr{T}^{\,\phantom{a}}_{\; {\bf k},\,{\bf k}}\hspace{0.03cm}
\bigl(\hspace{0.03cm}{\mathcal N}_{\hspace{0.02cm}{\bf k}}\,T^{\,e}\hspace{0.03cm}\bigr)^{a\hspace{0.03cm}a^{\prime}}
-\,
\mathscr{T}^{\hspace{0.03cm}\ast\,\phantom{a}}_{\; {\bf k},\,{\bf k}}
\hspace{0.03cm}
\bigl(\hspace{0.03cm}T^{\,e}{\mathcal N}_{\hspace{0.02cm}{\bf k}}\bigr)^{a\hspace{0.03cm}a^{\prime}}
\Bigr\}
\hspace{0.01cm}\bigl\langle\hspace{0.03cm}\mathcal{Q}^{\hspace{0.03cm}e}
\hspace{0.03cm}\bigr\rangle
-\!
\int\!d\hspace{0.02cm}{\bf k}_{1}\,
\bigl|\mathscr{T}_{{\bf k},\,{\bf k}_{1}}\bigr|^{\hspace{0.02cm}2}\,\times
\vspace{-0.25cm}
\label{eq:6t}
\end{equation}
\begin{align}
	\times\hspace{0.02cm}
	\biggl\{&\frac{1}{\Delta\hspace{0.02cm}\omega_{\hspace{0.04cm}{\bf k}, \hspace{0.03cm}{\bf k}_{1}}\! +\hspace{0.03cm} i\hspace{0.03cm}0}
	\Bigl(
	\bigl({\mathcal N}_{{\bf k}}\hspace{0.04cm} T^{\,d^{\hspace{0.03cm}\prime}}\!\hspace{0.01cm}{\mathcal N}_{{\bf k}_{1}} T^{\,d}\hspace{0.03cm}\bigr)^{a\hspace{0.03cm}a^{\prime}}
	\!f^{\hspace{0.03cm} d\, d^{\hspace{0.03cm}\prime}\hspace{0.02cm}e\hspace{0.03cm}}
	\bigl\langle\hspace{0.03cm}\mathcal{Q}^{\hspace{0.03cm}e}
	\hspace{0.03cm}\bigr\rangle
	+ i\hspace{0.04cm}
	\Bigl[
	\bigl({\mathcal N}_{{\bf k}}\hspace{0.04cm} T^{\,e}\hspace{0.03cm}T^{\,d}\hspace{0.03cm}\bigr)^{a\hspace{0.03cm}
	a^{\prime}}
	\!-\,
	\bigl(T^{\,e} {\mathcal N}_{{\bf k}_{1}} T^{\,d}\hspace{0.03cm}\bigr)^{a\hspace{0.03cm}a^{\prime}}
	\Bigr]
	\bigl\langle\hspace{0.03cm}\mathcal{Q}^{\hspace{0.03cm}d}
	\hspace{0.03cm}\bigr\rangle
	\bigl\langle\hspace{0.03cm}\mathcal{Q}^{\hspace{0.03cm}e}
	\hspace{0.03cm}\bigr\rangle
	\Bigr)
	\notag\vspace{0.12cm}\\[1ex]
	-\,
	&\frac{1}{\Delta\hspace{0.02cm}\omega_{\hspace{0.04cm}{\bf k},\hspace{0.03cm}{\bf k}_{1}} \!-\hspace{0.03cm} i\hspace{0.03cm}0}
	\Bigl(
	\bigl(T^{\,d^{\hspace{0.03cm}\prime}\!}{\mathcal N}_{{\bf k}_{1}} T^{\,d}{\mathcal N}_{{\bf k}}\hspace{0.03cm}\bigr)^{a\hspace{0.03cm}a^{\prime}}
	\!f^{\hspace{0.03cm} d\, d^{\hspace{0.03cm}\prime}\hspace{0.02cm}e\hspace{0.03cm}}
	\bigl\langle\hspace{0.03cm}\mathcal{Q}^{\hspace{0.03cm}e}
	\hspace{0.03cm}\bigr\rangle
	+ i\hspace{0.04cm}
	\Bigl[
	\bigl(T^{\,e}T^{\,d}{\mathcal N}_{{\bf k}} \hspace{0.03cm}\bigr)^{a\hspace{0.03cm}a^{\prime}}
	\!-
	\bigl(T^{\,e} {\mathcal N}_{{\bf k}_{1}} T^{\,d}\hspace{0.03cm}\bigr)^{a\hspace{0.03cm}a^{\prime}}
	\Bigr]
	\bigl\langle\hspace{0.03cm}\mathcal{Q}^{\hspace{0.03cm}d}
	\hspace{0.03cm}\bigr\rangle
	\bigl\langle\hspace{0.03cm}\mathcal{Q}^{\hspace{0.03cm}e}
	\hspace{0.03cm}\bigr\rangle
	\Bigr)\!\biggr\}.
	\notag
\end{align}
As opposed to our previous works \cite{markov_2020, markov_2023}, we literally do not gather the difference we need in the resulting expression
\begin{equation}
\frac{1}{\Delta\hspace{0.02cm}\omega_{\hspace{0.04cm}{\bf k},\hspace{0.03cm} {\bf k}_{1}}\! + i\hspace{0.03cm}0}
\,-\,
\frac{1}{\Delta\hspace{0.02cm}\omega_{\hspace{0.04cm}{\bf k},\hspace{0.03cm} {\bf k}_{1}}\! - i\hspace{0.03cm}0}\;
\big(\!\equiv
-\hspace{0.03cm}2\hspace{0.02cm}i\hspace{0.02cm}\pi\hspace{0.03cm}\delta(\Delta\hspace{0.03cm}\omega_{\hspace{0.04cm}{\bf k},\hspace{0.03cm}{\bf k}_{1}})\bigr),
\label{eq:6y}
\end{equation}
since here we have different arrangements of the plasmon number density matrices with different wave vectors: ${\mathcal N}_{{\bf k}}$ and ${\mathcal N}_{{\bf k}_{1}}$.


\section{First moment about color of the kinetic equation (\ref{eq:6t})}
\label{section_7}
\setcounter{equation}{0}

Let us consider the following color decomposition of the matrix function ${\mathcal N}^{\;a\hspace{0.03cm}a^{\prime}_{\phantom{1}}\!}_{\hspace{0.02cm}
{\bf k}}$:
\begin{equation}
{\mathcal N}^{\;a\hspace{0.03cm}a^{\prime}_{\phantom{1}}\!}_{\hspace{0.02cm}{\bf k}} 
= 
\delta^{\,a\hspace{0.02cm}a^{\prime}}\!\hspace{0.01cm} 
N^{\hspace{0.03cm}l}_{\bf k} + 
\bigl(T^{\,c}\bigr)^{a\hspace{0.02cm}a^{\prime}}\!\bigl\langle\hspace{0.01cm}\mathcal{Q}^{\hspace{0.03cm}c}\hspace{0.03cm}\bigr\rangle\hspace{0.03cm} W^{\hspace{0.03cm}l}_{\bf k}.
\label{eq:7q}
\end{equation}
In this section we define an equation for the colorless part of the plasmon number density, i.e. for $N^{\hspace{0.03cm}l}_{\bf k}$. For this purpose, we take the trace of the left- and right-hand sides of equation (\ref{eq:6t}) with respect to the color indices, i.e. we set $a = a^{\prime}$ and perform the summation over $a$. Using the color expansion (\ref{eq:7q}) and the formulae for the traces of the product of two and three color matrices (generators) in the adjoint representation in Appendix \ref{appendix_D}, Eqs.\,(\ref{ap:D2}) and (\ref{ap:D3}), we easily find for terms linear in the matrix functions ${\mathcal N}_{{\bf k}}$ and ${\mathcal N}_{{\bf k}_{1}}$ on the right-hand side of (\ref{eq:6t})
\begin{equation}
{\rm tr}\,{\mathcal N}_{\bf k} = (N^{\hspace{0.02cm}2}_{c} - 1)N^{\hspace{0.03cm}l}_{\bf k} 
\qquad
{\rm tr}\hspace{0.03cm}\bigl(T^{\,e} {\mathcal N}_{{\bf k}} \bigr) = N_{c}\hspace{0.04cm}\bigl\langle\hspace{0.01cm}\mathcal{Q}^{\hspace{0.03cm}e}\hspace{0.03cm}\bigr\rangle\hspace{0.03cm} W^{\hspace{0.03cm}l}_{\bf k},
\label{eq:7w}
\end{equation}
\[
{\rm tr}\bigl[
\bigl({\mathcal N}_{{\bf k}}\hspace{0.03cm}T^{\,e}\hspace{0.03cm} T^{\,d}\hspace{0.03cm}\bigr)
- 
\bigl(T^{\,e}{\mathcal N}_{{\bf k}_{1}} T^{\,d}\hspace{0.03cm}\bigr)\bigr]
=
\delta^{e\hspace{0.03cm}d}N_{c}\bigl(N^{\hspace{0.03cm}l}_{{\bf k}} - N^{\hspace{0.03cm}l}_{{\bf k}_{1}}\bigr)
-
\frac{1}{2}\,N_{c}\bigl(T^{\,c}\bigr)^{e\hspace{0.03cm}d}\bigl(\hspace{0.02cm}W^{\hspace{0.03cm}l}_{{\bf k}} + W^{\hspace{0.03cm}l}_{{\bf k}_{1}}\hspace{0.01cm}\bigr)
\bigl\langle\hspace{0.01cm}\mathcal{Q}^{\hspace{0.03cm}c}\hspace{0.03cm}\bigr\rangle,
\]
\[
{\rm tr}\bigl[
\bigl(T^{\,e}\hspace{0.03cm}T^{\,d} {\mathcal N}_{{\bf k}}\hspace{0.03cm}\bigr)
-
\bigl(T^{\,e} {\mathcal N}_{{\bf k}_{1}} T^{\,d}\hspace{0.03cm}\bigr) 
\bigr]
=
\delta^{e\hspace{0.03cm}d}N_{c}\bigl(N^{\hspace{0.03cm}l}_{{\bf k}} - N^{\hspace{0.03cm}l}_{{\bf k}_{1}}\bigr)
-
\frac{1}{2}\,N_{c}\bigl(T^{\,c}\bigr)^{e\hspace{0.03cm}d}\bigl(\hspace{0.02cm}W^{\hspace{0.03cm}l}_{{\bf k}} + W^{\hspace{0.03cm}l}_{{\bf k}_{1}}\hspace{0.01cm}\bigr)
\bigl\langle\hspace{0.01cm}\mathcal{Q}^{\hspace{0.03cm}c}\hspace{0.03cm}\bigr
\rangle.
\]
In contracting the last two expressions with $\bigl\langle\hspace{0.03cm}\mathcal{Q}^{\hspace{0.03cm}d}
\hspace{0.03cm}\bigr\rangle
\bigl\langle\hspace{0.03cm}\mathcal{Q}^{\hspace{0.03cm}e}
\hspace{0.03cm}\bigr\rangle$, the terms with a sum  $W^{\hspace{0.03cm}l}_{{\bf k}} + W^{\hspace{0.03cm}l}_{{\bf k}_{1}}$ on the right-hand sides turn to zero.\\
\indent The traces of terms quadratic in ${\mathcal N}_{{\bf k}}$ and ${\mathcal N}_{{\bf k}_{1}}$ in (\ref{eq:6t}) have a slightly more complicated structure. Thus, for example,
\begin{equation}
{\rm tr}\hspace{0.03cm}
\bigl(\hspace{0.03cm}{\mathcal N}_{{\bf k}}\hspace{0.04cm} T^{\,d^{\hspace{0.03cm}\prime}}\!\hspace{0.01cm}{\mathcal N}_{{\bf k}_{1}} T^{\,d}\hspace{0.03cm}\bigr)
=
N_{c}\hspace{0.04cm}\delta^{\hspace{0.03cm}d^{\hspace{0.03cm}\prime}d}
\hspace{0.02cm}N_{{\bf k}}\hspace{0.03cm}N_{{\bf k}_{1}}
\label{eq:7e}
\end{equation}
\[
+\,
\frac{i}{2}\,N_{c}\hspace{0.03cm}
f^{\hspace{0.03cm}d^{\hspace{0.03cm}\prime}d\,c\hspace{0.03cm}}
\bigl\langle\hspace{0.01cm}\mathcal{Q}^{\hspace{0.03cm}c}\hspace{0.03cm}\bigr
\rangle\hspace{0.03cm}W_{{\bf k}}\hspace{0.03cm}N_{{\bf k}_{1}}
-
\frac{i}{2}\,N_{c}\hspace{0.03cm}f^{\hspace{0.03cm} d^{\hspace{0.03cm}\prime}d\,c\hspace{0.03cm}}
\bigl\langle\hspace{0.01cm}\mathcal{Q}^{\hspace{0.03cm}c}\hspace{0.03cm}\bigr
\rangle\hspace{0.01cm}W_{{\bf k}_{1}}\hspace{0.03cm}N_{{\bf k}}
+
{\rm tr}\hspace{0.03cm}
\bigl(T^{\,c}\hspace{0.03cm} T^{\,d^{\hspace{0.03cm}\prime}}T^{\,f} T^{\,d}\hspace{0.03cm}\bigr)
\bigl\langle\hspace{0.01cm}\mathcal{Q}^{\hspace{0.03cm}c}\hspace{0.03cm}\bigr
\rangle
\bigl\langle\hspace{0.01cm}\mathcal{Q}^{\hspace{0.03cm}f}\hspace{0.03cm}\bigr
\rangle
\hspace{0.03cm}W_{{\bf k}}\hspace{0.03cm}W_{{\bf k}_{1}}.
\]
If we further contract this expression with the antisymmetric structure constants $f^{\hspace{0.03cm} d\, d^{\hspace{0.03cm}\prime}e}$, as it takes place in (\ref{eq:6t}), then the first term on the right-hand side of (\ref{eq:7e}) turns to zero. The last term also vanishes due to the symmetry property for the trace of the product of four generators (\ref{ap:D8}). As a result, we have
\[
f^{\hspace{0.03cm} d\, d^{\hspace{0.03cm}\prime}\hspace{0.01cm}e}\,
{\rm tr}\hspace{0.03cm}
\bigl({\mathcal N}_{{\bf k}}\hspace{0.04cm} T^{\,d^{\hspace{0.03cm}\prime}}\!\hspace{0.01cm}{\mathcal N}_{{\bf k}_{1}} T^{\,d}\hspace{0.03cm}\bigr)
=
-\hspace{0.03cm}\frac{i}{2}\,N^{\hspace{0.02cm}2}_{c}\hspace{0.03cm}
\bigl\langle\hspace{0.01cm}\mathcal{Q}^{\hspace{0.03cm}e}\hspace{0.03cm}
\bigr\rangle
\bigl(W^{\hspace{0.03cm}l}_{\bf k}\hspace{0.02cm}N^{\hspace{0.03cm}l}_{{\bf k}_{1}} 
-
N^{\hspace{0.03cm}l}_{\bf k}\hspace{0.03cm}W^{\hspace{0.03cm}l}_{{\bf k}_{1}}\bigr).
\] 
With the expressions obtained for the color traces, we can now write out the first moment about color for equation (\ref{eq:6t})
\begin{equation}
d_{A}\hspace{0.04cm}\frac{\partial\hspace{0.01cm} N^{\hspace{0.03cm}l}_{\bf k}}{\!\!\partial\hspace{0.03cm} t}
=
2\hspace{0.02cm}N_{c}\,{\mathfrak q}_{2}(t)\hspace{0.01cm} \bigl({\rm Im}\hspace{0.03cm}\mathscr{T}_{{\bf k},\hspace{0.03cm}{\bf k}}\bigr)\hspace{0.01cm}
W^{\hspace{0.03cm}l}_{\bf k}
\label{eq:7r}
\end{equation}
\[
-\hspace{0.03cm}
N_{c}\hspace{0.03cm}{\mathfrak q}_{2}(t)\!\!
\int\!d\hspace{0.02cm}{\bf k}_{1}\hspace{0.04cm}
\bigl|\mathscr{T}_{{\bf k},\hspace{0.03cm}{\bf k}_{1}}\bigr|^{\hspace{0.02cm}2}
\hspace{0.03cm}
\Bigl\{\bigl(N^{\hspace{0.03cm}l}_{{\bf k}} - N^{\hspace{0.03cm}l}_{{\bf k}_{1}}\bigr)
-
\frac{1}{2}\,N_{c}\hspace{0.03cm}\bigl(W^{\hspace{0.03cm}l}_{\bf k}\hspace{0.02cm}N^{\hspace{0.03cm}l}_{{\bf k}_{1}} 
-
N^{\hspace{0.03cm}l}_{\bf k}\hspace{0.03cm}W^{\hspace{0.03cm}l}_{{\bf k}_{1}}\bigr)\!\Bigr\}\hspace{0.03cm}
(2\pi)\,\delta(\omega^{\hspace{0.02cm}l}_{\hspace{0.02cm}{\bf k}} - \omega^{\hspace{0.02cm}l}_{\hspace{0.02cm}{\bf k}_{1}}\! - 
{\mathbf v}\cdot({\bf k} - {\bf k}_{1})).
\]
In deriving this expression, we have considered Sokhotsky's formula (\ref{eq:6y}) and introduced the notation for a colorless quadratic combination of the averaged color charge 
\begin{equation}
{\mathfrak q}_{2}(t) \equiv \bigl\langle\hspace{0.01cm}\mathcal{Q}^{\hspace{0.03cm}e}\hspace{0.03cm}
\bigr\rangle
\bigl\langle\hspace{0.01cm}\mathcal{Q}^{\hspace{0.03cm}e}\hspace{0.03cm}
\bigr\rangle.
\label{eq:7t}
\end{equation}
The coefficient $d_{A}\equiv N^{\hspace{0.02cm}2}_{c} - 1$ on the left-hand side of (\ref{eq:7r}) is an invariant for the group $SU(N_c)$. 


\section{Second moment about color of the kinetic equation (\ref{eq:6t})}
\label{section_8}
\setcounter{equation}{0}

Let us return to the matrix kinetic equation (\ref{eq:6t}). We now contract the left- and right-hand sides of this equation with the color matrix $(\hspace{0.02cm}T^{\,s})^{\hspace{0.02cm}a^{\prime}\hspace{0.01cm}a}$. As a result we get
\begin{equation}
\frac{\partial\,{\rm tr}\hspace{0.03cm}\bigl(T^{\,s} {\mathcal N}_{{\bf k}} \bigr)}{\!\!\partial\hspace{0.03cm} t}
=
-i\,\Bigl\{
{\rm tr}\hspace{0.03cm}\bigl(T^{\,e}\hspace{0.02cm}T^{\,s} {\mathcal N}_{{\bf k}}\hspace{0.02cm}\bigr)\hspace{0.03cm}
\mathscr{T}_{{\bf k},\hspace{0.03cm}{\bf k}}
-\,
{\rm tr}\hspace{0.03cm}\bigl(T^{\,s}T^{\,e} {\mathcal N}_{{\bf k}} \bigr)\hspace{0.03cm}
\mathscr{T}^{\,\ast}_{{\bf k},\hspace{0.03cm}{\bf k}}\Bigr\}\hspace{0.03cm}
\bigl\langle\hspace{0.01cm}\mathcal{Q}^{\hspace{0.03cm}e}\hspace{0.03cm}\bigr
\rangle 
\label{eq:8q}
\end{equation}
\[
-\!\int\!d\hspace{0.02cm}{\bf k}_{1}\hspace{0.04cm}
\bigl|\mathscr{T}_{{\bf k},\hspace{0.03cm} 
{\bf k}_{1}}\bigr|^{\hspace{0.02cm}2}\hspace{0.03cm}
\bigl\langle\hspace{0.03cm}\mathcal{Q}^{\hspace{0.03cm}e}
\hspace{0.03cm}\bigr\rangle
\,\times
\vspace{-0.15cm}
\]
\begin{align}
	\times\hspace{0.02cm}
	\biggl\{&\frac{1}{\Delta\hspace{0.02cm}\omega_{\hspace{0.03cm}{\bf k}, \hspace{0.03cm}{\bf k}_{1}}\! +\hspace{0.03cm} i\hspace{0.03cm}0}
	\Bigl({\rm tr}\hspace{0.03cm}
	\bigl({\mathcal N}_{{\bf k}}\hspace{0.03cm} T^{\,d^{\hspace{0.03cm}\prime}}\!{\mathcal N}_{{\bf k}_{1}} T^{\,d}\hspace{0.02cm}T^{\,s}\hspace{0.03cm}\bigr)
	f^{\hspace{0.03cm} d\, d^{\prime}\hspace{0.01cm}e\hspace{0.03cm}}
	+ i\hspace{0.04cm}
	\Bigl[{\rm tr}\hspace{0.03cm}
	\bigl(T^{\,s}{\mathcal N}_{{\bf k}}\hspace{0.02cm} T^{\,e}\hspace{0.03cm}T^{\,d}\hspace{0.03cm}\bigr)
	\!-\, {\rm tr}\hspace{0.03cm}
	\bigl(T^{\,s}\hspace{0.03cm}T^{\,e} {\mathcal N}_{{\bf k}_{1}} T^{\,d}\hspace{0.03cm}\bigr)
	\Bigr]
	\bigl\langle\hspace{0.03cm}\mathcal{Q}^{\hspace{0.03cm}d}
	\hspace{0.03cm}\bigr\rangle
	\notag\vspace{0.12cm}\\[1ex]
	-\,
	&\frac{1}{\Delta\hspace{0.02cm}\omega_{\hspace{0.03cm}{\bf k},\hspace{0.03cm}{\bf k}_{1}} \!-\hspace{0.03cm} i\hspace{0.03cm}0}
	\Bigl(
	{\rm tr}\hspace{0.03cm}
	\bigl({\mathcal N}_{{\bf k}_{1}}\hspace{0.01cm}T^{\,d}{\mathcal N}_{{\bf k}}\hspace{0.02cm}T^{\,s}\hspace{0.03cm} T^{\,d^{\hspace{0.03cm}\prime}}\hspace{0.03cm}\bigr)
	f^{\hspace{0.03cm} d\, d^{\prime}\hspace{0.01cm}e\hspace{0.03cm}}
	+ i\hspace{0.04cm}
	\Bigl[{\rm tr}\hspace{0.03cm}
	\bigl(T^{\,s}\hspace{0.03cm}T^{\,e}\hspace{0.03cm}T^{\,d}{\mathcal N}_{{\bf k}} \hspace{0.03cm}\bigr)
	\!- {\rm tr}\hspace{0.03cm}
	\bigl(T^{\,s}\hspace{0.03cm}T^{\,e} {\mathcal N}_{{\bf k}_{1}} T^{\,d}\hspace{0.03cm}\bigr)
	\Bigr]
	\bigl\langle\hspace{0.03cm}\mathcal{Q}^{\hspace{0.03cm}d}
	\hspace{0.03cm}\bigr\rangle
	\Bigr)\!\biggr\}.
	\notag
\end{align}
Here on the left-hand side, by virtue of the decomposition (\ref{eq:7q}) we have again
\begin{equation}
	{\rm tr}\hspace{0.03cm}\bigl(T^{\,s} {\mathcal N}_{{\bf k}} \bigr) = N_{c}\hspace{0.03cm}\bigl\langle\hspace{0.01cm}
	\mathcal{Q}^{\hspace{0.03cm}s}\hspace{0.03cm}\bigr\rangle\hspace{0.03cm} W^{\hspace{0.03cm}l}_{\bf k}.
	\label{eq:8w}
\end{equation}
Let us consider the traces in the first term on the right-hand side of (\ref{eq:8q}). By taking into account the representation (\ref{eq:7q}), a simple calculation leads us to
\[
{\rm tr}\bigl(T^{\,e}\hspace{0.03cm}T^{\,s} {\mathcal N}_{{\bf k}}\hspace{0.03cm}\bigr) 
=
\delta^{\hspace{0.02cm}e\hspace{0.02cm}s}N_{c}\hspace{0.03cm}N^{\hspace{0.03cm}l}_{\bf k}
+
\frac{i}{2}\,N_{c}\hspace{0.03cm}f^{\,e\hspace{0.03cm} s\hspace{0.03cm}c}\hspace{0.03cm}\bigl\langle\hspace{0.01cm}\mathcal{Q}^{\hspace{0.03cm}c}\hspace{0.03cm}\bigr\rangle\hspace{0.03cm} W^{\hspace{0.03cm}l}_{\bf k},
\quad
{\rm tr}\bigl(T^{\,s}\hspace{0.03cm}T^{\,e} {\mathcal N}_{{\bf k}} \hspace{0.03cm}\bigr) 
=
\delta^{\hspace{0.02cm}e\hspace{0.02cm}s}N_{c}\hspace{0.03cm}N^{\hspace{0.03cm}l}_{\bf k}
-
\frac{i}{2}\,N_{c}\hspace{0.03cm}f^{\,e\hspace{0.03cm} s\hspace{0.03cm}c}\hspace{0.03cm}\bigl\langle\hspace{0.01cm}\mathcal{Q}^{\hspace{0.03cm}c}\hspace{0.03cm}\bigr\rangle\hspace{0.03cm} W^{\hspace{0.03cm}l}_{\bf k}.
\]
The imaginary part in these two expressions turns to zero by contracting with the color charge 
$\bigl\langle\hspace{0.01cm}\mathcal{Q}^{\hspace{0.03cm}e}\hspace{0.03cm}
\bigr\rangle $ and finally the first term on the right-hand side (\ref{eq:8q}) takes the following form:
\begin{equation}
-\hspace{0.03cm}i\,\Bigl\{
{\rm tr}\hspace{0.03cm}\bigl(T^{\,e}\hspace{0.02cm}T^{\,s}{\mathcal N}_{{\bf k}}\hspace{0.02cm}\bigr)\hspace{0.03cm}
\mathscr{T}_{\hspace{0.03cm}{\bf k},\hspace{0.03cm}{\bf k}}
-\,
{\rm tr}\hspace{0.03cm}\bigl(T^{\,s}\hspace{0.02cm}T^{\,e} {\mathcal N}_{{\bf k}} \bigr)\hspace{0.03cm}
\mathscr{T}^{\,\ast}_{\hspace{0.03cm}{\bf k},\hspace{0.03cm}{\bf k}}\Bigr\}\hspace{0.03cm}
\bigl\langle\hspace{0.01cm}\mathcal{Q}^{\hspace{0.03cm}e}\hspace{0.03cm}\bigr
\rangle 
=
2\hspace{0.02cm}N_{c}\hspace{0.03cm}
\bigl({\rm Im}\hspace{0.03cm}\mathscr{T}_{\hspace{0.03cm}{\bf k},\hspace{0.03cm}{\bf k}}\bigr)
\hspace{0.01cm}N^{\hspace{0.03cm}l}_{\bf k}\,
\bigl\langle\hspace{0.01cm}\mathcal{Q}^{\hspace{0.03cm}s}\hspace{0.03cm}\bigr\rangle.
\label{eq:8e}
\end{equation}
\indent We now examine the traces of terms quadratic in ${\mathcal N}_{{\bf k}}$ and ${\mathcal N}_{{\bf k}_{1}}$ in (\ref{eq:8q}). Allowing for the decomposition (\ref{eq:7q}), we easily find for the first trace there 
\begin{equation}
{\rm tr}\hspace{0.03cm}
\bigl(\hspace{0.01cm} {\mathcal N}_{{\bf k}}\hspace{0.03cm} T^{\,d^{\hspace{0.03cm}\prime}}\!{\mathcal N}_{{\bf k}_{1}} T^{\,d}\hspace{0.035cm}T^{\,s}\hspace{0.01cm}\bigr)
=
{\rm tr}\hspace{0.03cm}\bigl(T^{\,d^{\hspace{0.03cm}\prime}} T^{\,d}\hspace{0.03cm}T^{\,s}\hspace{0.03cm}\bigr)
N_{{\bf k}}N_{{\bf k}_{1}}
+
{\rm tr}\hspace{0.03cm}\bigl(T^{\,c}\hspace{0.03cm} T^{\,d^{\hspace{0.03cm}\prime}}T^{\,d}\hspace{0.03 cm} T^{\,s}\hspace{0.03cm}\bigr)
\bigl\langle\hspace{0.01cm}\mathcal{Q}^{\hspace{0.03cm}c}\hspace{0.03cm}\bigr\rangle
W_{{\bf k}}N_{{\bf k}_{1}}
\label{eq:8r}
\end{equation}
\[
+\,
{\rm tr}\hspace{0.03cm}\bigl(T^{\,d^{\hspace{0.03cm}\prime}}\hspace{0.02cm} T^{\,c}\hspace{0.03cm}T^{\,d}\hspace{0.02cm}T^{\,s}\hspace{0.03cm}\bigr)
\bigl\langle\hspace{0.01cm}\mathcal{Q}^{\hspace{0.03cm}c}\hspace{0.03cm}\bigr\rangle
N_{{\bf k}}W_{{\bf k}_{1}}
+
{\rm tr}\hspace{0.03cm}\bigl(T^{\,d}\hspace{0.03cm} T^{\,s}\hspace{0.03cm}T^{\,c}\hspace{0.03cm}T^{\,d^{\hspace{0.03cm}\prime}} T^{\,f}\hspace{0.02cm} 
\bigr)
\bigl\langle\hspace{0.01cm}\mathcal{Q}^{\hspace{0.03cm}c}\hspace{0.03cm}\bigr\rangle
\bigl\langle\hspace{0.01cm}\mathcal{Q}^{\hspace{0.03cm}f}\hspace{0.03cm}\bigr\rangle
W_{{\bf k}}W_{{\bf k}_{1}}.
\]
Let us contract this expression with $f^{\hspace{0.03cm} d\, d^{\hspace{0.03cm}\prime}e\hspace{0.02cm}}\bigl\langle\hspace{0.01cm}\mathcal{Q}^{\hspace{0.03cm}e}\hspace{0.03cm}\bigr\rangle$, as it takes place in (\ref{eq:8q}). For the first term on the right-hand side of (\ref{eq:8r}), by virtue of formula (\ref{ap:D3}), we have 
\[
{\rm tr}\hspace{0.03cm}\bigl(T^{\,d^{\hspace{0.03cm}\prime}} T^{\,d}\hspace{0.03cm}T^{\,s}\hspace{0.03cm}\bigr)
N_{{\bf k}}N_{{\bf k}_{1}}
f^{\hspace{0.03cm} d\hspace{0.03cm} d^{\hspace{0.03cm}\prime}\hspace{0.01cm}e\hspace{0.03cm}}
\bigl\langle\hspace{0.01cm}\mathcal{Q}^{\hspace{0.03cm}e}\hspace{0.03cm}\bigr\rangle
=
-\hspace{0.03cm}\frac{i}{2}\,N^{\hspace{0.03cm}2}_{c}\hspace{0.03cm}
\bigl\langle\hspace{0.01cm}
\mathcal{Q}^{\hspace{0.03cm}s}\hspace{0.03cm}\bigr\rangle
N_{{\bf k}}N_{{\bf k}_{1}}.
\]
The third term in (\ref{eq:8r}), containing the trace of the product of four generators turns to zero due to the symmetry property (\ref{ap:D8}). The trace in the second term in (\ref{eq:8r}) is easily computed as follows
\[
{\rm tr}\hspace{0.03cm}\bigl(T^{\,c}\hspace{0.03cm} T^{\,d^{\hspace{0.03cm}\prime}}T^{\,d}\hspace{0.02cm} T^{\,s}\hspace{0.03cm}\bigr)
f^{\hspace{0.03cm} d\, d^{\hspace{0.03cm}\prime}\hspace{0.01cm}e\hspace{0.03cm}}
\]
\[
\equiv
\frac{1}{2}\,
{\rm tr}\hspace{0.03cm}\bigl(T^{\,c}\hspace{0.02cm}\bigl[\hspace{0.02cm} T^{\,d^{\hspace{0.03cm}\prime}}\!, T^{\,d}\hspace{0.02cm}\bigr] T^{\,s}\hspace{0.03cm}\bigr)
f^{\hspace{0.03cm} d\, d^{\hspace{0.03cm}\prime}\hspace{0.01cm}e}
=
\frac{i}{2}\,
{\rm tr}\hspace{0.03cm}\bigl(T^{\,c}\hspace{0.02cm}\hspace{0.02cm} T^{\,\lambda}\hspace{0.02cm} T^{\,s}\hspace{0.03cm}\bigr)
f^{\hspace{0.03cm} d^{\hspace{0.03cm}\prime}d\,\lambda\hspace{0.03cm}}
f^{\hspace{0.03cm} d\, d^{\hspace{0.03cm}\prime}\hspace{0.01cm}e}
=
-\!\left(\hspace{0.03cm}\frac{i}{2}\hspace{0.03cm}\right)^{\!\!2}\!N_{c}
\hspace{0.03cm}
f^{\hspace{0.03cm} c\, e\hspace{0.03cm}s\hspace{0.03cm}}.
\]
When constructing this expression with $\bigl\langle\hspace{0.01cm}\mathcal{Q}^{\hspace{0.03cm}e}\hspace{0.03cm}\bigr\rangle
\bigl\langle\hspace{0.01cm}\mathcal{Q}^{\hspace{0.03cm}c}\hspace{0.03cm}\bigr\rangle$ it turns to zero. Thus, instead of (\ref{eq:8r}), we now have
\begin{equation}
{\rm tr}\hspace{0.03cm}
\bigl(\hspace{0.01cm} {\mathcal N}_{{\bf k}}\hspace{0.03cm} T^{\,d^{\hspace{0.03cm}\prime}}\!{\mathcal N}_{{\bf k}_{1}} T^{\,d}\hspace{0.035cm}T^{\,s}\hspace{0.01cm}\bigr)
f^{\hspace{0.03cm} d\, d^{\hspace{0.03cm}\prime}\hspace{0.03cm}e\hspace{0.03cm}}
\bigl\langle\hspace{0.01cm}\mathcal{Q}^{\hspace{0.01cm}e}\bigr\rangle
\label{eq:8t}
\end{equation}
\[
=
-\hspace{0.03cm}\frac{i}{2}\,N^{\hspace{0.03cm}2}_{c}\hspace{0.03cm}
\bigl\langle\hspace{0.01cm}
\mathcal{Q}^{\hspace{0.03cm}s}\hspace{0.03cm}\bigr\rangle\hspace{0.03cm}
N_{{\bf k}}N_{{\bf k}_{1}}
+\,
{\rm tr}\hspace{0.03cm}\bigl(T^{\,d}\hspace{0.02cm} T^{\,s}\hspace{0.03cm}T^{\,c}\hspace{0.03cm}T^{\,d^{\hspace{0.03cm}\prime}} T^{\,f}\hspace{0.02cm} 
\bigr)
f^{\hspace{0.03cm} d\, d^{\hspace{0.03cm}\prime}\hspace{0.01cm}e\hspace{0.03cm}}
\bigl\langle\hspace{0.01cm}\mathcal{Q}^{\hspace{0.03cm}e}\hspace{0.03cm}\bigr\rangle
\bigl\langle\hspace{0.01cm}\mathcal{Q}^{\hspace{0.03cm}c}\hspace{0.03cm}\bigr\rangle
\bigl\langle\hspace{0.01cm}\mathcal{Q}^{\hspace{0.03cm}f}\hspace{0.03cm}\bigr\rangle \hspace{0.03cm}
W_{{\bf k}}W_{{\bf k}_{1}}.
\]
\indent We still need to consider the remaining trace of the fifth order. Due to the presence of the factor $\bigl\langle\hspace{0.01cm}\mathcal{Q}^{\hspace{0.03cm}c}\hspace{0.03cm}
\bigr\rangle
\bigl\langle\hspace{0.01cm}\mathcal{Q}^{\hspace{0.03cm}f}\hspace{0.03cm}\bigr\rangle$, which is symmetric with respect to the indices $c$ and $f$, here it is easiest to use the relation (\ref{ap:D10}) from Appendix \ref{appendix_D}. Using this relation, we immediately get
\[
{\rm tr}\hspace{0.03cm}\bigl(T^{\,d}\hspace{0.03cm} T^{\,s}\hspace{0.03cm}T^{\,c}\hspace{0.03cm} T^{\,d^{\hspace{0.03cm}\prime}} T^{\,f}\hspace{0.02cm} 
\bigr)
f^{\hspace{0.03cm} d\, d^{\hspace{0.03cm}\prime}\hspace{0.01cm}e\hspace{0.03cm}}
\bigl\langle\hspace{0.01cm}\mathcal{Q}^{\hspace{0.03cm}e}\hspace{0.03cm}\bigr\rangle
\bigl\langle\hspace{0.01cm}\mathcal{Q}^{\hspace{0.03cm}c}\hspace{0.03cm}\bigr\rangle
\bigl\langle\hspace{0.01cm}\mathcal{Q}^{\hspace{0.03cm}f}\hspace{0.03cm}\bigr\rangle
=
-\hspace{0.03cm}\frac{i}{2}\,
f^{\hspace{0.03cm}s\hspace{0.03cm}d\hspace{0.04cm}b}\,
{\rm tr}\hspace{0.03cm}\bigl(T^{\,b}\hspace{0.03cm} T^{\,f}\hspace{0.03cm} T^{\,d^{\hspace{0.03cm}\prime}} T^{\,c}\hspace{0.02cm} 
\bigr)
f^{\hspace{0.03cm} d\, d^{\hspace{0.03cm}\prime}\hspace{0.01cm}e\hspace{0.03cm}}
\bigl\langle\hspace{0.01cm}\mathcal{Q}^{\hspace{0.03cm}e}\hspace{0.03cm}\bigr\rangle
\bigl\langle\hspace{0.01cm}\mathcal{Q}^{\hspace{0.03cm}c}\hspace{0.03cm}\bigr\rangle
\bigl\langle\hspace{0.01cm}\mathcal{Q}^{\hspace{0.03cm}f}\hspace{0.03cm}\bigr\rangle.
\]
For the fourth-order trace on the right part, we apply formula (\ref{ap:D4}). Then, after simple algebraic transformations, the right-hand side of the previous expression can be reduced to the following form:
\begin{equation}
-\hspace{0.03cm}\frac{i}{2}\,
f^{\hspace{0.03cm}s\hspace{0.03cm}d\hspace{0.04cm}b\hspace{0.04cm}}
{\rm tr}\hspace{0.03cm}\bigl(T^{\,b}\hspace{0.03cm} T^{\,f}\hspace{0.03cm} T^{\,d^{\hspace{0.03cm}\prime}} T^{\,c}\hspace{0.02cm} 
\bigr)
f^{\hspace{0.03cm} d\, d^{\hspace{0.03cm}\prime}\hspace{0.01cm}e\hspace{0.03cm}}
\bigl\langle\hspace{0.01cm}\mathcal{Q}^{\hspace{0.03cm}e}\hspace{0.03cm}\bigr\rangle
\bigl\langle\hspace{0.01cm}\mathcal{Q}^{\hspace{0.03cm}c}\hspace{0.03cm}\bigr\rangle
\bigl\langle\hspace{0.01cm}\mathcal{Q}^{\hspace{0.03cm}f}\hspace{0.03cm}\bigr\rangle
\label{eq:8y}
\end{equation}
\[
= 
-\hspace{0.03cm}\frac{i}{2}\,\biggl\{\frac{1}{2}\,N_{c}\hspace{0.03cm}
\delta^{\hspace{0.02cm}f\hspace{0.02cm}c}\hspace{0.03cm}
\delta^{\hspace{0.02cm}s\hspace{0.03cm}e}
-\hspace{0.03cm}
\frac{1}{4}\,N_{c}\Bigl[\hspace{0.03cm}
{\rm tr}\hspace{0.03cm}\bigl(\hspace{0.02cm}T^{\,e}\hspace{0.03cm} T^{\,s}\hspace{0.03cm}T^{\,c}\hspace{0.03cm}T^{\,f}\hspace{0.02cm}\bigr)
-
{\rm tr}\hspace{0.03cm}\bigl(\hspace{0.02cm}T^{\,e}\hspace{0.03cm} T^{\,s}\hspace{0.01cm}D^{\,c}\hspace{0.01cm} D^{\,f}\hspace{0.02cm}\bigr)\Bigr]\!\biggr\}
\bigl\langle\hspace{0.01cm}\mathcal{Q}^{\hspace{0.03cm}e}\hspace{0.03cm}\bigr\rangle
\bigl\langle\hspace{0.01cm}\mathcal{Q}^{\hspace{0.03cm}c}\hspace{0.03cm}\bigr\rangle
\bigl\langle\hspace{0.01cm}\mathcal{Q}^{\hspace{0.03cm}f}\hspace{0.03cm}\bigr\rangle
\]
\[
-\hspace{0.03cm}\frac{i}{4}\,N_{c}\hspace{0.03cm}
\biggl\{\hspace{0.01cm}
\delta^{\hspace{0.02cm}f\hspace{0.02cm}c}\hspace{0.03cm}
\delta^{\hspace{0.02cm}s\hspace{0.03cm}e}
-\hspace{0.03cm}
\frac{1}{2}\,\Bigl(\delta^{\hspace{0.02cm}e\hspace{0.02cm}f}\hspace{0.03cm}
\delta^{\hspace{0.02cm}s\hspace{0.03cm}c}
+ 
\delta^{\hspace{0.02cm}e\hspace{0.02cm}c}\hspace{0.03cm}
\delta^{\hspace{0.02cm}f\hspace{0.03cm}s}
\Bigr)\!\biggr\}
\bigl\langle\hspace{0.01cm}\mathcal{Q}^{\hspace{0.03cm}e}\hspace{0.03cm}\bigr\rangle
\bigl\langle\hspace{0.01cm}\mathcal{Q}^{\hspace{0.03cm}c}\hspace{0.03cm}\bigr\rangle
\bigl\langle\hspace{0.01cm}\mathcal{Q}^{\hspace{0.03cm}f}\hspace{0.03cm}\bigr\rangle
= 0.
\]
Here we have used the formulas for the fourth-order traces (\ref{ap:D4}) and (\ref{ap:D5}). Thus the coefficient of the product $W_{{\bf k}}W_{{\bf k}_{1}}$ on the right-hand side of (\ref{eq:8t}) is zero and this contribution falls out of consideration.\\
\indent For the trace ${\rm tr}\hspace{0.03cm}
\bigl({\mathcal N}_{{\bf k}_{1}}T^{\,d}{\mathcal N}_{{\bf k}}\hspace{0.02cm}T^{\,s}\hspace{0.02cm}T^{\,d^{\hspace{0.03cm}\prime}}\hspace{0.03cm}\bigr)$ in (\ref{eq:8q}), we obtain similar result. In summary, for the quadratic in ${\mathcal N}_{{\bf k}}$ and ${\mathcal N}_{{\bf k}_{1}}$ terms in (\ref{eq:8q}), using the Sohotsky formula (\ref{eq:6y}) we get
\begin{equation}
\left\{\frac{{\rm tr}\hspace{0.03cm}
\bigl(\hspace{0.01cm} {\mathcal N}_{{\bf k}}\hspace{0.03cm} T^{\,d^{\hspace{0.03cm}\prime}}\!\hspace{0.02cm}{\mathcal N}_{{\bf k}_{1}} T^{\,d}\hspace{0.03cm}T^{\,s}\hspace{0.03cm}\bigr)}{\Delta\hspace{0.02cm}
\omega_{\hspace{0.03cm}{\bf k},\hspace{0.03cm}{\bf k}_{1}} +\, i\hspace{0.03cm}0} 
\,-\,
\frac{{\rm tr}\hspace{0.03cm}
\bigl({\mathcal N}_{{\bf k}_{1}}
T^{\,d}{\mathcal N}_{{\bf k}}\hspace{0.03cm}T^{\,s} \hspace{0.03cm}T^{\,d^{\hspace{0.03cm}\prime}}\hspace{0.03cm}\bigr)}
{\Delta\hspace{0.02cm}\omega_{\hspace{0.03cm}{\bf k},\hspace{0.03cm}{\bf k}_{1}} -\, i\hspace{0.03cm}0} 
\right\}\!
f^{\hspace{0.03cm} d\, d^{\hspace{0.03cm}\prime}\hspace{0.01cm}e\hspace{0.03cm}}
\bigl\langle\hspace{0.01cm}\mathcal{Q}^{\hspace{0.03cm}e}\hspace{0.03cm}\bigr\rangle
\label{eq:8u}
\end{equation}
\[
=
\frac{1}{2}\,i\hspace{0.03cm}N^{\hspace{0.03cm}2}_{c} 
N_{{\bf k}}\hspace{0.02cm}N_{{\bf k}_{1}}
\bigl\langle\hspace{0.01cm}\mathcal{Q}^{\hspace{0.03cm}s}\hspace{0.03cm}\bigr\rangle\hspace{0.03cm}
(2\pi)\,\delta(\omega^{l}_{\hspace{0.02cm}{\bf k}} - \omega^{l}_{\hspace{0.02cm}{\bf k}_{1}} - {\mathbf v}\cdot({\bf k} - {\bf k}_{1})).
\]
\indent Let us consider the remaining traces or, more precisely, differences of the traces that are linear in ${\mathcal N}_{{\bf k}}$ and ${\mathcal N}_{{\bf k}_{1}}$ in square brackets in (\ref{eq:8q}). Taking into account (\ref{eq:7q}) for the first difference of the traces, we find 
\[
\Bigl[{\rm tr}\hspace{0.03cm}
\bigl(T^{\,s}{\mathcal N}_{{\bf k}}\hspace{0.02cm} T^{\,e}\hspace{0.03cm}T^{\,d}\hspace{0.03cm}\bigr)
\!-\, {\rm tr}\hspace{0.03cm}
\bigl(T^{\,s}\hspace{0.03cm}T^{\,e} {\mathcal N}_{{\bf k}_{1}} T^{\,d}\hspace{0.03cm}\bigr)
\Bigr]
\bigl\langle\hspace{0.03cm}\mathcal{Q}^{\hspace{0.03cm}d}
\hspace{0.03cm}\bigr\rangle
\bigl\langle\hspace{0.03cm}\mathcal{Q}^{\hspace{0.03cm}e}
\hspace{0.03cm}\bigr\rangle
=
{\rm tr}\hspace{0.03cm}\bigl(T^{\,s} \hspace{0.03cm}T^{\,e}\hspace{0.03cm}T^{\,d} \hspace{0.02cm}\bigr)	
\bigl(N^{\,l}_{{\bf k}} - N^{\,l}_{{\bf k}_{1}}\!\hspace{0.03cm}\bigr)
\bigl\langle\hspace{0.03cm}\mathcal{Q}^{\hspace{0.03cm}d}
\hspace{0.03cm}\bigr\rangle
\bigl\langle\hspace{0.03cm}\mathcal{Q}^{\hspace{0.03cm}e}
\hspace{0.03cm}\bigr\rangle
\] 
\[
+\,
{\rm tr}\hspace{0.03cm}\bigl(\hspace{0.02cm}T^{\,e}\hspace{0.03cm} T^{\,s}\hspace{0.03cm}T^{\,c}\hspace{0.03cm}T^{\,f}\hspace{0.02cm}\bigr)
\bigl(W^{\,l}_{{\bf k}} - W^{\,l}_{{\bf k}_{1}}\!\hspace{0.03cm}\bigr)
\bigl\langle\hspace{0.03cm}\mathcal{Q}^{\hspace{0.03cm}c}
\hspace{0.03cm}\bigr\rangle
\bigl\langle\hspace{0.03cm}\mathcal{Q}^{\hspace{0.03cm}d}
\hspace{0.03cm}\bigr\rangle
\bigl\langle\hspace{0.03cm}\mathcal{Q}^{\hspace{0.03cm}e}
\hspace{0.03cm}\bigr\rangle
\vspace{0.1cm}
\]
\[
=
\Bigl\{2\hspace{0.03cm}\delta^{\hspace{0.02cm}s\hspace{0.02cm}d}\hspace{0.03cm}
\delta^{\hspace{0.02cm}c\hspace{0.03cm}e} 
+ \frac{1}{4}\,N_{c}\hspace{0.04cm}
d^{\hspace{0.03cm}s\hspace{0.03cm}d\hspace{0.03cm}\lambda}
\hspace{0.01cm}d^{\hspace{0.03cm}c\hspace{0.03cm}e\hspace{0.03cm}\lambda}
\Bigr\}\hspace{0.03cm}
\bigl(W^{\,l}_{{\bf k}} - W^{\,l}_{{\bf k}_{1}}\!\hspace{0.03cm}\bigr)
\bigl\langle\hspace{0.03cm}\mathcal{Q}^{\hspace{0.03cm}c}
\hspace{0.03cm}\bigr\rangle
\bigl\langle\hspace{0.03cm}\mathcal{Q}^{\hspace{0.03cm}d}
\hspace{0.03cm}\bigr\rangle
\bigl\langle\hspace{0.03cm}\mathcal{Q}^{\hspace{0.03cm}e}
\hspace{0.03cm}\bigr\rangle.
\]
Here we again used the formulae for the traces (\ref{ap:D3}) and (\ref{ap:D4}). The term with the difference $N^{\,l}_{{\bf k}} - N^{\,l}_{{\bf k}_{1}}$ has turned to zero. The coefficient of the survive term with the difference $W^{\,l}_{{\bf k}} - W^{\,l}_{{\bf k}_{1}}$ contains the contraction of two symmetric structure constants, which in the general case of arbitrary $N_{c}$ cannot be reduced to an expansion in terms of the Kronecker deltas.\\
\indent Taking into account the above obtained expression and  (\ref{eq:8w}), (\ref{eq:8e}), and (\ref{eq:8u}), instead of (\ref{eq:8q}) we find a kinetic equation for the second ``color'' part of the plasmon number density of soft gluon plasma excitations $W^{\,l}_{{\bf k}}$  
\begin{equation}
N_{c}\hspace{0.03cm}\frac{\partial\hspace{0.03cm}  \bigl(\bigl\langle\hspace{0.01cm}\mathcal{Q}^{\hspace{0.03cm}s}\hspace{0.03cm}\bigr\rangle\hspace{0.02cm} W^{\hspace{0.03cm}l}_{\bf k}\hspace{0.03cm} \bigr)}{\!\!\partial\hspace{0.03cm} t}
=
2\hspace{0.015cm}N_{c}\hspace{0.03cm}
\bigl({\rm Im}\hspace{0.03cm}\mathscr{T}_{\hspace{0.02cm}{\bf k},\hspace{0.03cm}{\bf k}}\bigr)
\hspace{0.03cm}N^{\hspace{0.03cm}l}_{\bf k}\,
\bigl\langle\hspace{0.01cm}\mathcal{Q}^{\hspace{0.03cm}s}\hspace{0.03cm}\bigr\rangle
\label{eq:8i}
\end{equation}
\[
-\!\int\!d\hspace{0.02cm}{\bf k}_{1}\,
\bigl|\mathscr{T}_{\hspace{0.02cm}{\bf k},\hspace{0.03cm}
{\bf k}_{1}}\bigr|^{\hspace{0.02cm}2}\hspace{0.04cm}
\Bigl\{2\hspace{0.03cm}\delta^{\hspace{0.02cm}s\hspace{0.02cm}d}
\hspace{0.03cm}
\delta^{\hspace{0.02cm}c\hspace{0.03cm}e} 
+ \frac{1}{4}\,N_{c}\hspace{0.05cm}
d^{\hspace{0.03cm}s\hspace{0.03cm}d\hspace{0.03cm}\lambda}
\hspace{0.01cm}d^{\hspace{0.03cm}c\hspace{0.03cm}e\hspace{0.03cm}\lambda}
\Bigr\}\hspace{0.03cm}
\bigl(W^{\,l}_{{\bf k}} - W^{\,l}_{{\bf k}_{1}}\bigr)
\bigl\langle\hspace{0.03cm}\mathcal{Q}^{\hspace{0.03cm}c}
\hspace{0.03cm}\bigr\rangle
\bigl\langle\hspace{0.03cm}\mathcal{Q}^{\hspace{0.03cm}d}
\hspace{0.03cm}\bigr\rangle
\bigl\langle\hspace{0.03cm}\mathcal{Q}^{\hspace{0.03cm}e}
\hspace{0.03cm}\bigr\rangle 
\]
\[
\times\,
(2\pi)\hspace{0.03cm}\delta(\omega^{\hspace{0.02cm}l}_{\hspace{0.03cm}{\bf k}} - \omega^{\hspace{0.02cm}l}_{\hspace{0.03cm}{\bf k}_{1}} - {\mathbf v}\cdot({\bf k} - {\bf k}_{1}))
\]
\[
+\;
\frac{1}{2}\,N^{\hspace{0.03cm}2}_{c}\!\int\!d\hspace{0.02cm}{\bf k}_{1}\hspace{0.04cm}
\bigl|\mathscr{T}_{\hspace{0.02cm}{\bf k},\hspace{0.03cm}{\bf k}_{1}}\bigr|^{\hspace{0.02cm}2}\,
N^{\,l}_{{\bf k}}\hspace{0.02cm}N^{\,l}_{{\bf k}_{1}}
\bigl\langle\hspace{0.01cm}\mathcal{Q}^{\hspace{0.03cm}s}\hspace{0.03cm}\bigr\rangle\hspace{0.03cm}
(2\pi)\,\delta(\omega^{\hspace{0.02cm}l}_{\hspace{0.03cm}{\bf k}} - \omega^{\hspace{0.02cm}l}_{\hspace{0.03cm}{\bf k}_{1}} - {\mathbf v}\cdot({\bf k} - {\bf k}_{1})).
\]
We can simplify the contraction of two symmetric structure constants in (\ref{eq:8i}) in two cases. The first ``trivial'' case relates to $N_{c} = 2$, when $d^{\hspace{0.03cm}a\hspace{0.03cm}b\hspace{0.03cm}c}\equiv 0$. The second case refers to $N_{c} = 3$ and is connected with the availability of a well-known accidental cancellation expressed by the relation (\ref{ap:D13}) from Appendix \ref{appendix_D}. We will discuss this point in more detail in section \ref{section_10}.


\section{\bf Equation for the expected value of color charge ${\mathcal Q}^{\hspace{0.03cm}d}$}
\label{section_9}
\setcounter{equation}{0}

In this section we consider the derivation of the equation of motion for the expected value of color charge ${\mathcal Q}^{\hspace{0.03cm}d}$. Our first step is to average equation (\ref{eq:5e}). Thus, we obtain
\begin{equation}
\frac{d \hspace{0.01cm}\langle\hspace{0.01cm}\mathcal{Q}^{\,d}\hspace{0.03cm}
\rangle} {d\hspace{0.03cm}t}
=
i\hspace{0.02cm}
f^{\hspace{0.03cm}d\hspace{0.03cm}d^{\hspace{0.02cm}\prime}\hspace{0.01cm}e}
\!\!\int\!d\hspace{0.02cm}{\bf k}_{1}\hspace{0.02cm}
d\hspace{0.02cm}{\bf k}_{2}\, 
\mathscr{T}^{\,a_{1}\hspace{0.03cm}a_{2}\,d^{\hspace{0.02cm}\prime}}_{\; {\bf k}_{1},\, {\bf k}_{2}} 
\bigl\langle c^{\hspace{0.03cm}\ast\,a_{1}}_{{\bf k}_{1}}c^{\hspace{0.03cm}a_{2}}_{{\bf k}_{2}}\hspace{0.03cm}\mathcal{Q}^{\,e}\hspace{0.02cm}\bigr\rangle,
\label{eq:9q}
\end{equation}
where the fourth-order correlation function on the right-hand side, due to (\ref{eq:6q}), has the following form 
\begin{equation}
\bigl\langle\hspace{0.03cm}c^{\ast\ \!\!a_{1}}_{{\bf k}_{1}}\hspace{0.01cm} c^{\phantom{\ast}\!\!a_{2}}_{{\bf k}_{1}}\hspace{0.02cm}
\mathcal{Q}^{\,e}\hspace{0.03cm}\bigr\rangle
\hspace{0.03cm}\simeq\hspace{0.03cm}
\delta({\bf k}_{1} - {\bf k}_{2})\,
{\mathcal N}^{\;a_{1}\hspace{0.03cm}a^{\phantom{\prime}}_{2}\!}_{\hspace{0.02cm}
{\bf k}}\,
\bigl\langle\hspace{0.03cm}\mathcal{Q}^{\,e}
\hspace{0.03cm}\bigr\rangle
\vspace{0.15cm}
\label{eq:9w}
\end{equation}
\[
+\;
\frac{1}{\Delta\hspace{0.02cm}\omega_{\,{\bf k}_{1},\hspace{0.03cm}
{\bf k}_{2}}\! + i\hspace{0.03cm}0}\,
\mathscr{T}^{\hspace{0.03cm}\ast\,\phantom{a}}_{\; {\bf k}_{1},
\hspace{0.03cm}{\bf k}_{2}}
\bigl[\hspace{0.02cm}
\bigl({\mathcal N}_{{\bf k}_{1}}\hspace{0.03cm}T^{\,e^{\prime}}\bigr)^{\,a_{1}\hspace{0.02cm}a_{2}}
-
\bigl(T^{\,e^{\prime}}\!\hspace{0.02cm}{\mathcal N}_{{\bf k}_{2}}\bigr)^{a_{1}\hspace{0.02cm}a_{2}}
\hspace{0.03cm}\bigr]\hspace{0.03cm} 
\bigl\langle\hspace{0.03cm}\mathcal{Q}^{\hspace{0.03cm}e}
\hspace{0.03cm}\bigr\rangle
\hspace{0.03cm}\bigl\langle\hspace{0.03cm}\mathcal{Q}^{\hspace{0.03cm}
e^{\prime}}\hspace{0.03cm}\bigr\rangle
\]
\[
-\;
\frac{i}{\Delta\hspace{0.02cm}\omega_{\,{\bf k}_{1},\hspace{0.03cm}{\bf k}_{2}}\! + i\hspace{0.03cm}0}\,
\mathscr{T}^{\hspace{0.03cm}\ast\,\phantom{a}}_{\; {\bf k}_{1}, \hspace{0.03cm} {\bf k}_{2}}
\hspace{0.03cm}
\bigl({\mathcal N}_{{\bf k}_{1}}\hspace{0.01cm}T^{\,d^{\hspace{0.03cm}\prime\prime}\!}{\mathcal N}_{{\bf k}_{2}}\bigr)^{a_{1}\hspace{0.02cm}a_{2}}
\hspace{0.03cm}
\bigl\langle\hspace{0.03cm}\mathcal{Q}^{\hspace{0.03cm}e^{\prime}}
\hspace{0.03cm}\bigr\rangle\hspace{0.02cm}
f^{\hspace{0.03cm}e\hspace{0.03cm}d^{\hspace{0.02cm}\prime\prime}
\hspace{0.01cm}e^{\prime}}.
\]
Let us substitute the first term of the correlation function into the right-hand side of (\ref{eq:9q}). Then, by taking into account the color factorization of the total effective amplitude, Eq.\,(\ref{eq:5i}), we find 
\[
i\hspace{0.02cm}
f^{\hspace{0.03cm}d\hspace{0.03cm}d^{\hspace{0.02cm}\prime}\hspace{0.01cm}e}
\!\!\int\!d\hspace{0.02cm}{\bf k}_{1}\hspace{0.02cm}
d\hspace{0.02cm}{\bf k}_{2}\hspace{0.04cm} 
\mathscr{T}_{\; {\bf k}_{1},\, {\bf k}_{2}}\hspace{0.04cm} 
{\rm tr}\bigl(\hspace{0.03cm}T^{\,d^{\hspace{0.02cm}\prime}}\!\hspace{0.02cm}
{\mathcal N}_{{\bf k}_{1}}\bigr)\hspace{0.03cm}
\bigl\langle\hspace{0.03cm}\mathcal{Q}^{\,e}
\hspace{0.02cm}\bigr\rangle\hspace{0.03cm}.
\]
By virtue of the equality ${\rm tr}\bigl(\hspace{0.03cm}T^{\,d^{\hspace{0.03cm}\prime}}\!\hspace{0.02cm}
{\mathcal N}_{{\bf k}_{1}}\bigr) = N_{c}\hspace{0.03cm} W^{\hspace{0.03cm}l}_{{\bf k}_{1}}\bigl\langle\hspace{0.03cm}
\mathcal{Q}^{\,d^{\hspace{0.03cm}\prime}}
\hspace{0.02cm}\bigr\rangle$, the expression above is zero. Furthermore, we substitute the second term from the correlation function (\ref{eq:9w}) into (\ref{eq:9q})
\[
f^{\hspace{0.03cm}d\hspace{0.03cm}d^{\hspace{0.02cm}\prime}\hspace{0.01cm}e}
\!\!\int\!d\hspace{0.02cm}{\bf k}_{1}\hspace{0.02cm}
d\hspace{0.02cm}{\bf k}_{2}\, 
|\mathscr{T}_{\; {\bf k}_{1},\, {\bf k}_{2}}|^{\hspace{0.02cm}2}\,
\frac{1}{\Delta\hspace{0.02cm}\omega_{\,{\bf k}_{1},\hspace{0.03cm}{\bf k}_{2}}\! + i\hspace{0.03cm}0}\,
\bigl[
{\rm tr}\bigl(\hspace{0.03cm}T^{\,d^{\prime}}\!\hspace{0.02cm}
{\mathcal N}_{{\bf k}_{1}}T^{\,e^{\prime}}\bigr)
-
{\rm tr}\bigl(\hspace{0.03cm}T^{\,d^{\hspace{0.02cm}\prime}}T^{\,e^{\prime}}
\!\hspace{0.02cm}{\mathcal N}_{{\bf k}_{2}}\bigr)
\bigr]
\hspace{0.01cm}\bigl\langle\hspace{0.03cm}\mathcal{Q}^{\hspace{0.03cm}e}
\hspace{0.03cm}\bigr\rangle
\hspace{0.03cm}\bigl\langle\hspace{0.03cm}
\mathcal{Q}^{\hspace{0.03cm}e^{\prime}}\hspace{0.03cm}\bigr\rangle.	
\]
Using the corresponding formula for the traces in (\ref{eq:7w}) for the last expression we easily find
\[
f^{\hspace{0.03cm}d\hspace{0.03cm}d^{\hspace{0.02cm}\prime}e}
\!\!\int\!d\hspace{0.02cm}{\bf k}_{1}\hspace{0.02cm}
d\hspace{0.02cm}{\bf k}_{2}\, 
|\mathscr{T}_{\; {\bf k}_{1},\, {\bf k}_{2}}|^{\hspace{0.02cm}2}\,
\frac{1}{\Delta\hspace{0.02cm}\omega_{\,{\bf k}_{1},\hspace{0.03cm}
{\bf k}_{2}}\! + i\hspace{0.03cm}0}\,
\]
\[
\times\hspace{0.03cm}\bigl[\hspace{0.03cm}
\delta^{\hspace{0.03cm}d^{\hspace{0.03cm}\prime}e^{\prime}\!}N_{c}
\hspace{0.03cm}\bigl(N^{\hspace{0.03cm}l}_{{\bf k}_{1}} - N^{\hspace{0.03cm}l}_{{\bf k}_{2}}\bigr)
+
\frac{1}{2}\,N_{c}\bigl(T^{\,c}\bigr)^{d^{\hspace{0.03cm}\prime}
\hspace{0.02cm}e^{\hspace{0.01cm}\prime}\!}\bigl(W^{\hspace{0.03cm}l}_{{\bf k}_{2}} + W^{\hspace{0.03cm}l}_{{\bf k}_{1}}\hspace{0.03cm}\bigr)
\bigl\langle\hspace{0.01cm}\mathcal{Q}^{\hspace{0.03cm}c}\hspace{0.03cm}\bigr
\rangle\bigr]
\hspace{0.01cm}\bigl\langle\hspace{0.03cm}\mathcal{Q}^{\hspace{0.03cm}e}
\hspace{0.03cm}\bigr\rangle
\hspace{0.03cm}\bigl\langle\hspace{0.03cm}
\mathcal{Q}^{\hspace{0.03cm}e^{\prime}}\hspace{0.03cm}\bigr\rangle.	
\]
By virtue of the antisymmetry of the structural constant $f^{\hspace{0.03cm}a\hspace{0.03cm}b\hspace{0.03cm}c}$ this contribution also turns to zero.\\
\indent As the last step, we substitute the third term of the correlation function (\ref{eq:9w}) into the right-hand side of (\ref{eq:9q}). As a result, we get
\[
-\hspace{0.03cm}if^{\hspace{0.03cm}d\hspace{0.03cm}d^{\hspace{0.02cm}\prime}\hspace{0.01cm}e}
f^{\hspace{0.03cm}e\hspace{0.03cm}d^{\hspace{0.02cm}\prime\prime}
e^{\prime}}
\!\!\int\!d\hspace{0.02cm}{\bf k}_{1}\hspace{0.02cm}
d\hspace{0.02cm}{\bf k}_{2}\, 
|\mathscr{T}_{\; {\bf k}_{1},\, 
{\bf k}_{2}}|^{\hspace{0.02cm}2}\hspace{0.03cm}
\frac{1}{\Delta\hspace{0.02cm}\omega_{\,{\bf k}_{1},\hspace{0.03cm}{\bf k}_{2}}\! + i\hspace{0.03cm}0}\,
{\rm tr}\bigl(\hspace{0.03cm}T^{\,d^{\hspace{0.02cm}\prime}}\!{\mathcal N}_{{\bf k}_{1}}T^{\,d^{\hspace{0.03cm}\prime\prime}}\!{\mathcal N}_{{\bf k}_{2}}\bigr)
\hspace{0.03cm}\bigl\langle\hspace{0.03cm}
\mathcal{Q}^{\hspace{0.03cm}e^{\prime}}\hspace{0.03cm}\bigr\rangle.
\]
According to formula (\ref{eq:7e}) for the trace in the integrand we have  
\begin{equation}
{\rm tr}\hspace{0.03cm}
\bigl(\hspace{0.03cm}{\mathcal N}_{{\bf k}_{1}}\hspace{0.03cm} T^{\,d^{\hspace{0.03cm}\prime}}\!{\mathcal N}_{{\bf k}_{2}} T^{\,d^{\hspace{0.03cm}\prime\prime}}\hspace{0.03cm}\bigr)
=
N_{c}\hspace{0.04cm}\delta^{\hspace{0.03cm}d^{\hspace{0.03cm}\prime}
d^{\hspace{0.03cm}\prime\prime}}\!\hspace{0.025cm}
N^{\hspace{0.03cm}l}_{{\bf k}_{1}}N^{\hspace{0.03cm}l}_{{\bf k}_{2}}
\label{eq:9e}
\end{equation}
\[
+\,
\frac{i}{2}\,N_{c}\hspace{0.03cm}f^{\hspace{0.03cm} d^{\hspace{0.03cm}\prime}\hspace{0.01cm}d^{\hspace{0.03cm}\prime\prime}\hspace{0.01cm}c\hspace{0.03cm}}
\bigl\langle\hspace{0.01cm}\mathcal{Q}^{\hspace{0.03cm}c}\hspace{0.03cm}\bigr
\rangle\hspace{0.03cm}W^{\hspace{0.03cm}l}_{{\bf k}_{1}}N^{\hspace{0.03cm}l}_{{\bf k}_{2}}
-
\frac{i}{2}\,N_{c}\hspace{0.03cm}f^{\hspace{0.03cm} d^{\hspace{0.03cm}\prime}\hspace{0.01cm}d^{\hspace{0.03cm}\prime\prime}\hspace{0.01cm}c\hspace{0.03cm}}
\bigl\langle\hspace{0.01cm}\mathcal{Q}^{\hspace{0.03cm}c}\hspace{0.04cm}
\bigr\rangle\hspace{0.01cm}W^{\hspace{0.03cm}l}_{{\bf k}_{2}}N^{\hspace{0.03cm}l}_{{\bf k}_{1}}
+
{\rm tr}\hspace{0.03cm}
\bigl(T^{\,c}\hspace{0.03cm} T^{\,d^{\hspace{0.03cm}\prime}}T^{\,f} T^{\,d^{\hspace{0.03cm}\prime\prime}}\hspace{0.03cm}\bigr)
\bigl\langle\hspace{0.01cm}\mathcal{Q}^{\hspace{0.03cm}c}\hspace{0.03cm}\bigr
\rangle
\bigl\langle\hspace{0.01cm}\mathcal{Q}^{\hspace{0.03cm}f}\hspace{0.03cm}\bigr
\rangle\hspace{0.03cm}W^{\hspace{0.03cm}l}_{{\bf k}_{1}}
W^{\hspace{0.03cm}l}_{{\bf k}_{2}}.
\]
For the first term on the right-hand side, we have the following color factor 
\[
f^{\hspace{0.03cm}d\hspace{0.03cm}d^{\hspace{0.02cm}\prime}\hspace{0.01cm}e}
f^{\hspace{0.03cm}e\hspace{0.03cm}d^{\hspace{0.02cm}\prime\prime}
e^{\prime}}
\delta^{\hspace{0.03cm}d^{\hspace{0.03cm}\prime}d^{\hspace{0.03cm}\prime\prime}}
\bigl\langle\hspace{0.03cm}\mathcal{Q}^{\,e^{\prime}}\hspace{0.03cm}\bigr\rangle
=
-N_{c}\hspace{0.04cm}\delta^{\hspace{0.03cm}d\hspace{0.03cm}e^{\hspace{0.03cm}\prime}}
\hspace{0.01cm}\bigl\langle\hspace{0.03cm}\mathcal{Q}^{\,e^{\prime}}
\hspace{0.03cm}\bigr\rangle
=
-N_{c}\hspace{0.03cm}\bigl\langle\hspace{0.03cm}\mathcal{Q}^{\,d}\hspace{0.03cm}\bigr\rangle.
\]
Next, the color factor for the second term in (\ref{eq:9e}) is 
\[
f^{\hspace{0.03cm}d\hspace{0.03cm}d^{\hspace{0.02cm}\prime}\hspace{0.01cm}e}
f^{\hspace{0.03cm}e\hspace{0.03cm}d^{\hspace{0.02cm}\prime\prime}
e^{\prime}}\!
f^{\hspace{0.03cm} d^{\hspace{0.03cm}\prime}\hspace{0.01cm}d^{\hspace{0.03cm}\prime\prime}
\hspace{0.01cm}c\hspace{0.03cm}}
\hspace{0.01cm}\bigl\langle\hspace{0.03cm}\mathcal{Q}^{\,e^{\prime}}\hspace{0.03cm}\bigr\rangle
\hspace{0.01cm}\bigl\langle\hspace{0.03cm}\mathcal{Q}^{\,c}\hspace{0.03cm}\bigr\rangle
=
i\,{\rm tr}\hspace{0.03cm}
\bigl(T^{\,d}\hspace{0.03cm}T^{\,e^{\hspace{0.03cm}\prime}}T^{\,c} \hspace{0.03cm}\bigr)
\hspace{0.01cm}\bigl\langle\hspace{0.03cm}\mathcal{Q}^{\,e^{\prime}}\hspace{0.03cm}\bigr\rangle
\hspace{0.01cm}\bigl\langle\hspace{0.03cm}\mathcal{Q}^{\,c}\hspace{0.03cm}\bigr\rangle
=
-\hspace{0.03cm}\frac{1}{2}\,N_{c}\hspace{0.03cm}
f^{\hspace{0.03cm}d\hspace{0.03cm}e^{\hspace{0.02cm}\prime}\hspace{0.01cm}c}
\hspace{0.01cm}\bigl\langle\hspace{0.03cm}\mathcal{Q}^{\,e^{\prime}}\hspace{0.03cm}\bigr\rangle
\hspace{0.01cm}\bigl\langle\hspace{0.03cm}\mathcal{Q}^{\,c}\hspace{0.03cm}\bigr\rangle \equiv 0.
\]
Thus, the contribution with the product $W_{{\bf k}_{1}}\hspace{0.01cm}N_{{\bf k}_{2}}$ in (\ref{eq:9e}) equals zero.
By arguments that are completely analogous to the previous ones, the contribution with the product $W_{{\bf k}_{2}}\hspace{0.01cm}N_{{\bf k}_{1}}$ vanishes. Finally, the color factor for the last term (\ref{eq:9e}) has the form 
\[
f^{\hspace{0.03cm}d\hspace{0.03cm}d^{\hspace{0.02cm}\prime}\hspace{0.01cm}e}
f^{\hspace{0.03cm}e\hspace{0.03cm}d^{\hspace{0.02cm}\prime\prime}
\hspace{0.01cm}e^{\prime}}
{\rm tr}\hspace{0.03cm}
\bigl(T^{\,c}\hspace{0.03cm} T^{\,d^{\hspace{0.03cm}\prime}}T^{\,f} T^{\,d^{\hspace{0.03cm}\prime\prime}}\hspace{0.03cm}\bigr)
\hspace{0.01cm}\bigl\langle\hspace{0.03cm}\mathcal{Q}^{\,e^{\prime}}\hspace{0.03cm}\bigr\rangle
\bigl\langle\hspace{0.01cm}\mathcal{Q}^{\hspace{0.03cm}c}\hspace{0.03cm}\bigr
\rangle
\bigl\langle\hspace{0.01cm}\mathcal{Q}^{\hspace{0.03cm}f}\hspace{0.03cm}\bigr
\rangle.
\]
This contribution up to a multiplier and redefinition of the color indices coincides with the expression (\ref{eq:8y}) of the previous section. There we have shown that it is zero and thus the contribution proportional to $W_{{\bf k}_{1}}W_{{\bf k}_{2}}$ also drops out and, thus, only the first term survives from the whole expression (\ref{eq:9e}). As a result, for equation (\ref{eq:9q}) we have 
\begin{equation}
\frac{d \hspace{0.01cm}\langle\hspace{0.01cm}\mathcal{Q}^{\,d}\hspace{0.03cm}
\rangle} {d\hspace{0.03cm}t}
=
i\hspace{0.03cm}N^{\hspace{0.03cm}2}_{c}\! \int\!
d\hspace{0.02cm}{\bf k}_{1}\hspace{0.02cm}d\hspace{0.02cm}{\bf k}_{2}\, 
|\mathscr{T}_{\; {\bf k}_{1},\,{\bf k}_{2}}|^{\hspace{0.02cm}2}\,
\frac{1}{\Delta\hspace{0.02cm}\omega_{\,{\bf k}_{1},\hspace{0.03cm}
{\bf k}_{2}}\! + i\hspace{0.03cm}0}\,N^{\hspace{0.03cm}l}_{{\bf k}_{1}}N^{\hspace{0.03cm}l}_{{\bf k}_{2}}
\bigl\langle\hspace{0.01cm}\mathcal{Q}^{\hspace{0.03cm}d}\hspace{0.03cm}
\bigr \rangle.
\label{eq:9r}
\end{equation}
However, it is not the end of the story. In the final step, we use Sohotsky's formula 
\[
\frac{1}{\Delta\hspace{0.02cm}\omega_{\,{\bf k}_{1},\hspace{0.03cm}
{\bf k}_{2}}\! + i\hspace{0.03cm}0}
 =
\mathcal{P}\frac{1}{\Delta\hspace{0.02cm}\omega_{\,{\bf k}_{1},\hspace{0.03cm}{\bf k}_{2}}}
-
\hspace{0.02cm}i\hspace{0.02cm}\pi\hspace{0.03cm}\delta(\Delta\hspace{0.03cm}\omega_{\hspace{0.03cm}{\bf k}_{1},\hspace{0.03cm}{\bf k}_{2}}),
\]
where the symbol $\mathcal{P}$ denotes the principle value. The square of the amplitude modulus $|\mathscr{T}_{\; {\bf k}_{1},\hspace{0.03cm}{\bf k}_{2}}|^{\hspace{0.03cm}2}$, by virtue of property (\ref{eq:5o}), is an even function relative to the exchange ${\bf k}_{1} \rightleftarrows {\bf k}_{2}$, whereas the function $\Delta\hspace{0.02cm}\omega_{\,{\bf k}_{1},\hspace{0.03cm}{\bf k}_{2}}$ by virtue of the definition (\ref{eq:4a}) is an odd function with respect to the same exchange, that is, $\Delta\hspace{0.02cm}\omega_{\,{\bf k}_{1},\hspace{0.03cm}{\bf k}_{2}} = - \Delta\hspace{0.02cm}\omega_{\,{\bf k}_{2},\hspace{0.03cm}{\bf k}_{1}}$. Therefore, in the integrand of (\ref{eq:9r}) only the second term in the Sohotsky representation is kept and thus, finally, instead of (\ref{eq:9r}), we can write
\begin{equation}
\frac{d \hspace{0.01cm}\langle\hspace{0.01cm}\mathcal{Q}^{\,d}\hspace{0.03cm}
\rangle} {d\hspace{0.03cm} t}
=
N^{\hspace{0.03cm}2}_{c}\!\int\!d\hspace{0.02cm}
{\bf k}_{1}\hspace{0.02cm}d\hspace{0.02cm}{\bf k}_{2}\, 
|\mathscr{T}_{\; {\bf k}_{1},\, 
{\bf k}_{2}}|^{\hspace{0.03cm}2}\hspace{0.04cm}
\pi\hspace{0.03cm}\delta(\omega^{\hspace{0.02cm}l}_{\hspace{0.02cm}{\bf k}_{1}} - \omega^{\hspace{0.02cm}l}_{\hspace{0.02cm}{\bf k}_{2}} - {\mathbf v}\cdot({\bf k}_{1} - {\bf k}_{2}))
\hspace{0.03cm}N^{\hspace{0.03cm}l}_{{\bf k}_{1}}N^{\hspace{0.03cm}l}_{{\bf k}_{2}}
\bigl\langle\hspace{0.01cm}\mathcal{Q}^{\hspace{0.03cm}d}\hspace{0.03cm}
\bigr\rangle.
\label{eq:9t}
\end{equation}
\indent Let us introduce the notation 
\begin{equation}
A(t) \equiv 
N^{\hspace{0.03cm}2}_{c}\!\int\!d\hspace{0.02cm}{\bf k}_{1}\hspace{0.02cm} d\hspace{0.02cm}{\bf k}_{2}\,
\bigl|\mathscr{T}_{{\bf k}_{1},\hspace{0.02cm}
{\bf k}_{2}}\bigr|^{\hspace{0.02cm}2}
\hspace{0.03cm} N^{\hspace{0.03cm}l}_{{\bf k}_{1}}N^{\hspace{0.03cm}l}_{{\bf k}_{2}}\hspace{0.03cm}
(2\pi)\hspace{0.03cm}\delta(\omega^{\hspace{0.02cm}l}_{\hspace{0.02cm}{\bf k}} - \omega^{\hspace{0.02cm}l}_{\hspace{0.02cm}{\bf k}_{1}} - {\mathbf v}\cdot({\bf k} - {\bf k}_{1}))
\label{eq:9y}
\end{equation}
and rewrite equation (\ref{eq:9t}) in a visual form
\begin{equation}
\frac{d\hspace{0.04cm}\bigl\langle\hspace{0.01cm}{\mathcal Q}^{\hspace{0.03cm}d}\hspace{0.03cm}\bigr\rangle}{d\hspace{0.03cm} t}
=
\frac{1}{2}\,A(t)\langle\hspace{0.01cm}{\mathcal Q}^{\hspace{0.03cm}d}\hspace{0.03cm}\bigr\rangle,
\quad
\langle\hspace{0.01cm}{\mathcal Q}^{\hspace{0.03cm}d}\hspace{0.03cm}
\bigr\rangle|_{t\hspace{0.02cm}=\hspace{0.02cm}t_{0}}
=
{\mathcal Q}^{\hspace{0.03cm}d}_{\hspace{0.03cm}0},
\label{eq:9u}
\end{equation}
where ${\mathcal Q}^{\hspace{0.03cm}d}_{\hspace{0.03cm}0}$ is some fixed (nonrandom) vector in the internal color space that the high-energy color-charged particle possessed at the initial moment of time. We are interested in the time dependence of the colorless quadratic combination ${\mathfrak q}_{\hspace{0.02cm}2}(t)$ for the averaged color charge, Eq.\,(\ref{eq:7t}), as well as the colorless combination of the fourth order
\begin{equation}
{\mathfrak q}_{\hspace{0.02cm}4}(t) =
{\mathfrak q}^{\hspace{0.03cm}a}_{\hspace{0.02cm}2}(t)\hspace{0.03cm}
{\mathfrak q}^{\hspace{0.03cm}a}_{\hspace{0.02cm}2}(t),
\label{eq:9i}
\end{equation}
where
\[
{\mathfrak q}^{\hspace{0.03cm}a}_{\hspace{0.02cm}2}(t) \equiv
d^{\hspace{0.03cm}a\hspace{0.03cm}b\hspace{0.03cm}c}
\langle\hspace{0.01cm}{\mathcal Q}^{\hspace{0.03cm}b}\hspace{0.03cm}\bigr\rangle
\langle\hspace{0.01cm}{\mathcal Q}^{\hspace{0.03cm}c}\hspace{0.03cm}\bigr\rangle.
\]
From (\ref{eq:9u}), we immediately find the desired time dependence of these combinations as nonlinear functionals of the colorless part $N^{\hspace{0.03cm}l}_{{\bf k}}$ of the plasmon number density
\begin{equation}
{\mathfrak q}_{\hspace{0.02cm}2}(t) = {\mathfrak q}_{\hspace{0.02cm}2}(t_{0}) \exp\left\{\int^{\hspace{0.03cm}t}_{t_{0}}\!A(\tau)
\hspace{0.03cm} d\hspace{0.02cm}\tau\right\},
\qquad
{\mathfrak q}_{\hspace{0.02cm}4}(t) = {\mathfrak q}_{\hspace{0.02cm}4}(t_{0}) \exp\left\{2\!\int^{\hspace{0.03cm}t}_{t_{0}}\!A(\tau)
\hspace{0.03cm}d\hspace{0.02cm}\tau\right\}.
\label{eq:9o}
\end{equation}
Thus, the square of the averaged color charge ${\mathfrak q}_{\hspace{0.02cm}2}(t)$ of a hard particle is not conserved in the interaction with the random soft bosonic excitations of a hot gluon plasma.


\section{\bf System of kinetic equations for soft gluon excitations}
\label{section_10}
\setcounter{equation}{0}

In this section, we write out once more the kinetic equations for the scalar plasmon number densities $N^{\hspace{0.03cm}l}_{\bf k}$ and $W^{\hspace{0.03cm}l}_{\bf k}$ obtained in the previous sections and analyze some of their simplest consequences. The equation for the colorless part $N^{\hspace{0.03cm}l}_{\bf k}$ according to (\ref{eq:7r}) has the following form: 
\begin{equation}
d_{A}\hspace{0.03cm}\frac{\partial N^{\hspace{0.03cm}l}_{\bf k}}{\!\!\partial\hspace{0.04cm} t}
=
2\hspace{0.03cm}{\mathfrak q}_{2}(t)\hspace{0.02cm}N_{c}\hspace{0.03cm}
\bigl({\rm Im}\hspace{0.03cm}\mathscr{T}_{\,{\bf k},\hspace{0.03cm}{\bf k}}\bigr)\hspace{0.01cm}
W^{\hspace{0.03cm}l}_{\bf k}
\label{eq:10q}
\end{equation}
\[
-\,
{\mathfrak q}_{2}(t)\hspace{0.03cm}N_{c}\!\!\hspace{0.03cm}
\int\!d\hspace{0.02cm}{\bf k}_{1}\hspace{0.04cm}\bigl|\mathscr{T}_{\,{\bf k}, 
\hspace{0.03cm}{\bf k}_{1}}\bigr|^{\hspace{0.02cm}2}\hspace{0.03cm}
\Bigl\{\bigl(N^{\hspace{0.03cm}l}_{{\bf k}} - N^{\hspace{0.03cm}l}_{{\bf k}_{1}}\bigr)
-
\frac{1}{2}\,N_{c}\hspace{0.03cm}\bigl(W^{\hspace{0.03cm}l}_{\bf k}\hspace{0.02cm}N^{\hspace{0.03cm}l}_{{\bf k}_{1}} 
-
N^{\hspace{0.03cm}l}_{\bf k}\hspace{0.03cm}W^{\hspace{0.03cm}l}_{{\bf k}_{1}}\bigr)\!\Bigr\}
\]
\[
\times\hspace{0.04cm}
(2\pi)\,\delta(\omega^{\hspace{0.02cm}l}_{\hspace{0.02cm}{\bf k}} - \omega^{\hspace{0.02cm}l}_{\hspace{0.02cm}{\bf k}_{1}} - {\mathbf v}\cdot({\bf k} - {\bf k}_{1})).
\]
Let us introduce the total number of ``colorless'' plasmons setting by the definition
\begin{equation}
\mathbbm{N}^{\hspace{0.03cm}l}(t) \equiv 
\!\int\!d\hspace{0.02cm}{\bf k}\,N^{\hspace{0.03cm}l}_{\bf k}(t).
\label{eq:10w}
\end{equation}
With the help of (\ref{eq:10q}) and the fact that the square of the amplitude modulus is an even function with respect to the permutation of momenta ${\bf k}$ and ${\bf k}_{1}$, we easily find 
\begin{equation}
\frac{d\hspace{0.04cm}\mathbbm{N}^{\hspace{0.03cm}l}(t)}{d\hspace{0.02cm}t} 
=
2\,\frac{N_{c}}{d_{A}}\,{\mathfrak q}_{2}(t)
\!\int\!d\hspace{0.02cm}{\bf k}\hspace{0.04cm} 
\bigl({\rm Im}\hspace{0.03cm}\mathscr{T}_{\,{\bf k},
\hspace{0.03cm}{\bf k}}\bigr)\hspace{0.01cm}
W^{\hspace{0.03cm}l}_{\bf k}(t),
\label{eq:10e}
\end{equation}
i.e. this total number will be conserved only in the absence of dissipative effects when ${\rm Im}\hspace{0.03cm}\mathscr{T}_{\,{\bf k},
\hspace{0.03cm}{\bf k}} = 0$ (see the next section).\\
\indent Furthermore, if we introduce into consideration the total energy and momentum of the wave system of colorless plasmons
\begin{equation}
\mathbbm{E}^{\hspace{0.03cm}l}(t) \equiv \!\int\!d\hspace{0.02cm}{\bf k}\;
\omega^{\hspace{0.02cm}l}_{\hspace{0.02cm}{\bf k}}\hspace{0.03cm} N^{\hspace{0.03cm}l}_{\bf k}(t)
\quad\mbox{and}\quad
{\bf K}^{\hspace{0.02cm}l}(t) \equiv \!\int\!d\hspace{0.03cm}{\bf k}\,
{\bf k}\hspace{0.03cm} N^{\hspace{0.03cm}l}_{\bf k}(t),
\label{eq:10r}
\end{equation}
then by virtue of the same kinetic equation we have
\begin{equation}
\frac{d\hspace{0.02cm}\bigl(\mathbbm{E}^{\hspace{0.03cm}l}- {\bf v}\cdot{\bf K}^{\hspace{0.02cm}l}\bigr)(t)}{d\hspace{0.03cm}t} 
=
2\,\frac{N_{c}}{d_{A}}\,{\mathfrak q}_{2}(t)
\!\int\!d\hspace{0.02cm}{\bf k}\, 
(\omega^{\hspace{0.02cm}l}_{\hspace{0.02cm}{\bf k}} - {\bf v}\cdot{\bf k})
\bigl({\rm Im}\hspace{0.03cm}\mathscr{T}_{\,{\bf k}, 
\hspace{0.03cm}{\bf k}}\bigr)\hspace{0.01cm}
W^{\hspace{0.03cm}l}_{\bf k}(t). 
\label{eq:10t}
\end{equation}
Here, we see that the quantity $\bigl(\mathbbm{E}^{\hspace{0.03cm}l}- {\bf v}\cdot{\bf K}^{\hspace{0.02cm}l}\bigr)(t)$ is also not conserved due to the dissipative effects.\\
\indent Finally, let us consider the time dependence of the entropy of the colorless longitudinally polarized soft gluon excitations 
\[
\mathbbm{S}^{\hspace{0.03cm}l}(t) = \!\int\!d\hspace{0.02cm}{\bf k}\hspace{0.04cm}\ln\hspace{0.03cm}\!N^{\hspace{0.03cm}l}_{\bf k}(t).
\]
The time derivative by virtue of (\ref{eq:10q}) gives us
\begin{equation}
\frac{d\hspace{0.03cm} \mathbbm{S}^{\hspace{0.03cm}l}(t)} {\!\!d\hspace{0.03cm} t}
=
2\,\frac{N_{c}}{d_{A}}\,{\mathfrak q}_{2}(t)
\!\int\!d\hspace{0.02cm}{\bf k}\hspace{0.04cm} 
\bigl({\rm Im}\hspace{0.03cm}\mathscr{T}_{\,{\bf k},
\hspace{0.03cm}{\bf k}}\bigr)\hspace{0.03cm}
\frac{W^{\hspace{0.03cm}l}_{\bf k}}{N^{\hspace{0.03cm}l}_{\bf k}}
\label{eq:10y}
\end{equation}
\vspace{-0.6cm}
\begin{align}
+\;
{\mathfrak q}_{2}(t)N_{c}\!
&\int\!d\hspace{0.03cm}{\bf k}\hspace{0.03cm}d\hspace{0.03cm}{\bf k}_{1} \hspace{0.04cm}
\bigl|\mathscr{T}_{\,{\bf k}, 
\hspace{0.03cm}{\bf k}_{1}}\bigr|^{\hspace{0.02cm}2}\,
\frac{\bigl(N^{\hspace{0.03cm}l}_{{\bf k}} - N^{\hspace{0.03cm}l}_{{\bf k}_{1}}\bigr)^{\!\hspace{0.02cm}2}}{N^{\hspace{0.03cm}l}_{\bf k}\hspace{0.02cm}N^{\hspace{0.03cm}l}_{{\bf k}_{1}}}
\hspace{0.03cm}
(2\pi)\,\delta(\omega^{\hspace{0.02cm}l}_{\hspace{0.02cm}{\bf k}}\! - \omega^{\hspace{0.02cm}l}_{\hspace{0.02cm}{\bf k}_{1}}\! - {\mathbf v}\cdot({\bf k} - {\bf k}_{1}))
\notag\\[1.5ex]
-\,
\frac{1}{2}\,{\mathfrak q}_{2}(t)\hspace{0.03cm}N^{\hspace{0.03cm}2}_{c}
\!\!
&\int\!d\hspace{0.03cm}{\bf k}\hspace{0.03cm}
d\hspace{0.03cm}{\bf k}_{1}\hspace{0.04cm}
\bigl|\mathscr{T}_{\,{\bf k}, 
\hspace{0.03cm}{\bf k}_{1}}\bigr|^{\hspace{0.02cm}2}\hspace{0.03cm}
\biggl(\frac{1}{N^{\hspace{0.03cm}l}_{\bf k}}
-
\frac{1}{N^{\hspace{0.03cm}l}_{{\bf k}_{1}}}\biggr)
\bigl(W^{\hspace{0.03cm}l}_{\bf k}\hspace{0.02cm}N^{\hspace{0.03cm}l}_{{\bf k}_{1}}\! 
-
N^{\hspace{0.03cm}l}_{\bf k}\hspace{0.03cm}W^{\hspace{0.03cm}l}_{{\bf k}_{1}}\bigr)
\hspace{0.03cm}
(2\pi)\,\delta(\omega^{\hspace{0.02cm}l}_{\hspace{0.02cm}{\bf k}} - \omega^{\hspace{0.02cm}l}_{\hspace{0.02cm}{\bf k}_{1}} - {\mathbf v}\cdot({\bf k} - {\bf k}_{1})).
\notag
\end{align}
While the second term is positive, the third term mixed in $N^{\hspace{0.03cm}l}_{\bf k}$ and $W^{\hspace{0.03cm}l}_{\bf k}$, is indefinite. This can be seen more clearly from the following rewriting
\[
\biggl(\frac{1}{N^{\hspace{0.03cm}l}_{\bf k}}
-
\frac{1}{N^{\hspace{0.03cm}l}_{{\bf k}_{1}}}\biggr)
\bigl(W^{\hspace{0.03cm}l}_{\bf k}\hspace{0.02cm}N^{\hspace{0.03cm}l}_{{\bf k}_{1}} 
-
N^{\hspace{0.03cm}l}_{\bf k}\hspace{0.03cm}W^{\hspace{0.03cm}l}_{{\bf k}_{1}}\bigr)
=
N^{\hspace{0.03cm}l}_{\bf k}\hspace{0.02cm}N^{\hspace{0.03cm}l}_{{\bf k}_{1}}
\biggl(\frac{1}{N^{\hspace{0.03cm}l}_{\bf k}}
-
\frac{1}{N^{\hspace{0.03cm}l}_{{\bf k}_{1}}}\biggr)
\biggl(\frac{W^{\hspace{0.03cm}l}_{\bf k}}{N^{\hspace{0.03cm}l}_{\bf k}}
-
\frac{W^{\hspace{0.03cm}l}_{{\bf k}_{1}}}{N^{\hspace{0.03cm}l}_{{\bf k}_{1}}}\biggr).
\]
As is known \cite{zakharov_book_1992}, because the kinetic equations obtained after the averaging procedure describe an irreversible evolution toward thermodynamic equilibrium, the entropy of a {\it closed} wave system can only increase. The mathematical statement of irreversibility is the theorem of entropy growth which is similar to Boltzmann's H-theorem for gas kinetics \cite{balescu_book_1975}. In our case, even in the absence of dissipation, due to the presence of the indefinite term in (\ref{eq:10y}), we cannot state unambiguously that
\[
\frac{d\hspace{0.03cm} \mathbbm{S}^{\hspace{0.03cm}l}(t)} {\!\!d\hspace{0.03cm} t} > 0.
\]
The presence of an external hard particle in the wave system under consideration breaks its closure.\\
\indent Furthermore, the equation for the color part $W^{\hspace{0.03cm}l}_{\bf k}$ of the plasmon number density is given by the expression (\ref{eq:8i}). Let us contract the left- and right-hand sides of this equation with  $\bigl\langle\hspace{0.01cm}{\mathcal Q}^{\hspace{0.03cm}s}\hspace{0.03cm}\bigr\rangle$. Taking into account the evolution equation for the averaged color charge, Eq.\,(\ref{eq:9u}), the  definition of colorless charge combinations $\mathfrak{q}_{2}(t)$ and $\mathfrak{q}_{4}(t)$, Eqs.\,(\ref{eq:7t}) and (\ref{eq:9i}), and canceling the left- and right-hand sides by the product $N_{c}\hspace{0.04cm}{\mathfrak q}_{2}(t)$, the equation for function $W^{\hspace{0.03cm}l}_{\bf k}$ can be cast in
\begin{equation}
\frac{\partial\hspace{0.03cm} W^{\hspace{0.03cm}l}_{\bf k}}{\!\!\partial\hspace{0.03cm} t}
=
-\hspace{0.03cm}\frac{1}{2}\,A(t)\hspace{0.03cm}W^{\hspace{0.03cm}l}_{\bf k}
+
2\hspace{0.03cm}\hspace{0.03cm}
\bigl({\rm Im}\hspace{0.03cm}\mathscr{T}_{{\bf k}, {\bf k}}\bigr)
\hspace{0.03cm}N^{\hspace{0.03cm}l}_{\bf k}
\label{eq:10u}
\end{equation}
\[
-\!
\int\!d\hspace{0.02cm}{\bf k}_{1}\,
\bigl|\mathscr{T}_{\,{\bf k},\hspace{0.03cm}
{\bf k}_{1}}\bigr|^{\hspace{0.02cm}2}\,
\Bigl\{\uprho\hspace{0.04cm}{\mathfrak q}_{2}(t)
\bigl(W^{\,l}_{{\bf k}} - W^{\,l}_{{\bf k}_{1}}\bigr)
-
\frac{1}{2}\,N_{c}\hspace{0.04cm}
N^{\,l}_{{\bf k}}\hspace{0.02cm}N^{\,l}_{{\bf k}_{1}}\Bigr\}
\hspace{0.03cm}
(2\pi)\,\delta(\omega^{l}_{{\bf k}} - \omega^{l}_{{\bf k}_{1}} - {\mathbf v}\cdot({\bf k} - {\bf k}_{1})),
\]
where the coefficient $\uprho$ in braces by virtue of representations for the functions $\mathfrak{q}_{2}(t)$ and $\mathfrak{q}_{4}(t)$, Eq.\,(\ref{eq:9o}), has the form
\[
\uprho \equiv 
\frac{2}{N_{c}}
+ \frac{1}{4}\,\frac{{\mathfrak q}_{4}(t)}{({\mathfrak q}_{2}(t))^{\hspace{0.02cm}2}}
=
\frac{2}{N_{c}}
+ \frac{1}{4}\frac{{\mathfrak q}_{4}(t_{0})}{({\mathfrak q}_{2}(t_{0}))^{\hspace{0.02cm}2}}.
\]
Thus, the coefficient $\uprho$ is a constant that depends in a complex way on the initial value of the color charge ${\mathcal Q}^{\hspace{0.03cm}a}_{\hspace{0.03cm}0}$. The latter, we recall, is a given deterministic vector in the effective color space. For the special case $N_{c} = 2$, when $d^{\hspace{0.03cm}a\hspace{0.03cm}b\hspace{0.03cm}c}\equiv 0$ and, as a consequence ${\mathfrak q}_{4}(t_{0})\equiv 0$, we have
\[
\uprho = 1.
\]
In another, more nontrivial special case $N_{c} = 3$, by virtue of the definition (\ref{eq:9i}) and the property (\ref{ap:D13}), we find
\[
{\mathfrak q}_{4}(t_{0}) = 
d^{\hspace{0.03cm}a\hspace{0.03cm}b\hspace{0.03cm}c}
d^{\hspace{0.03cm}a\hspace{0.03cm}b^{\hspace{0.02cm}\prime}
c^{\hspace{0.02cm}\prime}}
\langle\hspace{0.01cm}{\mathcal Q}^{\hspace{0.03cm}b}_{\hspace{0.03cm}0}\hspace{0.03cm}\bigr\rangle
\langle\hspace{0.01cm}{\mathcal Q}^{\hspace{0.03cm}c}_{\hspace{0.03cm}0}\hspace{0.03cm}\bigr\rangle
\langle\hspace{0.01cm}{\mathcal Q}^{\hspace{0.03cm}b^{\hspace{0.02cm}\prime}}_{\hspace{0.03cm}0}\hspace{0.03cm}\bigr\rangle
\langle\hspace{0.01cm}{\mathcal Q}^{\hspace{0.03cm}c^{\hspace{0.02cm}\prime}}_{\hspace{0.03cm}0}\hspace{0.03cm}\bigr\rangle
=
\frac{1}{3}\,\bigl(
\langle\hspace{0.01cm}\mathcal{Q}^{\hspace{0.03cm}a}_{\hspace{0.03cm}0}
\hspace{0.03cm}\bigr\rangle
\langle\hspace{0.01cm}\mathcal{Q}^{\hspace{0.03cm}a}_{\hspace{0.03cm}0}
\hspace{0.03cm}\bigr\rangle\bigr)^{2}
\equiv
\frac{1}{3}\,({\mathfrak q}_{2}(t_{0}))^{\hspace{0.02cm}2}
\]
and as a result one obtains
\[
\uprho = \frac{2}{3} + \frac{1}{12} = \frac{3}{4}.
\]
Let us write these two cases together 
\[
\uprho = \left\{\!
\begin{array}{rl}
1, & \mbox{if}\; N_{c} = 2 \\[1ex]
3/4, & \mbox{if}\; N_{c} = 3.
\end{array}
\right.	
\]
Only for these particular values the quantity $\uprho$ does not depend on the initial value of the color charge. Equation (\ref{eq:10u}) must be added to (\ref{eq:10q}), and thus we obtain a self-consistent system of kinetic equations defining the evolution of the colorless and color parts of the plasmon number density in a hot gluon plasma, during the interaction of soft boson excitations with a hard color-charged particle.\\
\indent In conclusion of this section we note that for the kinetic equation (\ref{eq:10u}) there are no conservation laws (in the absence of dissipative effects) similar to (\ref{eq:10e}) or (\ref{eq:10t}). However, there is one nontrivial relation of another type here. Let us introduce into consideration the quantity
\begin{equation}
\mathbbm{W}^{\hspace{0.02cm}l}(t) 
\equiv 
\!\int\!d\hspace{0.02cm}{\bf k}\,W^{\hspace{0.02cm}l}_{\bf k}(t).
\label{eq:10o}
\end{equation}
We drop the term with the imaginary part of the scattering amplitude in equation (\ref{eq:10u}) and integrate it over ${\bf k}$. The term with the difference $W^{\,l}_{{\bf k}} - W^{\,l}_{{\bf k}_{1}}$ turns to zero, and we are left with
\[
\frac{d\hspace{0.04cm}\mathbbm{W}^{\hspace{0.03cm}l}(t)}{d\hspace{0.02cm}t} 
=
-\hspace{0.03cm}\frac{1}{2}\,A(t)\hspace{0.03cm}\mathbbm{W}^{\hspace{0.03cm}l}_{\bf k}
\hspace{0.03cm}+\hspace{0.03cm}
\frac{1}{2}\,N_{c}\!
\int\!d\hspace{0.03cm}{\bf k}\hspace{0.04cm}
d\hspace{0.03cm}{\bf k}_{1}\hspace{0.03cm}
\bigl|\mathscr{T}_{\,{\bf k},\hspace{0.03cm}
	{\bf k}_{1}}\bigr|^{\hspace{0.02cm}2}\hspace{0.02cm}
N^{\hspace{0.03cm}l}_{{\bf k}}\hspace{0.02cm}N^{\hspace{0.03cm}l}_{{\bf k}_{1}}
\hspace{0.03cm}
(2\pi)\,\delta(\omega^{l}_{{\bf k}} - \omega^{l}_{{\bf k}_{1}} - {\mathbf v}\cdot({\bf k} - {\bf k}_{1}))
\] 
or by virtue of the definition of the function $A(t)$, Eq.\,(\ref{eq:9y}),
\[
\frac{d\hspace{0.04cm}\mathbbm{W}^{\hspace{0.03cm}l}(t)}{d\hspace{0.02cm}t} 
=
\frac{1}{2\hspace{0.01cm}N_{c}}\,A(t)\hspace{0.03cm}\bigl\{1 - N_{c}\hspace{0.03cm}\mathbbm{W}^{\hspace{0.03cm}l}(t)\bigr\}.
\]
It then follows trivially from equation (\ref{eq:9t}) and the definition of the function ${\mathfrak q}_{2}(t)$ that 
\[
A(t) = \frac{d\hspace{0.01cm}\ln{\mathfrak q}_{2}(t)}{\!\!d\hspace{0.03cm}t}
\] 
and therefore, the previous equation can also be represented as
\[
-\hspace{0.02cm}2\hspace{0.03cm}d\ln\hspace{0.02cm}\bigl\{1 - N_{c}\hspace{0.03cm}\mathbbm{W}^{\hspace{0.03cm}l}(t)\bigr\}
=
d\ln{\mathfrak q}_{2}(t). 
\] 
A simple integration results in the desired relation
\begin{equation}
\biggl(\frac{1 - N_{c}\hspace{0.03cm}\mathbbm{W}^{\hspace{0.02cm}l}(t_{0})}
{\!1 - N_{c}\hspace{0.04cm}\mathbbm{W}^{\hspace{0.03cm}l}\hspace{0.01cm}(t)}\biggr)^{\!\!2}
=
\frac{\!{\mathfrak q}_{2}(t)}{{\mathfrak q}_{\hspace{0.02cm}2}(t_{0})}.
\label{eq:10p}
\end{equation}
This relation connects the behavior of the colorless quadratic combination ${\mathfrak q}_{2}(t)$ of the averaged color charge of a hard particle with the behavior of the integral characteristic (\ref{eq:10o}) of the color part $W^{\hspace{0.02cm}l}_{\bf k}$ of the number density of longitudinal soft gluon excitations.\\
\indent The physical meaning of the expression (\ref{eq:10p}) will become more obvious if we introduce, by analogy with the works of \cite{luscher_1978, reinhardt_2011, cruz_2020, gonzo_2021}, the notion of a color charge for soft gluon excitations, which in terms of the normal field variables reads
\[
\mathfrak{Q}^{\hspace{0.03cm}a}
\equiv
i\hspace{0.03cm}
f^{\hspace{0.03cm}a\hspace{0.02cm}b\hspace{0.03cm}c}\!
\int\!d\hspace{0.02cm}{\bf k}\, 
c^{\hspace{0.03cm}\ast\hspace{0.03cm}b}_{\hspace{0.02cm}{\bf k}}\hspace{0.03cm} c^{\!\!\phantom{\ast}c}_{{\bf k}}.
\]
Taking into account the definition of the correlation function (\ref{eq:5r}), the decomposition (\ref{eq:7q}) and the definition (\ref{eq:10o}), we obtain for the averaged value of the color charge of the gauge field
\[
\langle\hspace{0.02cm}\mathfrak{Q}^{\hspace{0.03cm}a}\hspace{0.01cm}
\bigr\rangle
=
N_{c}\hspace{0.04cm}\bigl\langle\hspace{0.01cm}\mathcal{Q}^{\hspace{0.03cm}a}\hspace{0.03cm}\bigr\rangle\hspace{0.03cm}\mathbbm{W}^{\hspace{0.02cm}l}(t).
\]
If we define, similar to (\ref{eq:7t}), the colorless quadratic combination of the averaged color charge 
\[
\mathbbm{q}_{2}(t) \equiv \langle\hspace{0.02cm}\mathfrak{Q}^{\hspace{0.03cm}a}\hspace{0.01cm}
\bigr\rangle
\langle\hspace{0.02cm}\mathfrak{Q}^{\hspace{0.03cm}a}\hspace{0.01cm}
\bigr\rangle,
\]
then the relation (\ref{eq:10p}) can be rewritten in somewhat different form
\[
\bigl(\sqrt{{\mathfrak q}_{2}(t)} - \sqrt{\mathbbm{q}_{2}(t)}\,\bigr)^{2} = \mbox{const}.
\]
This expression can be interpreted as a sort of averaged color charge conservation of the constituent system: a hot gluon plasma and an external hard color particle.


\section{\bf Interaction of infinitely narrow packets}
\label{section_11}
\setcounter{equation}{0}

To get some understanding of the behavior of the solution of the system of kinetic equations (\ref{eq:10q}) and (\ref{eq:10u}), we consider the model problem of the interaction of two infinitely narrow packets with typical wavevectors ${\bf k}_{\hspace{0.02cm}0}$ and ${\bf k}^{\hspace{0.02cm}\prime}_{\hspace{0.02cm}0}$.
We have already considered the problem of this type earlier in the works \cite{markov_2000, markov_2001}. The kinetic equation defining the process of elastic scattering of colorless plasmons off each other was analyzed in the paper \cite{markov_2000}. A self-consistent system of two kinetic equations defining nonlinear interaction between plasminos and plasmons was considered in \cite{markov_2001}. Due to the relative simplicity of the initial equations written out in these papers, exact solutions of the reduced kinetic equations were obtained. This allowed us to explicitly define the kinematic relations between the wavevectors of excitations determining a direction of the effective spectral pumping of the plasma excitation energy across the spectrum towards small wavenumbers with complete conservation of the excitation energy for the collision process among plasmons and the effective spectral pumping from the fermion branch of plasma excitations to the boson branch and vice versa for the process of scattering of plasminos and plasmons off each other. We note that the approach of this type for determining a direction of the effective spectral pumping of excitation energy is used in the weak wave turbulence theory of an Abelian (i.e electron-ion) plasma.\\
\indent Let us introduce the scalar plasmon number densities $N^{\hspace{0.03cm}l}_{{\bf k}}$ and $W^{\hspace{0.03cm}l}_{{\bf k}}$ as follows
\begin{equation}
\begin{split}
&N^{\hspace{0.03cm}l}_{\bf k}(t) = N_{1}(t)\hspace{0.03cm} 
\delta({\bf k} - {\bf k}_{\hspace{0.02cm}0}) +
N_{2}(t)\hspace{0.03cm}\delta({\bf k} - 
{\bf k}^{\hspace{0.02cm}\prime}_{\hspace{0.02cm}0}), 
\\[1.5ex] 
&W^{\hspace{0.03cm}l}_{\bf k}(t) = W_{1}(t)\hspace{0.03cm} 
\delta({\bf k} -{\bf k}_{\hspace{0.02cm}0}) +
W_{2}(t)\hspace{0.03cm}\delta({\bf k} - {\bf k}^{\hspace{0.02cm}\prime}_{\hspace{0.02cm}0})
\end{split}
\label{eq:11q}
\end{equation}
at that ${\bf k}_{\hspace{0.02cm}0}\!\neq\!{\bf k}^{\hspace{0.02cm}\prime}_{\hspace{0.02cm}0}$, i.e. the longitudinal wave packets do not overlap. In particular, from this representation for the function $N^{\hspace{0.03cm}l}_{\bf k}(t)$ and from the definitions (\ref{eq:10w}) and (\ref{eq:10r}), immediately follows
\begin{equation}
\begin{split}
&\mathbbm{N}^{\hspace{0.03cm}l}(t) =  N_{1}(t) +  N_{2}(t),\\[1.5ex] 	
\bigl(\mathbbm{E}^{\hspace{0.03cm}l} - 
{\bf v}\cdot{\bf K}^{\hspace{0.03cm}l}\bigr)(t)
&=
(\omega^{\hspace{0.02cm}l}_{\hspace{0.02cm}{\bf k}_{\hspace{0.02cm}0}} - {\bf v}\cdot{\bf k}_{\hspace{0.02cm}0})N_{1}(t)
+
(\omega^{\hspace{0.02cm}l}_{\hspace{0.02cm}{\bf k}^{\hspace{0.02cm}\prime}_{\hspace{0.02cm}0}} - {\bf v}\cdot{\bf k}^{\hspace{0.02cm}\prime}_{\hspace{0.02cm}0})N_{2}(t).
\end{split}
\label{eq:11w}
\end{equation}
Let us substitute (\ref{eq:11q}) into the left- and right-hand sides of equations (\ref{eq:10q}) and (\ref{eq:10u}). We obtain the coupled nonlinear equations, which we write as follows: the first equation for the functions $N_{1}(t)$ and $N_{2}(t)$
\begin{equation}
\frac{d N_{1}(t)}{d\hspace{0.03cm}t}\,\delta({\bf k} - {\bf k}_{\hspace{0.02cm}0})
+
\frac{d N_{2}(t)}{d\hspace{0.03cm}t}\,\delta({\bf k} - {\bf k}^{\hspace{0.02cm}\prime}_{\hspace{0.02cm}0})
\label{eq:11e}
\end{equation}
\vspace{-0.6cm}
\begin{align}
=\; &\biggl\{A_{11}\hspace{0.02cm}N_{1} + A_{13}\hspace{0.02cm}W_{1}
- 
\frac{N_{c}}{d_{A}}\,{\mathfrak q}_{2}(t)B\hspace{0.03cm}\bigl(N_{1}W_{2} - N_{2}\hspace{0.03cm}W_{1}\bigr)\biggr\}\hspace{0.03cm}
\delta({\bf k} - {\bf k}_{\hspace{0.02cm}0})
\notag\\
+\,
&\biggl\{A_{22}\hspace{0.02cm}N_{2} + A_{24}\hspace{0.02cm}W_{2}
+ 
\frac{N_{c}}{d_{A}}\,{\mathfrak q}_{2}(t)B\hspace{0.03cm}\bigl(N_{1}W_{2} - N_{2}\hspace{0.03cm}W_{1}\bigr)\biggr\}\hspace{0.03cm}
\delta({\bf k} - {\bf k}^{\hspace{0.02cm}\prime}_{\hspace{0.02cm}0})
\notag
\end{align}
\[	
+\;\frac{N_{c}}{d_{A}}\,
{\mathfrak q}_{2}(t)\hspace{0.03cm}
\Bigl\{\hspace{0.03cm}\bigl|\mathscr{T}_{\hspace{0.04cm}{\bf  k},\hspace{0.03cm} 
	{\bf k}_{\hspace{0.02cm}0}}\bigr|^{\hspace{0.02cm}2}\,
(2\pi)\,\delta(\Delta\hspace{0.02cm}\omega_{\,{\bf k},
\hspace{0.03cm}{\bf k}_{\hspace{0.02cm}0}})\hspace{0.02cm}N_{1}
+
\bigl|\mathscr{T}_{\hspace{0.04cm}{\bf  k},\hspace{0.03cm} 
{\bf k}^{\hspace{0.02cm}\prime}_{\hspace{0.02cm}0}}\bigr|^{\hspace{0.02cm}2}\,
(2\pi)\,\delta(\Delta\hspace{0.02cm}\omega_{\,{\bf k},
\hspace{0.03cm}{\bf k}^{\hspace{0.02cm}\prime}_{\hspace{0.02cm}0}})\hspace{0.02cm}N_{2}\Bigr\},
\]
and the second equation for the functions $W_{1}(t)$ and $W_{2}(t)$
\begin{equation}
	\frac{d\hspace{0.03cm}W_{1}(t)}{d\hspace{0.03cm}t}\,\delta({\bf k} - {\bf k}_{\hspace{0.02cm}0})
	+
	\frac{d\hspace{0.03cm}W_{2}(t)}{d\hspace{0.03cm}t}\,\delta({\bf k} - {\bf k}^{\hspace{0.02cm}\prime}_{\hspace{0.02cm}0})
	\label{eq:11r}
\end{equation}
\vspace{-0.6cm}
\begin{align}
=\; &\biggl\{A_{31}\hspace{0.02cm}N_{1} + A_{33}\hspace{0.02cm}W_{1}
-\hspace{0.03cm}\displaystyle\frac{1}{2}\,A(t)\hspace{0.02cm}W_{1}
+ 
B\hspace{0.01cm}N_{1}N_{2}
+
\frac{1}{2}\,N_{c}\hspace{0.03cm}
\bigl|\mathscr{T}_{\hspace{0.04cm}{\bf  k}_{\hspace{0.02cm}0},\hspace{0.03cm} 
{\bf k}_{\hspace{0.02cm}0}}\bigr|^{\hspace{0.02cm}2}\,
(2\pi)\,\delta(\Delta\hspace{0.02cm}\omega_{\,{\bf k}_{\hspace{0.02cm}0},
\hspace{0.03cm}{\bf k}_{\hspace{0.02cm}0}})\hspace{0.02cm}N^{\hspace{0.03cm}2}_{1}	
\biggr\}\hspace{0.03cm}
\delta({\bf k} - {\bf k}_{\hspace{0.02cm}0})
\notag\\
+\,
&\biggl\{A_{42}\hspace{0.02cm}N_{2} + A_{44}\hspace{0.02cm}W_{2}
-\hspace{0.03cm}\displaystyle\frac{1}{2}\,A(t)\hspace{0.02cm}W_{2}
+ 
B\hspace{0.02cm}N_{1}N_{2}
+
\frac{1}{2}\,N_{c}\hspace{0.03cm}
\bigl|\mathscr{T}_{\hspace{0.04cm}{\bf  k}^{\hspace{0.02cm}\prime}_{\hspace{0.02cm}0},\hspace{0.03cm} 
{\bf k}^{\hspace{0.02cm}\prime}_{\hspace{0.02cm}0}}\bigr|^{\hspace{0.02cm}2}\,
(2\pi)\,\delta(\Delta\hspace{0.02cm}\omega_{\,{\bf k}^{\hspace{0.02cm}\prime}_{\hspace{0.02cm}0},
\hspace{0.03cm}{\bf k}^{\hspace{0.02cm}\prime}_{\hspace{0.02cm}0}})\hspace{0.02cm}N^{\hspace{0.03cm}2}_{2}	
\biggr\}\hspace{0.03cm}
\delta({\bf k} - {\bf k}^{\hspace{0.02cm}\prime}_{\hspace{0.02cm}0})
\notag
\end{align}
\[
+\, 
\uprho\hspace{0.05cm}
{\mathfrak q}_{2}(t)\hspace{0.03cm}
\Bigl\{\hspace{0.03cm}\bigl|\mathscr{T}_{\hspace{0.04cm}{\bf  k},\hspace{0.03cm} 
{\bf k}_{\hspace{0.02cm}0}}\bigr|^{\hspace{0.02cm}2}\,
(2\pi)\,\delta(\Delta\hspace{0.02cm}\omega_{\,{\bf k},\hspace{0.03cm}
{\bf k}_{\hspace{0.02cm}0}})\hspace{0.02cm}W_{1}
+
\bigl|\mathscr{T}_{\hspace{0.04cm}{\bf  k}, {\bf k}^{\hspace{0.02cm}\prime}_{\hspace{0.02cm}0}}\bigr|^{\hspace{0.02cm}2}\,
(2\pi)\,\delta(\Delta\hspace{0.02cm}\omega_{\,{\bf k}, \hspace{0.03cm}\hspace{0.03cm}
{\bf k}^{\hspace{0.02cm}\prime}_{\hspace{0.02cm}0}})\hspace{0.02cm}W_{2}\Bigr\},
\]
where, we recall
\begin{equation}
	\Delta\hspace{0.02cm}\omega_{\,{\bf k},\hspace{0.03cm}{\bf k}_{1}} 
	\equiv
	\omega^{\hspace{0.02cm}l}_{\hspace{0.03cm}{\bf k}} - 
	\omega^{\hspace{0.02cm}l}_{\hspace{0.03cm}{\bf k}_{1}} - {\bf v}\cdot({\bf k} - {\bf k}_{1}).
	\label{eq:11t}
\end{equation}
Nonzero ``matrix elements'' $A_{ij},\,i,j = 1,\ldots,4$ are defined by the following expressions 
\[
\begin{array}{lll}
&A_{31} = 2\hspace{0.03cm}\hspace{0.03cm}
\bigl({\rm Im}\hspace{0.03cm}\mathscr{T}_{\hspace{0.04cm}{\bf k}_{\hspace{0.02cm}0}, \hspace{0.03cm}{\bf k}_{\hspace{0.02cm}0}}\bigr),
\qquad\;
&A_{13} = \displaystyle\frac{N_{c}}{d_{A}}\,
{\mathfrak q}_{2}(t)\hspace{0.03cm}A_{31},\\[2ex]
&A_{42} = 2\hspace{0.03cm}\hspace{0.03cm}
\bigl({\rm Im}\hspace{0.03cm}\mathscr{T}_{\hspace{0.04cm}{\bf k}^{\hspace{0.02cm}\prime}_{\hspace{0.02cm}0}, 
\hspace{0.03cm}{\bf k}^{\hspace{0.02cm}\prime}_{\hspace{0.02cm}0}}\bigr),
\qquad\;
&A_{24} = \displaystyle\frac{N_{c}}{d_{A}}\,
{\mathfrak q}_{2}(t)\hspace{0.03cm}A_{42},\\[2ex]
&A_{11} = -\hspace{0.03cm}\displaystyle\frac{N_{c}}{d_{A}}\,{\mathfrak q}_{2}(t)\!
\int\!d\hspace{0.02cm}{\bf k}\,
\bigl|\mathscr{T}_{\hspace{0.04cm}{\bf k}_{\hspace{0.02cm}0},\hspace{0.03cm} 
{\bf k}}\bigr|^{\hspace{0.02cm}2}\hspace{0.03cm}
(2\pi)\,\delta(\Delta\hspace{0.02cm}\omega_{\,{\bf k}_{\hspace{0.02cm}0},
\hspace{0.03cm}{\bf k}}),
\qquad\;
&A_{33} = 
 \uprho\hspace{0.03cm}\displaystyle\frac{d_{A}}{N_{c}} A_{11},\\[2ex]
&A_{22} = -\hspace{0.03cm}\displaystyle\frac{N_{c}}{d_{A}}\,
{\mathfrak q}_{2}(t)\!
\int\!d\hspace{0.02cm}{\bf k}\,
\bigl|\mathscr{T}_{\hspace{0.04 cm}{\bf k}^{\hspace{0.02cm}\prime}_{\hspace{0.02cm}0},\hspace{0.03cm} 
{\bf k}}\bigr|^{\hspace{0.02cm}2}\hspace{0.03cm}
(2\pi)\,\delta(\Delta\hspace{0.02cm}\omega_{\,{\bf k}^{\hspace{0.02cm}\prime}_{\hspace{0.02cm}0},\hspace{0.03cm}{\bf k}}),
\qquad\;
&A_{44} = 
   \uprho\hspace{0.03cm}\displaystyle\frac{d_{A}}{N_{c}} A_{22}.
	\end{array}
\]
Finally, the coefficient $B$ is given by the following expression
\begin{equation}
B = \frac{1}{2}\,N_{c}\,
{\mathfrak q}_{2}(t)\hspace{0.03cm}
\bigl|\mathscr{T}_{\hspace{0.04cm}{\bf k}^{\phantom{\prime}}_{\hspace{0.02cm}0},
	\hspace{0.03cm} 
{\bf k}^{\hspace{0.02cm}\prime}_{\hspace{0.02cm}0}}\bigr|^{\hspace{0.02cm}2}
 \,
(2\pi)\,\delta(\Delta\hspace{0.02cm}\omega_{\,{\bf k}^{\phantom{\prime}}_{\hspace{0.02cm}0},\hspace{0.03cm}
{\bf k}^{\hspace{0.02cm}\prime}_{\hspace{0.02cm}0}}).
\label{eq:11u}
\end{equation}
Note, however, that this coefficient is, generally speaking, {\it a generalized} function. The notations for the matrix elements $A_{ij}$ are chosen to correspond to the usual matrix multiplication of the matrix $\mathcal{A}\equiv (A_{ij})$ by some effective vector ${\bf \Lambda}$, composed of the desired functions as follows 
\[
{\bf \Lambda}
=
\left(\!\!\!\!\!\!
\begin{array}{ll}
&\Lambda_{1}\\
&\Lambda_{2}\\
&\Lambda_{3}\\
&\Lambda_{4}	
\end{array}
\!\right)
\equiv
\left(\!\!\!\!\!\!
\begin{array}{ll}
	&N_{1}\\
	&N_{2}\\
	&W_{1}\\
	&W_{2}\\
\end{array}
\!\right).
\] 
From the structure of the right-hand sides of equations (\ref{eq:11e}) and (\ref{eq:11r}) we see that the only last terms in these equations do not allow us to separate the contributions proportional to the $\delta$-functions: $\delta({\bf k} - {\bf k}_{\hspace{0.02cm}0})$ and $\delta({\bf k} - {\bf k}^{\hspace{0.02cm}\prime}_{\hspace{0.02cm}0})$, as was the case, for example, in \cite{markov_2000, markov_2001} for similar problems of interaction of infinitely narrow packets. In addition, (\ref{eq:11r}) contains terms with badly defined products such as  $\bigl|\mathscr{T}_{\hspace{0.04cm}{\bf k}_{\hspace{0.02cm}0}, {\bf k}_{\hspace{0.02cm}0}}\bigr|^{\hspace{0.02cm}2}\,
(2\pi)\,\delta(\Delta\hspace{0.02cm}\omega_{\,{\bf k}_{\hspace{0.02cm}0},
{\bf k}_{\hspace{0.02cm}0}})$, which in further reasoning we simply discard.\\
\indent Then we can proceed as follows. Let us integrate expressions (\ref{eq:11e}) and (\ref{eq:11r}) over ${\bf k}$. By virtue of our requirement ${\bf k}_{\hspace{0.02cm}0} \neq {\bf k}^{\hspace{0.02cm}\prime}_{\hspace{0.02cm}0}$ we can choose the integration region around the wave vectors ${\bf k}_{\hspace{0.02cm}0}$ and ${\bf k}^{\hspace{0.02cm}\prime}_{\hspace{0.02cm}0}$ so that they do not overlap. In this case, from (\ref{eq:11e}) and (\ref{eq:11r}), we obtain a system of four nonlinear first-order ordinary differential equations
\begin{equation}
\begin{split}
&\frac{d N_{1}(t)}{d\hspace{0.03cm}t}
=
A_{13}\hspace{0.02cm}W_{1}
- 
\frac{N_{c}}{d_{A}}\,{\mathfrak q}_{2}(t)B\hspace{0.03cm}\bigl(N_{1}W_{2} - N_{2}\hspace{0.03cm}W_{1}\bigr),
\\[1ex]
&\frac{d N_{2}(t)}{d\hspace{0.03cm}t}
=
A_{24}\hspace{0.02cm}W_{2}
+  
\frac{N_{c}}{d_{A}}\,{\mathfrak q}_{2}(t)B\hspace{0.03cm}\bigl(N_{1}W_{2} - N_{2}\hspace{0.03cm}W_{1}\bigr),
\\[1ex]
&\frac{d\hspace{0.03cm}W_{1}(t)}{d\hspace{0.03cm}t}
=
A_{31}\hspace{0.02cm}N_{1} -\hspace{0.03cm}\displaystyle\frac{1}{2}\,A(t)\hspace{0.02cm}W_{1}
+ 
B\hspace{0.03cm}N_{1}N_{2},
\\[1ex]
&\frac{d\hspace{0.03cm}W_{2}(t)}{d\hspace{0.03cm}t}
=
A_{42}\hspace{0.02cm}N_{2} -\hspace{0.03cm}\displaystyle\frac{1}{2}\,A(t)\hspace{0.02cm}W_{2}
+ 
B\hspace{0.03cm}N_{1}N_{2}.
\end{split}
\label{eq:11i}
\end{equation}
Note that in this approach the contributions with the matrix elements $A_{11}, A_{22}, A_{33}$ and $A_{44}$ are exactly reduced with the last terms in (\ref{eq:11e}) and (\ref{eq:11r}). System (\ref{eq:11i}) can also be obtained in a slightly different way. Consider for concreteness the original expression (\ref{eq:11e}). Let us integrate it over ${\bf k}$ over all momentum space ${\bf R}^{3}$. Then we multiply (\ref{eq:11e}) by the function $(\omega^{l}_{{\bf k}} - {\mathbf v}\cdot{\bf k})$ and integrate again over ${\bf k}$. As a result, we obtain a system of two equations, which can be conveniently represented in matrix form:
\begin{equation}
\left(\!\!\!\!\!\!
\begin{array}{ccc}
	&1 &1\\[1ex]
	&\omega^{l}_{{\bf k}_{\hspace{0.02cm}0}} - {\mathbf v}\cdot{\bf k}_{\hspace{0.02cm}0} &\omega^{l}_{{\bf k}^{\hspace{0.02cm}\prime}_{\hspace{0.02cm}0}} - {\mathbf v}\cdot{\bf k}^{\hspace{0.02cm}\prime}_{\hspace{0.02cm}0}
\end{array}
\!\right)
\left(\!\!\!\!\!\!
\begin{array}{ll}
	&\dot{N}_{1}\\[1ex]
	&\dot{N}_{2}
	\end{array}
\!\right)
\label{eq:11o}
\end{equation}
\[
=
\left(\!\!\!\!\!\!
\begin{array}{ccc}
	&1 &1\\[1ex]
	&\omega^{l}_{{\bf k}_{\hspace{0.02cm}0}} - {\mathbf v}\cdot{\bf k}_{\hspace{0.02cm}0} &\omega^{l}_{{\bf k}^{\hspace{0.02cm}\prime}_{\hspace{0.02cm}0}} - {\mathbf v}\cdot{\bf k}^{\hspace{0.02cm}\prime}_{\hspace{0.02cm}0}
\end{array}
\!\right)
\left(\!\!\!\!\!\!
\begin{array}{ll}
	&A_{13}\hspace{0.03cm}W_{1}\\[1ex]
	&A_{24}\hspace{0.03cm}W_{2}
\end{array}
\!\right)
+
\left(\!\!\!\!\!\!
\begin{array}{cc}
	&0\\[1ex]
	&\displaystyle\frac{N_{c}}{d_{A}}\,{\mathfrak q}_{2}(t)B\hspace{0.03cm}
	\bigl(N_{1}W_{2} - N_{2}\hspace{0.03cm}W_{1}\bigr)
	\Delta\hspace{0.02cm}\omega_{\,{\bf k}_{\hspace{0.02cm}0},\hspace{0.03cm}{\bf k}^{\hspace{0.02cm}\prime}_{\hspace{0.02cm}0}} 
\end{array}
\!\right),
\] 
where the dot above $N_{1}$ and $N_{2}$ denotes differentiation with respect to $t$. Suppose that the determinant of the matrix on the left-hand side (\ref{eq:11o})
\[
\det\!\hspace{0.02cm} \left(\!\!\!\!\!\!
\begin{array}{ccc}
	&1 &1\\[1ex]
	&\omega^{l}_{{\bf k}_{\hspace{0.02cm}0}} - {\mathbf v}\cdot{\bf k}_{\hspace{0.02cm}0} &\omega^{l}_{{\bf k}^{\hspace{0.02cm}\prime}_{\hspace{0.02cm}0}} - {\mathbf v}\cdot{\bf k}^{\hspace{0.02cm}\prime}_{\hspace{0.02cm}0}
\end{array}
\!\right)
=
-\hspace{0.03cm}\Delta\hspace{0.02cm}
\omega_{\,{\bf k}_{\hspace{0.02cm}0},\hspace{0.03cm}{\bf k}^{\hspace{0.02cm}\prime}_{\hspace{0.02cm}0}} 
\]
is nonzero. Recall that we use the notation (\ref{eq:11t}). In this case, multiplying equation (\ref{eq:11o}) by the corresponding inverse matrix, we easily arrive at the first two equations of the system (\ref{eq:11i}). Note that in these equations, the factor $\Delta\hspace{0.02cm} \omega_{\,{\bf k}_{\hspace{0.02cm}0},\hspace{0.03cm}
{\bf k}^{\hspace{0.02cm}\prime}_{\hspace{0.02cm}0}}$ is completely reduced (it is only presented in the $B$ factor, Eq.\,(\ref{eq:11u}), as the argument of the $\delta$-function) and the requirement $\Delta\hspace{0.02cm} \omega_{\,{\bf k}_{\hspace{0.02cm}0},\hspace{0.03cm}
{\bf k}^{\hspace{0.02cm}\prime}_{\hspace{0.02cm}0}}\neq 0$ can be formally discarded. The second pair of equations in the system (\ref{eq:11i}) can be derived in a completely analogous way. It is easy to see that the first two equations in (\ref{eq:11i}) are in perfect agreement with the relations obtained in the previous section, namely, with (\ref{eq:10e}) and (\ref{eq:10t}) when using the representation (\ref{eq:11q}).\\
\indent Let us simplify the resulting system as much as possible. The matrix elements $A_{13}, A_{24}, A_{31}$ and $A_{42}$ in front of the linear terms on the right-hand sides of (\ref{eq:11i}) are proportional to the factors 
${\rm Im}\hspace{0.03cm}\mathscr{T}_{\hspace{0.04cm}{\bf k}_{\hspace{0.02cm}0}, \hspace{0.03cm}{\bf k}_{\hspace{0.02cm}0}}({\bf v})$ and 
${\rm Im}\hspace{0.03cm}\mathscr{T}_{\hspace{0.04cm}{\bf k}^{\hspace{0.02cm}\prime}_{\hspace{0.02cm}0}, \hspace{0.03cm}{\bf k}^{\hspace{0.02cm}\prime}_{\hspace{0.02cm}0}}({\bf v})$. These factors are actually related to the collisionless (Landau) damping of soft gluon oscillations and thus must contain the Dirac delta function which reflects the corresponding conservation laws for energy and momentum:
\[
{\rm Im}\hspace{0.03cm}\mathscr{T}_{\hspace{0.04cm}{\bf  k}_{\hspace{0.02cm}0},\,{\bf k}_{\hspace{0.02cm}0}}({\bf v})
\sim
\int\!\frac{d\vspace{0.4cm}\Omega_{{\bf v}^{\prime}}}{4\pi}\,
w_{{\bf v}^{\prime}}({\bf v};{\bf k}_{\hspace{0.02cm}0})\hspace{0.03cm}
(2\pi)\hspace{0.03cm}\delta(\omega^{l}_{{\bf k}_{\hspace{0.02cm}0}} - {\mathbf v}^{\prime}\cdot{\bf k}_{\hspace{0.02cm}0}),
\]
where $w_{{\bf v}^{\prime}}({\bf v}; {\bf k}_{\hspace{0.02cm}0})$ is the probability for the Landau damping process. An explicit form of this probability can be obtained by using the expressions for the scattering amplitude (\ref{eq:12f}), the three-point amplitude ${\mathcal V}_{{\bf k},\, {\bf k}_{1},\, {\bf k}_{2}}$, Eq.\,(\ref{eq:2j}), and the HTL-correction $\delta\hspace{0.025cm} \Gamma^{\mu\hspace{0.02cm}\nu\rho}(k, k_{1}, k_{2})$, Eq.\,(\ref{ap:A3}). However, as is well known, the {\it linear} Landau damping is kinematically forbidden in hot gluon plasma and therefore, these matrix elements can be set to zero, i.e.,
\[
A_{13} = A_{24} = A_{31} = A_{42} = 0.
\] 
Next, we consider the terms in the last two equations in (\ref{eq:11i}) containing the function $A(t)$. By virtue of the definition (\ref{eq:9y}), this function is quadratic in the (colorless part) plasmon number density, and thus, the terms $A(t)\hspace{0.02cm}W_{1}$ and $A(t)\hspace{0.02cm}W_{2}$ in (\ref{eq:11i}) are of the third order. In constructing the kinetic equations (\ref{eq:10q}) and (\ref{eq:10u}), we limited ourselves to linear and quadratic contributions of the plasmon number density. For this reason, within the accepted accuracy, in the last two equations in (\ref{eq:11i}) one should drop the contributions with the function $A(t)$, and in the first two equations, for the same reason, one should consider 
\[
{\mathfrak q}_{2}(t) \simeq {\mathfrak q}_{2}(t_{0})\equiv {\mathfrak q}^{0}_{2}.
\]
Taking all of the above-mentioned into account, instead of (\ref{eq:11i}) we now have
\[
\begin{split}
	&\frac{d N_{1}(t)}{d\hspace{0.03cm}t}
	=
	-\hspace{0.03cm} 
	\frac{N_{c}}{d_{A}}\,{\mathfrak q}^{0}_{2}\hspace{0.03cm}B\hspace{0.03cm}\bigl(N_{1}W_{2} - N_{2}\hspace{0.03cm}W_{1}\bigr),
	\\[1ex]
	&\frac{d N_{2}(t)}{d\hspace{0.03cm}t}
	=
	\frac{N_{c}}{d_{A}}\,{\mathfrak q}^{0}_{2}\hspace{0.03cm}B\hspace{0.03cm}\bigl(N_{1}W_{2} - N_{2}\hspace{0.03cm}W_{1}\bigr),
	\\[1ex]
	&\frac{d\hspace{0.03cm}W_{1}(t)}{d\hspace{0.03cm}t}
	=
	B\hspace{0.03cm}N_{1}N_{2},
	\\[1ex]
	&\frac{d\hspace{0.03cm}W_{2}(t)}{d\hspace{0.03cm}t}
	=
	B\hspace{0.03cm}N_{1}N_{2},
\end{split}
\] 
whence, in particular, the relations immediately follow
\begin{equation}
N_{1}(t) + N_{2}(t) = \mathcal{C}_{1}, 
\qquad 
W_{1}(t) - W_{2}(t) = \mathcal{C}_{2},
\label{eq:11p}
\end{equation}
where $\mathcal{C}_{1}$ and $\mathcal{C}_{2}$ are constants. These relations allow us to reduce the system of four equations to a system of two equations
\begin{equation}
\begin{split}
&\frac{d N_{1}(t)}{d\hspace{0.03cm}t}
=
\beta B\bigl[\hspace{0.01cm}N_{1}\hspace{0.02cm}(\mathcal{C}_{2} - W_{1})	
+ W_{1}\hspace{0.02cm}(\mathcal{C}_{1} - N_{1})\bigr],
		\\[1ex]
		&\frac{d\hspace{0.03cm}W_{1}(t)}{d\hspace{0.03cm}t}
		=
		B\hspace{0.01cm}N_{1}\hspace{0.02cm}(\mathcal{C}_{1} - N_{1}),
	\end{split}
\label{eq:11a}
\end{equation}
where for the sake of brevity, we have designated $\beta\equiv N_{c}\,{\mathfrak q}^{0}_{2}\hspace{0.02cm}/d_{A}$.
Obviously, this system has two stationary points, one of which is trivial:
$N_{1} = W_{1} = 0$, and the second one is
\[
N_{1} = \mathcal{C}_{1},\qquad W_{1} = \mathcal{C}_{2}.
\]
From (\ref{eq:11w}) and the first relation in (\ref{eq:11p}) it follows that $\mathbbm{N}^{\hspace{0.03cm}l}(t) = \mbox{const}$.\\
\indent At a certain relation between the constants $\mathcal{C}_{1}$ and $\mathcal{C}_{2}$, namely, at
\[
\mathcal{C}^{\hspace{0.03cm}2}_{2} = \frac{1}{2\hspace{0.02cm}\beta}\,\mathcal{C}^{\hspace{0.03cm}2}_{1},
\]
we can obtain the exact solution of the system (\ref{eq:11a}). For this purpose, the first step, due to the autonomy of the right-hand sides, is to reduce this system to a single equation
\[
\frac{d N_{1}}{d\hspace{0.03cm}W_{1}}
=
\beta\hspace{0.03cm}\biggl(\hspace{0.01cm}\frac{\mathcal{C}_{2} - W_{1}}{\mathcal{C}_{1} - N_{1}}
+
\frac{W_{1}}{N_{1}}\biggr)
\]
or, in a slightly different form,
\begin{equation}
\bigl[\hspace{0.02cm}(2\hspace{0.02cm}N_{1} - \mathcal{C}_{1})\hspace{0.02cm} W_{1} - 
\mathcal{C}_{2}\hspace{0.01cm}N_{1}\bigr]\hspace{0.03cm}
\frac{d\hspace{0.03cm}W_{1}} {d\hspace{0.02cm}N_{1}}
=
\frac{1}{\beta}\,(N_{1}^{\hspace{0.02cm}2} - \mathcal{C}_{1}\hspace{0.01cm}N_{1}), 	
\label{eq:11s}
\end{equation}
which defines $W_{1}$ as a function of $N_{1}$. The construction of the solution of equation (\ref{eq:11s}) and then of the original system (\ref{eq:11a}) is given in Appendix \ref{appendix_E}. Here, we provide at once an explicit form of the exact solution for the system (\ref{eq:11a}) in parametric form
\begin{equation}
\begin{split}
		&N_{1} = N_{1}(\tau,C) = \frac{1}{2}\,\mathcal{C}_{1} - \frac{1}{4\hspace{0.03cm}a}\,\mathcal{C}_{1}\hspace{0.03cm}
		\mathcal{C}_{2}\hspace{0.03cm}\frac{\tau}{f(\tau)},
		\\[1ex]
		&W_{1} = W_{1}(\tau,C) = a \biggl[\frac{1 + \tau}{\tau}\,
		f(\tau)
		-
		\frac{1}{2}\,\frac{\tau}{f(\tau)}
		\biggr]
		+ \frac{\mathcal{C}_{2}\hspace{0.03cm}N_{1}}
		{2\hspace{0.01cm}N_{1} - \mathcal{C}_{1}},
		\\[1ex]
		&t = t\hspace{0.03cm}(\tau, C, \widetilde{C}) =-\hspace{0.03cm}\frac{2a}{B\hspace{0.03cm}
		\mathcal{C}^{\hspace{0.03cm}2}_{1}}\int\!\frac{d\hspace{0.03cm}\tau }{(1 + \tau)\hspace{0.01cm}f(\tau)}
		\hspace{0.03cm}+\hspace{0.03cm}\widetilde{C},
	\end{split}
	\label{eq:11d}
\end{equation}
where $C$ and $\widetilde{C}$ are arbitrary integration constants and 
\[
a^{2} = \frac{1}{4\hspace{0.02cm}\beta}\,\mathcal{C}^{\hspace{0.03cm}2}_{1},
\qquad
f(\tau)\equiv\bigl(\tau - \ln|1 + \tau| - C\hspace{0.02cm}\bigr)^{1/2}.
\]
In Appendix \ref{appendix_E}, the integral over $\tau$ in the representation for time $t$ in (\ref{eq:11d}) is rewritten in terms of an integral of the famous Lambert $W$-\hspace{0.03cm}function \cite{corless_1996}.\\
\indent Although the problem considered above for the process of scattering of plasmons off a hard particle allows us to obtain in a particular case exact analytical solutions of the corresponding reduced system of kinetic equations (\ref{eq:10q}) and (\ref{eq:10u}), these solutions do not make it possible to determine unambiguously the directions of pumping energy of plasmon excitations both across the spectrum and between the colorless and color components of the plasmon number density. This is due to the fact that the solutions are obtained in a parametric form and to the fact that the time parametrization can be represented only in the form of an indefinite integral. Probably, numerical methods should be used here.


\section{Coefficient functions $\widetilde{V}^{\,(1)\,a_{1}\hspace{0.03cm}a_{2}\,a}_{\ {\bf k}_{1},\, 
{\bf k}_{2}}$ and $\widetilde{V}^{\,(2)\,a_{1}\hspace{0.03cm}a_{2}\,a}_{\ {\bf k}_{1},\, {\bf k}_{2}}$ of the cano\-nical transformation (\ref{eq:3t})}
\label{section_12}
\setcounter{equation}{0}

\begin{figure}[hbtp]
	\begin{center}
		\includegraphics*[height=3.4cm]{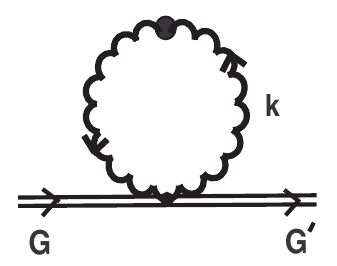}
	\end{center}
	\caption{\small Influence of the medium on a free-moving particle}
	\label{fig4}
\end{figure}

In this section we consider the problem of defining an explicit form of the third-order coefficient functions 
$\widetilde{V}^{\hspace{0.03cm}(1)\,a\,a_{1}\hspace{0.03cm}a_{2}}_{\ {\bf k},\, {\bf k}_{1}}$ and $\widetilde{V}^{\hspace{0.03cm}(2)\, a\, a_{1}\hspace{0.03cm}a_{2}}_{\ {\bf k},\, {\bf k}_{1}}$ entering into the canonical transformation of the bosonic normal variable $a^{\,a}_{\hspace{0.03cm}{\bf k}}$, Eq.\,(\ref{eq:3t}). Let us return once again to the fourth-order interaction Hamilto\-nian (\ref{eq:2d}), more exactly to its terms
\[
\frac{1}{2}\int\!d\hspace{0.03cm}{\bf k}\,d\hspace{0.03cm}{\bf k}_{1}\,
\Bigl\{\hspace{0.02cm}T^{\,\ast\,(1)\, a\, a_{1}\hspace{0.03cm}a_{2}}_{\,{\bf k},\,
{\bf k}_{1}}\hspace{0.03cm}a^{\,a}_{\hspace{0.03cm}{\bf k}}\hspace{0.04cm}
a^{\,a_{1}}_{\hspace{0.03cm}{\bf k}_{1}}\hspace{0.03cm}Q^{\,a_{2}}
\,+\,
T^{\,(1)\,a\,a_{1}\hspace{0.03cm}a_{2}}_{\,{\bf k},\, {\bf k}_{1}}\, {a}^{\,\ast\, a}_{\hspace{0.03cm}{\bf k}}\hspace{0.01cm}
a^{\,\ast\,a_{1}}_{\hspace{0.03cm}{\bf k}_{1}} Q^{\,a_{2}}\hspace{0.01cm} 
\Bigr\}.
\]
It is obvious that even if the nonresonant fourth-order contributions with the vertex functions $T^{\,(1)\,a\,a_{1}\hspace{0.03cm}a_{2}}_{\,{\bf k},\, {\bf k}_{1}}$ and $T^{\,\ast\,(1)\,a\,a_{1}\hspace{0.03cm}a_{2}}_{\,{\bf k},\, {\bf k}_{1}}$ are zero by virtue of the properties of the system under consideration, still they will inevitably be generated by the canonical transformations (\ref{eq:3t}) and (\ref{eq:3y}) from the free field and third- and fourth-order Hamiltonians, Eqs.\,(\ref{eq:2p}), (\ref{eq:2s}) and (\ref{eq:2d}), respectively. We can determine the third-order coefficient function $\widetilde{V}^{\hspace{0.03cm}(1)\, a\, a_{1}\hspace{0.03cm}a_{2}}_{\ {\bf k},\, {\bf k}_{1}}$ in (\ref{eq:3t}) from the requirement of vanishing these ``induced'' contributions.\\
\indent The first step is to find all the contributions proportional to the products 
$c^{\,a_{1}}_{\hspace{0.03cm}{\bf k}_{1}}\hspace{0.03cm}c^{\,a_{2}}_{\hspace{0.03cm}{\bf k}_{2}}\hspace{0.03cm} \mathcal{Q}^{\,a}$ and $c^{\,\ast\, a_{1}}_{\hspace{0.03cm}{\bf k}_{1}} \hspace{0.03cm}a^{\,\ast\,a_{2}}_{\hspace{0.03cm}{\bf k}_{2}} \mathcal{Q}^{\,a}$ from the free-field Hamiltonian $H^{(0)}$ given by Eq.\,(\ref{eq:2p}) under the canonical transfor\-mation (\ref{eq:3t}). In the case of the product $c^{\,\ast\, a_{1}}_{\hspace{0.03cm}{\bf k}_{1}} \hspace{0.01cm}c^{\,\ast\, a_{2}}_{\hspace{0.03cm}{\bf k}_{2}}  \mathcal{Q}^{\,a}$ we obtain the following contributions from $H^{(0)}$:
\begin{equation}
\int\!d\hspace{0.03cm}{\bf k}_{1}\hspace{0.03cm}d\hspace{0.03cm}{\bf k}_{2}\;
c^{\,\ast\,a_{1}}_{\hspace{0.03cm}{\bf k}_{1}}\hspace{0.03cm}
c^{\,\ast\,a_{2}}_{\hspace{0.03cm}{\bf k}_{2}}\hspace{0.03cm} \mathcal{Q}^{\,a}
\hspace{0.03cm}\biggl\{\frac{1}{2}\,
\Bigl[\hspace{0.03cm}
\bigl(\omega^{\hspace{0.03cm}l}_{\hspace{0.03cm}{\bf k}_{1}} - {\mathbf v}\cdot {\mathbf k}_{1}\bigr)\,
\widetilde{V}^{\,(1)\,a_{1}\hspace{0.03cm}a_{2}\,a}_{\ {\bf k}_{1},\, {\bf k}_{2}}
+
\bigl(\omega^{\hspace{0.03cm}l}_{\hspace{0.03cm}{\bf k}_{2}} - {\mathbf v}\cdot {\mathbf k}_{2}\bigr)\,
\widetilde{V}^{\,(1)\,a_{2}\,a_{1}\hspace{0.03cm}a}_{\ {\bf k}_{2},\, {\bf k}_{1}}
\Bigr]
\label{eq:12q}
\end{equation}
\[
+\!
\int\!d\hspace{0.03cm}{\bf k}\, 
\bigl(\omega^{l}_{\hspace{0.03cm}{\bf k}} - {\mathbf v}\cdot {\mathbf k}\bigr)\hspace{0.03cm}
\Bigl[F^{\,\ast}_{\hspace{0.03cm}\bf k}\hspace{0.04cm}
V^{\hspace{0.03cm}(3)\, a\, a_{1}\hspace{0.03cm}a_{2}}_{\ {\bf k},\, {\bf k}_{1},\, {\bf k}_{2}}
+
F^{\phantom{\ast}}_{\hspace{0.03cm}\bf k}\hspace{0.03cm}
V^{\,\ast\hspace{0.03cm}(1)\, a\, a_{2}\, a_{1}}_{\ {\bf k},\, {\bf k}_{2},\, {\bf k}_{1}}
\Bigr]\biggr\}.
\]
Further, we determine the desired fourth-order contributions from the Hamiltonian $H^{(3)}$, Eq.\,(\ref{eq:2s}). Performing, when needed, the relevant symmetrization, in the case of the product $c^{\,\ast\, a_{1}}_{\hspace{0.03cm}{\bf k}_{1}} \hspace{0.01cm}c^{\,\ast\, a_{2}}_{\hspace{0.03cm}{\bf k}_{2}} \mathcal{Q}^{\,a}$, we obtain
\begin{equation}
\int\!d\hspace{0.03cm}{\bf k}_{1}\hspace{0.03cm}d\hspace{0.03cm}{\bf k}_{2}\;
c^{\,\ast\, a_{1}}_{\hspace{0.03cm}{\bf k}_{1}} \hspace{0.01cm}
c^{\,\ast\, a_{2}}_{\hspace{0.03cm}{\bf k}_{2}} \mathcal{Q}^{\,a}
\hspace{0.03cm}\biggl\{	
\frac{1}{2}\,
\Bigl[\hspace{0.03cm}\upphi^{\,\ast}_{\hspace{0.03cm}{\bf k}_{1}}\hspace{0.01cm}
M^{\,\ast\hspace{0.03cm}a_{1}\hspace{0.03cm}a_{2}\,a}_{\; {\bf k}_{2}}	
\,+\,
\upphi^{\,\ast}_{\hspace{0.03cm}{\bf k}_{2}}\hspace{0.02cm}
M^{\,\ast\hspace{0.03cm}a_{2}\,a_{1}\hspace{0.03cm}a}_{\; {\bf k}_{1}}
\Bigr]
\label{eq:12w}
\end{equation}
\[
+\!
\int\!d\hspace{0.03cm}{\bf k}\, 
\Bigl[\upphi^{\phantom{\ast}}_{\hspace{0.03cm}\bf k}\hspace{0.03cm}
V^{\hspace{0.03cm}(3)\,a\,a_{1}\hspace{0.03cm}a_{2}}_{\ {\bf k},\,
{\bf k}_{1},\, {\bf k}_{2}}
+
\upphi^{\,\ast}_{\hspace{0.03cm}\bf k}\,
V^{\,\ast\hspace{0.03cm}(1)\,a\,a_{2}\,a_{1}}_{\ {\bf k},\, {\bf k}_{2},\, {\bf k}_{1}}
\Bigr]
\]
\[
+\!
\int\!d\hspace{0.03cm}{\bf k}\,
\Bigl[\hspace{0.03cm}{\mathcal V}^{\,\ast\, a\, a_{1}\hspace{0.03cm}a_{2}}_{\,{\bf k},\, {\bf k}_{1},\, {\bf k}_{2}}\, F^{\phantom{\ast}}_{\hspace{0.03cm}\bf k}
\hspace{0.04cm}
\delta({\bf k} - {\bf k}_{1} - {\bf k}_{2})
\,+\,
{\mathcal U}^{\,*\,a\,a_{1}\hspace{0.03cm}a_{2}}_{\,{\bf k},\, {\bf k}_{1},\, {\bf k}_{2}}\, 
F^{\,\ast}_{\hspace{0.03cm}\bf k}\hspace{0.04cm}
\delta({\bf k} + {\bf k}_{1} + {\bf k}_{2})
\hspace{0.03cm}\Bigr]\biggr\}.
\]
In deriving (\ref{eq:12q}) and (\ref{eq:12w}) we considered that the coefficient functions $V^{(1)\, a\, a_{1}\hspace{0.03cm}a_{2}}_{\ {\bf k},\, {\bf k}_{1},\, {\bf k}_{2}}$ and $V^{(3)\, a\, a_{1}\hspace{0.03cm}a_{2}}_{\ {\bf k},\, {\bf k}_{1},\, {\bf k}_{2}}$ are proportional to the corresponding Dirac delta functions depending on the sum or difference of the wave vectors\footnote{\hspace{0.02cm}We have also taken this circumstance into account in deriving the canonicity conditions given in Appendix~\ref{appendix_B}, namely, the canonicity conditions (\ref{ap:B1}), (\ref{ap:B2}) and (\ref{ap:B4}), also containing the coefficient functions $V^{(1)\, a\, a_{1}\hspace{0.03cm}a_{2}}_{\ {\bf k},\, {\bf k}_{1},\, {\bf k}_{2}}$ and $V^{(3)\, a\, a_{1}\hspace{0.03cm}a_{2}}_{\ {\bf k},\, {\bf k}_{1},\, {\bf k}_{2}}$.}, by virtue of the relations (\ref{eq:4r}). Finally, from the Hamiltonian $H^{(4)}$, Eq.\,(\ref{eq:2d}), we easily find the relevant contribution 
\begin{equation}
\frac{1}{2} 
\int\!d\hspace{0.03cm}{\bf k}_{1}\hspace{0.03cm}d\hspace{0.03cm}{\bf k}_{2}\;
T^{\,(1)\ a_{1}\hspace{0.03cm}a_{2}\, a}_{\,{\bf k}_{1},\, {\bf k}_{2}}\hspace{0.03cm}c^{\,\ast\, a_{1}}_{\hspace{0.03cm}{\bf k}_{1}} \hspace{0.02cm}
c^{\,\ast\,a_{2}}_{\hspace{0.03cm}{\bf k}_{2}} \hspace{0.03cm} Q^{\,a}\hspace{0.01cm}.
\label{eq:12e}
\end{equation}
\indent Let us analyze the expressions (\ref{eq:12q}) and (\ref{eq:12w}). In the first expression (\ref{eq:12q}) we eliminate the coefficient function $\widetilde{V}^{\,(1)\, a_{2}\, a_{1}\hspace{0.03cm}a}_{\ {\bf k}_{2},\, {\bf k}_{1}}$. For this purpose, we make use of the canonicity condition (\ref{eq:3p}) connecting the functions $\widetilde{V}^{\,(1)\, a_{2}\, a_{1}\hspace{0.03cm}a}_{\ {\bf k}_{2},\, {\bf k}_{1}}$ and $\widetilde{V}^{\,(1)\, a_{1}\hspace{0.03cm}a_{2}\, a}_{\ {\bf k}_{1},\, {\bf k}_{2}}$ among themselves. Further we notice that in the sum of the expressions (\ref{eq:12q}) and (\ref{eq:12w}) the terms with the coefficient functions $V^{\hspace{0.03cm}(1,3)\, a\, a_{1}\hspace{0.03cm}a_{2}}_{\ {\bf k},\, {\bf k}_{1},\, {\bf k}_{2}}$ exactly reduce each other when considering the relation (\ref{eq:4w}). For the coefficient function $M^{\,a\,a_{1}\,a_{2}}_{\; {\bf k}_{1}}$ in (\ref{eq:12w}) we use the relation (\ref{eq:4e}) relating this function to the vertex function $\upphi^{\phantom{\ast}}_{\hspace{0.03cm}{\bf k}_{1}}$.\\
\indent Finally, in the last line of the expression (\ref{eq:12w}), it is necessary to integrate over ${\bf k}$ with the $\delta$-functions and go from the coefficient function $F^{\phantom{\ast}}_{\hspace{0.03cm}{\bf k}}$ to the ``physical'' function $\upphi^{\phantom{\ast}}_{\hspace{0.03cm}\bf k}$ by the rule (\ref{eq:4w}).
By summing the expressions (\ref{eq:12q})\,--\,(\ref{eq:12e}) and considering all the above-mentioned, we lead to the following fourth-order expression in the new variables $c^{\,\ast\, a_{1}}_{\hspace{0.03cm}{\bf k}_{1}} $, $c^{\,\ast\, a_{2}}_{\hspace{0.03cm}{\bf k}_{2}}$ and $\mathcal{Q}^{\,a}$, which must be added to the effective Hamiltonian (\ref{eq:4y})
\begin{equation}
	\frac{1}{2}
	\int\!d\hspace{0.03cm}{\bf k}_{1}\hspace{0.03cm}d\hspace{0.03cm}
	 {\bf k}_{2}\;
	c^{\,\ast\, a_{1}}_{\hspace{0.03cm}{\bf k}_{1}} \hspace{0.01cm}c^{\,\ast\, a_{2}}_{\hspace{0.03cm}{\bf k}_{2}} \mathcal{Q}^{\,a}
	\hspace{0.03cm}\biggl\{\hspace{0.01cm}
	\bigl[\hspace{0.03cm}
	\omega^{\hspace{0.03cm}l}_{\hspace{0.03cm}{\bf k}_{1}} + \omega^{\hspace{0.03cm}l}_{\hspace{0.03cm}{\bf k}_{2}}
	- {\mathbf v}\cdot ({\mathbf k}_{1} + {\mathbf k}_{2})\bigr]\hspace{0.03cm}
	\widetilde{V}^{\,(1)\,a_{1}\hspace{0.03cm}a_{2}\,a}_{\ {\bf k}_{1},\, {\bf k}_{2}}
	+\,
	\widetilde{T}^{\,(1)\,a_{1}\hspace{0.03cm}a_{2}\,a}_{\,{\bf k}_{1},\, {\bf k}_{2}}\biggr\},
\label{eq:12r}
\end{equation}
where the effective amplitude $\widetilde{T}^{\,(1)\,a_{1}\hspace{0.03cm}a_{2}\, a}_{\,{\bf k}_{1},\, 
{\bf k}_{2}}$ has the following structure:
\begin{equation}
\widetilde{T}^{\,(1)\,a_{1}\hspace{0.03cm}a_{2}\,a}_{\,{\bf k}_{1},\, {\bf k}_{2}}
=
T^{\,(1)\,a_{1}\hspace{0.03cm}a_{2}\,a}_{\,{\bf k}_{1},\,{\bf k}_{2}}
\label{eq:12t}
\end{equation}
\vspace{0.15cm}
\[
-\,
f^{\hspace{0.03cm}a_{1}\hspace{0.03cm}a_{2}\,a}\hspace{0.03cm}
\biggl\{i\,
\frac{{\upphi}^{\hspace{0.03cm}\ast}_{\,{\bf k}_{1}}
	\hspace{0.01cm}{\upphi}^{\hspace{0.03cm}\ast}_{\,{\bf k}_{2}}}
{\omega^{\hspace{0.02cm} l}_{\hspace{0.03cm}{\bf k}_{2}} - {\bf v}\cdot {\bf k}_{2}}
+
2\hspace{0.04cm}\biggl(\,
\frac{{\mathcal V}^{\,\ast}_{\,{\bf k}_{1} + {\bf k}_{2},\,{\bf k}_{1},\, {\bf k}_{2}} 
	{\upphi}^{\hspace{0.03cm}{\ast}}_{\,{\bf k}_{1} + {\bf k}_{2}}}
{\omega^{\hspace{0.03cm}l}_{\hspace{0.03cm}{\bf k}_{1} + {\bf k}_{2}}\! - {\bf v}\cdot ({\bf k}_{1} + {\bf k}_{2})}
\,+\,
\frac{{\mathcal U}^{\,{\ast}}_{\,-{\bf k}_{1} - {\bf k}_{2},\,{\bf k}_{1},\, {\bf k}_{2}} 
	{\upphi}^{\phantom{\ast}}_{\,-{\bf k}_{1} - {\bf k}_{2}}}
{\omega^{\hspace{0.03cm}l}_{-{\bf k}_{1} - {\bf k}_{2}}\! + {\bf v}\cdot ({\bf k}_{1} + {\bf k}_{2})}\biggr)\!\biggr\}.
\vspace{0.2cm}
\]
Note that the amplitude $\widetilde{T}^{\,(1)\,a_{1}\hspace{0.03cm}a_{2}\, a}_{\,{\bf k}_{1},\, 
{\bf k}_{2}}$ itself does not possess the symmetry property
\[
\widetilde{T}^{\,(1)\,a_{1}\hspace{0.03cm}a_{2}\, a}_{\,{\bf k}_{1},\, 
{\bf k}_{2}}
=
\widetilde{T}^{\,(1)\,a_{2}\,a_{1}\hspace{0.03cm}a}_{\,{\bf k}_{2},\, 
{\bf k}_{1}},
\]
as it would seem to imply (\ref{eq:12r}). As shown in Appendices \ref{appendix_B} and \ref{appendix_C}, the presence of the term breaking this symmetry is important in determining the explicit form of the coefficient function $M^{\hspace{0.025cm}(1)\hspace{0.03cm}a\,a_{1}\hspace{0.03cm}a_{2}\,a_{3}}_{\; {\bf k}_{1},\, {\bf k}_{2}}$ in the canonical transformation (\ref{eq:3y}).
The requirement of vanishing the expression (\ref{eq:12r}) uniquely defines the explicit form of the desired higher-order coefficient function $\widetilde{V}^{\,(1)\,a_{1}\hspace{0.03cm}a_{2}\, a}_{\ {\bf k}_{1},\, 
{\bf k}_{2}}$ for the canonical transformation (\ref{eq:3t}):
\begin{equation}
	\widetilde{V}^{\,(1)\,a_{1}\hspace{0.03cm}a_{2}\,a}_{\ {\bf k}_{1},\, {\bf k}_{2}}
	=
	- \frac{1}{\omega^{\hspace{0.03cm}l}_{\hspace{0.03cm}{\bf k}_{1}} + \omega^{\hspace{0.03cm}l}_{\hspace{0.03cm}{\bf k}_{2}}
		- {\mathbf v}\cdot ({\mathbf k}_{1} + {\mathbf k}_{2})}\,
	\widetilde{T}^{\,(1)\,a_{1}\hspace{0.03cm}a_{2}\,a}_{\,{\bf k}_{1},\, {\bf k}_{2}}.
	\label{eq:12y}
\end{equation}
\indent Now we turn to the determination of the coefficient function $\widetilde{V}^{\,(2)\,a_{1}\hspace{0.03cm}a_{2}\,a}_{\ {\bf k}_{1},\, 
{\bf k}_{2}}$. We recall that this coefficient function enters into the expression for the effective fourth-order Hamiltonian describing the elastic scattering process of plasmon off hard color-charged test particle: 
\begin{equation}
{\mathcal H}^{(4)}_{g\hspace{0.02cm}G\hspace{0.02cm}\rightarrow
\hspace{0.02cm} g\hspace{0.02cm}G} 
=
i\!\int\!d\hspace{0.03cm}{\bf k}_{1}\hspace{0.03cm}d\hspace{0.03cm}{\bf k}_{2}\,
\mathscr{T}^{\hspace{0.03cm}(2)\hspace{0.03cm} a\,a_{1}\hspace{0.03cm}a_{2}}_{\; {\bf k}_{1},\, {\bf k}_{2}}\,
c^{\ast\ \!\!a_{1}}_{\hspace{0.03cm}{\bf k}_{1}} c^{\hspace{0.03cm}a_{2}}_{\hspace{0.03cm}{\bf k}_{2}}
\mathcal{Q}^{\,a}.
\label{eq:12u}
\end{equation}
Here the complete effective amplitude is defined by the following expression
\begin{equation}
	\mathscr{T}^{\hspace{0.03cm}(2)\hspace{0.03cm} a\,a_{1}\hspace{0.03cm} a_{2}}_{\; {\bf k}_{1},\, {\bf k}_{2}}
	=
	-\hspace{0.02cm}i\,
	\bigl[\hspace{0.03cm}\omega^{\hspace{0.03cm}l}_{\hspace{0.03cm}{\bf k}_{1}} - \omega^{\hspace{0.03cm}l}_{\hspace{0.03cm}{\bf k}_{2}}
	-
	{\mathbf v}\cdot ({\mathbf k}_{1} - {\mathbf k}_{2})\bigr]\,
	\widetilde{V}^{\,(2)\,a_{1}\hspace{0.03cm}a_{2}\,a}_{\ {\bf k}_{1},\, {\bf k}_{2}}
	+\,
	\widetilde{T}^{\,(2)\,a_{1}\hspace{0.03cm}a_{2}\,a}_{\,{\bf k}_{1},\, {\bf k}_{2}}, 
	\label{eq:12i}
\end{equation}
where, in turn, the effective amplitude $\widetilde{T}^{\,(2)\,a_{1}\hspace{0.03cm}a_{2}\,a}_{\,{\bf k}_{1},\, 
{\bf k}_{2}} = f^{\hspace{0.03cm}a_{1}\hspace{0.02cm}a_{2}\hspace{0.03cm}a}\hspace{0.04cm}
\widetilde{T}^{\,(2)}_{\,{\bf k}_{1},\, {\bf k}_{2}}$ has the following structure:
\begin{equation}
\widetilde{T}^{\,(2)}_{\,{\bf k}_{1},\, {\bf k}_{2}}
=
T^{\hspace{0.03cm}(2)}_{\; {\bf k}_{1},\, {\bf k}_{2}}
+
\biggl\{\frac{{\upphi}^{\hspace{0.03cm}\ast}_{\,{\bf k}_{1}}\hspace{0.03cm}{\upphi}^{\phantom{\ast}}_{\,{\bf k}_{2}}}
{\omega^{\hspace{0.03cm}l}_{\hspace{0.03cm}{\bf k}_{2}} - {\bf v}\cdot {\bf k}_{2}}
\,+\,
2\hspace{0.03cm}i\hspace{0.02cm}\biggl(\,
\frac{{\mathcal V}^{\phantom{\ast}}_{\,{\bf k}_{1},\, {\bf k}_{2},\, {\bf k}_{1} - {\bf k}_{2}} 
	{\upphi}^{\hspace{0.03cm}{\ast}}_{\,{\bf k}_{1} - {\bf k}_{2}}}
{\omega^{\hspace{0.03cm}l}_{\hspace{0.03cm}{\bf k}_{1} - {\bf k}_{2}}\! - {\bf v}\cdot ({\bf k}_{1} - {\bf k}_{2})}
\,-\,
\frac{{\mathcal V}^{\,{\ast}}_{\,{\bf k}_{2},\, {\bf k}_{1},\, {\bf k}_{2} - {\bf k}_{1}} 
	{\upphi}^{\phantom{\ast}}_{\,{\bf k}_{2} - {\bf k}_{1}}}
{\omega^{\hspace{0.03cm}l}_{\hspace{0.03cm}{\bf k}_{2} - {\bf k}_{1}}\! - {\bf v}\cdot ({\bf k}_{2} - {\bf k}_{1})}\biggr)\!\biggr\}.
\label{eq:12o}
\end{equation}
However, an explicit form of the coefficient function $\widetilde{V}^{\,(2)\, a_{1}\hspace{0.03cm}a_{2}\, a}_{\ {\bf k}_{1},\, {\bf k}_{2}}$ as opposed to the previous one, will no longer be so unambiguous. This is because the canonical transformations admit a certain freedom (in the case of the coefficient $\widetilde{V}^{\,(1)\, a_{1}\hspace{0.03cm}a_{2}\, a}_{\ {\bf k}_{1},\, {\bf k}_{2}}$ this freedom is limited by the condition of exclusion of the nonresonant term from ${\mathcal H}^{(4)}$). For the coefficient function $\widetilde{V}^{\,(2)\,a_{1}\hspace{0.03cm}a_{2}\, a}_{\ {\bf k}_{1},\, {\bf k}_{2}}$ we have one canonicity condition (\ref{eq:3a}), which we write out here once again:
\[
\widetilde{V}^{\hspace{0.03cm}(2)\,a\,a_{1}\hspace{0.03cm}a_{2}}_{\ {\bf k}_{1},\, {\bf k}_{2}} 
+
\widetilde{V}^{\,\ast\hspace{0.03cm}(2)\, a\, a_{2}\, a_{1}}_{\ {\bf k}_{2},\, {\bf k}_{1}} 
=
-\hspace{0.03cm}
i\hspace{0.03cm}f^{\,a\,a_{1}\hspace{0.03cm}a_{2}}
F^{\hspace{0.03cm}\phantom{\ast}}_{\, {\bf k}_{1}}
F^{\hspace{0.03cm} \ast}_{\, {\bf k}_{2}} 
\]
or, if we factorize the color and momentum dependence  $\widetilde{V}^{\hspace{0.03cm}(2)\,a\,a_{1}\hspace{0.03cm}a_{2}}_{\ {\bf k}_{1},\, {\bf k}_{2}} = f^{\,a\,a_{1}\hspace{0.03cm} a_{2}}\hspace{0.03cm}\widetilde{V}^{\hspace{0.03cm}(2)}_{\ {\bf k}_{1},\, {\bf k}_{2}}$, 
\begin{equation}
	\widetilde{V}^{\hspace{0.03cm}(2)}_{\ {\bf k}_{1},\, {\bf k}_{2}} 
	-
	\widetilde{V}^{\,\ast\hspace{0.03cm}(2)}_{\ {\bf k}_{2},\, {\bf k}_{1}} 
	= -\hspace{0.03cm}
	 i\hspace{0.03cm}F^{\hspace{0.03cm}\phantom{\ast}}_{\, {\bf k}_{1}} F^{\hspace{0.03cm} \ast}_{\, {\bf k}_{2}}. 
\label{eq:12p}
\end{equation}
The last relation can be considered as a functional equation for the coefficient function $\widetilde{V}^{\hspace{0.03cm}(2)}_{\ {\bf k}_{1},\, {\bf k}_{2}}$. We can find its solution by writing down the general solution of the associated homogeneous equation and the particular solution of the nonhomogeneous one. Searching for a partial solution in the form $\alpha\hspace{0.03cm}F^{\hspace{0.03cm}\phantom{\ast}}_{\, {\bf k}_{1}} F^{\hspace{0.03cm} \ast}_{\, {\bf k}_{2}}$, where $\alpha$ is some complex number and passing then from the coefficient function $F^{\phantom{\ast}}_{\hspace{0.03cm}\bf k}$ to the vertex function $\upphi^{\phantom{\ast}}_{\hspace{0.03cm}{\bf k}}$ by the rule (\ref{eq:4w}), we get from (\ref{eq:12p})
\[
\Bigl(\widetilde{V}^{\hspace{0.03cm}(2)}_{\ {\bf k}_{1},\, {\bf k}_{2}}\Bigr)_{\rm inhom}\! 
=
\biggl(\mathrm{Re}\,\alpha - \frac{i}{2}\hspace{0.03cm}\biggr)
\hspace{0.03cm} 
\frac{{\upphi}^{\hspace{0.03cm}\ast}_{\,{\bf k}_{1}}\hspace{0.03cm}
{\upphi}^{\phantom{\ast}}_{\,{\bf k}_{2}}}
{\bigl(\omega^{\hspace{0.03cm}l}_{\hspace{0.03cm}{\bf k}_{1}} - {\bf v}\cdot {\bf k}_{1}\bigr)
\bigl(\omega^{\hspace{0.03cm}l}_{\hspace{0.03cm}{\bf k}_{2}} - {\bf v}\cdot {\bf k}_{2}\bigr)},
\]
where $\mathrm{Re}\,\alpha$ is an arbitrary numerical parameter.\\
\indent We take the general solution of the homogeneous equation in the following form
\[
\Bigl(\widetilde{V}^{\hspace{0.03cm}(2)}_{\ {\bf k}_{1},\, {\bf k}_{2}}\Bigr)_{\rm hom} 
=
\Lambda^{\hspace{0.03cm}(2)}_{\ {\bf k}_{1},\, {\bf k}_{2}}
\]
\[
+\,
\frac{{\mathcal V}^{\phantom{\ast}}_{\,{\bf k}_{1},\, {\bf k}_{2},\, {\bf k}_{1} - {\bf k}_{2}} 
	{\upphi}^{\hspace{0.03cm}{\ast}}_{\hspace{0.03cm}{\bf k}_{1} - 
	{\bf k}_{2}}}
{\bigl(\omega^{\hspace{0.03cm}l}_{\hspace{0.03cm}{\bf k}_{1}} - \omega^{\hspace{0.03cm}l}_{\hspace{0.03cm}{\bf k}_{2}} - \omega^{\hspace{0.03cm}l}_{\hspace{0.03cm}{\bf k}_{1} - {\bf k}_{2}}\bigr)
	\bigl(\omega^{\hspace{0.03cm}l}_{\hspace{0.03cm}{\bf k}_{1} - {\bf k}_{2}}\! - {\bf v}\cdot ({\bf k}_{1} - {\bf k}_{2})\bigr)}
\,+\,
\frac{{\mathcal V}^{\,{\ast}}_{\,{\bf k}_{2},\, {\bf k}_{1},\, {\bf k}_{2} - {\bf k}_{1}} 
	{\upphi}^{\phantom{\ast}}_{\hspace{0.03cm}{\bf k}_{2} - {\bf k}_{1}}}
{\bigl(\omega^{\hspace{0.03cm}l}_{\hspace{0.03cm}{\bf k}_{2}} - \omega^{\hspace{0.03cm}l}_{\hspace{0.03cm}{\bf k}_{1}} - \omega^{\hspace{0.03cm}l}_{\hspace{0.03cm}{\bf k}_{2} - {\bf k}_{1}}\bigr)
	\bigl(\omega^{\hspace{0.03cm}l}_{\hspace{0.03cm}{\bf k}_{2} - {\bf k}_{1}}\! - {\bf v}\cdot ({\bf k}_{2} - {\bf k}_{1})\bigr)},
\]
where $\Lambda^{\hspace{0.03cm}(2)}_{\ {\bf k}_{1},\, {\bf k}_{2}}$ is an arbitrary function satisfying the condition
\[
\Lambda^{\hspace{0.03cm}(2)}_{\, {\bf k}_{1},\, {\bf k}_{2}} 
=
\Lambda^{\ast\hspace{0.03cm}(2)}_{\, {\bf k}_{2},\, {\bf k}_{1}}.
\]
Summing the partial and general solutions and assuming for the sake of definiteness
\[
\mathrm{Re}\,\alpha \equiv 0,
\qquad
\Lambda^{\hspace{0.03cm}(2)}_{\, {\bf k}_{1},\, {\bf k}_{2}} 
\equiv 0,
\]
we obtain the required function $\widetilde{V}^{\hspace{0.03cm}(2)}_{\ {\bf k}_{1},\, {\bf k}_{2}}$
\begin{equation}
\widetilde{V}^{\hspace{0.03cm}(2)}_{\ {\bf k}_{1},\, {\bf k}_{2}}
= 
-\hspace{0.03cm}\frac{i}{2}\,
\frac{{\upphi}^{\hspace{0.03cm}\ast}_{\hspace{0.03cm}{\bf k}_{1}}\hspace{0.03cm}{\upphi}^{\phantom{\ast}}_{\hspace{0.03cm}{\bf k}_{2}}}
{\bigl(\omega^{\hspace{0.03cm}l}_{\hspace{0.03cm}{\bf k}_{1}} - {\bf v}\cdot {\bf k}_{1}\bigr)
	\bigl(\omega^{\hspace{0.02cm} l}_{\hspace{0.03cm}{\bf k}_{2}} - {\bf v}\cdot {\bf k}_{2}\bigr)}
\label{eq:12a}
\end{equation}
\[
+\,
\frac{{\mathcal V}^{\phantom{\ast}}_{\,{\bf k}_{1},\, {\bf k}_{2},\, {\bf k}_{1} - {\bf k}_{2}} 
	{\upphi}^{\hspace{0.03cm}{\ast}}_{\hspace{0.03cm}{\bf k}_{1} - {\bf k}_{2}}}
{\bigl(\omega^{\hspace{0.03cm}l}_{\hspace{0.03cm}{\bf k}_{1}} - \omega^{\hspace{0.03cm}l}_{\hspace{0.03cm}{\bf k}_{2}} - \omega^{\hspace{0.03cm}l}_{\hspace{0.03cm}{\bf k}_{1} - {\bf k}_{2}}\bigr)
	\bigl(\omega^{\hspace{0.03cm}l}_{\hspace{0.03cm}{\bf k}_{1} - {\bf k}_{2}}\! - {\bf v}\cdot ({\bf k}_{1} - {\bf k}_{2})\bigr)}
\,+\,
\frac{{\mathcal V}^{\,{\ast}}_{\,{\bf k}_{2},\, {\bf k}_{1},\, {\bf k}_{2} - {\bf k}_{1}} 
	{\upphi}^{\phantom{\ast}}_{\hspace{0.03cm}{\bf k}_{2} - {\bf k}_{1}}}
{\bigl(\omega^{\hspace{0.03cm}l}_{\hspace{0.03cm}{\bf k}_{2}} - \omega^{\hspace{0.03cm}l}_{\hspace{0.03cm}{\bf k}_{1}} - \omega^{\hspace{0.03cm}l}_{\hspace{0.03cm}{\bf k}_{2} - {\bf k}_{1}}\bigr)
	\bigl(\omega^{\hspace{0.03cm}l}_{\hspace{0.03cm}{\bf k}_{2} - 
	{\bf k}_{1}}\! - {\bf v}\cdot ({\bf k}_{2} - {\bf k}_{1})\bigr)}.
\vspace{0.15cm}
\]
\indent We need to consider in more detail the practical implication of the coefficient function $\widetilde{V}^{\hspace{0.03cm}(2)}$ for the Hamilton formalism under consideration when the resonance frequency difference (\ref{eq:4a}) is different from zero. We recall for this purpose that the function $T^{\,(2)\,a\,a_{1}\hspace{0.03cm}a_{2}}_{\,{\bf k},\,{\bf k}_{1}}$ in the initial fourth-order interaction Hamiltonian $H^{(4)}$, Eq.\,(\ref{eq:2d}), satisfies the requirement of reality of this Hamiltonian 
\begin{equation}
	T^{\,(2)\,a\,a_{1}\hspace{0.03cm}a_{2}}_{\,{\bf k},\, {\bf k}_{1}}
	=
	- T^{\,\ast\,(2)\,a_{1}\hspace{0.03cm}a\, a_{2}}_{\,{\bf k}_{1},\, {\bf k}}. 
\label{eq:12s}
\end{equation}
\indent Let us now consider the effective amplitude $\widetilde{T}^{\,(2)\, a_{1}\hspace{0.03cm}a_{2}\,a}_{\,{\bf k}_{1},\,{\bf k}_{2}}$, which is defined by the expression (\ref{eq:12o}). If one does not use the resonance condition  
\begin{equation}
	\omega^{\hspace{0.03cm}l}_{\hspace{0.03cm}{\bf k}_{1}} - \omega^{\hspace{0.03cm}l}_{\hspace{0.03cm}{\bf k}_{2}}
	-
	{\mathbf v}\cdot ({\mathbf k}_{1} - {\mathbf k}_{2}) = 0,
\label{eq:12d}
\end{equation}
then it is not difficult to verify that in contrast to (\ref{eq:12s}), we have
\[
\widetilde{T}^{\,(2)\,a_{1}\hspace{0.03cm}a_{2}\,a}_{\,{\bf k}_{1},\, 
{\bf k}_{2}}
\neq
-\hspace{0.04cm}
\widetilde{T}^{\,\ast\,(2)\,a_{2}\,a_{1}\hspace{0.03cm}a}_{\,{\bf k}_{2},\, {\bf k}_{1}}.
\]
\indent Substituting the functions $\widetilde{V}^{\hspace{0.03cm}(2)}_{\ {\bf k}_{1},\, {\bf k}_{2}}$ and $\widetilde{T}^{\,(2)}_{\,{\bf k}_{1},\, {\bf k}_{2}}$, Eqs.\,(\ref{eq:12a}) and (\ref{eq:12o}), into (\ref{eq:12i}) and performing simple algebraic transformations, we define an explicit form of the {\it complete effective amplitude} $\mathscr{T}^{\hspace{0.03cm}(2)\hspace{0.03cm}a\,a_{1}\hspace{0.03cm} a_{2}}_{\; {\bf k}_{1},\, {\bf k}_{2}} = f^{\hspace{0.03cm}a\,a_{1}\hspace{0.02cm}a_{2}}\hspace{0.01cm}
\mathscr{T}^{\hspace{0.03cm}(2)}_{\; {\bf k}_{1},\, {\bf k}_{2}}$ with
\begin{equation}
\mathscr{T}^{\hspace{0.03cm}(2)}_{\; {\bf k}_{1},\, {\bf k}_{2}}
=
T^{\,(2)}_{\,{\bf k}_{1},\, {\bf k}_{2}}
+
\frac{1}{2}\,\biggl(\frac{1}
{\omega^{\hspace{0.03cm}l}_{\hspace{0.03cm}{\bf k}_{1}} - {\bf v}\cdot {\bf k}_{1}}
+
\frac{1}
{\omega^{\hspace{0.03cm}l}_{\hspace{0.03cm}{\bf k}_{2}} - {\bf v}\cdot {\bf k}_{2}}\biggr)
\hspace{0.03cm}{\upphi}^{\hspace{0.03cm}\ast}_{\hspace{0.03cm}{\bf k}_{1}}\hspace{0.03cm}
{\upphi}^{\phantom{\ast}}_{\hspace{0.03cm}{\bf k}_{2}}
	\label{eq:12f}
	\vspace{-0.3cm}
\end{equation}
\begin{align}
	+\;i\,\Biggl[
	\Biggl(&\frac{1}
	{\omega^{\hspace{0.03cm}l}_{\hspace{0.03cm}{\bf k}_{1} - {\bf k}_{2}}\! - {\bf v}\cdot ({\bf k}_{1} - {\bf k}_{2})}
	+
	\frac{1}
	{\omega^{\hspace{0.03cm}l}_{\hspace{0.03cm}{\bf k}_{1} - {\bf k}_{2}} -\omega^{\hspace{0.03cm}l}_{\hspace{0.03cm}{\bf k}_{1}} + \omega^{\hspace{0.03cm}l}_{\hspace{0.03cm}{\bf k}_{2}}}		
	\Biggr)\hspace{0.03cm}
	{\mathcal V}^{\phantom{\ast}}_{\,{\bf k}_{1},\, {\bf k}_{2},\, {\bf k}_{1} - {\bf k}_{2}} 
	{\upphi}^{\hspace{0.03cm}{\ast}}_{\hspace{0.03cm}{\bf k}_{1} - {\bf k}_{2}}
	\notag\\[1.5ex]
-\,
	\Biggl(&\frac{1}
	{\omega^{\hspace{0.03cm}l}_{\hspace{0.03cm}{\bf k}_{2} - {\bf k}_{1}}\! - {\bf v}\cdot ({\bf k}_{2} - {\bf k}_{1})}
	+
	\frac{1}
	{\omega^{\hspace{0.03cm}l}_{\hspace{0.03cm}{\bf k}_{2} - {\bf k}_{1}} - \omega^{\hspace{0.03cm}l}_{\hspace{0.03cm}{\bf k}_{2}} + \omega^{\,l}_{\hspace{0.03cm}{\bf k}_{1}}}		
	\Biggr)\hspace{0.03cm}
	{\mathcal V}^{\,{\ast}}_{\,{\bf k}_{2},\, {\bf k}_{1},\, {\bf k}_{2} - {\bf k}_{1}} 
	{\upphi}^{\phantom{\ast}}_{\hspace{0.03cm}{\bf k}_{2} - {\bf k}_{1}}
	\Biggr].
	\notag
\end{align}
The presented form (\ref{eq:12f}) of the complete effective amplitude $\mathscr{T}^{\,(2)}$ makes the validity of the requirement of reality of the Hamiltonian (\ref{eq:12u}) practically obvious 
\[
\mathscr{T}^{\hspace{0.03cm}(2)\hspace{0.03cm}a\,a_{1}\hspace{0.03cm} a_{2}}_{\; {\bf k}_{1},\, {\bf k}_{2}}
=
-\hspace{0.03cm}
\mathscr{T}^{\hspace{0.03cm}\ast\,(2)\hspace{0.03cm}a\, a_{2}\hspace{0.03cm} a_{1}}_{\; {\bf k}_{2},\, {\bf k}_{1}}. 
\]
Thus, the role of the coefficient function $\widetilde{V}^{\,(2)\,a_{1}\hspace{0.03cm} a_{2}\, a}_{\ {\bf k}_{1},\, {\bf k}_{2}}$ is reduced to the total symmetrization of the effective amplitude $\widetilde{T}^{\,(2)\,a_{1}\hspace{0.03cm}a_{2}\, a}_{\,{\bf k}_{1},\, {\bf k}_{2}}$. This involves the fulfillment of the necessary symmetry condition without any use of the resonance condition (\ref{eq:12d}). In other words, the effective amplitude $\widetilde{T}^{\,(2)\,a_{1}\hspace{0.03cm}a_{2}\,a}_{\,{\bf k}_{1},\,{\bf k}_{2}}$ satisfies the symmetry condition of the (\ref{eq:12s}) type only on the resonance surface described by Eq.\,(\ref{eq:12d}), while accounting for the contribution with the function $\widetilde{V}^{\,(2)\,a_{1}\hspace{0.03cm}a_{2}\,a}_{\ {\bf k}_{1},\,{\bf k}_{2}}$ makes it possible to extend the symmetry condition throughout the space of the vectors ${\bf k}_{1}$ and ${\bf k}_{2}$. In this case the resonance frequency difference $\Delta\hspace{0.02cm}\omega_{\hspace{0.03cm}
{\mathbf k}_{1},\hspace{0.03cm}{\mathbf k}_{2}}$, Eq.\,(\ref{eq:4a}), can be arbitrary. Note that the explicit expressions we derived for the coefficient functions $\widetilde{V}^{\,(1)\,a_{1}\hspace{0.03cm}a_{2}\,a}_{\ {\bf k}_{1},\,{\bf k}_{2}}$ and  $\widetilde{V}^{\,(2)\,a_{1}\hspace{0.03cm}a_{2}\,a}_{\ {\bf k}_{1},\,{\bf k}_{2}}$, Eqs.\,(\ref{eq:12y}) and (\ref{eq:12a}), will be obtained by another more rigorous approach in our second part \cite{markov_II_2023}, which serves to further confirm their validity.\\
\indent To complete the picture, in conclusion of this section we consider the remaining fourth-order term in the initial fourth-order interaction Hamiltonian (\ref{eq:2d}), namely  
\begin{equation}
\Xi\hspace{0.03cm}Q^{\hspace{0.03cm}a}\hspace{0.02cm}Q^{\hspace{0.03cm}a}.
\label{eq:12g}
\end{equation}
By virtue of the Wong equation (\ref{eq:1r}), the colorless combination $Q^{\hspace{0.03cm}a}\hspace{0.02cm}Q^{\hspace{0.03cm}a}$ is a conserved quantity, i.e. 
\[
Q^{\hspace{0.03cm}a}(t)\hspace{0.03cm}Q^{\hspace{0.03cm}a}(t)
=
Q^{\hspace{0.03cm}a}_{0}\hspace{0.03cm}Q^{\hspace{0.03cm}a}_{0},
\]
where $Q^{\hspace{0.03cm}a}_{\hspace{0.03cm}0}\equiv	Q^{\hspace{0.03cm}a}(t_{0})$ and so the contribution (\ref{eq:12g}) is actually just some constant. However, we cannot assume the coefficient $\Xi$ to be zero. The value for $\Xi$ is determined from the condition of vanishing the corresponding fourth-order contributions 
that are generated by the canonical transformations (\ref{eq:3t}) and (\ref{eq:3y}) from the free field and third-order Hamiltonians, Eqs.\,(\ref{eq:2p}) and (\ref{eq:2s}), respectively
\[
H^{(0)} + H^{(3)}
=
\int\!d\hspace{0.03cm}{\bf k}\,
\bigl[
(\omega^{\hspace{0.03cm}l}_{\hspace{0.03cm}{\bf k}} - {\mathbf v}\cdot {\mathbf k})\hspace{0.03cm}
F^{\,\ast}_{\hspace{0.03cm}\bf k}F^{\,\phantom{\ast}}_{\hspace{0.03cm} \bf k} 
+
{\upphi}^{\phantom{\ast}}_{\hspace{0.03cm}{\bf k}}
F^{\,\phantom{\ast}}_{\hspace{0.03cm} \bf k}
+
{\upphi}^{\hspace{0.02cm}\ast}_{\hspace{0.03cm}{\bf k}}
F^{\,\ast}_{\hspace{0.03cm}\bf k}
\bigr]\hspace{0.03cm} 
{\mathcal Q}^{\hspace{0.03cm}a}\hspace{0.02cm}
{\mathcal Q}^{\hspace{0.03cm}a}.
\]
We add this expression to (\ref{eq:12g}), where we make the replacement $Q^{\hspace{0.03cm}a}\rightarrow {\mathcal Q}^{\hspace{0.03cm}a}$, and require that the resulting sum be turned to zero. Taking into consideration the connection (\ref{eq:4w}), we obtain an explicit form of the desired coefficient $\Xi$:
\[
\Xi = \!\int\!d\hspace{0.02cm}{\bf k}\; 
\frac{|{\upphi}_{\hspace{0.03cm}{\bf k}}|^{2}}
{\omega^{\hspace{0.03cm}l}_{\hspace{0.03cm}{\bf k}} - {\mathbf v}\cdot {\mathbf k}}\,.
\]
As we will discuss in the Conclusion, we associate the quadratic in color charge ${\mathcal Q}^{\hspace{0.03cm}a}$ terms with particular ``one-loop'' corrections to the process of tree-level elastic scattering of plasmon off a hard particle. However, because the external soft gluon lines are absent here, a contribution of the type (\ref{eq:12g}) can be diagrammatically interpreted as depicted in fig.\,\ref{fig4}.
\\
\indent It should be noted, however, that the situation with the contribution (\ref{eq:12g}) (as well as with contributions such as $d^{\hspace{0.03cm}a\hspace{0.03cm}b\hspace{0.03cm}c}Q^{\hspace{0.03cm}a}
\hspace{0.02cm}Q^{\hspace{0.03cm}b}\hspace{0.02cm}Q^{\hspace{0.03cm}c}$ for more complicated interaction Hamiltonians) is not quite clear. The point is that we must consider all coefficient functions in the integrands of the original interaction Hamiltonians $H^{(3)}$ and $H^{(4)}$, Eqs.\,(\ref{eq:2s}) and (\ref{eq:2d}), as given functions reflecting the properties of the real physical medium. In the case of the contribution (\ref{eq:12g}) the situation is different. Here the $\Xi$ coefficient itself is defined from the vanishing requirement of such a contribution of a new effective fourth-order Hamiltonian ${\mathcal H}^{(4)}_{g\hspace{0.01cm}G\hspace{0.02cm}\rightarrow\hspace{0.02cm} g\hspace{0.01cm}G}$.


\section{Conclusion}
\label{section_13}
\setcounter{equation}{0}

\begin{figure}[hbtp]
	\begin{center}
		\includegraphics*[height=3.4cm]{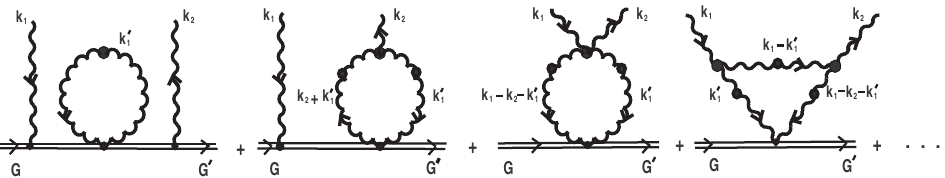}
	\end{center}
	\caption{\small Some soft one-loop corrections to the process of tree-level elastic scattering of plasmon off a hard particle depicted in fig.\,\ref{fig2}. The third diagram here contains the effective four-gluon vertex function defined by the formulas  (\ref{ap:A4})\,--\,(\ref{ap:A6}) of Appendix \ref{appendix_A}}
	\label{fig5_red}
\end{figure}

In this paper, we have set up the classical Hamiltonian formalism needed to describe the process of nonlinear interaction between soft and hard excitation modes in a high-temperature gluon plasma. The canonical transformations for the bosonic normal variable $a^{\hspace{0.03cm}a}_{\hspace{0.03cm}{\bf k}}$ and the color charge $Q^{\hspace{0.03cm}a}$ of a hard test particle, Eqs.\,(\ref{eq:3t}) and (\ref{eq:3y}), are constructed in an explicit form. We sought these transformations in the form of an integro-power series in a new normal variable $c^{\hspace{0.03cm}a}_{{\bf k}}$ and a new color charge $\mathcal{Q}^{\hspace{0.03cm}a}$. The canonical transformations enabled us to exclude the third-order interaction Hamiltonian $H^{(3)}$, Eq.\,(\ref{eq:2s}). This in turn allowed us to define a new effective interaction Hamiltonian ${\mathcal H}^{(4)}_{g\hspace{0.01cm}G\hspace{0.02cm}\rightarrow\hspace{0.02cm} g\hspace{0.01cm}G}$, Eq.\,(\ref{eq:4y}), with the gauge-covariant scattering amplitude $\mathscr{T}^{\hspace{0.03cm}(2)\hspace{0.03cm} a\, a_{1}\hspace{0.03cm} a_{2}}_{\; {\bf k}_{1},\, {\bf k}_{2}}$, Eq.\,(\ref{eq:12f}). This amplitude describes the elastic scattering 
of plasmon off a hard color-charged particle in the tree approximation. On the basis of the canonicity conditions (\ref{eq:3e}) and (\ref{eq:3r}) we  determined a complete system of independent relations of the algebraic and integral types connecting the lowest and highest coefficient functions in the integrands of various terms in the canonical transfor\-mations (\ref{eq:3t}) and (\ref{eq:3y}).\\
\indent In the present work we have restricted ourselves to the detailed consideration of only the simplest process of nonlinear interaction of soft purely collective excitations in the gluon plasma: elastic scattering of plasmon off hard particle of the type 2 $\rightarrow 2$, i.e.
\begin{equation}
{\rm g}^{\ast}_{\hspace{0.02cm}1} + {\rm G} \rightleftharpoons
{\rm g}^{\ast}_{\hspace{0.02cm}2} + {\rm G}^{\hspace{0.02cm}\prime},
\label{eq:14q}
\end{equation}
where ${\rm g}^{\ast}_{1}$ and ${\rm g}^{\ast}_{2}$ are plasmon collective excitations and ${\rm G},\, {\rm G}^{\prime}$ are excitations with characteristic momenta of order the temperature $T$ and above. At least for the weakly-excited system correspon\-ding to the level of thermal fluctuations, the given process is dominant. The approach developed allows us to consider more complicated scattering processes, for example, the simplest scattering processes of the ``inelastic'' type
\[
{\rm g}^{\ast}_{\hspace{0.02cm}1} + {\rm g}^{\ast}_{\hspace{0.02cm}2} + 
{\rm G}\rightleftharpoons
{\rm g}^{\ast}_{\hspace{0.02cm}3} + {\rm G}^{\hspace{0.02cm}\prime},
\]
or
\[
{\rm g}^{\ast}_{\hspace{0.02cm}1} + {\rm G}\rightleftharpoons
{\rm g}^{\ast}_{\hspace{0.02cm}2} + {\rm g}^{\ast}_{\hspace{0.02cm}3} + 
{\rm G}^{\hspace{0.02cm}\prime}.
\]
The corresponding Hamiltonian of this scattering process has the form  
\[
{\mathcal H}^{(5)}_{2g\hspace{0.02cm}G\hspace{0.02cm}\rightarrow\hspace{0.02cm} g\hspace{0.02cm}G} 
=
\int\!\frac{d\hspace{0.03cm}{\bf x}_{0}}{(2\pi)^{3}}\!
\int\!\prod_{i=1}^{3} d\hspace{0.03cm}{\bf k}_{i}\,
\mathscr{T}^{\hspace{0.03cm}(3)\hspace{0.03cm}a_{1}\hspace{0.03cm}a_{2}\, a_{3}\,a_{4}}_{\; {\bf k}_{1},\, {\bf k}_{2},\, {\bf k}_{3}}\hspace{0.03cm}
c^{\ast\ \!\!a_{1}}_{{\bf k}_{1}} c^{\ast\ \!\!a_{2}}_{{\bf k}_{2}}
c^{\hspace{0.03cm}a_{3}}_{{\bf k}_{3}} 
\hspace{0.03cm} \mathcal{Q}^{\,a_{4}}
\,+\,
(\mathrm{compl.\,conj.}).
\]
Although the calculation of the effective amplitude $\mathscr{T}^{\hspace{0.03cm}(3)\hspace{0.03cm}a_{1}\, a_{2}\, a_{3}\, a_{4}}_{\; {\bf k}_{1},\, {\bf k}_{2},\, {\bf k}_{3}}$ becomes considerably more cumbersome here but is still quite manageable. For more complicated scattering processes the proposed approach becomes computationally inefficient and here it is necessary to develop other methods in the spirit, for example, of the paper \cite{markov_2004}, where the procedure of practically automatic calculation at least of the scattering amplitudes of any complexity of processes is proposed.\\
\indent However, even for the simplest elastic scattering process (\ref{eq:14q}) the story does not end here. Generally, the most general form of the effective fourth-order Hamiltonian describing the elastic scattering of a plasmon off a hard color-charged test particle is, instead of (\ref{eq:12u}), 
\[
	{\mathcal H}^{(4)}_{g\hspace{0.02cm}G\hspace{0.02cm}\rightarrow\hspace{0.02cm} g\hspace{0.02cm}G} 
	=
	i\!
	\int\!\frac{d\hspace{0.03cm}{\bf x}_{0}}{(2\pi)^{3}}\!
	\int\!d\hspace{0.03cm}{\bf k}_{1}\hspace{0.03cm}d\hspace{0.03cm}{\bf k}_{2}\,
	\mathscr{T}^{\hspace{0.03cm}(2)\hspace{0.03cm}a_{1}\hspace{0.03cm}a_{2}}_{\; {\bf k}_{1},\, {\bf k}_{2}}(\mathcal{Q})\,
	c^{\ast\ \!\!a_{1}}_{{\bf k}_{1}} c^{\hspace{0.03cm}a_{2}}_{{\bf k}_{2}}	,
\]
where the effective amplitude $\mathscr{T}^{\hspace{0.03cm}(2)\hspace{0.03cm}a_{1}\, a_{2}}_{\; {\bf k}_{1},\, {\bf k}_{2}}(\mathcal{Q})$ is generally speaking an arbitrary function of the color vector $\mathcal{Q} = (\mathcal{Q}^{\,a})$. We can represent this amplitude in the form of an expansion in the color charge $\mathcal{Q}^{\,a}$ of an energetic particle
\begin{equation}
\mathscr{T}^{\hspace{0.03cm}(2)\hspace{0.03cm}a_{1}\hspace{0.03cm}a_{2}}_{\; {\bf k}_{1},\, {\bf k}_{2}}(\mathcal{Q})
=
\mathscr{T}^{\hspace{0.03cm}(2)\hspace{0.03cm}a_{1}\hspace{0.03cm}a_{2}
\hspace{0.04cm}a_{3}}_{\; {\bf k}_{1},\, {\bf k}_{2}}\mathcal{Q}^{\,a_{3}}
+
\frac{1}{2}\,
\mathscr{T}^{\hspace{0.03cm}(2)\hspace{0.03cm}a_{1}\hspace{0.03cm}a_{2} \hspace{0.04cm}a_{3}\hspace{0.03cm}a_{4}}_{\; {\bf k}_{1},\, 
{\bf k}_{2}}\mathcal{Q}^{\,a_{3}} \mathcal{Q}^{\,a_{4}}
+\hspace{0.03cm} \ldots\,.
\label{eq:14w}
\end{equation}
Thus, our expression for the effective amplitude (\ref{eq:12i}) represents only the first term of this expansion. The remaining terms of the expansion are tacitly assumed to be suppressed by the coupling constant. However, what physical meaning or interpretation do they have? At least the second term in the expansion (\ref{eq:14w}) can be calculated explicitly within the framework of our approach. An example of calculating such a function can be found in \cite{markov_2004}. The diagrammatic interpretation of some terms in $\mathscr{T}^{\hspace{0.03cm}(2)\hspace{0.03cm}a_{1}\hspace{0.03cm}a_{2}
\hspace{0.04cm}a_{3}\, a_{4}}_{\; {\bf k}_{1},\, {\bf k}_{2}}$ is presented in fig.\,\ref{fig5_red}.
From the view of these diagrams, we can conclude that here we are concerned with soft ``one-loop'' corrections to the elastic scattering process of plasmon off the hard color-charged particle that is suppressed by a power of $g^{2}$ compared with the tree approximation (\ref{eq:12u}), (\ref{eq:12i}). We note that the region of integration in the loops is restricted by cone $v\cdot k^{\hspace{0.03cm}\prime}_{1} = 0$. ``Radiation'' corrections of this kind have been considered, for example, in the theory of wave turbulence \cite{gurarie_1995, rosenhaus_2023} in the construction of the kinetic equations for weak excitations to next-to-leading order. There they were interpreted as fluctuations about the stationary wave turbulent state. We can also suggest that the higher terms in the expansion (\ref{eq:14w}) are a kind of fluctuational deviation from the elastic scattering process (\ref{eq:14q}) in the tree approximation. In particular, the first diagram in fig.\ref{fig5_red} represents the one-loop correction to the propagator of the hard particle, defined by the expression (\ref{eq:4aa}).


\section*{\bf Acknowledgment}

The research was funded by the Ministry of Education and Science of the Russian Federation within the framework of the project ``Analytical and numerical methods of mathematical physics in problems of tomography, quantum field theory, and fluid and gas mechanics'' (no. of state registration: 121041300058-1).

\begin{appendices}
\numberwithin{equation}{section}


\numberwithin{equation}{section}
\section{Effective gluon vertices and gluon propa\-ga\-tor}
\numberwithin{equation}{section}
\label{appendix_A}

In this Appendix, we have provided the explicit form of the vertex functions and gluon propagator in the hard thermal loop (HTL) approximation \cite{blaizot_2002, ghiglieri_2020, braaten_1990}.\\
\indent Effective three-gluon vertex
\begin{equation}
	\,^{\ast} \Gamma^{\hspace{0.03cm}\mu\hspace{0.02cm} \nu  \rho}(k, k_{1}, k_{2}) \equiv
	\Gamma^{\hspace{0.03cm}\mu\hspace{0.02cm} \nu  \rho}(k, k_{1}, k_{2}) +
	\delta\hspace{0.025cm} \Gamma^{\hspace{0.03cm}\mu\hspace{0.02cm} \nu  \rho}(k, k_{1}, k_{2})
	\label{ap:A1}
\end{equation}
is the sum of bare three-gluon vertex
\begin{equation}
	\Gamma^{\hspace{0.03cm}\mu\hspace{0.02cm}\nu\hspace{0.02cm}\rho}(k, k_{1}, k_{2}) =
	g^{\hspace{0.03cm}\mu\hspace{0.02cm}\nu} (k - k_{1})^{\rho} + g^{\hspace{0.03cm}\nu\hspace{0.02cm}\rho} (k_{1} - k_{2})^{\mu} +
	g^{\hspace{0.03cm}\mu\hspace{0.02cm}\rho} (k_{2} - k)^{\nu}
	\label{ap:A2}
\end{equation}
and the corresponding HTL correction
\begin{equation}
	\delta\hspace{0.025cm} \Gamma^{\hspace{0.03cm}\mu\hspace{0.02cm} \nu  \rho}(k, k_{1}, k_{2}) =
	3\hspace{0.035cm}\omega^{\hspace{0.02cm} 2}_{\rm pl}\!\int\!\frac{d\hspace{0.035cm}\Omega}{4 \pi} \,
	\frac{v^{\hspace{0.03cm}\mu}\hspace{0.02cm} v^{\hspace{0.03cm}\nu} v^{\hspace{0.03cm}\rho}}{v\cdot k + i\hspace{0.025cm}\epsilon} \,
	\Biggl(\frac{\omega_{2}}{v\cdot k_{2} - i\epsilon} -
	\frac{\omega_1}{v\cdot k_{1} - i\epsilon}\Biggr),
	\quad \epsilon\rightarrow +\hspace{0.02cm}0,
	\label{ap:A3}
\end{equation}
where $v^{\hspace{0.03cm}\mu} = (1,{\bf {\bf v}})$, $k^{\hspace{0.03cm}\mu} = (\omega, {\bf k})$ is a gluon four-momentum with $k  + k_{1} + k_{2} = 0$, $d\hspace{0.035cm}\Omega$ is a differential solid angle and $\omega_{\rm pl}^{\hspace{0.02cm}2} = g^{\hspace{0.03cm}2}N_{c}T^{\hspace{0.02cm}2}/\hspace{0.02cm}9$ is the gluon plasma frequency squared. We present below useful properties of the three-gluon HTL-resumed vertex function for complex conjugation and permutation of momenta:
\[
	\left(\!\,^{\ast}\Gamma_{\mu\hspace{0.02cm} \mu_{1} \mu_{2}}(-k_{1} - k_{2}, k_{1}, k_{2})\right)^{\ast} =
	-\!\,^{\ast}\Gamma_{\mu\hspace{0.02cm} \mu_{1} \mu_{2}}(k_{1} + k_{2}, -k_{1}, -k_{2}) 
	= \!\,^{\ast}\Gamma_{\mu\hspace{0.02cm} \mu_{2}\mu_{1}}(k_{1} + k_{2}, -k_{2}, -k_{1}).
\]	
%
Furthermore, the effective four-gluon vertex
\begin{equation}
	^{\ast} \Gamma^{\hspace{0.03cm}\mu\hspace{0.02cm} \nu  \lambda \sigma}(k,k_1,k_2,k_3) \equiv
	\Gamma^{\hspace{0.03cm}\mu\hspace{0.02cm} \nu  \lambda \sigma}(k,k_1,k_2,k_3) +
	\delta\hspace{0.025cm} \Gamma^{\hspace{0.03cm}\mu\hspace{0.02cm} \nu  \lambda \sigma}(k,k_1,k_2,k_3)
	\label{ap:A4}
\end{equation}
is the sum of bare four-gluon vertex
\begin{equation}
	\Gamma^{\hspace{0.03cm}\mu\hspace{0.02cm}\nu \lambda\sigma} =
	2\hspace{0.02cm}g^{\hspace{0.03cm}\mu\hspace{0.02cm}\nu }g^{\hspace{0.03cm}\lambda\sigma} - g^{\hspace{0.03cm}\mu \sigma}g^{\hspace{0.03cm}\nu\lambda} -
	g^{\hspace{0.03cm}\mu \lambda}g^{\hspace{0.03cm}\sigma\nu}
	\label{ap:A5}
\end{equation}
and the corresponding HTL correction
\begin{equation}
	\delta\hspace{0.025cm}\Gamma^{\hspace{0.03cm}\mu\hspace{0.02cm}\nu  \lambda\sigma}(k, k_1, k_2,k_3) 
	= 
	3\hspace{0.035cm}\omega^2_{\rm pl}\!\int\!\frac{d\hspace{0.035cm}\Omega}{4 \pi} \, \frac{v^{\hspace{0.03cm}\mu}\hspace{0.02cm}v^{\hspace{0.03cm}\nu}
	\hspace{0.02cm} v^{\hspace{0.03cm}\lambda}v^{\hspace{0.03cm}\sigma}}{v\cdot k + i\hspace{0.025cm}\epsilon}
	\label{ap:A6}
\end{equation}
\[
\times\Biggl[
\,\frac{1}{v\cdot (k + k_1) + i\epsilon}\, 
\Biggl(\frac{\omega_{2}}{v\cdot k_2 - i \epsilon} - \frac{\omega_3}{v\cdot k_3 - i \epsilon} \Biggr)
- \frac{1}{v\cdot (k + k_3) + i \epsilon}\,
\Biggl(\frac{\omega_{1}}{v\cdot k_1 - i \epsilon} - \frac{\omega_2}{v\cdot k_2 - i \epsilon} \Biggr)\Biggr].
\]
Finally, the expression
\begin{equation}
^{\ast}\widetilde{\cal D}_{\mu\hspace{0.02cm} \nu }(k) = 
- P_{\mu\hspace{0.02cm} \nu }(k) \,^{\ast}\!\Delta^t(k) - \widetilde{Q}_{\mu\hspace{0.02cm} \nu }(k) \,^{\ast}\!\Delta^l(k)
- \xi_{0}\ \!\frac{k^{2}}{(k\cdot u)^{2}}\ \!D_{\mu\hspace{0.02cm} \nu }(k)
\label{ap:A7}
\end{equation}
is a gluon (retarded) propagator in the $A_0$\hspace{0.02cm}-\hspace{0.02cm}gauge, which is modified by effects of the medium. Here, the ``scalar'' transverse and longitudinal propagators have the following form
\begin{equation}
\hspace{-1cm}\,^{\ast}\!\Delta^{t}(k) = \frac{1}{k^2 - \Pi^{\hspace{0.025cm}t}(k)},
\qquad\quad\;
\,^{\ast}\!\Delta^{l}(k) = \frac{1}{k^{2} - \Pi^{\hspace{0.025cm}l}(k)},
\label{ap:A8}
\end{equation}
where
\[
\Pi^{\hspace{0.025cm}t}(k) = \frac{1}{2}\, \Pi^{\hspace{0.02cm}\mu\nu}(k) P_{\mu\nu}(k),
\qquad
\Pi^{\hspace{0.025cm} l}(k) = \Pi^{\hspace{0.02cm}\mu\nu}(k) \widetilde{Q}_{\mu\nu}(k).
\hspace{0.2cm}
\]
The polarization tensor $\Pi_{\mu\hspace{0.02cm} \nu }(k)$ in the HTL approximation takes the form
\[
\Pi^{\hspace{0.025cm}\mu\hspace{0.02cm}\nu }(k) = 3\hspace{0.035cm}\omega_{\rm pl}^{2}\!\hspace{0.02cm}
\left(u^{\hspace{0.03cm}\mu}u^{\hspace{0.03cm}\nu} - \omega\!\int\!\frac{d\hspace{0.035cm}\Omega}{4 \pi}
\,\frac{v^{\hspace{0.03cm}\mu}\hspace{0.02cm} v^{\hspace{0.03cm}\nu}}{v\cdot k + i \epsilon} \right)
\]
and the longitudinal and transverse projectors are defined by the expressions
\begin{equation}
	\begin{split}
		&\widetilde{Q}_{\mu\hspace{0.02cm}\nu }(k) =
		\frac{\tilde{u}_{\hspace{0.02cm}\mu}(k)\hspace{0.02cm} \tilde{u}_{\hspace{0.03cm}\nu}(k)}{\bar{u}^2(k)}\,,\\[0.7ex]
		&P_{\mu\nu}(k) = g_{\hspace{0.02cm}\mu\nu} - u_{\hspace{0.02cm}\mu}u_{\hspace{0.03cm}\nu}
		- \widetilde{Q}_{\mu\hspace{0.02cm} \nu }(k)\hspace{0.03cm}\frac{(k\cdot u)^{2}}{k^{2}}\, ,
	\end{split}
	\label{ap:A9}
\end{equation}
respectively. Two four-vectors 
\begin{equation}
	\tilde{u}_{\mu} (k) = \frac{k^2}{(k\cdot u)}\ \! \Bigl(k_{\mu} - u_{\mu}(k\cdot u)\Bigr)
	\quad \mbox{and} \quad
	\bar{u}_{\mu}(k) = k^{2}u_{\mu} - k_{\hspace{0.02cm}\mu}(k\cdot u)
	\label{ap:A10}
\end{equation}
are the projectors onto the longitudinal direction of wavevector ${\bf k}$ written in the Lorentz-invariant form in the Hamilton and Lorentz gauges, respectively. Here, $u^{\hspace{0.02cm}\mu}$ is the four-velocity of the medium. In the rest frame of the system $u^{\hspace{0.02cm}\mu}=(1,0,0,0)$.

\section{System of the canonicity conditions}
\numberwithin{equation}{section}
\label{appendix_B}

Here, we present a system of the canonicity conditions, which follows from the Lie-Poisson brackets (\ref{eq:3ea})\hspace{0.03cm}--\,(\ref{eq:3ec}). This system connects the coefficient functions of the second, third and so on orders in the integrands of the canonical transformations (\ref{eq:3t}) and (\ref{eq:3y}) among themselves. Substituting the transformation (\ref{eq:3t}) into equations (\ref{eq:3ea}) and (\ref{eq:3eb}), equating the coefficients of the same powers of a product of the functions $c^{\hspace{0.025cm}a}_{\bf k},\,c^{\hspace{0.03cm}\ast\,a}_{{\bf k}}$ and color charge ${\mathcal Q}^{\hspace{0.03cm}a}$ to zero, and taking into account the relations (\ref{eq:3i}), we obtain the required canonicity conditions. The first of them has the following form:
\begin{align}
	W^{\hspace{0.025cm}\ast\hspace{0.025cm}(2)\hspace{0.03cm}a\,a_{1}\hspace{0.02cm}a_{2}
	\, a_{3}}_{\ {\bf k},\, {\bf k}_{1},\, {\bf k}_{2}}
	&+
	2\hspace{0.03cm}W^{\hspace{0.025cm}(1)\hspace{0.03cm}a_{2}\,a_{1}
	\hspace{0.02cm}a\, a_{3}}_{\ {\bf k}_{2},\, {\bf k}_{1},\, {\bf k}}
	\,
	+
	i\hspace{0.03cm}\Bigl(F^{\phantom{(2)}}_{{\bf k}_{2}}\!\!
	\widetilde{V}^{\hspace{0.025cm}\ast\hspace{0.025cm}(1)\hspace{0.03cm}a\,a_{1}
	\hspace{0.02cm} a^{\prime}}_{\ \ {\bf k},\, {\bf k}_{1}}\!
	f^{\,a_{2}\,a^{\prime}\hspace{0.02cm}a_{3}}
	+
	F^{\,\ast\phantom{(1)}}_{{\bf k}}\!\!\!\!
	\widetilde{V}^{\hspace{0.025cm}(2)\hspace{0.03cm}a_{2}\, a_{1}\hspace{0.03cm}a^{\prime}}_{\ \ {\bf k}_{2},\, {\bf k}_{1}}\! 
	f^{\,a^{\prime}\hspace{0.02cm}a\,a_{3}}\Bigr) 
	\label{ap:B1}\\[1ex]
	&- 2\!\int\!d\hspace{0.03cm}{\bf k}^{\hspace{0.015cm}\prime}\,\Bigl[\hspace{0.02cm}
	\widetilde{V}^{\,(2)\hspace{0.03cm}a_{2}\,a^{\prime}\hspace{0.02cm}a_{3}}_{\ \ {\bf k}_{2},\, {\bf k}^{\prime}}\,
	V^{\hspace{0.03cm}(1)\hspace{0.03cm} a^{\prime}\hspace{0.02cm}a_{1}\hspace{0.03cm}a}_{\ \ {\bf k}^{\prime},\, {\bf k}_{1},\, {\bf k}} 
	-
	\widetilde{V}^{\hspace{0.025cm}\ast\hspace{0.025cm}(2)\hspace{0.03cm} a\, a^{\prime}\hspace{0.02cm}a_{3}}_{\ \  {\bf k},\, {\bf k}^{\prime}}\,
	V^{\hspace{0.03cm}\hspace{0.025cm}(1)\hspace{0.03cm}a_{2}\,a^{\prime}\,a_{1}}_{\ \ {\bf k}_{2},\, {\bf k}^{\prime},\, {\bf k}_{1}}
	+ 
	\widetilde{V}^{\,(1)\hspace{0.03cm}a_{2}\,a^{\prime}\hspace{0.02cm}a_{3}}_{\ \ {\bf k}_{2},\, {\bf k}^{\prime}}\,
	V^{\hspace{0.03cm}\hspace{0.025cm}\ast\hspace{0.025cm}(3)\hspace{0.03cm}a\,a_{1}\hspace{0.02cm}
	a^{\prime}}_{\ \ {\bf k},\, {\bf k}_{1},\, {\bf k}^{\prime}}
	\notag\\[1ex]
	&\hspace{9cm}-
	\widetilde{V}^{\hspace{0.025cm}\ast\hspace{0.025cm}(1)\hspace{0.03cm}a\, a^{\prime}\hspace{0.02cm}a_{3}}_{\ \ {\bf k},\, {\bf k}^{\prime}}
	\,
	V^{\hspace{0.03cm}\hspace{0.025cm}\ast\hspace{0.025cm}(1)\hspace{0.03cm}a_{1}\hspace{0.03cm}
	a^{\prime}\, a_{2}}_{\ \  {\bf k}_{1},\, {\bf k}^{\prime},\, {\bf k}_{2}}\Bigr] 
	= 0.
	\notag
\end{align}
This relation defines the connection of the two higher functions $W^{\hspace{0.025cm} (1)\hspace{0.03cm} a\, a_{1}\, a_{2}\, a_{3}}_{\ {\bf k},\, {\bf k}_{1},\, {\bf k}_{2}}$ and $W^{\hspace{0.025cm} (2)\hspace{0.03cm} a\, a_{1}\, a_{2}\, a_{3}}_{\ {\bf k},\, {\bf k}_{1},\, {\bf k}_{2}}$ to each other. It can be considered as a generalization of the first relation in (\ref{eq:3i}) to higher coefficient functions. The corresponding generalization of the second relation in (\ref{eq:3i}) should be considered as two relations of the form 
\begin{subequations} 
	\label{ap:B2}
	\begin{align}
		W^{\hspace{0.025cm}(2)\hspace{0.03cm}a\,a_{1}\hspace{0.03cm}a_{2}\,a_{3}}_{\ {\bf k},\, {\bf k}_{1},\, {\bf k}_{2}}
		&-
		W^{\hspace{0.025cm} (2)\hspace{0.03cm}a_{1}\hspace{0.03cm}a\, a_{2}\,a_{3}}_{\ {\bf k}_{1},\, {\bf k},\, {\bf k}_{2}}
		\,+
		i\hspace{0.04cm}\Bigl(F^{\phantom{(1)}}_{{\bf k}_{1}}\!\!
		\widetilde{V}^{\hspace{0.025cm}(2)\hspace{0.03cm}a\,a_{2}\,a^{\prime}}_{\ \ {\bf k},\, {\bf k}_{2}}\!
		f^{\,a_{1}\hspace{0.03cm}a^{\prime}\hspace{0.02cm}a_{3}}
		+
		F^{\hspace{0.03cm}\phantom{(1)}}_{{\bf k}}\!\!\!\!\hspace{0.03cm}
		\widetilde{V}^{\hspace{0.025cm}(2)\hspace{0.03cm}a_{1}\hspace{0.03cm}a_{2}\, a^{\prime}}_{\ \ {\bf k}_{1},\, {\bf k}_{2}}\! 
		f^{\,a^{\prime}\hspace{0.02cm}a\,a_{3}}\Bigr) 
		\label{ap:B2a}\\[1ex]
		&- 2\!\int\!d\hspace{0.03cm}{\bf k}^{\hspace{0.015cm}\prime}\,\Bigl[\hspace{0.02cm}
		\widetilde{V}^{(2)\hspace{0.03cm}a_{1}\hspace{0.03cm}a^{\prime}\hspace{0.02cm}
		a_{3}}_{\ \ {\bf k}_{1},\, {\bf k}^{\prime}}\,
		V^{\hspace{0.03cm}\hspace{0.03cm}\ast\hspace{0.03cm}(1)\hspace{0.03cm}a_{2}\,a^{\prime}
		\hspace{0.03cm}a}_{\ \ {\bf k}_{2},\, {\bf k}^{\prime},\, {\bf k}} 
		-
		\widetilde{V}^{\hspace{0.025cm}(1)\hspace{0.03cm}a\,a^{\prime}\hspace{0.02cm}
		a_{3}}_{\ \  {\bf k},\, {\bf k}^{\prime}}\,
		V^{\hspace{0.03cm}\hspace{0.025cm}(1)\hspace{0.03cm}a_{1}\hspace{0.03cm}a^{\prime}\hspace{0.02cm}a_{2}}_{\ \ {\bf k}_{1},\, {\bf k}^{\prime},\, {\bf k}_{2}}
		+ 
		\widetilde{V}^{(1)\hspace{0.03cm}a_{1}\hspace{0.03cm}a^{\prime}\hspace{0.02cm}
		a_{3}}_{\ \ {\bf k}_{1},\, {\bf k}^{\prime}}\,
		V^{\hspace{0.03cm}\hspace{0.025cm}(1)\hspace{0.03cm}a\,a^{\prime}\hspace{0.02cm}a_{2}}_{\ \  {\bf k},\, {\bf k}^{\prime},\, {\bf k}_{2}}
		\notag\\[1ex]
		&\hspace{9cm}-
		\widetilde{V}^{\hspace{0.025cm}(2)\hspace{0.03cm}a\, a^{\prime}\hspace{0.02cm}a_{3}}_{\ \ {\bf k},\, {\bf k}^{\prime}}
		\,
		V^{\hspace{0.03cm}\hspace{0.025cm}\ast\hspace{0.025cm}(1)\hspace{0.03cm}a_{2}\,a^{\prime}
		\hspace{0.03cm} a_{1}}_{\ \ {\bf k}_{2},\, {\bf k}^{\prime},\, {\bf k}_{1}}\Bigr] 
		= 0, 
		\notag\\[1.5ex]
		W^{\hspace{0.025cm}(3)\hspace{0.03cm}a\,a_{1}\hspace{0.03cm}a_{2}\,a_{3}}_{\ {\bf k},\, {\bf k}_{1},\, {\bf k}_{2}}
		&-
		W^{\hspace{0.025cm} (3)\hspace{0.03cm}a_{1}\hspace{0.03cm}a\, a_{2}\,a_{3}}_{\ {\bf k}_{1},\, {\bf k},\, {\bf k}_{2}}
		\,+
		\frac{i}{2}\,\Bigl(F^{\phantom{(1)}}_{{\bf k}_{1}}\!\!
		\widetilde{V}^{\hspace{0.025cm}(1)\hspace{0.03cm} a\, a_{2}\, a^{\prime}}_{\ \ {\bf k},\, {\bf k}_{2}}\!
		f^{\,a_{1}\hspace{0.03cm}a^{\prime}\hspace{0.02cm}a_{3}}
		+
		F^{\hspace{0.03cm}\phantom{(1)}}_{{\bf k}}\!\!\!\!\hspace{0.03cm}
		\widetilde{V}^{\hspace{0.025cm}(1)\hspace{0.03cm}a_{1}\hspace{0.03cm}a_{2}\, a^{\prime}}_{\ \ {\bf k}_{1},\, {\bf k}_{2}}\! 
		f^{\,a^{\prime}\hspace{0.02cm}a\,a_{3}}\Bigr) 
		\label{ap:B2b}\\[1ex]
		&+ \!\int\!d\hspace{0.03cm}{\bf k}^{\hspace{0.015cm}\prime}\,\Bigl[\hspace{0.02cm}
		\widetilde{V}^{(2)\hspace{0.03cm}a_{1}\hspace{0.03cm}a^{\prime}\hspace{0.02cm}
		a_{3}}_{\ \ {\bf k}_{1},\, {\bf k}^{\prime}}\,
		V^{\hspace{0.03cm}\hspace{0.03cm}(3)\hspace{0.03cm}a\,\hspace{0.03cm}a_{2}\, a^{\prime}}_{\ \ {\bf k},\, {\bf k}_{2},\, {\bf k}^{\prime}} 
		-
		\widetilde{V}^{\hspace{0.025cm}(1)\hspace{0.03cm}a\,a^{\prime}\hspace{0.02cm}
		a_{3}}_{\ \  {\bf k},\, {\bf k}^{\prime}}\,
		V^{\hspace{0.03cm}\hspace{0.025cm}\ast\hspace{0.025cm}(1)\hspace{0.03cm}a^{\prime}\hspace{0.02cm}a_{2}\, a_{1}}_{\ \ {\bf k}^{\prime},\, {\bf k}_{2},\, {\bf k}_{2}}
		+ 
		\widetilde{V}^{(1)\hspace{0.03cm}a_{1}\hspace{0.03cm}a^{\prime}\hspace{0.02cm}
		a_{3}}_{\ \ {\bf k}_{1},\, {\bf k}^{\prime}}\,
		V^{\hspace{0.03cm}\hspace{0.025cm}\ast\hspace{0.025cm}(1)\hspace{0.03cm}a^{\prime}\hspace{0.02cm}a_{2}\, a}_{\ \ {\bf k}^{\prime},\, {\bf k}_{2},\, {\bf k}}
		\notag\\[1ex]
		&\hspace{9cm}-
		\widetilde{V}^{\hspace{0.025cm}(2)\hspace{0.03cm}a\,a^{\prime}\hspace{0.02cm}
		a_{3}}_{\ \ {\bf k},\, {\bf k}^{\prime}}
		\,
		V^{\hspace{0.03cm}\hspace{0.025cm}(3)\hspace{0.03cm}a_{1}\hspace{0.03cm}a_{2}\,a^{\prime}}_{\ \ {\bf k}_{1},\, {\bf k}_{2},\, {\bf k}^{\prime}}\Bigr] 
		= 0, 
		\notag
	\end{align}
\end{subequations} 
and a generalization of the canonicity conditions (\ref{eq:3p}) and (\ref{eq:3a}) should be considered as the following relations for the fifth-order coefficient functions from canonical transformation (\ref{eq:3t})
\begin{subequations} 
	\label{ap:B3}
	\begin{align}
		\widetilde{W}^{\hspace{0.025cm}(1)\hspace{0.03cm}a\,a_{1}\hspace{0.03cm}a_{2}\, a_{3}}_{\ {\bf k},\, {\bf k}_{1}}
		&-
		\widetilde{W}^{\hspace{0.025cm}(1)\hspace{0.03cm}a_{1}\hspace{0.03cm}a\, a_{2}\, a_{3}}_{\ {\bf k}_{1},\, {\bf k}}
		\,+
		i\hspace{0.03cm}\Bigl[\Bigl(F^{\phantom{(1)}}_{{\bf k}_{1}}\!\! G^{\hspace{0.03cm}\,a\,a_{3}\,a^{\prime}}_{{\bf k}}\!
		f^{\,a_{1}\hspace{0.03cm}a^{\prime}\hspace{0.02cm}a_{2}}
		+
		F^{\phantom{(1)}}_{{\bf k}}\!\!\!
		G^{\hspace{0.03cm}\,a_{1}\hspace{0.03cm}a_{2}\, a^{\prime}}_{{\bf k}_{1}}\! 
		f^{\,a^{\prime}\hspace{0.02cm}a\,a_{3}}\Bigr) + (2 \leftrightarrow 3)\Bigr]
		\label{ap:B3a}\\[1ex]
		+\,\frac{1}{2}&\int\!d\hspace{0.02cm}
		{\bf k}^{\hspace{0.015cm}\prime}\hspace{0.03cm}
		\Bigl[\hspace{0.02cm}\Bigl(
		\widetilde{V}^{\hspace{0.03cm}(2)\hspace{0.03cm}a_{1}\hspace{0.03cm}a^{\prime}
		\hspace{0.03cm}
		a_{2}}_{\ \ {\bf k}_{1},\, {\bf k}^{\prime}}\,
		\widetilde{V}^{\hspace{0.03cm}(1)\hspace{0.03cm}a\,a^{\prime}\hspace{0.02cm}a_{3}}_{\ \  {\bf k},\, {\bf k}^{\prime}} 
		+
		\widetilde{V}^{\hspace{0.025cm}(2)\hspace{0.03cm}a_{1}\hspace{0.03cm}a^{\prime}
		\hspace{0.02cm}a_{3}}_{\ \ {\bf k}_{1},\, {\bf k}^{\prime}}\,
		\widetilde{V}^{\hspace{0.025cm}(1)\hspace{0.03cm}a\,a^{\prime}\hspace{0.02cm}
		a_{2}}_{\ \ {\bf k},\, {\bf k}^{\prime}}
		\Bigr)
		- 
		\Bigl(
		\widetilde{V}^{\hspace{0.03cm}(1)\hspace{0.03cm}a_{1}
		\hspace{0.03cm}a^{\prime}\hspace{0.02cm}
		a_{3}}_{\ \ {\bf k}_{1},\, {\bf k}^{\prime}}\,
		\widetilde{V}^{\hspace{0.03cm}(2)\hspace{0.03cm}a\,a^{\prime}\hspace{0.02cm}a_{2}}_{\ \  {\bf k},\, {\bf k}^{\prime}} 
		\notag\\[1ex]
		&\hspace{9cm}+
		\widetilde{V}^{\hspace{0.025cm}(1)\hspace{0.03cm}a_{1}\hspace{0.03cm}a^{\prime}
		\hspace{0.03cm}a_{2}}_{\ \ {\bf k}_{1},\, {\bf k}^{\prime}}\,
		\widetilde{V}^{\hspace{0.025cm}(2)\hspace{0.03cm}a\,a^{\prime}\hspace{0.02cm}a_{3}}_{\ \ {\bf k},\, {\bf k}^{\prime}}\Bigl)\Bigr] 
		= 0, 
		\notag\\[1.5ex]
		\widetilde{W}^{\hspace{0.025cm}\ast\hspace{0.025cm}(2)\hspace{0.03cm}a\,a_{1}
		\hspace{0.03cm}a_{2}\,a_{3}}_{\ {\bf k},\, {\bf k}_{1}}
		&+
		\widetilde{W}^{\hspace{0.025cm}(2)\hspace{0.03cm}a_{1}\hspace{0.03cm}a\,a_{2}\,a_{3}}_{\ {\bf k}_{1},\, {\bf k}}
		\,+
		i\hspace{0.03cm}\Bigl[\Bigl(F^{\phantom{(1)}}_{{\bf k}_{1}}\!\!
		G^{\hspace{0.03cm}\hspace{0.025cm}\ast\hspace{0.025cm} a\, a_{3}\, a^{\prime}}_{{\bf k}}\!f^{\,a_{1}\hspace{0.03cm}a^{\prime}
		\hspace{0.03cm}a_{2}}
		+
		F^{\hspace{0.03cm}\ast\phantom{(1)}}_{{\bf k}}\!\!\!\!\!\hspace{0.02cm} 
		G^{\hspace{0.03cm}\,a_{1}\hspace{0.03cm}a_{2}\,a^{\prime}}_{{\bf k}_{1}}\! 
		f^{\,a^{\prime}\hspace{0.02cm}a\,a_{3}}\Bigr) + (2 \leftrightarrow 3)\Bigr]
		\label{ap:B3b}\\[1ex]
		+\,\frac{1}{2}&\int\!d\hspace{0.03cm}{\bf k}^{\hspace{0.015cm}\prime}\,\Bigl[\hspace{0.02cm}\Bigl(
		\widetilde{V}^{\hspace{0.03cm}(2)\hspace{0.03cm}a_{1}\hspace{0.03cm}
		a^{\prime}\hspace{0.02cm}
		a_{2}}_{\ \ {\bf k}_{1},\, {\bf k}^{\prime}}\,
		\widetilde{V}^{\hspace{0.03cm}\ast\hspace{0.03cm}(2)\hspace{0.03cm}a
		\hspace{0.03cm}a^{\prime}\, a_{3}}_{\ \ {\bf k},\, {\bf k}^{\prime}} 
		+
		\widetilde{V}^{\hspace{0.025cm}(2)\hspace{0.03cm}a_{1}\hspace{0.03cm}a^{\prime}
		\hspace{0.03cm}a_{3}}_{\ \ {\bf k}_{1},\, {\bf k}^{\prime}}\,
		\widetilde{V}^{\hspace{0.025cm}\ast\hspace{0.025cm}(2)\hspace{0.03cm}a\, a^{\prime}\hspace{0.02cm}a_{2}}_{\ \ {\bf k},\, {\bf k}^{\prime}}
		\Bigr)
		- 
		\Bigl(
		\widetilde{V}^{\hspace{0.03cm}(1)\hspace{0.03cm}a_{1}\hspace{0.03cm}
		a^{\prime}\hspace{0.02cm}
		a_{2}}_{\ \ {\bf k}_{1},\, {\bf k}^{\prime}}\,
		\widetilde{V}^{\hspace{0.03cm}\ast\hspace{0.03cm}(1)\hspace{0.03cm}a\, a^{\prime}\, a_{3}}_{\ \  {\bf k},\, {\bf k}^{\prime}} 
		\notag\\[1ex]
		&\hspace{9cm}+
		\widetilde{V}^{\hspace{0.025cm}(1)\hspace{0.03cm}a_{1}\hspace{0.03cm}a^{\prime}
		\hspace{0.03cm}a_{3}}_{\ \ {\bf k}_{1},\, {\bf k}^{\prime}}\,
		\widetilde{V}^{\hspace{0.025cm}\ast\hspace{0.025cm}(1)\hspace{0.03cm} a\, a^{\prime}\hspace{0.02cm}a_{2}}_{\ \ {\bf k},\, 
		{\bf k}^{\prime}}\Bigl)\Bigr] 
		= 0. 
		\notag
	\end{align}
\end{subequations} 
The systems (\ref{ap:B2}) and (\ref{ap:B3}) define the rearrangement rules for the color indices $a$ and $a_{1}$, and for the corresponding momentum arguments ${\bf k}$ and ${\bf k}_{1}$ of the fourth- and fifth-order coefficient functions $W^{\hspace{0.025cm}(2)\hspace{0.03cm}a\,a_{1}\hspace{0.03cm}a_{2}\,a_{3}}_{\ {\bf k},\, {\bf k}_{1},\, {\bf k}_{2}}$, $W^{\hspace{0.025cm} (3)\hspace{0.03cm}a\, a_{1}\hspace{0.03cm}a_{2}\,a_{3}}_{\ {\bf k},\, {\bf k}_{1},\, {\bf k}_{2}}$, $\widetilde{W}^{\hspace{0.025cm} (1)\hspace{0.03cm}a\,a_{1}\hspace{0.03cm}a_{2}\,a_{3}}_{\ {\bf k},\, {\bf k}_{1}}$ and $\widetilde{W}^{\hspace{0.025cm} (2)\hspace{0.03cm}a\,a_{1}\hspace{0.03cm}a_{2}\, a_{3}}_{\ {\bf k},\, {\bf k}_{1}}$.\\
\indent Further substituting the canonical transformations (\ref{eq:3t}) and (\ref{eq:3y}) into equation (\ref{eq:3ec}), equating the coefficients of the same powers of a product of the functions $c^{\hspace{0.025cm}a}_{\bf k},\,c^{\ast\,a}_{{\bf k}}$ and color charge ${\mathcal Q}^{\hspace{0.03cm}a}$ to zero, and taking into account the relations (\ref{eq:3i}), we obtain the other canonicity conditions for the higher coefficient functions in (\ref{eq:3y}). For the fourth-order functions, we have
\begin{subequations} 
\label{ap:B4}
\begin{align}
		M^{\hspace{0.025cm}(1)\hspace{0.03cm}a\,a_{1}\hspace{0.03cm}a_{2}\,a_{3}}_{\; {\bf k}_{1},\, {\bf k}_{2}}
		-\!
		\int\!d\hspace{0.03cm}{\bf k}^{\hspace{0.015cm}\prime}\,
		\Bigl[\hspace{0.02cm}M^{\hspace{0.03cm}\ast\,a\,a^{\prime}\, a_{3}}_{\hspace{0.03cm}{\bf k}^{\prime}}\,
		&V^{\hspace{0.03cm}\hspace{0.03cm}\ast\hspace{0.025cm}(3)\hspace{0.03cm}a_{1}\hspace{0.03cm}a_{2}\, a^{\prime}}_{\ \ {\bf k}_{1},\, {\bf k}_{2},\, {\bf k}^{\prime}}
		+
		M^{\,a\,a^{\prime}\hspace{0.02cm}a_{3}}_{\hspace{0.03cm}{\bf k}^{\prime}}\,
		V^{\hspace{0.03cm}\hspace{0.025cm}(1)\hspace{0.03cm}a^{\prime}\hspace{0.02cm}a_{2}\,a_{1}}_{\ \  {\bf k}^{\prime},\, {\bf k}_{2},\, {\bf k}_{1}}
		\Bigr]
		\label{ap:B4a}\\[1ex]
		&+
		\frac{i}{2}\,\Bigl(\widetilde{V}^{\hspace{0.025cm}\ast
		\hspace{0.025cm}
		(1 )\hspace{0.03cm}a_{1}\hspace{0.03cm}a_{2}\,a^{\prime}}_{\ \ {\bf k}_{1},\, {\bf k}_{2}}\!f^{\,a\,a^{\prime}\hspace{0.02cm}a_{3}}
		+
		F^{\hspace{0.03cm}\ast\phantom{(1)}}_{\hspace{0.03cm}{\bf k}_{1}}\!\!\!\!M^{\, a\, a_{2}\, a^{\prime}}_{\hspace{0.03cm}{\bf k}_{2}}\! 
		f^{\,a^{\prime}\hspace{0.02cm}a_{1}\hspace{0.03cm}a_{3}}\Bigr) = 0, 
		\notag\\[1ex]
		{M}^{\hspace{0.025cm}(2)\hspace{0.03cm}a\,a_{1}\hspace{0.03cm}a_{2}\,a_{3}}_{\; {\bf k}_{1},\, {\bf k}_{2}}
		+2\!
		\int\!d\hspace{0.03cm}{\bf k}^{\hspace{0.015cm}\prime}
		\,\Bigl[\hspace{0.02cm}
		M^{\,a\,a^{\prime}\hspace{0.02cm}a_{3}}_{\hspace{0.03cm}{\bf k}^{\prime}}\,
		&V^{\hspace{0.03cm}\hspace{0.03cm}\ast\hspace{0.025cm}(1)\hspace{0.03cm}a_{2}\,a^{\prime}
		\hspace{0.03cm} a_{1}}_{\ \ {\bf k}_{2},\, {\bf k}^{\prime},\, {\bf k}_{1}}
		+
		M^{\hspace{0.03cm}\ast\,a\,a^{\prime}\hspace{0.02cm}a_{3}}_{\hspace{0.03cm}{\bf k}^{\prime}}\,
		V^{\hspace{0.03cm}\hspace{0.025cm}(1)\hspace{0.03cm}a_{1}\hspace{0.03cm}a^{\prime}\hspace{0.02cm}a_{2}}_{\ \ {\bf k}_{1},\, {\bf k}^{\prime},\, {\bf k}_{2}}
		\Bigr]
		\label{ap:B4b}\\[1ex]
		&-
		i\hspace{0.04cm}\Bigl(\widetilde{V}^{\hspace{0.025cm}(2)
		\hspace{0.03cm}a_{1}
		\hspace{0.03cm} a_{2}\, a^{\prime}}_{\ \ {\bf k}_{1},\, 
		{\bf k}_{2}}\!f^{\,a\,a^{\prime}\hspace{0.02cm}a_{3}}
		+
		F^{\phantom{(1)}}_{\hspace{0.03cm}{\bf k}_{1}}\!M^{\, a\, a_{2}\, a^{\prime}}_{\hspace{0.03cm}
		{\bf k}_{2}}\!f^{\,a^{\prime}\hspace{0.02cm}a_{1}\hspace{0.03cm}a_{3}}
	    \Bigr) = 0. 
		\notag
	\end{align}
\end{subequations} 
Correspondingly, for the fifth-order coefficient function we obtain
\begin{align}
	&\widetilde{M}^{\,a\,a_{1}\hspace{0.03cm}a_{2}\,a_{3}}_{\; {\bf k}_{1}}
	+
	\frac{1}{2}
	\int\!d\hspace{0.03cm}{\bf k}^{\hspace{0.015cm}\prime}
	\,\Bigl[\hspace{0.02cm}\Bigl(\!
	M^{\,a\,a^{\prime}\hspace{0.02cm}a_{3}}_{\hspace{0.03cm}{\bf k}^{\prime}}\,
	\widetilde{V}^{\hspace{0.03cm}\ast\hspace{0.025cm}(2)\hspace{0.03cm} a_{1}\hspace{0.03cm}a^{\prime}\hspace{0.02cm}a_{2}}_{\ \ {\bf k}_{1},\, {\bf k}^{\prime}}
	+
	M^{\,a\,a^{\prime}\hspace{0.02cm}a_{2}}_{\hspace{0.03cm}{\bf k}^{\prime}}\,
	\widetilde{V}^{\hspace{0.025cm}\ast\hspace{0.025cm}(2)\hspace{0.03cm} a_{1}\hspace{0.03cm}a^{\prime}\hspace{0.02cm}a_{3}}_{\ \ {\bf k}_{1},\, {\bf k}^{\prime}}
	\Bigr)
	\notag\\[1ex]
	&\hspace{7.3cm}-
	\Bigl(\!
	M^{\hspace{0.03cm}\ast\,a\,a^{\prime}\hspace{0.02cm}
	a_{3}}_{\hspace{0.03cm}{\bf k}^{\prime}}\,
	\widetilde{V}^{\hspace{0.03cm}\ast\hspace{0.025cm}(1)\hspace{0.03cm} a_{1}\hspace{0.03cm}a^{\prime}\hspace{0.02cm}a_{2}}_{\ \ {\bf k}_{1},\, {\bf k}^{\prime}}
	+
	M^{\hspace{0.03cm}\ast\,a\,a^{\prime}\hspace{0.02cm}
	a_{2}}_{\hspace{0.03cm}{\bf k}^{\prime}}\,
	\widetilde{V}^{\hspace{0.025cm}\ast\hspace{0.025cm}(1)\hspace{0.03cm}
	a_{1}\hspace{0.03cm}a^{\prime}\hspace{0.02cm}a_{3}}_{\ \ {\bf k}_{1},\, 
	{\bf k}^{\prime}}
	\Bigr)\Bigr]
	\notag\\[1ex]
	&+\,
	i\hspace{0.04cm}\Bigl[F^{\hspace{0.03cm}\ast\phantom{(1)}}_{{\bf k}_{1}}\!\!\!\!\Bigl(
	F^{\hspace{0.025cm} a\, a_{2}\, a^{\prime}}\!
	f^{\,a^{\prime}\hspace{0.02cm}a_{1}\hspace{0.03cm}a_{3}}
	+
	F^{\, a\, a_{3}\, a^{\prime}}\! 
	f^{\,a^{\prime}\hspace{0.02cm}a_{1}\hspace{0.03cm}a_{2}}\Bigr) 
	- 
	\Bigl(G^{\hspace{0.03cm}\hspace{0.025cm}\ast\hspace{0.03cm}a_{1}\hspace{0.03cm}a_{2}\, a^{\prime}}_{{\bf k}_{1}}\!
	f^{\,a^{\prime}\hspace{0.02cm}a\,a_{3}}
	+
	G^{\hspace{0.03cm}\hspace{0.03cm}\ast\hspace{0.025cm}a_{1}\hspace{0.03cm}a_{3}\, a^{\prime}}_{{\bf k}_{1}}\! f^{\, a^{\prime}\, a\,a_{2}}\Bigr) \Bigr] = 0. 
	\notag
\end{align}
The last three conditions represent explicit definitions of higher coefficient functions $M^{\hspace{0.025cm}(1)\hspace{0.03cm} a\, a_{1}\, a_{2}\, a_{3}}_{\ {\bf k}_{1},\, {\bf k}_{2}}$, $M^{\hspace{0.025cm}(2)\hspace{0.03cm} a\, a_{1}\, a_{2}\, a_{3}}_{\ {\bf k}_{1},\, {\bf k}_{2}}$ and $\widetilde{M}^{\, a\, a_{1}\, a_{2}\, a_{3}}_{\ {\bf k}_{1},\, {\bf k}_{2}}$ through the lower ones, as is also the case for the function $M^{\,a\,a_{1}\hspace{0.03cm}a_{2}}_{\; {\bf k}}$, Eq.\,(\ref{eq:3o}). The explicit form of the first two of these is given in the following Appendix.\\
\indent The second system of equations (\ref{eq:3r}) does not give any new canonicity conditions compared to those written out above. All relations derived from (\ref{eq:3r}) either turn into an identity or reproduce those obtained earlier. Note only that on the right-hand part of (\ref{eq:3rc}) instead of the colored charge $Q^{\hspace{0.03cm}a}$, one must substitute the integro-power series (\ref{eq:3y}).


\numberwithin{equation}{section}
\section{Explicit expressions for the coefficient func\-tions $M^{\hspace{0.03cm}(1)\hspace{0.03cm}a\, a_{1}\hspace{0.03cm}a_{2}\,a_{3}}_{\ {\bf k}_{1},\, {\bf k}_{2}}$ and $M^{\hspace{0.03cm}(2)\hspace{0.03cm}a\,a_{1}\hspace{0.03cm}a_{2}\,a_{3}}_{\ {\bf k}_{1},\, {\bf k}_{2}}$}
\numberwithin{equation}{section}
\label{appendix_C}

Here, we write out an explicit form of the higher-order coefficient functions $M^{(n)\hspace{0.03cm}a\,a_{1}\hspace{0.03cm}a_{2}\,a_{3}}_{\ 
{\bf k}_{1},\, {\bf k}_{2}}$ for $n = 1$ and 2, which enter into the integrands of the canonical transformation of the color charge $Q^{\,a}$, Eq.\,(\ref{eq:3y}). The form of these functions can be easily restored from the conditions of canonicity (\ref{ap:B4a}) and (\ref{ap:B4b}). Substituting the explicit form of the coefficient functions $F_{\bf k}$, $M^{\,a\,a_{1}\hspace{0.03cm}a_{2}}_{\; {\bf k}},\,V^{\hspace{0.03cm}\,(1,\hspace{0.03cm}3)\hspace{0.03cm}a\,a_{1}\hspace{0.03cm}a_{2}}_{\ {\bf k},\, {\bf k}_{1},\, {\bf k}_{2}}$ and $\widetilde{V}^{\,(1,\hspace{0.03cm}2)\hspace{0.03cm} a_{1}\hspace{0.03cm}a_{2}\,a}_{\ {\bf k}_{1},\, {\bf k}_{2}}$, Eqs.\,(\ref{eq:4w}), (\ref{eq:4e}), (\ref{eq:4r}), (\ref{eq:12y}) and (\ref{eq:12a}), respectively, we find for the first of them
\begin{equation}
M^{\hspace{0.03cm}(1)\hspace{0.03cm}a\,a_{1}\hspace{0.03cm}a_{2}\,a_{3}}_{\ {\bf k}_{1},\, 
{\bf k}_{2}}
=
	-\hspace{0.03cm}\frac{1}{4}\,
	\bigl(f^{\hspace{0.03cm}a\,a_{1}\hspace{0.03cm}e}\hspace{0.02cm}
	f^{\hspace{0.03cm}e\,a_{2}\hspace{0.03cm}a_{3}}
	+
	f^{\hspace{0.03cm}a\,a_{2}\hspace{0.03cm}e}\hspace{0.02cm}
	f^{\hspace{0.03cm}e\,a_{1}\hspace{0.03cm}a_{3}}\bigr)\hspace{0.03cm}
	\frac{{\upphi}^{\phantom{\ast}}_{\hspace{0.03cm}{\bf k}_{1}}\hspace{0.02cm}
	{\upphi}^{\phantom{\ast}}_{\hspace{0.03cm}{\bf k}_{2}}}
	{\bigl(\omega^{\hspace{0.02cm}l}_{\hspace{0.02cm}{\bf k}_{1}} - 
	{\bf v}\cdot {\bf k}_{1}\bigr)
	\bigl(\omega^{\hspace{0.02cm}l}_{\hspace{0.02cm}{\bf k}_{2}} - 
	{\bf v}\cdot {\bf k}_{2}\bigr)}
\label{ap:C1}
\end{equation}
\[
+\,f^{\hspace{0.03cm}a_{1}\hspace{0.03cm}a_{2}\hspace{0.03cm}e}\hspace{0.02cm}
f^{\hspace{0.03cm}e\, a\hspace{0.03cm} a_{3}}
\,\frac{1}{\bigl(\omega^{\hspace{0.02cm} l}_{\hspace{0.02cm}{\bf k}_{1}} + \omega^{\hspace{0.02cm}l}_{\hspace{0.02cm}{\bf k}_{2}} - {\bf v}\cdot ({\bf k}_{1} + {\bf k}_{2})\bigr)}
\]
\[
\times\,\biggl\{-\hspace{0.03cm}\frac{1}{4}\,
{\upphi}_{\hspace{0.03cm}{\bf k}_{1}}\hspace{0.02cm}
{\upphi}_{\hspace{0.03cm}{\bf k}_{2}}
\biggl(\frac{1}
{\omega^{\hspace{0.02cm}l}_{\hspace{0.02cm}{\bf k}_{1}} - {\bf v}\cdot 
{\bf k}_{1}}
\,-\,
\frac{1}
{\omega^{\hspace{0.02cm}l}_{\hspace{0.02cm}{\bf k}_{2}} - {\bf v}\cdot 
{\bf k}_{2}}
\biggr)
\]
\vspace{0.2cm}
\[
+\,i\,\biggl(\hspace{0.02cm}
\frac{{\mathcal U}_{\,-{\bf k}_{1} - {\bf k}_{2},\, {\bf k}_{1},\, 
{\bf k}_{2}}\hspace{0.02cm} 
	{\upphi}^{\ast}_{\hspace{0.03cm}-{\bf k}_{1} - {\bf k}_{2}}}
{\omega^{\hspace{0.02cm} l}_{-{\bf k}_{1} - {\bf k}_{2}} + \omega^{\hspace{0.02cm} l}_{\hspace{0.02cm}{\bf k}_{1}}\! + \omega^{\hspace{0.02cm} l}_{{\bf k}_{2}}}
\notag\\[1.5ex]
\,+\,
\frac{{\mathcal V}_{\, {\bf k}_{1} + {\bf k}_{2},\, {\bf k}_{1},\, {\bf k}_{2}}\hspace{0.02cm} 
	{\upphi}^{\phantom{\ast}}_{\hspace{0.03cm}{\bf k}_{1} + {\bf k}_{2}}}
{\omega^{\hspace{0.02cm}l}_{\hspace{0.02cm}{\bf k}_{1} + {\bf k}_{2}} - \omega^{\hspace{0.02cm}l}_{\hspace{0.02cm}{\bf k}_{1}}\! - \omega^{\hspace{0.02cm}l}_{\hspace{0.02cm}{\bf k}_{2}}}\biggr)\!\biggr\},
\]
and for the second coefficient function, we get correspondingly
\begin{equation}
	M^{\hspace{0.03cm}(2)\hspace{0.03cm}a\,a_{1}\hspace{0.03cm}a_{2}\,a_{3}}_{\ {\bf k}_{1},\, 
	{\bf k}_{2}}
	=
\frac{1}{2}\,
\bigl(f^{\hspace{0.03cm}a\,a_{2}\hspace{0.03cm}e}\hspace{0.02cm}
f^{\hspace{0.03cm}e\,a_{1}\hspace{0.03cm}a_{3}}
+
f^{\hspace{0.03cm}a\,a_{1}\hspace{0.03cm}e}\hspace{0.02cm}
f^{\hspace{0.03cm} e\, a_{2}\hspace{0.03cm} a_{3}}\bigr)\hspace{0.03cm}
	\frac{{\upphi}^{\ast}_{\hspace{0.03cm}{\bf k}_{1}}\hspace{0.03cm}
	{\upphi}^{\phantom{\ast}}_{\hspace{0.03cm}{\bf k}_{2}}}
	{\bigl(\omega^{\hspace{0.02cm}l}_{\hspace{0.02cm}{\bf k}_{1}} - {\bf v}\cdot {\bf k}_{1}\bigr)
	\bigl(\omega^{\hspace{0.02cm}l}_{\hspace{0.02cm}{\bf k}_{2}} - {\bf v}\cdot {\bf k}_{2}\bigr)}
\label{ap:C2}
\end{equation}
\begin{align}
+\,i\hspace{0.02cm}
f^{\hspace{0.03cm}a_{1}\hspace{0.02cm}a_{2}\hspace{0.03cm}e}\hspace{0.02cm}
f^{\hspace{0.03cm}e\,a\hspace{0.03cm}a_{3}}\hspace{0.03cm}
\Biggl\{
&\frac{{\mathcal V}_{\,{\bf k}_{1}, {\bf k}_{2},\, {\bf k}_{1} - {\bf k}_{2}} 
\hspace{0.02cm}{\upphi}^{\hspace{0.03cm}\ast}_{\hspace{0.03cm}{\bf k}_{1} - {\bf k}_{2}}}
{\bigl(\omega^{\hspace{0.02cm}l}_{\hspace{0.02cm}{\bf k}_{1}}\! - \omega^{\hspace{0.02cm}l}_{\hspace{0.02cm}{\bf k}_{2}}\! -
	\omega^{\hspace{0.02cm}l}_{\hspace{0.02cm}{\bf k}_{1} - 
	{\bf k}_{2}}\bigr)
	\bigl(\omega^{\hspace{0.02cm}l}_{\hspace{0.02cm}{\bf k}_{1} - {\bf k}_{2}}\! - {\bf v}\cdot ({\bf k}_{1} - {\bf k}_{2})\bigr)}
\notag\\[1ex]
+\;
&\frac{{\mathcal V}^{\hspace{0.04cm}\ast}_{\, {\bf k}_{2}, {\bf k}_{1}, {\bf k}_{2} - {\bf k}_{1}}\hspace{0.02cm} 
	{\upphi}^{\phantom{\ast}}_{\hspace{0.03cm}{\bf k}_{2} - {\bf k}_{1}}}
{\bigl(\omega^{\hspace{0.02cm}l}_{\hspace{0.02cm}{\bf k}_{2}}\! - \omega^{\hspace{0.02cm} l}_{{\bf k}_{1}}\! -
	\omega^{\hspace{0.02cm}l}_{\hspace{0.02cm}{\bf k}_{2} - 
	{\bf k}_{1}}\bigr)
	\bigl(\omega^{\hspace{0.02cm} l}_{\hspace{0.02cm}{\bf k}_{2} - 
	{\bf k}_{1}}\! - {\bf v}\cdot ({\bf k}_{2} - {\bf k}_{1})\bigr)}\Biggr\}.
\notag
\end{align}
In deriving these expressions we used the symmetry properties of effective vertex functions ${\mathcal V}_{\, {\bf k},\, {\bf k}_{1},\, {\bf k}_{2}}
$ and ${\mathcal U}_{\, {\bf k},\, {\bf k}_{1},\, {\bf k}_{2}}$, Eq.\,(\ref{eq:2l}). It is easy to verify that the coefficient function (\ref{ap:C1}) satisfies the corresponding natural symmetry condition in (\ref{eq:3u})
\begin{equation}
M^{(1)\,a\,a_{1}\hspace{0.03cm}a_{2}\,a_{3}}_{\ {\bf k}_{1},\, {\bf k}_{2}} 
= 
M^{(1)\,a\,a_{2}\hspace{0.03cm}a_{1}\hspace{0.03cm}a_{3}}_{\ {\bf k}_{2},\, {\bf k}_{1}}.
\label{ap:C3}
\end{equation}
As mentioned in section \ref{section_12}, the coefficient function $\widetilde{V}^{\hspace{0.025cm}(1)\hspace{0.03cm}a_{1}	\hspace{0.03cm} a_{2}\, a^{\prime}}_{\ \ {\bf k}_{1},\, {\bf k}_{2}}$, which enters into the expression (\ref{ap:B4a}), by virtue of its definition (\ref{eq:12t}) and (\ref{eq:12y}) does not possess a symmetry relation similar to (\ref{ap:C3}).
However, the presence of the last term   $F^{\hspace{0.03cm}\ast}_{\hspace{0.03cm}{\bf k}_{1}}\!\hspace{0.03cm}M^{\, a\, a_{2}\, a^{\prime}}_{\hspace{0.03cm}
{\bf k}_{2}}\!f^{\,a^{\prime}\hspace{0.02cm}a_{1}\hspace{0.03cm}a_{3}}$ in
(\ref{ap:B4a}) allows us to recover this property.\\
\indent Furthermore, the coefficient function (\ref{ap:C2}) is represented in such a form that the fulfillment of the first condition of a reality of the color charges $Q^{\hspace{0.03cm}a}$ and $\mathcal{Q}^{\hspace{0.03cm}a}$ under the canonical transformation (\ref{eq:3y}), namely
\[
M^{\hspace{0.03cm}\ast\hspace{0.03cm}(2)\,a\,a_{1}\hspace{0.03cm}a_{2}\, a_{3}}_{\ {\bf k}_{1},\, {\bf k}_{2}}
=
M^{\hspace{0.03cm}(2)\,a\,a_{2}\,a_{1}\hspace{0.03cm}a_{3}}_{\ {\bf k}_{2},\, {\bf k}_{1}},
\]
is almost obvious.


\section{Traces for generators in the adjoint repre\-sentation}
\numberwithin{equation}{section}
\label{appendix_D}

In this Appendix, we have provided the traces for generators in the adjoint representation, which we use in the text of this paper. A comprehensive list of the various traces, relations and identities for the color matrices in the adjoint representation can be found in \cite{kaplan_1967, macfarlane_1968, azcarraga_1998, fadin_2005, nikolaev_2005, haber_2021}. The original definition of matrices $T^{\,a}$ is:
\[
\bigl(T^{\,a}\bigr)^{\hspace{0.01cm}b\hspace{0.03cm}c} \equiv -\hspace{0.02cm}i\hspace{0.03cm}
f^{\hspace{0.03cm}a\hspace{0.02cm}b\hspace{0.03cm}c},
\]
where $f^{\hspace{0.03cm}a\hspace{0.02cm}b\hspace{0.03cm}c}$ are the antisymmetric structure constants in the $\mathfrak{su}(N_{c})$ Lie algebra. These matrices are traceless, i.e.
\[
{\rm tr}\,T^{\,a} = 0
\]
and satisfy the following commutation relation
\[
\bigl[\hspace{0.02cm}T^{\,a},T^{\,b}\hspace{0.02cm}\bigr] = i\hspace{0.02cm}f^{\hspace{0.03cm}a\hspace{0.02cm}b\hspace{0.03cm}c}
\,T^{\,c}.
\]
The traces of a product of two and three generators are
\begin{align}
&{\rm tr}\hspace{0.03cm}
\bigl(T^{\,a}\hspace{0.03cm}T^{\,b}\bigr) = N_{c}\hspace{0.04cm}\delta^{\hspace{0.03cm}a\hspace{0.03cm}b},
\label{ap:D2}\\[1ex]
&{\rm tr}\hspace{0.03cm}
\bigl(T^{\,a}\hspace{0.03cm}T^{\,b}\hspace{0.03cm}T^{\,c}\bigr) = \displaystyle\frac{i}{2}\,N_{c}\hspace{0.03cm}
f^{\hspace{0.03cm}a\hspace{0.02cm}b\hspace{0.03cm}c}.
\label{ap:D3}
\end{align}
We will need two traces of four generators 
\begin{align}
&{\rm tr}\hspace{0.03cm}\bigl(T^{\,a}\hspace{0.02cm} T^{\,b}\hspace{0.02cm} T^{\,c}\hspace{0.02cm} 
T^{\,d}\hspace{0.03cm}\bigr)
=
\delta^{\hspace{0.02cm}a\hspace{0.02cm}d}\hspace{0.03cm}\delta^{\hspace{0.02cm}b\hspace{0.03cm}c}
+
\frac{1}{2}\,\bigl(\hspace{0.02cm}
\delta^{\hspace{0.02cm}a\hspace{0.02cm}b}\hspace{0.03cm}\delta^{\hspace{0.02cm}c\hspace{0.03cm}d}
+
\delta^{\hspace{0.02cm}a\hspace{0.02cm}c}\hspace{0.03cm}
\delta^{\hspace{0.02cm}b\hspace{0.03cm}d}\hspace{0.02cm}\bigr)
+
\frac{1}{4}\,N_{c}\hspace{0.02cm}\bigl(\hspace{0.02cm}
f^{\hspace{0.03cm}a\hspace{0.03cm}d\hspace{0.03cm}e}
\hspace{0.01cm}f^{\hspace{0.03cm}b\hspace{0.03cm}c\hspace{0.03cm}e}
\!+
d^{\hspace{0.04cm}a\hspace{0.03cm}d\hspace{0.03cm}e}
\hspace{0.02cm}d^{\hspace{0.04cm}b\hspace{0.03cm}c\hspace{0.03cm}e}
\hspace{0.02cm}\bigr),
\label{ap:D4} \\[1ex]
&{\rm tr}\hspace{0.03cm}\bigl(T^{\,a}\hspace{0.02cm} T^{\,b}\hspace{0.02cm} D^{\,c}\hspace{0.02cm}D^{\,d}\hspace{0.03cm}\bigr)
=
\frac{1}{2}\,
\bigl(\delta^{\hspace{0.02cm}a\hspace{0.02cm}b}\delta^{\hspace{0.02cm}c\hspace{0.02cm}d}
\!-
\delta^{\hspace{0.02cm}a\hspace{0.02cm}c}\delta^{\hspace{0.02cm}b\hspace{0.02cm}d}\bigr)
+\,
\biggl(\frac{N^{\hspace{0.02cm}2}_{c} - 8}{4\hspace{0.02cm}N_{c}}\biggr)
f^{\hspace{0.03cm}a\hspace{0.03cm}d\hspace{0.03cm}e}
\hspace{0.01cm}f^{\hspace{0.03cm}b\hspace{0.03cm}c\hspace{0.03cm}e}
+
\frac{1}{4}\,N_{c}\hspace{0.04cm}
d^{\hspace{0.04cm}a\hspace{0.03cm}d\hspace{0.03cm}e}
\hspace{0.02cm}d^{\hspace{0.04cm}b\hspace{0.03cm}c\hspace{0.03cm}e},
\label{ap:D5} 
\end{align}
where $d^{\hspace{0.04cm}a\hspace{0.03cm}b\hspace{0.03cm}c}$ are the completely symmetric structure constants in the $\mathfrak{su}(N_{c})$ Lie algebra and $\bigl(D^{\,a}\bigr)^{\hspace{0.01cm}b\hspace{0.03cm}c} \equiv 
d^{\hspace{0.03cm}a\hspace{0.02cm}b\hspace{0.03cm}c}$. If one uses the relation  	
\[
f^{\hspace{0.03cm}a\hspace{0.03cm}b\hspace{0.03cm}e}
\hspace{0.01cm}f^{\hspace{0.03cm}c\hspace{0.03cm}d\hspace{0.03cm}e}
=
\frac{2}{N_{c}}\,\bigl(\hspace{0.02cm}
\delta^{\hspace{0.02cm}a\hspace{0.02cm}c}\hspace{0.03cm}
\delta^{\hspace{0.02cm}b\hspace{0.03cm}d}
-
\delta^{\hspace{0.02cm}a\hspace{0.02cm}d}\hspace{0.03cm}
\delta^{\hspace{0.02cm}b\hspace{0.03cm}c}\hspace{0.02cm}\bigr)
+
\bigl(\hspace{0.02cm}d^{\hspace{0.03cm}a\hspace{0.03cm}
c\hspace{0.03cm}e}
\hspace{0.01cm}d^{\hspace{0.03cm}b\hspace{0.03cm}d\hspace{0.03cm}e}
\!-
d^{\hspace{0.04cm}b\hspace{0.03cm}c\hspace{0.03cm}e}
\hspace{0.03cm}d^{\hspace{0.04cm}a\hspace{0.03cm}d\hspace{0.03cm}e}
\hspace{0.02cm}\bigr),
\]
%
then the trace (\ref{ap:D4}) can also be presented in a slightly different form
\begin{equation}
{\rm tr}\hspace{0.03cm}\bigl(T^{\,a}\hspace{0.02cm} T^{\,b}\hspace{0.02cm} T^{\,c}\hspace{0.02cm} 
T^{\,d}\hspace{0.03cm}\bigr)
=
\delta^{\hspace{0.02cm}a\hspace{0.02cm}b}\hspace{0.03cm}
\delta^{\hspace{0.02cm}c\hspace{0.03cm}d}
+		\delta^{\hspace{0.02cm}a\hspace{0.02cm}d}\hspace{0.03cm}
\delta^{\hspace{0.02cm}b\hspace{0.03cm}c}
+
\frac{1}{4}\,N_{c}\hspace{0.02cm}\bigl(\hspace{0.02cm}
d^{\hspace{0.03cm}a\hspace{0.03cm}b\hspace{0.03cm}e}
\hspace{0.01cm}d^{\hspace{0.03cm}c\hspace{0.03cm}d\hspace{0.03cm}e}
\!+
d^{\hspace{0.04cm}a\hspace{0.03cm}d\hspace{0.03cm}e}
\hspace{0.03cm}d^{\hspace{0.04cm}b\hspace{0.03cm}c\hspace{0.03cm}e}
-
d^{\hspace{0.04cm}a\hspace{0.03cm}c\hspace{0.03cm}e}
\hspace{0.03cm}d^{\hspace{0.04cm}b\hspace{0.03cm}d\hspace{0.03cm}e}
\hspace{0.02cm}\bigr).
\label{ap:D7}
\end{equation}
The trace (\ref{ap:D7}) is written in such a form, which shows its symmetry with respect to the permutation of indices $a$ and $c$ (and, correspondingly, $b$ and $d$\hspace{0.03cm}), i.e.
\begin{equation}
{\rm tr}\hspace{0.03cm}\bigl(T^{\,a}\hspace{0.02cm} T^{\,b}\hspace{0.02cm} T^{\,c}\hspace{0.02cm} 
T^{\,d}\hspace{0.03cm}\bigr)
=
{\rm tr}\hspace{0.03cm}\bigl(T^{\,c}\hspace{0.02cm} T^{\,b}\hspace{0.02cm} T^{\,a}\hspace{0.02cm}T^{\,d}\hspace{0.03cm}\bigr)
=
{\rm tr}\hspace{0.03cm}\bigl(T^{\,a}\hspace{0.02cm} T^{\,d}\hspace{0.02cm} T^{\,c}\hspace{0.02cm} 
T^{\,b}\hspace{0.03cm}\bigr).
\label{ap:D8}
\end{equation}
\indent The trace of five generators $T^{\,a}$ can be represented as a linear combination of the traces of four generators \cite{ritbergen_1999}\hspace{0.03cm}\footnote{\hspace{0.03cm}In the cited paper in formula (45) for the trace of five generators in one of the terms on the right-hand side, two indices are incorrectly placed.}
\begin{equation}
{\rm tr}\hspace{0.03cm}\bigl(T^{\,a_{1}}\hspace{0.02cm} T^{\,a_{2}}\hspace{0.02cm} T^{\,a_{3}}\hspace{0.02cm} 
T^{\,a_{4}}\hspace{0.03cm}T^{\,a_{5}}\hspace{0.02cm}\bigr)
\label{ap:D9}
\end{equation}
\[
\begin{split}
=
-\hspace{0.03cm}\frac{i}{2}\,\Bigl\{ 
&f^{\hspace{0.03cm}a_{4}\hspace{0.02cm}a_{3}\hspace{0.02cm} b}\hspace{0.05cm}
{\rm tr}\hspace{0.03cm}\bigl(T^{\,a_{1}}\hspace{0.02cm} T^{\,a_{2}}\hspace{0.02cm} T^{\,b}\hspace{0.02cm} 
T^{\,a_{5}}\hspace{0.03cm}\bigr)
+
 f^{\hspace{0.03cm}a_{5}\hspace{0.02cm}a_{3}\hspace{0.02cm} b}\hspace{0.05cm}
{\rm tr}\hspace{0.03cm}\bigl(T^{\,a_{1}}\hspace{0.02cm} T^{\,a_{2}}\hspace{0.02cm} 
T^{\,a_{4}}\hspace{0.02cm}T^{\,b}\hspace{0.03cm}\bigr)\\[0.7ex]
+\,
&f^{\hspace{0.03cm}a_{5}\hspace{0.02cm}a_{4}\hspace{0.02cm} b}\hspace{0.05cm}
{\rm tr}\hspace{0.03cm}\bigl(T^{\,a_{1}}\hspace{0.02cm}T^{\,a_{2}}
\hspace{0.02cm} T^{\,b}\hspace{0.02cm} 
T^{\,a_{3}}\hspace{0.03cm}\bigr)
+
f^{\hspace{0.03cm}a_{2}\hspace{0.02cm}a_{1}\hspace{0.02cm} b}\hspace{0.05cm}
{\rm tr}\hspace{0.03cm}\bigl(T^{\,b}\hspace{0.02cm} T^{\,a_{5}}\hspace{0.02cm} T^{\,a_{4}}\hspace{0.02cm} 
T^{\,a_{3}}\hspace{0.03cm}\bigr)\Bigr\}.
\end{split}
\]
This expression is a consequence of the sign reversal property of the permutation of matrices $T^{\,a}$ under the trace sign in inverse order 
\[
{\rm tr}\hspace{0.03cm}\bigl(T^{\,a_{1}}\hspace{0.02cm} T^{\,a_{2}}\hspace{0.02cm} T^{\,a_{3}}\hspace{0.02cm} 
T^{\,a_{4}}\hspace{0.03cm}T^{\,a_{5}}\hspace{0.02cm}\bigr)
=
-{\rm tr}\hspace{0.03cm}\bigl(T^{\,a_{5}}\hspace{0.02cm} T^{\,a_{4}}\hspace{0.02cm} T^{\,a_{3}}\hspace{0.02cm} 
T^{\,a_{2}}\hspace{0.03cm}T^{\,a_{1}}\hspace{0.02cm}\bigr),
\]
which, in turn, is a trivial consequence of the identity 
\[
{\rm tr}\hspace{0.03cm}\bigl(T^{\,a_{1}}\hspace{0.02cm} T^{\,a_{2}}\hspace{0.02cm} T^{\,a_{3}}\hspace{0.02cm} 
T^{\,a_{4}}\hspace{0.03cm}T^{\,a_{5}}\hspace{0.02cm}\bigr)
=
-\hspace{0.03cm}2\hspace{0.03cm}
{\rm tr}\hspace{0.03cm}\bigl(t^{\,b}\hspace{0.02cm}\bigl[\hspace{0.03cm}
t^{\,a_{1}},\bigl[\hspace{0.02cm} t^{\,a_{2}},\bigl[\hspace{0.02cm} t^{\,a_{3}},\bigl[\hspace{0.02cm}t^{\,a_{4}},\bigl[\hspace{0.03cm}t^{\,a_{5}},t^{b}\bigr]\bigr]\bigr]\bigr]\bigr]\hspace{0.02cm}\bigr),
\]
where $t^{\,a}$ are the $N^{\hspace{0.03cm}2}_{c} - 1$ generators in the defining representation of the $\mathfrak{su}(N_{c})$ Lie algebra.\\ 
\indent Based on the representation (\ref{ap:D9}) and considering the symmetry property (\ref{ap:D8}), it is easy to see that the following useful relation exists for the fifth-order trace under permutations of indices $a_{3}$ and $a_{5}$:
\begin{equation}	
{\rm tr}\hspace{0.03cm}\bigl(T^{\,a_{1}}\hspace{0.02cm} T^{\,a_{2}}\hspace{0.02cm} T^{\,a_{3}}\hspace{0.02cm} 
T^{\,a_{4}}\hspace{0.03cm}T^{\,a_{5}}\hspace{0.02cm}\bigr)
+
{\rm tr}\hspace{0.03cm}\bigl(T^{\,a_{1}}\hspace{0.02cm} T^{\,a_{2}}\hspace{0.02cm} T^{\,a_{5}}\hspace{0.02cm} 
T^{\,a_{4}}\hspace{0.03cm}T^{\,a_{3}}\hspace{0.02cm}\bigr)
=
-\hspace{0.03cm}if^{\hspace{0.03cm}a_{2}\hspace{0.02cm}a_{1}
\hspace{0.01cm}b}\hspace{0.04cm}
{\rm tr}\hspace{0.03cm}\bigl(T^{\,b}\hspace{0.03cm} T^{\,a_{5}}\hspace{0.02cm} T^{\,a_{4}}\hspace{0.02cm} 
T^{\,a_{3}}\hspace{0.03cm}\bigr).
\label{ap:D10}
\end{equation}	
We use this relation in section \ref{section_8} to calculate the fifth-order trace.\\ 
\indent There are also two additional identities for the special case $N_{c} = 3$ \cite{macfarlane_1968, haber_2021}. We will just need one of these 
\begin{equation}
d^{\,a\hspace{0.02cm}b\hspace{0.03cm}e}\hspace{0.02cm} d^{\,c\hspace{0.02cm}d\hspace{0.03cm}e}
+
d^{\,a\hspace{0.02cm}c\hspace{0.03cm}e}\hspace{0.02cm} d^{\,b\hspace{0.02cm}d\hspace{0.03cm}e}
+
d^{\,a\hspace{0.02cm}d\hspace{0.03cm}e}\hspace{0.02cm} d^{\,b\hspace{0.02cm}c\hspace{0.03cm}e}
= 
\frac{1}{3}\,\bigl(\hspace{0.02cm}
\delta^{\hspace{0.02cm}a\hspace{0.02cm}b}\delta^{\hspace{0.02cm}
c\hspace{0.02cm} d} 
+
\delta^{\hspace{0.02cm}a\hspace{0.02cm}c}\delta^{\hspace{0.02cm}
b\hspace{0.02cm}d}
+
\delta^{\hspace{0.02cm}a\hspace{0.02cm}d}\delta^{\hspace{0.02cm}
b\hspace{0.02cm} c}\hspace{0.02cm}\bigr).
\label{ap:D13}
\end{equation}


\numberwithin{equation}{section}
\section{Construction of the exact solution of the system (\ref{eq:11a})}
\numberwithin{equation}{section}
\label{appendix_E}

Let us rewrite the system (\ref{eq:11a}) and equation (\ref{eq:11s}) in a slightly different form, introducing the notations generally accepted in the theory of differential equations. We set
\begin{equation}
y \equiv W_{1}, \qquad  x\equiv N_{1},
\label{ap:E1}
\end{equation}
then instead of (\ref{eq:11a}), we have
\begin{equation}
\left\{\;
\begin{split}
&\frac{d\hspace{0.03cm}y(t)}{d\hspace{0.03cm}t}
=
x\hspace{0.03cm}(\mathcal{C}_{\hspace{0.02cm}1} - x)		,
\\[1ex]
&\frac{d\hspace{0.02cm}x(t)}{d\hspace{0.03cm}t}
=
\beta\hspace{0.02cm}\bigl[x\hspace{0.02cm}(\mathcal{C}_{\hspace{0.02cm}2} - y)	
+ y\hspace{0.02cm}(\mathcal{C}_{\hspace{0.02cm}1} - x)\bigr],
\end{split}
\right.
\label{ap:E2}
\end{equation}
and instead of (\ref{eq:11s}), in turn,
\begin{equation}
\bigl[\hspace{0.02cm}(2\hspace{0.02cm}x - \mathcal{C}_{\hspace{0.02cm}1})\hspace{0.03cm}y - 
\mathcal{C}_{\hspace{0.02cm}2}\hspace{0.02cm}x\bigr]\hspace{0.03cm}
\frac{d\hspace{0.03cm}y} {d\hspace{0.02cm}x}
=
\frac{1}{\beta}\,x\hspace{0.02cm}(x - \mathcal{C}_{\hspace{0.02cm}1})	
\label{ap:E3}
\end{equation}
or in more standard notations \cite{polyanin_1995}
\begin{equation}
\bigl[\hspace{0.02cm}g_{1}(x)\hspace{0.03cm}y + 
g_{0}(x)\bigr]\hspace{0.03cm}
\frac{d\hspace{0.03cm}y}{d\hspace{0.02cm}x}
=
f_{0}(x), 	
\label{ap:E4}
\end{equation}
where 
\[
g_{0}(x) \equiv -\hspace{0.03cm}\mathcal{C}_{\hspace{0.02cm}2}\hspace{0.03cm}x, \qquad g_{1}(x) \equiv 2\hspace{0.02cm}x - \mathcal{C}_{\hspace{0.02cm}1}, \qquad 
f_{0}(x) \equiv \frac{1}{\beta}\,x\hspace{0.02cm}(x - \mathcal{C}_{\hspace{0.02cm}1}).
\]
In the system (\ref{ap:E2}) we eliminated the parameter $B$, formally redefining the time 
\begin{equation}
t \rightarrow t/B.	
\label{ap:E5}
\end{equation}
Recall that the given parameter, by virtue of the definition (\ref{eq:11u}), is a generalized function of wave vectors ${\bf k}_{0}$ and ${\bf k}^{\hspace{0.02cm}\prime}_{0}$. Equation (\ref{ap:E4}) belongs to the class of the Abel equations of the second kind. The first step is to reduce it to the ``normal'' form. We perform replacement of the unknown function 
\[
w = y + \frac{g_{0}(x)}{g_{1}(x)}.
\]
This transformation reduces the original equation (\ref{ap:E3}) to the form: 
\begin{equation}
w\hspace{0.03cm}\frac{d\hspace{0.02cm}w}{d\hspace{0.02cm}x} = F_{1}(x)\hspace{0.02cm}w +  F_{0}(x),
\label{ap:E6}
\end{equation}
where 
\[
F_{1}(x) = \frac{d}{d\hspace{0.02cm}x}\hspace{0.02cm}\biggl(\frac{g_{0}(x)}{g_{1}(x)}
\biggr) 
= \frac{\mathcal{C}_{\hspace{0.02cm}1}\hspace{0.02cm}\mathcal{C}_{\hspace{0.02cm}2}}
{(2\hspace{0.02cm}x - \mathcal{C}_{\hspace{0.02cm}1})^{\hspace{0.02cm}2}},
\qquad
F_{0}(x) = \frac{f_{0}(x)}{g_{1}(x)} = \frac{1}{\beta}\,
\frac{x\hspace{0.02cm}(x - \mathcal{C}_{\hspace{0.02cm}1})}{2\hspace{0.02cm}x - \mathcal{C}_{\hspace{0.02cm}1}}.
\]
Next, the replacement of the argument of the function
\[
\xi = \!\int\!F_{1}(x)\hspace{0.03cm}d\hspace{0.02cm}x
=
-\frac{1}{2}\,\frac{\mathcal{C}_{\hspace{0.02cm}1}\hspace{0.02cm}\mathcal{C}_{\hspace{0.02cm}2}}
{2\hspace{0.02cm}x - \mathcal{C}_{\hspace{0.02cm}1}},
\quad\mbox{or}\quad
x = \frac{1}{2}\,\mathcal{C}_{\hspace{0.02cm}1} - \frac{1}{4\hspace{0.03cm}\xi}\,\mathcal{C}_{\hspace{0.02cm}1}\hspace{0.03cm}\mathcal{C}_{\hspace{0.02cm}2}
\]
allows us to bring equation (\ref{ap:E6}) to the final form
\begin{equation}
w\hspace{0.03cm}\frac{d\hspace{0.02cm}w}{d\hspace{0.02cm}\xi} = w +  F(\xi),
\label{ap:E7}
\end{equation}
where 
\[
F(\xi) = \frac{F_{0}(x)}{F_{1}(x)} = 
\frac{\mathcal{C}^{\hspace{0.03cm}2}_{1}}{8\hspace{0.02cm}\beta}\,
\biggl(\frac{1}{\xi} 
\hspace{0.03cm}-\hspace{0.03cm} \frac{\mathcal{C}^{\hspace{0.03cm}2}_{2}}{4\hspace{0.02cm}
\xi^{\hspace{0.03cm}3}}\biggr).
\]
For equation (\ref{ap:E7}), at a certain ratio between the parameters 
$\mathcal{C}_{\hspace{0.02cm}1}$ and $\mathcal{C}_{\hspace{0.02cm}2}$, namely, for
\begin{equation}
\mathcal{C}^{\hspace{0.03cm}2}_{2} = \frac{1}{2\hspace{0.02cm}\beta}\,\mathcal{C}^{\hspace{0.03cm}2}_{1},
\label{ap:E9}
\end{equation}
there is an exact solution in the parametric form (Eq.\,7 in subsection 1.3.1. of \cite{polyanin_1995})
\begin{align}
&\xi = \frac{a}{\tau}\,\bigl(\tau - \ln|1 + \tau| - C\bigr)^{1/2},
\notag\\[1ex]
&w = a \biggl[\frac{1 + \tau}{\tau}\,\bigl(\tau - \ln|1 + \tau| - C\bigr)^{1/2}
-
\frac{1}{2}\,\tau \bigl(\tau - \ln|1 + \tau| - C\bigr)^{-1/2}
\biggr],
\notag
\end{align}
where $\tau$ is a parameter, $C$ is an arbitrary constant and 
\begin{equation}
a^{2} = \frac{1}{4\hspace{0.02cm}\beta}\,\mathcal{C}^{\hspace{0.03cm}2}_{1}.
\label{ap:E10}
\end{equation}
Returning to the original function $y$ and to its argument $x$, we determine the exact solution for the equation (\ref{ap:E3})
\begin{subequations} 
\label{ap:E11}
\begin{align}
&x = x(\tau,C) = \frac{1}{2}\,\mathcal{C}_{\hspace{0.02cm}1} - \frac{1}{4\hspace{0.03cm}a}\,\mathcal{C}_{\hspace{0.02cm}1}\hspace{0.03cm}
\mathcal{C}_{\hspace{0.02cm}2}\hspace{0.03cm}\frac{\tau}{f(\tau)},
\label{ap:E11a}\\[1ex]
&y = y(\tau,C) = a \biggl[\frac{1 + \tau}{\tau}\,f(\tau)
-
\frac{1}{2}\,\frac{\tau}{f(\tau)}\biggr] + \frac{\mathcal{C}_{\hspace{0.02cm}2}\hspace{0.03cm}x}
{2\hspace{0.02cm}x - \mathcal{C}_{\hspace{0.02cm}1}}.
\label{ap:E11b}
\end{align}
\end{subequations}
Here, we introduce the notation
\begin{equation}
f(\tau)\equiv\bigl(\tau - \ln|1 + \tau| - C\bigr)^{1/2}.
\label{ap:E12}
\end{equation}
The solution of the initial dynamical system (\ref{ap:E2}) is determined by the formulae 
\begin{equation}
x = x(\tau,C), \quad y = y(\tau,C),\quad
t = \frac{1}{\beta}\int\!\frac{\dot{x}_{\tau}\hspace{0.03cm}
	d\hspace{0.02cm}\tau}{[\hspace{0.03cm}\mathcal{C}_{\hspace{0.02cm}1} - 2\hspace{0.02cm}x(\tau,C)]\hspace{0.03cm}y(\tau,C) + 
	\mathcal{C}_{\hspace{0.02cm}2}\hspace{0.02cm}x(\tau,C)} 
\hspace{0.03cm}+\hspace{0.03cm} \widetilde{C}.
\label{ap:E13}
\end{equation}
Here, $\dot{x}_{\tau}\equiv dx(\tau,C)/d\tau$ and $\widetilde{C}$ is another arbitrary constant. The latter relation defines the implicit dependence of parameter $\tau$ on time $t$: $\tau = \tau(t, C, \widetilde{C})$. Using the formulae (\ref{ap:E11a}) and (\ref{ap:E11b}), we can find the dependence of $x$ and $y$ on $t$. Note that instead of the last expression in (\ref{ap:E13}), we could equally well use the formula
\[
t = \int\!\frac{\dot{y}_{\hspace{0.02cm}\tau}\hspace{0.03cm}d\hspace{0.02cm}\tau}{x(\tau,C)\hspace{0.03cm}
(\mathcal{C}_{\hspace{0.02cm}1} - x(\tau,C))} \hspace{0.03cm}+\hspace{0.03cm} \widetilde{C}.
\] 
Unfortunately, the direct use of both parametric representations for time $t = t\hspace{0.03cm}(\tau,C,\widetilde{C})$ leads to rather cumbersome expressions, so here we will proceed somewhat differently.\\
\indent In fact, the task here is to determine a function $\mathcal{F}(\tau)$, such that the system of equations is satisfied
\begin{equation}
\left\{\;
\begin{split}
&\frac{d\hspace{0.03cm}y(\tau,C)}{d\hspace{0.03cm}\tau}
=
\mathcal{F}(\tau)\hspace{0.02cm}x\hspace{0.03cm}(\mathcal{C}_{\hspace{0.02cm}1} - x)		,
\\[1ex]
&\frac{d\hspace{0.02cm}x(\tau,C)}{d\hspace{0.03cm}\tau}
=
\beta\hspace{0.01cm}\mathcal{F}(\tau)\hspace{0.01cm}
\bigl[\hspace{0.02cm}(\mathcal{C}_{\hspace{0.02cm}1} - 2\hspace{0.02cm}x)\hspace{0.03cm}y + 
\mathcal{C}_{\hspace{0.02cm}2}\hspace{0.02cm}x\bigr].
\end{split}
\right.
\label{ap:E14}
\end{equation}
Here, on the right-hand side for brevity, we have suppressed the dependence on $\tau$ (and $C$) functions $x$ and $y$. The connection with time in this case has the form
\begin{equation}
t = \int\!\mathcal{F}(\tau)\hspace{0.03cm}d\hspace{0.03cm}\tau 
\hspace{0.03cm}+\hspace{0.03cm}\widetilde{C}.
\label{ap:E15}
\end{equation}
Next, we express $f(\tau)$ as a function of $x$ from the parametric solution (\ref{ap:E11a}), substitute it into $y$, the solution (\ref{ap:E11b}), and differentiate with respect to $\tau$. As a result, we have
\[
\dot{y}_{\hspace{0.02cm}\tau} = \frac{1}{2}\,\frac{\mathcal{C}_{\hspace{0.02cm}1}\hspace{0.03cm}
\mathcal{C}_{\hspace{0.02cm}2}}{\mathcal{C}_{\hspace{0.02cm}1} - 2\hspace{0.02cm}x}
\hspace{0.03cm}+\hspace{0.03cm}
\mathcal{C}_{\hspace{0.02cm}1}\hspace{0.03cm}
\mathcal{C}_{\hspace{0.02cm}2}\,\frac{\tau\hspace{0.01cm}\dot{x}_{\tau}}{(\mathcal{C}_{\hspace{0.02cm}1} - 2\hspace{0.02cm}x)^{2}}
\hspace{0.03cm}+\hspace{0.03cm}
\frac{2\hspace{0.02cm}a^{2}}{\mathcal{C}_{\hspace{0.02cm}1}\hspace{0.03cm}\mathcal{C}_{\hspace{0.02cm}2}}\,\dot{x}_{\tau}.
\]
Recall that the coefficient $a^{2}$ was defined by the equality (\ref{ap:E10}). The main peculiarity of the last expression is the absence of the derivative $\dot{x}_{\tau}$ in the first term on the right-hand side. The last step is to substitute the corresponding right-hand sides from (\ref{ap:E14}) into the previous expression instead of the derivatives $\dot{y}_{\hspace{0.02cm}\tau}$ and $\dot{x}_{\hspace{0.01cm}\tau}$, and then to substitute the exact solutions (\ref{ap:E11}) instead of $y$ and $x$. After simple but somewhat cumbersome algebraic transformations, considering the relations (\ref{ap:E9}) and (\ref{ap:E10}), we find the required function 
\begin{equation}
\mathcal{F}(\tau) = -\frac{2a}{\mathcal{C}^{\hspace{0.03cm}2}_{1}}\,
\frac{1}{(1 + \tau)f(\tau)}.
\label{ap:E16}
\end{equation}
The advantage of this approach is that most of the terms are mutually reduced and we are left with a rather simple expression. Substituting $\mathcal{F}(\tau)$ into (\ref{ap:E15}) and recalling the definition of (\ref{ap:E12}), we finally find the desired time parameterization
\begin{equation}
t = t\hspace{0.03cm}(\tau, C, \widetilde{C}) =-\frac{2\hspace{0.02cm}a}{\mathcal{C}^{\hspace{0.03cm}2}_{1}}\int\!
\frac{d\hspace{0.02cm}\tau} 
{(1 + \tau)\bigl(\tau - \ln|1 + \tau| - C\bigr)^{1/2}}
\hspace{0.03cm}+\hspace{0.03cm} \widetilde{C}.
\label{ap:E17}
\end{equation}
By directly substituting (\ref{ap:E16}) and the exact solutions (\ref{ap:E11}) into the system of equations (\ref{ap:E14}), we verify that they are turning to the identity. Coming to the original functions $N_{1}$ and $W_{1}$ according to (\ref{ap:E1}) and restoring the parameter $B$ for time, Eq.\,(\ref{ap:E5}), we thus determine the exact solution of the system (\ref{eq:11a}) as defined by the expression (\ref{eq:11d}).\\
\indent Let us analyze the structure of the integral in (\ref{ap:E17}) in more detail. For this purpose, let us replace the integration variable
\[
\ln\hspace{0.03cm}|\hspace{0.02cm}1 + \tau\hspace{0.02cm}| = \zeta,
\]
which gives 
\begin{equation}
\int\!\frac{d\hspace{0.02cm}\tau}{(1 + \tau)\bigl(\tau - \ln|1 + \tau| - C\bigr)^{1/2}}
=
\int\!\frac{d\hspace{0.02cm}\zeta}{\bigl[\hspace{0.03cm}\pm\hspace{0.03cm}
{\rm e}^{\hspace{0.03cm}\zeta} - \zeta - \bigl(1 + C\bigr)\bigr]^{1/2}},
\label{ap:E18}
\end{equation}
where on the right-hand side in the integrand we have
\[
\left\{\!\!\!\!\!\!\!
\begin{array}{ll}
&+\hspace{0.04cm}{\rm e}^{\hspace{0.03cm}\zeta},\; \mbox{for}\; \tau > -1,\\[1ex]
&-\hspace{0.03cm}{\rm e}^{\hspace{0.03cm}\zeta},\; \mbox{for}\; \tau < -1.
\end{array} 
\right.
\]
Let us again perform the replacement of the integration variable
\begin{equation}
\pm\hspace{0.03cm}{\rm e}^{\hspace{0.03cm}\zeta} - \zeta - \bigl(1 + C\bigr) = \lambda.
\label{ap:E19}
\end{equation}
The solution of this expression with respect to the new variable $\lambda$ can be represented in the following form:
\[
\zeta = \zeta(\lambda) = - \lambda - \bigl(1 + C\bigr) - W\bigl(\mp\hspace{0.04cm}{\rm e}^{-(1 + C)}\hspace{0.03cm}{\rm e}^{-\lambda}\bigr), 
\]
where $W(x)$ is the Lambert $W$ function \cite{corless_1996}. Using the rule of differentiation for this function \cite{corless_1996, corless_1997}, we find further
\[
d\hspace{0.03cm}\zeta = -\hspace{0.03cm}d\hspace{0.03cm}\lambda
+
\frac{W\bigl(\mp\hspace{0.04cm}{\rm e}^{-(1 + C)}\hspace{0.03cm}{\rm e}^{-\lambda}\bigr)}{1 + W\bigl(\mp\hspace{0.04cm}{\rm e}^{-(1 + C)}\hspace{0.03cm}{\rm e}^{-\lambda}\bigr)}\,d\hspace{0.03cm}\lambda
\equiv
-\hspace{0.03cm}
\frac{1}{1 + W\bigl(\mp\hspace{0.04cm}{\rm e}^{-(1 + C)}\hspace{0.03cm}{\rm e}^{-\lambda}\bigr)}\,d\hspace{0.03cm}\lambda.
\]
Substituting the replacement of the variable (\ref{ap:E19}) and the differential $d\hspace{0.03cm}\zeta$ into (\ref{ap:E18}), we find, instead of the last integral in (\ref{ap:E18}),
\[
-\!
\int\!
\frac{d\hspace{0.03cm}\lambda}{\lambda^{1/2}\hspace{0.03cm}\bigl[1 + W\bigl(\mp\hspace{0.04cm}{\rm e}^{-\hspace{0.03cm}(1 + C)}\hspace{0.03cm}{\rm e}^{-\lambda}\bigr)\bigr]}
\]
or
\begin{equation}
\hspace{1.3cm}	
-\hspace{0.03cm}
2\!\int\!
\frac{d\hspace{0.03cm}\xi}{1 + W\bigl(\mp\hspace{0.04cm}{\rm e}^{-(1 + C)}\hspace{0.03cm}{\rm e}^{-\hspace{0.02cm}\xi^{\hspace{0.02cm}2}}\bigr)},
\quad 
\xi \equiv \lambda^{1/2}.
\label{ap:E20}
\end{equation}
Thus, we reduced the integral involving the transcendental function in (\ref{ap:E17}) to the integral of the Lambert $W$ function.\\
\indent However, such an integral cannot be calculated directly either. The difficulty is that the Lambert $W$ function depends on the variable $\xi$ as a function ${\rm e}^{-\hspace{0.02cm}\xi^{\hspace{0.02cm}2}}$. Here, we can use the particular integral representation for the Lambert $W$ function given in \cite{kalugin_2011} (Eq.\,(45)). In our case it will look as follows
\begin{equation}
\int\!
\frac{d\hspace{0.03cm}\xi}{1 + W\bigl(\mp\hspace{0.04cm}{\rm e}^{-(1 + C)}\hspace{0.03cm}{\rm e}^{-\hspace{0.02cm}\xi^{\hspace{0.02cm}2}}\bigr)}
=
\frac{1}{\pi}\int\limits^{\pi}_{0}dv\!
\int\!
\frac{d\hspace{0.03cm}\xi}{1 \mp {\rm e}^{-\hspace{0.03cm}(1 + C)}\hspace{0.03cm}
{\rm e}^{-\hspace{0.02cm}\xi^{\hspace{0.02cm}2}}{\rm e}^{\hspace{0.03cm}v\cot v}\sin v/v}.
\label{ap:E21}
\end{equation}
Further we present the integrand as a series expansion
\[
\frac{1}{1 \mp {\rm e}^{-(1 + C)}\hspace{0.03cm}
{\rm e}^{-\hspace{0.02cm}\xi^{\hspace{0.02cm}2}}{\rm e}^{\hspace{0.03cm}v\cot v}\sin v/v}
=
1 + \sum\limits^{\infty}_{\nu\hspace{0.03cm}=\hspace{0.03cm}1}
(\pm 1)^{\hspace{0.02cm}\nu}\hspace{0.02cm}
{\rm e}^{-\hspace{0.03cm}\nu\hspace{0.03cm}(1 + C)}\hspace{0.03cm}
{\rm e}^{-\hspace{0.03cm}\nu\hspace{0.03cm}\xi^{\hspace{0.02cm}2}}
\left\{{\rm e}^{v\cot v}\,\frac{\sin v}{v}\right\}^{\!\nu}.
\]
Let us write the integral over $\xi$ through the Gauss error function
\[
\int\limits^{\xi}_{0}\!
{\rm e}^{-\hspace{0.02cm}\nu\hspace{0.03cm}\xi^{\hspace{0.02cm}2}}d\hspace{0.03cm}\xi
=
\frac{1}{2}\,\sqrt{\frac{\pi}{\nu}}\;{\rm erf}(\sqrt{\nu}\hspace{0.03cm}\xi),
\] 
and for the integration over $v$ we use Corollary 3.4. from \cite{kalugin_2011}
\[
\int\limits^{\pi}_{0}\!\left\{{\rm e}^{v\cot v}\,\frac{\sin v}{v}\right\}^{\!\nu}\!d\hspace{0.02cm}v
=
\frac{\pi\hspace{0.02cm}\nu^{\hspace{0.03cm}\nu}}{\nu\hspace{0.02cm}!}. 
\]
As a result, for the integral (\ref{ap:E21}) we find a representation in the form of a series 
\[
\int\!
\frac{d\hspace{0.03cm}\xi}{1 + W\bigl(\mp\hspace{0.04cm}{\rm e}^{-(1 + C)}\hspace{0.03cm}{\rm e}^{-\xi^{\hspace{0.02cm}2}}\bigr)}
= 
\xi + \frac{\sqrt{\pi}}{2}\,
\sum\limits^{\infty}_{\nu\hspace{0.03cm}=\hspace{0.03cm}1}
\frac{(\pm 1)^{\hspace{0.02cm}\nu}}
{\nu\hspace{0.02cm}!\hspace{0.03cm}\sqrt{\nu}}\,
\bigl({\rm e}^{-\hspace{0.03cm}\hspace{0.03cm}(1 + C)}\nu\bigr)^{\nu}\hspace{0.03cm}
\hspace{0.03cm}{\rm erf}(\sqrt{\nu}\hspace{0.03cm}\xi).
\]
Substituting this representation into (\ref{ap:E20}) and returning to the original variable $\tau$ (we simply replace $\xi$ with $f(\tau)$), we find the following representation for the time parameterization (\ref{ap:E17})
\[
t = \frac{4\hspace{0.02cm}a}{\mathcal{C}^{\hspace{0.03cm}2}_{1}}
\,\biggl\{f(\tau)
+
\frac{\sqrt{\pi}}{2}\,
\sum\limits^{\infty}_{\nu\hspace{0.03cm}=\hspace{0.03cm}1}
\frac{(\pm 1)^{\hspace{0.02cm}\nu}}{\nu\hspace{0.02cm}!\hspace{0.03cm}\sqrt{\nu}}\,
\bigl({\rm e}^{-\hspace{0.03cm}\hspace{0.03cm}(1 + C)}\nu\bigr)^{\nu}\hspace{0.03cm}
\,{\rm erf}\bigl(\sqrt{\nu}\hspace{0.03cm}f(\tau)\bigr)
\!\hspace{0.02cm}\biggr\}
\hspace{0.03cm}+\hspace{0.03cm}\widetilde{C}.
\]
Here, we recall that under the sum sign we choose  $(+1)^{\hspace{0.02cm}\nu}\equiv 1$ if $\tau > -1$ and $(-1)^{\hspace{0.02cm}\nu}$, if $\tau < -1$.
Perhaps this representation is more convenient for approximate expressions of time $t$ as a function of the parameter $\tau$, using, for example, several first terms of the series or, vice versa, using asymptotic approximation at $\nu\rightarrow\infty$ for the terms of this series.

\end{appendices}

\newpage

\end{document}